\documentclass[english]{IEEEtran}
\usepackage[T1]{fontenc}
\usepackage[latin9]{inputenc}
\usepackage{array}
\usepackage{float}
\usepackage{textcomp}
\usepackage{bm}
\usepackage{amsmath}
\usepackage{amssymb}
\usepackage{graphicx}
\usepackage{threeparttable}
\usepackage{epstopdf}
\makeatletter

\newcommand{\lyxmathsym}[1]{\ifmmode\begingroup\def\b@ld{bold}
  \text{\ifx\math@version\b@ld\bfseries\fi#1}\endgroup\else#1\fi}

\providecommand{\tabularnewline}{\\}

\usepackage[acronym]{glossaries}
\usepackage{babel}
\usepackage{algorithm}
\usepackage{algorithmic}
\usepackage[unicode=true,bookmarks=false,breaklinks=false,pdfborder={0 0 1},,colorlinks=false]{hyperref}\hypersetup{colorlinks}

\@ifundefined{showcaptionsetup}{}{%
 \PassOptionsToPackage{caption=false}{subfig}}
\usepackage{subfig}
\makeatother

\begin{document}
\title{A Tutorial on Fluid Antenna System for 6G Networks: Encompassing Communication Theory, Optimization Methods and Hardware Designs}
\author{Wee Kiat New, \textit{Member, IEEE}, 
            Kai-Kit Wong, \textit{Fellow, IEEE}, 
            Hao Xu, \textit{Member, IEEE,}\\
            Chao Wang, \textit{Senior Member, IEEE,} 
            Farshad Rostami Ghadi, \textit{Member, IEEE,} 
            Jichen Zhang, \textit{Graduate Student Member, IEEE,} 
            Junhui Rao, \textit{Graduate Student Member, IEEE,} 
            Ross Murch, \textit{Fellow, IEEE},\\
            Pablo Ram\'irez-Espinosa, 
            David Morales-Jimenez, \textit{Senior Member, IEEE},\\ 
            Chan-Byoung Chae, \textit{Fellow, IEEE}, 
            and Kin-Fai Tong, \textit{Fellow, IEEE}

\thanks{The work of W. K. New, K. K. Wong, H. Xu, F. Rostami Ghadi, and K. F. Tong is supported by the Engineering and Physical Sciences Research Council (EPSRC) under grant EP/W026813/1.}
\thanks{The work of C. Wang was supported by the National Natural Science Foundation of China under Grant 62371357, the Funds for International Cooperation and Exchange of the National Natural Science Foundation of China under Grant 12411530120, and the International Postdoctoral Exchange Fellowship Program under Grant PC2021060.}
\thanks{The work of R. Murch is supported by the Hong Kong Research Grants Council Area of Excellence grant AoE/E-601/22-R.}
\thanks{The work of P. Ram\'{i}rez-Espinosa has been funded by the European Union under the Marie Sklodowska-Curie grant agreement No. 101109529.}
\thanks{The work of D. Morales-Jimenez is also supported in part by the AEI of Spain and the European Social Fund under grant RYC2020-030536-I and by MICIU/AEI/10.13039/501100011033 and FEDER/UE under grant PID2023-149975OB-I00 (COSTUME).}
\thanks{The work of C.-B. Chae was supported by the Institute for Information and Communication Technology Planning and Evaluation (IITP)/NRF grant funded by the Ministry of Science and ICT (MSIT), South Korea, under Grant RS-2024-00428780 and 2022R1A5A1027646.}

\thanks{W. K. New, K. K. Wong, H. Xu, F. Rostami Ghadi, and K. F. Tong are with the Department of Electronic and Electrical Engineering, University College London, London WC1E7JE, United Kingdom. K. K. Wong is also affiliated with Yonsei Frontier Laboratory, Yonsei University, Seoul, 03722, Korea.}
\thanks{C. Wang is with the Integrated Service Networks Lab, Xidian University, Xi'an 710071, China.}
\thanks{J. Zhang, J. Rao and R. Murch are with the Department of Electronic and Computer Engineering and Institute for Advanced Study (IAS), The Hong Kong University of Science and Technology, Clear Water Bay, Hong Kong SAR, China.}
\thanks{P. Ram\'irez-Espinosa and D. Morales-Jimenez are with the Department of Signal Theory, Networking and Communications, Universidad de Granada, Granada 18071, Spain.}
\thanks{C.-B. Chae are with the School of Integrated Technology, Yonsei University, Seoul, 03722, Korea.}
\thanks{Corresponding author: Kai-Kit Wong (e-mail: $\rm kai\text{-}kit.wong@ucl.ac.uk$).}

}

\maketitle

\newacronym{1D}{1D}{one-dimensional} 

\newacronym{2D}{2D}{two-dimensional} 

\newacronym{3D}{3D}{three-dimensional} 

\newacronym{5G}{5G}{fifth generation} 

\newacronym{6G}{6G}{sixth generation} 

\newacronym{AI}{AI}{Artificial intelligence} 

\newacronym{AoA}{AoA}{angle-of-arrivals} 

\newacronym{AoD}{AoD}{angle-of-departures} 

\newacronym{AWGN}{AWGN}{additive white Gaussian noise} 

\newacronym{BLAST}{BLAST}{Bell Laboratories Layered Space-Time}

\newacronym{BS}{BS}{base station} 

\newacronym{CAP}{CAP}{continuous aperture} 

\newacronym{CDF}{CDF}{cumulative distribution function} 

\newacronym{CNN}{CNN}{Convolutional Neural Networks} 

\newacronym{CPU}{CPU}{central processing unit} 

\newacronym{CSCG}{CSCG}{circularly symmetric complex Gaussian} 

\newacronym{CSI}{CSI}{channel state information} 

\newacronym{CUMA}{CUMA}{compact ultra massive antenna array} 

\newacronym{DFT}{DFT}{discrete Fourier transform} 

\newacronym{DMT}{DMT}{Diversity and Multiplexing Tradeoff} 

\newacronym{DTDR}{DTDR}{distributed transmission and directional reception} 

\newacronym{eMBB}{eMBB}{enhanced Mobile Broadband} 

\newacronym{EWOD}{EWOD}{electrowetting-on-dielectric} 

\newacronym{FAS}{FAS}{fluid antenna system} 

\newacronym{FD}{FD}{full-duplex} 

\newacronym{FFT}{FFT}{fast Fourier transform} 

\newacronym{FAMA}{FAMA}{fluid antenna multiple access} 

\newacronym{gdof}{gdof}{generalized degrees of freedom} 

\newacronym{GNN}{GNN}{Graph Neural Networks} 

\newacronym{HD}{HD}{half-duplex} 

\newacronym{HK}{HK}{Han-Kobayashi} 

\newacronym{i.i.d.}{i.i.d.}{independent and identically distributed} 

\newacronym{IMT}{IMT}{International Mobile Telecommunications} 

\newacronym{IoT}{IoT}{Internet-of-Things} 

\newacronym{ISAC}{ISAC}{Integrated sensing and communication} 

\newacronym{ITU-T}{ITU-T}{International Telecommunication Union Telecommunication Standardization Sector} 

\newacronym{L3SCR}{L3SCR}{low-sample-size sparse channel reconstruction} 

\newacronym{LoS}{LoS}{line-of-sight} 

\newacronym{LSTM}{LSTM}{long short-term memory} 

\newacronym{mMTC}{mMTC}{massive Machine Type Communication} 

\newacronym{MRC}{MRC}{maximum ratio combining} 

\newacronym{MRT}{MRT}{maximum ratio transmission} 

\newacronym{MSE}{MSE}{mean squared error} 

\newacronym{ML}{ML}{machine learning} 

\newacronym{MLP}{MLP}{multi-layer perceptron} 

\newacronym{MIMO}{MIMO}{multiple-input multiple-output} 

\newacronym{MISO}{MISO}{multiple-input single-output} 

\newacronym{NGMA}{NGMA}{next-generation multiple access} 

\newacronym{NLoS}{NLoS}{non line-of-sight} 

\newacronym{NMSE}{NMSE}{normalized mean squared error} 

\newacronym{NOMA}{NOMA}{non-orthogonal multiple access} 

\newacronym{NTN}{NTN}{non-terrestrial networks} 

\newacronym{OMA}{OMA}{orthogonal multiple access} 

\newacronym{OMP}{OMP}{orthogonal matching pursuit} 

\newacronym{ORTHO}{ORTHO}{orthogonalization} 

\newacronym{PDF}{PDF}{probability density function} 

\newacronym{PLS}{PLS}{physical layer security} 

\newacronym{RF}{RF}{radio frequency} 

\newacronym{RIS}{RIS}{reconfigurable intelligent surfaces} 

\newacronym{RRQR}{RRQR}{rank-revealing QR} 

\newacronym{RSMA}{RSMA}{rate-splitting multiple access} 

\newacronym{RZF}{RZF}{regularized zero forcing}

\newacronym{SIC}{SIC}{successive interference cancellation} 

\newacronym{SINR}{SINR}{signal-to-interference-plus-noise ratio} 

\newacronym{SIMO}{SIMO}{single-input multiple-output} 

\newacronym{SISO}{SISO}{single-input single-output} 

\newacronym{SNR}{SNR}{signal-to-noise ratio}

\newacronym{SPO}{SPO}{Smart ``Predict, then Optimize''}

\newacronym{SVD}{SVD}{singular value decomposition} 

\newacronym{TAS}{TAS}{traditional antenna system} 

\newacronym{THz}{THz}{Terahertz} 

\newacronym{TIN}{TIN}{treating interference as noise} 

\newacronym{UAMA}{UAMA}{unified asymmetric masked autoencoder} 

\newacronym{URLLC}{URLLC}{ultra reliable and low-latency communication} 

\newacronym{XL}{XL}{extremely large-scale} 

\begin{abstract}
The advent of the sixth-generation (6G) networks presents another round of revolution for the mobile communication landscape, promising an immersive experience, robust reliability, minimal latency, extreme connectivity, ubiquitous coverage, and capabilities beyond communication, including intelligence and sensing. To achieve these ambitious goals, it is apparent that 6G networks need to incorporate the state-of-the-art technologies. One of the technologies that has garnered rising interest is fluid antenna system (FAS) which represents any software-controllable fluidic, conductive, or dielectric structure capable of dynamically changing its shape and position to reconfigure essential radio-frequency (RF) characteristics. Compared to traditional antenna systems (TASs) with fixed-position radiating elements, the core idea of FAS revolves around the unique flexibility of reconfiguring the radiating elements within a given space. One recent driver of FAS is the recognition of its position-flexibility as a new degree of freedom (dof) to harness diversity and multiplexing gains. In this paper, we provide a comprehensive tutorial, covering channel modeling, signal processing and estimation methods, information-theoretic insights, new multiple access techniques, and hardware designs. Moreover, we delineate the challenges of FAS and explore the potential of using FAS to improve the performance of other contemporary technologies. By providing insights and guidance, this tutorial paper serves to inspire researchers to explore new horizons and fully unleash the potential of FAS.
\end{abstract}


\section{Introduction}
\subsection{Definition and Motivation of FAS}
\IEEEPARstart{As}{ the landscape} of mobile communication continues to evolve, the anticipation surrounding the advent of the \gls{6G} networks is steadily growing. Envisioned as the natural successor to the \gls{5G} networks, \gls{6G} is expected to usher in another wave of revolutionary advancements, offering enormous data speeds, robust reliability, minimal latency, extreme connectivity, ubiquitous coverage, intelligence and sensing capabilities. These qualities are instrumental in enabling innovative applications that were previously unimaginable, and hence potentially fostering new business opportunities and creating employment prospects. With that, academia and industry professionals are gearing up to meet highly ambitious key performance index requirements, which include achieving a peak rate of $1$ Tbps, an end-to-end latency of $1$ ms, and a connection density of $10^{7}$ devices per km$^{2}$, and more \cite{rajatheva2020white,Zhang-2019,Saad-2020,Tariq-2020}.

\begin{figure}
\centering \includegraphics[scale=0.6]{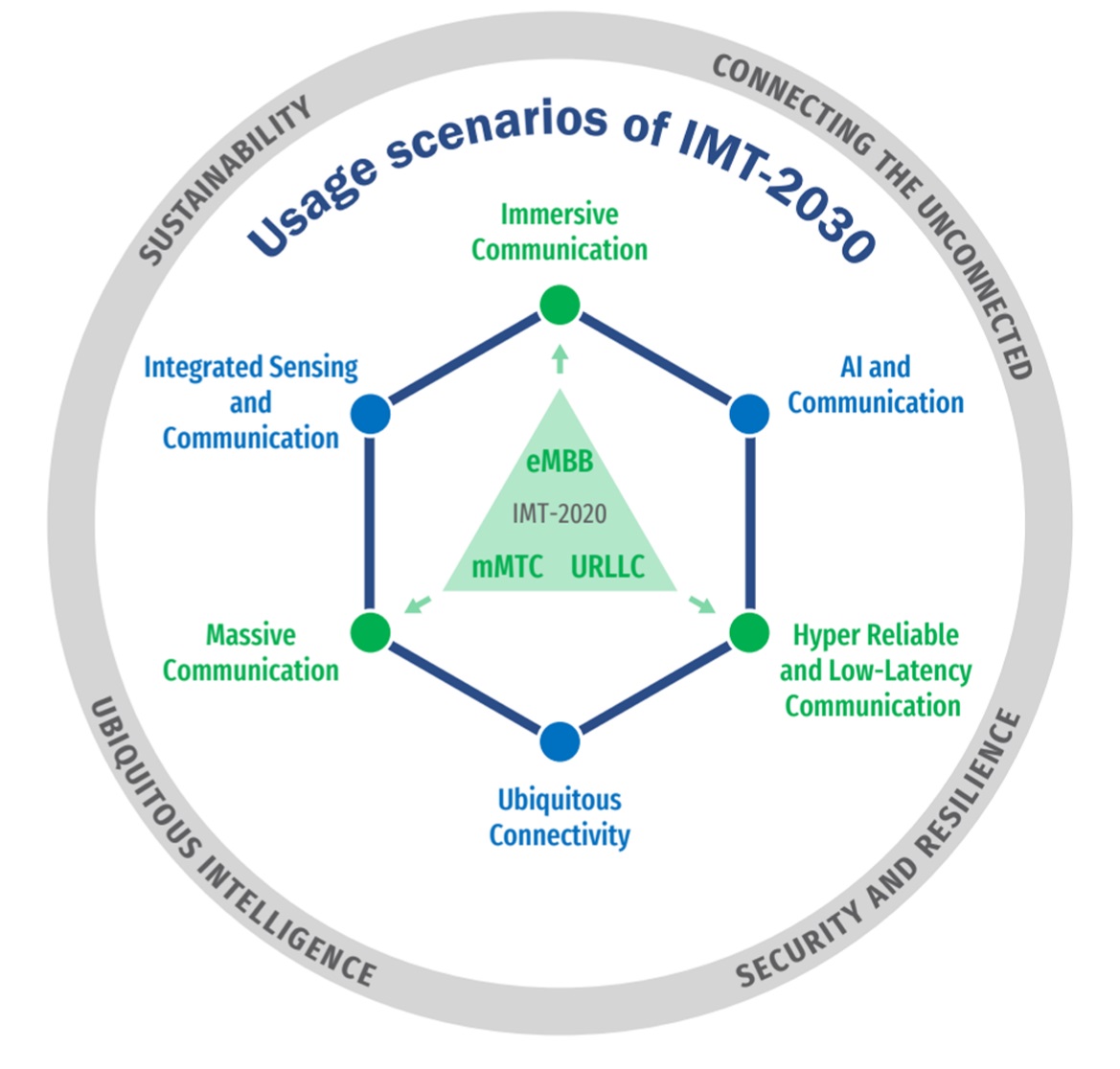} \caption{Usage scenarios of IMT-2030 \cite{ITUwhite}.}\label{IMT2030}
\end{figure}

In recent developments, the \gls{ITU-T} has reached a consensus on the anticipated usage scenarios for \gls{6G}, as shown in Fig.~\ref{IMT2030}. These include \cite{ITUwhite}:
\begin{itemize}
\item Immersive communication;
\item Hyper reliable, low-latency communication (HRLLC);
\item Massive communication;
\item Artificial intelligence (AI) and communication;
\item Ubiquitous connectivity;
\item Integrated sensing and communication (ISAC).
\end{itemize}

Immersive communication advances and broadens the capability of enhanced mobile broadband (eMBB) of \gls{IMT}-2020, to deal with the use cases that provide rich and interactive mobile services, including interactions with machine interfaces, communication for immersive extended reality, remote multi-sensory telepresence, and holographic communication as well as supporting a range of environments, such as hotspots, urban and rural \cite{Du-2020}. HRLLC builds upon the ultra reliable and low-latency communication (URLLC) of \gls{IMT}-2020, catering to specialized use cases with stringent requirements on reliability and latency \cite{Viswanathan-2020,Tao-2023}. Well-known examples include the time-synchronized operations for emergency services, telemedicine, and monitoring in electrical power transmission and distribution. On the other hand, massive communication extends the concept of massive machine-type communication (mMTC) from \gls{IMT}-2020, involving the connection of a huge number of long-lasting devices or sensors for various \gls{IoT} applications \cite{Chen-2021,Liu-2018}. 

While the aforesaid usage scenarios are progressive continuation of the 5G use cases, 6G is boosted by three new usage scenarios. The first usage is AI and communication, addressing distributed computing and AI applications such as automated driving \cite{Chen-2021iot}, autonomous collaboration between devices for medical assistance \cite{Mucchi-2020}, offloading heavy computational operations across devices and networks, and the creation and prediction with digital twins \cite{Sun-2020,Lu-2021}. Ubiquitous connectivity is another new scenario, looking to address currently underserved areas, especially rural, remote, and sparsely populated regions. Finally, ISAC makes up the last new usage scenario, aiming to facilitate applications and services requiring sensing capabilities such as movement tracking, environmental monitoring, and providing sensing data or information for AI, extended reality, and digital twin applications \cite{Zhang-2021,Liu-2022}.

It goes without saying that realizing 6G comes with fundamental technical challenges and a single technology will prove to be inadaquate. Over the past generations, it is apparent that every generational leap needs a major technology push in the physical layer. From analogue to digital communications, from circuit switching to packet switching, then from single-antenna to multi-antenna systems, and most recently from single-user to multiuser signal processing, we always rely on advances in the physical layer to take mobile communications to the next level. Undoubtedly, 6G should see the degree of freedom (dof) in the physical layer increase further to meet the demands.


For this very reason, researchers from around the world are constantly exploring different cutting-edge technologies, such as the integration of AI or \gls{ML} \cite{8808168}, \gls{NGMA} \cite{9693417,9806417}, \gls{NTN} \cite{9275613}, \gls{RIS} \cite{Huang-2019,8796365,Renzo-2020}, \gls{THz} communications \cite{8732419,Petrov-2020} and \gls{XL} \gls{MIMO} \cite{10098681} or \gls{CAP}-\gls{MIMO} \cite{Wan-2023}, just to name a few. There are also the LiFi (light fidelity) \cite{Haas-2016} and quantum technologies \cite{Nawaz-2019,Wang-2022} that are in the conversation of the physical layer for 6G. A combination of these technologies are expected to be key for achieving the 6G targets \cite{10298069}.

With the desire to increase the dof in the physical layer, it is worth taking notice the emerging technology of \gls{FAS} which represents the next generation reconfigurable antenna technologies. According to \cite{https://doi.org/10.1049/el.2020.2788,wong2022bruce}, 
\begin{quote}
{\em ``Fluid antenna is a radical approach that advocates software-controlled position-flexible shape-flexible antenna.''} \dots and \dots {\em ``FAS represents any software-controllable fluidic, conductive or dielectric structure that can alter its shape and position to reconfigure the gain, radiation pattern, operating frequency, and other characteristics.''}
\end{quote}

A \gls{FAS} device can leverage the entire spatial diversity within a predefined space by finely reconfiguring the positions of the radiating elements. This introduces additional dofs, resulting in significant performance gains. In contrast to massive \gls{MIMO} or antenna selection techniques that normally require a multitude of fixed physical antennas and/or \gls{RF}-chains \cite{1341263,Molisch-2004,Gao-2019,Kong-2023}, \gls{FAS} distinguishes itself by offering the potential to reduce hardware costs and power consumption. This is achieved with fewer antennas and/or \gls{RF}-chains, all while maintaining the spatial diversity inherent in the designated space. Besides, \gls{FAS} can dynamically adjust the dimensions of radiating elements to optimally serve different frequencies, rotate the orientations of the radiating elements to create optimal polarization, and perform directional beamforming without relying on analog or digital signal processing. These advancements are made possible through recent breakthroughs in utilizing flexible materials such as liquid materials \cite{liquid_metals,paracha2019liquid,9388928,9982508}, reconfigurable RF pixels \cite{1367557,5991914,6678331,hoang2021single,9785489}, stepper motors \cite{1353484,8060521} and metamaterials \cite{7031374,Yurduseven:18} for antennas. FAS is independent of the other cutting-edge technologies and can combine with them to a great effect.

The original concept of FAS was first introduced by Wong {\em et al.}~in \cite{9131873,9264694}. The inspiration behind resonates with the wisdom shared by Bruce Lee, who articulated the philosophy of Jeet Kune Do, a martial art that he created \cite{wong2022bruce}. Specifically, Bruce Lee had the following famous quote \cite{lee2020water}: 
\begin{quote}
{\em ``Be like water making its way through cracks. Do not be assertive, but adjust to the object, and you shall find a way around or through it. If nothing within you stays rigid, outward things will disclose themselves.\\
\vspace{-3mm}

Empty your mind, be formless. Shapeless, like water. If you put water into a cup, it becomes the cup. You put water into a bottle and it becomes the bottle. You put it in a teapot, it becomes the teapot. Now, water can flow or it can crash. Be water, my friend.''}
\end{quote}

Applying this philosophy to communications, it is observed that antennas can also be formless, shapeless like fluid, and adjusted to different situations. This gives rise to fluid antennas, providing the ultimate reconfigurability and agility for signal and information processing \cite{9131873,9264694}. It is worth noting that the term `fluid' in \gls{FAS} is used to highlight the dynamic nature of the antennas and the motivation behind this idea; however, the antennas may not necessarily be made of liquid or gas. In short, \gls{FAS} encompasses all forms of flexible reconfigurable antennas that fall within its definition \cite{zhu2024historical}.

\subsection{Comparison to State-of-the-Art Technologies}
As myriads of cutting-edge technologies emerge to facilitate \gls{6G}, this subsection aims to highlight the key similarities and differences between \gls{FAS} and some related technologies.

\subsubsection{XL-MIMO and Antenna Selection}
In 1994, Paulraj and Kailath filed a patent for an apparatus designed to increase wireless system capacity through \gls{DTDR} \cite{paulraj1994increasing}. This innovation captures the essence of contemporary \gls{MIMO} technology, establishing Paulraj as the {\em father} of \gls{MIMO}. Despite this pioneering conceptualization, \gls{MIMO} did not come to fruition until the seminal work of Foschini and Gans in 1998 \cite{foschini1998limits}. Their groundbreaking contribution introduced the \gls{BLAST} architecture, showcasing the capacity benefits of \gls{MIMO} and transforming the third-generation (3G) wireless systems \cite{Adjoudani-2003}. Since then, \gls{MIMO} systems have been upgraded to multiuser MIMO, first in \cite{904670,Wong-2002} and then in the classical results \cite{Vishwanath-2003,Spencer-2004}. Massive \gls{MIMO} became the latest version in charge of providing the bandwidth for massive connectivity in 5G \cite{Marzetta-2010,Larsson-2014}. It has the ability to scale up spectral efficiency from space, address small-scale fading with channel hardening, and mitigate interference through favorable propagation. 

Researchers are naturally encouraged to explore \gls{XL}-\gls{MIMO}, where the basic idea is to deploy an extremely large number of antennas in a compact space in which the antenna aperture could be discrete or continuous \cite{10379539}. The latter design is also referred to as \gls{CAP}-\gls{MIMO} \cite{Wan-2023}. Compared to \gls{XL}-\gls{MIMO}, \gls{FAS} may require much fewer physical radiating elements or \gls{RF}-chains. For example, a \gls{FAS} device built on the liquid antenna design may only require one RF-chain and one active radiating element. Nevertheless, its performance may match that of \gls{XL}-\gls{MIMO} if the same number of \gls{RF}-chains is considered. On the other hand, \gls{FAS} built on the pixel-based antenna design can be made similar to \gls{XL}-\gls{MIMO} and vice versa, but with a dense arrangement of antennas. In this case, \gls{FAS} can be viewed as an alternative architecture of compact \gls{XL}-\gls{MIMO} with antenna selection, as will be discussed later in Section \ref{sec:Promising-Research-Directions}. 

In contrast to antenna selection in \gls{TAS}, in which the radiating elements are separated by half a wavelength, a \gls{FAS} device can precisely configure the positions of the radiating elements to utilize the entire channel variation within a designated space, even in spaces smaller than half a wavelength. This flexibility significantly reduces the outage probability and power consumption. Additionally, in multiuser \gls{MIMO} scenarios, each user can be assigned a spatial signature to maximize its received signal and/or suppress the multiuser interference. Contrary to \gls{TAS}, \gls{FAS} can reconfigure the steering vector and the antenna positions to mitigate the traditional tradeoff between signal maximization and interference nulling \cite{10278220,10382559}. Lastly, by intelligently reconfiguring the positions of the radiating elements, \gls{FAS} can achieve the same as or a stronger level of channel hardening effect than the \gls{TAS}, but with much fewer radiating elements \cite{zheng2023flexibleposition}. This improves the overall spectral efficiency and energy efficiency.

\subsubsection{Non-Orthogonal Multiple Access (NOMA) and Rate-Splitting Multiple Access (RSMA)}
Existing systems, including the \gls{5G} networks, have primarily relied on \gls{OMA} techniques such as time-division multiple access (TDMA), frequency-division multiple access (FDMA) and code-division multiple access (CDMA) to serve multiple users. In \gls{OMA}, users are allocated orthogonal radio resources to prevent multiuser interference. While \gls{OMA} provides certain reliability, it faces limitations in allowing massive connectivity when radio resources are scarce. To facilitate massive connectivity, recent interest has therefore turned to non-orthogonal techniques such as \gls{NOMA} and \gls{RSMA}. 

In NOMA, users share the same radio resource unit using superposition coding and \gls{SIC} \cite{7973146,8114722}. For instance, in the downlink two-user power-domain NOMA, the transmitter superimposes the user signals with dedicated power and broadcasts the signal to the wireless channel. During reception, the weaker NOMA user treats the interference as noise, while the stronger NOMA user performs \gls{SIC} to decode its interference signal, then subtract it. On the other hand, RSMA employs rate-splitting and \gls{SIC} \cite{10038476,mao2018rate}. In a downlink two-user RSMA, the transmitter splits the user messages into common and private streams, each of which is allocated at a different rate. Each (user) receiver then decodes the common and its private streams to retrieve its messages, allowing RSMA to partially cancel multiuser interference and partially treat the interference as noise. These multiple access schemes permit aggressive sharing of the spectrum amongst multiple users and are often viewed as an enabler for massive connectivity. Clearly, these schemes can also be implemented in \gls{FAS} to enhance user performance greatly when compared to \gls{TAS} \cite{FAS_NOMA}. Specifically, the positions of the radiating elements and beamforming vector in \gls{FAS} can be reconfigured to create correlated or semi-correlated channels between users, thereby enhancing the performance of non-orthogonal techniques.

Nonetheless, both NOMA and RSMA require \gls{CSI} at the transmitter and \gls{SIC} at the receiver, leading to higher system and protocol complexity. To address this, one key research direction in \gls{FAS} considers no \gls{CSI} at the transmitter, and no \gls{SIC} at the receiver for multiple access. In this context, researchers in \gls{FAS} have explored the concept of \gls{FAMA} \cite{9650760,10066316}, considering opportunistic \gls{FAMA} \cite{10078147,Waqar-2024}, \gls{CUMA} \cite{10318083} and other variations to achieve the goal of TK$\mu$ extreme connectivity. The TK$\mu$ extreme connectivity is a performance target statement for 6G originally proposed in \cite{10183795}, which aims to achieve date rate of $1~{\rm Tbps}$, spectral efficiency of $1~{\rm kbps/Hz}$ and latency of $1~\mu{\rm s}$. Instead of relying on cell-free massive \gls{MIMO}, photonics-assisted \gls{THz} wireless access, and spatio-temporal channel coding, \gls{FAMA} and its variants leverage the fact that by reconfiguring the FAS at the user side to operate where interference undergoes deep fade, the interference can be mitigated without \gls{CSI} at the transmitter and \gls{SIC} at the receiver. \gls{FAMA} offers an exciting possibility for (transmit) \gls{CSI}-free/\gls{CSI}-less massive connectivity.

\subsubsection{Movable Antenna System}
This is an emerging wireless system where the antenna positions are flexibly adjusted within a spatial region to enhance channel conditions and communication performance \cite{10286328,10318061}. This system was inspired by \gls{FAS} and can be regarded as a specific design within the field of \gls{FAS}, where only position reconfiguration is considered. As a majority of studies in \gls{FAS} focus on position reconfiguration, the results are often applicable to both systems and can thus be interchangeable due to the similarity in concepts \cite{zhu2024historical}. 

Nevertheless, it is important to recognize that certain findings in movable antenna systems may not be directly applicable to \gls{FAS}, especially concerning the pixel-based fluid antenna design and other general reconfiguration capabilities. For example, pixel-based antennas are more closely related to \gls{XL}-\gls{MIMO}, where many antennas or pixels are deployed within a given surface.\footnote{Note that there is difference between a pixel and an antenna. An antenna is typically composed of several pixels with optimized connections. More details will be given in Section \ref{ssec:pixel}.} This design introduces mutual coupling issues, even if the antennas or pixels are non-active. Consequently, antenna and circuit theories play pivotal roles in determining the performance of \gls{FAS}. Nonetheless, mutual coupling is less severe in mechanical movable or liquid-based antenna designs as the effect only occurs between the active radiating elements in that case. Current research in movable antenna systems also does not explore the reconfiguration of shape or size in radiating elements. These reconfigurations are of great interest in \gls{FAS} too since they play crucial roles in applications such as cognitive radio networks \cite{8105795,8762117}, \gls{IoT}, body area networks \cite{9982508} and etc., despite less progress in these areas.

\subsection{Main Contributions}

\begin{table*}[h]
\caption{\label{Comparison}The summary of different review papers.}

\begin{centering}
\begin{tabular}{|c|>{\centering}p{15cm}|}
\hline 
Ref. & Summary\tabularnewline[\doublerulesep]
\hline 
\hline 
\cite{wong2022bruce} & \begin{raggedright}
a) Simplified channel model and theoretical performance of \gls{FAS}
and \gls{FAMA} were discussed. \\
b) Two hardware designs to enable \gls{FAS} were described. \\
c) Six research topics were outlined, including \gls{RIS}+\gls{FAMA},
\gls{MIMO}-\gls{FAMA}, \gls{MIMO}-\gls{FAS}, port selection and
utility interference,
\par\end{raggedright}
\raggedright{}\hspace{1em}implementation of \gls{FAS}, and low-latency
high security \gls{FAMA}.\tabularnewline[\doublerulesep]
\hline 
\cite{liquid_metals} & \raggedright{}a) Gallium-based liquid metals were reviewed.\\
b) Their physical, chemical and biological properties were discussed.
\\
c) The compatibility of Gallium-based liquid metals for different
applications were highlighted.\tabularnewline[\doublerulesep]
\hline 
\cite{paracha2019liquid} & \raggedright{}a) Comprehensive materials used in liquid metals were
discussed. \\
b) Various methods for fabricating the antennas were investigated.\\
c) The performance of frequency, polarization and pattern reconfiguration
of the antennas were reviewed.\\
d) The potential applications of liquid metals in wearable devices,
\gls{IoT} and wireless power transfer were highlighted.\tabularnewline[\doublerulesep]
\hline 
\cite{9388928} & \raggedright{}a) Metallic liquid and non-metallic liquid antennas
were reviewed.\\
b) State-of-art designs were presented.\\
c) Challenges of making the liquid-based antennas for real-world applications
were identified.\tabularnewline[\doublerulesep]
\hline 
\cite{9982508}$^{\ddagger}$ & \raggedright{}a) Different liquid antenna technologies that can be
used to build arrays were presented.\\
b) Challenges in integrating liquid antenna for wireless communications
were enumerated.\\
c) Some potential solutions to address the challenges were discussed.\tabularnewline[\doublerulesep]
\hline 
\cite{zheng2023flexibleposition}$^{\ddagger}$ & \raggedright{}a) Various hardware designs were elaborated including
traditional approaches.\\
b) The benefits of flexible-position \gls{MIMO} in channel hardening,
spectral efficiency and energy efficiency were explained.\\
c) Optimizing the trajectory of the antenna positions via \gls{ML}
techniques was demonstrated.\\
d) Future directions were outlined.\tabularnewline[\doublerulesep]
\hline 
\cite{10286328}$^{\ddagger}$ & \raggedright{}a) The general hardware design of movable antenna system
was discussed.\\
b) The opportunities of movable antenna systems in improved signal
power, suppressed interference, flexible beamforming, and enhanced
spatial multiplexing performance were demonstrated.\\
c) The challenges in channel estimations and antenna position optimization
were highlighted.\tabularnewline[\doublerulesep]
\hline 
\cite{9770295} & \raggedright{}a) Conventional active and semi-active \gls{RIS} were
discussed.\\
b) Surface-wave based \gls{RIS} was introduced.\\
c) The performance of \gls{FAS} and \gls{FAMA} were examined.\\
d) The potential of combining \gls{MIMO}, \gls{RIS}, and \gls{FAS} was envisioned.\tabularnewline[\doublerulesep]
\hline 
\cite{10146274}$^{\dagger}$ & \raggedright{}a) More recent channel models and theoretical performance of \gls{MIMO}-\gls{FAS} and \gls{FAMA} were summarized.\tabularnewline[\doublerulesep]
\hline 
\cite{10146286}$^{\dagger}$ & \raggedright{}a) Few research opportunities in \gls{FAS} were outlined. These include \gls{CAP}-\gls{MIMO}, \gls{MIMO}-\gls{FAMA}, FAS-assisted wireless power transfer and \gls{FAS}-assisted \gls{PLS}.\tabularnewline[\doublerulesep]
\hline 
\cite{10146262}$^{\dagger}$ & \raggedright{}a) Investigated the use of \gls{RIS}s as distributed artificial scattering surfaces to produce a rich scattering environment that enables \gls{FAS} to prevent multiuser interference at each user.\tabularnewline[\doublerulesep]
\hline 
This paper & \raggedright{}a) The similarities and differences of \gls{FAS} and other state-of-the-art technologies are discussed.\\
b) Several existing channel models were introduced and key factors that shape the system models are highlighted.\\
c) State-of-the-art methods for channel estimation and system optimization, including mathematical approaches and \gls{ML} techniques are presented.\\
d) The superiority of \gls{FAS} as compared to \gls{TAS} is demonstrated from an information-theoretic perspective.\\
e) New methods to perform multiple access, exclusive to \gls{FAS}, are introduced.\\
f) Various hardware designs are elaborated, including their strengths and weaknesses.\\
g) The new challenges and potential synergy of \gls{FAS} and other contemporary technologies are thoroughly discussed.\tabularnewline[\doublerulesep]
\hline 
\end{tabular}
\par\end{centering}
$^{\dagger}$Letters,$^{\ddagger}$Magazine articles
\end{table*}

\gls{FAS} is an interdisciplinary subject and this paper's aim is to provide a comprehensive tutorial on this topic. Some review articles have briefly addressed this matter. For instance, \cite{9770295} discussed the synergy between \gls{MIMO}, \gls{RIS} and \gls{FAS} whereas \cite{wong2022bruce} outlined six research topics and highlighted the potential of \gls{FAS} under simplified channel models. Moreover, liquid-based antenna designs were reviewed in \cite{liquid_metals,paracha2019liquid,9388928,9982508} but the emphasis was on the RF design and implementation and the communication-theoretic aspects of \gls{FAS} were not covered. A recent three-part paper provided insights into the preliminaries \cite{10146274}, research opportunities \cite{10146286}, as well as a paradigm of using many RISs as distributed artificial scattering surfaces for massive connectivity \cite{10146262} from the information-theoretic perspective of \gls{FAS}. Furthermore, \cite{10286328} outlined the opportunities and challenges of the movable antenna system, and there was a short article touching on the origins of the various names used in \gls{FAS} \cite{zhu2024historical}. Additionally, \cite{zheng2023flexibleposition} provided a more general coverage on the fundamentals, challenges, and future research directions of flexible-position \gls{MIMO} systems. 

The interest of FAS is evident. Although there have been articles providing some insightful discussions, their scopes are quite limited. Summaries of these review papers are given in Table \ref{Comparison}. This article serves as the first comprehensive tutorial of \gls{FAS}, encompassing state-of-the-art communication theory, estimation and optimization methods, and hardware designs, with emphasis on position reconfiguration. This tutorial can be easily understood with basic knowledge in communication, circuit, and antenna theories, as well as AI and optimization. The main contributions are summarized as follows:
\begin{itemize}
\item This tutorial paper introduces existing channel models in the field of \gls{FAS} and highlights various key factors that influence the system models. Readers will gain insights into adopting or developing system models tailored to their research applications effectively.
\item Additionally, this tutorial covers state-of-the-art methods for channel estimation and system optimization, including both mathematical approaches and \gls{ML} techniques. Also, it delves into the fundamentals of \gls{FAS}, providing information-theoretic insights into its superiority compared to \gls{TAS} across various setups and cases.
\item This paper introduces new multiple access methods exclusive to \gls{FAS}, offering a new perspective to achieve the \gls{6G} TK$\mu$ extreme connectivity. Utilizing information theory, approximation techniques, and symbol-level coding, the tutorial aims to facilitate readers' comprehension of concepts ranging from simple to sophisticated.
\item An overview of various hardware designs, including mechanical movable antennas, liquid-based antennas, pixel-based antennas, and hybrid antennas, is provided. These insights contribute to progress in hardware development and offer valuable information for antenna experts.
\item Lastly, this tutorial paper outlines promising research directions in \gls{FAS}, inspiring further studies to unlock the full potential of \gls{6G} networks. Specifically, we describe new challenges within \gls{FAS} and explore potential synergies with other recent technologies, enabling researchers to harness the full benefits of \gls{FAS}.
\end{itemize}

\subsection{Organization of the Tutorial}
\begin{figure*}[!pht]
\centering{}\includegraphics[width=17cm]{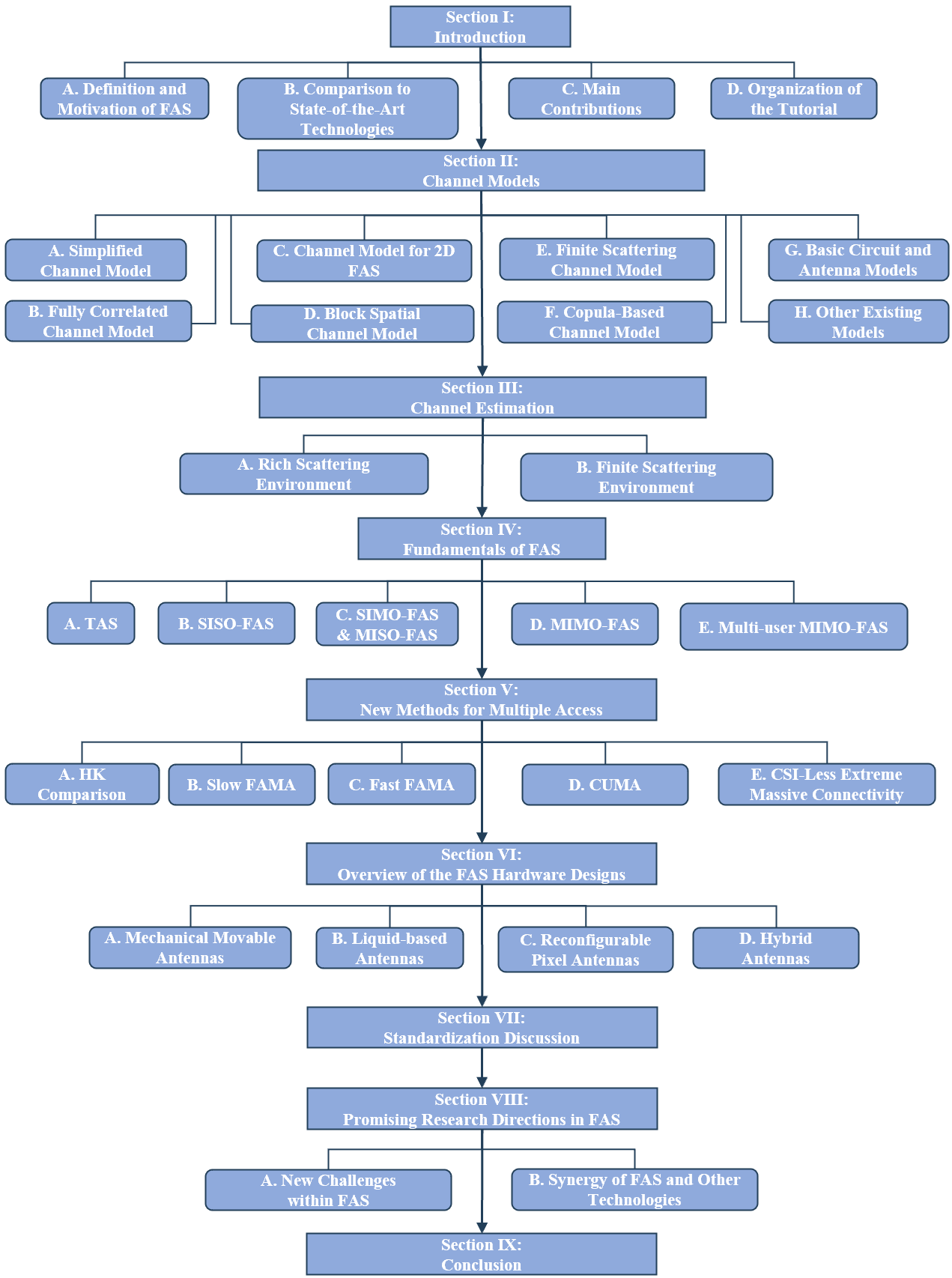} \caption{Organization of the tutorial.}
\label{Organization}
\end{figure*}

As shown in Fig.~\ref{Organization}, this paper is organized into sections with the aim of providing a comprehensive understanding of \gls{FAS}. In Section \ref{sec:intro}, we introduce different system models essential in characterizing the performance of FAS, while Section \ref{sec:ce} discusses various estimation methods to operate \gls{FAS}. Section \ref{sec:funda} delves into the fundamentals of \gls{FAS}, providing readers information-theoretic insights into its superior characteristics compared to \gls{TAS} across typical setups and cases. Section \ref{sec:fama} explores new multiple access methods exclusive to \gls{FAS}, presenting a new perspective for massive communications. In Section \ref{sec:hardware}, an overview of various hardware designs related to \gls{FAS} is discussed. Section \ref{sec:standard} briefly discusses the implication regarding standardization in light of \gls{FAS}. Section \ref{sec:Promising-Research-Directions} outlines the new challenges of \gls{FAS} and promising research directions.
 This tutorial concludes in Section \ref{sec:conclude}, where essential remarks and key takeaways are provided to encapsulate the tutorial's insights and findings. To facilitate the readers, key abbreviations are summarized in Table \ref{Abbreviations}. 

\begin{table*}
\begin{centering}
\caption{\label{Abbreviations}Key abbreviations.}
\par\end{centering}
\begin{centering}
\begin{tabular}{|c|c|c|c|}
\hline 
Abbreviation & Definition & Abbreviation & Definition\tabularnewline
\hline 
\hline 
AoA & Angle-of-Arrivals & MRC & Maximum Ratio Combining\tabularnewline
\hline 
AoD & Angle-of-Departure & NGMA & Next Generation Multiple Access\tabularnewline
\hline 
AWGN & Additive White Gaussian Noise & NMSE & Normalized Mean Squared Error\tabularnewline
\hline 
BS & Base Station & NLoS & Non Line-of-Sight\tabularnewline
\hline 
CAP-MIMO & Continuous Aperture MIMO & NOMA & Non-Orthogonal Multiple Access\tabularnewline
\hline 
CSCG & Circularly Symmetric Complex Gaussian & OMA & Orthogonal Multiple Access\tabularnewline
\hline 
CSI & Channel State Information & OMP & Orthogonal Matching Pursuit\tabularnewline
\hline 
CUMA & Compact Ultra Massive Antenna Array & RF & Radio Frequency\tabularnewline
\hline 
DFT & Discrete Fourier Transform & RIS & Reconfigurable Intelligent Surfaces\tabularnewline
\hline 
DMT & Diversity and Multiplexing Tradeoff & RSMA & Rate-Splitting Multiple Access\tabularnewline
\hline 
EWOD & Electrowetting-on-Dielectric & RZF & Regularized Zero-Forcing\tabularnewline
\hline 
FAS & Fluid Antenna System & SIC & Successive Interference Cancellation\tabularnewline
\hline 
FAMA & Fluid Antenna Multiple Access & SINR & Signal-to-Interference-plus-Noise Ratio\tabularnewline
\hline 
FFT & Fast Fourier Transform & SIMO & Single-Input Multiple-Output\tabularnewline
\hline 
HK & Han-Kobayashi & SISO & Single-Input Single-Output\tabularnewline
\hline 
L3SCR & Low-Sample-Size Sparse Channel Reconstruction & SNR & Signal-to-Noise Ratio\tabularnewline
\hline 
LoS & Line-of-Sight & TAS & Traditional Antenna System\tabularnewline
\hline 
MIMO & Multiple-Input Multiple-Output & TIN & Treating Interference as Noise\tabularnewline
\hline 
MISO & Multiple-Input Single-Output & UAMA & Unified Asymmetric Masked Autoencoder\tabularnewline
\hline 
ML & Machine Learning & XL-MIMO & Extremely Large-Scale MIMO\tabularnewline
\hline 
\end{tabular}
\par\end{centering}
\centering{}
\end{table*}

\section{Channel Models}\label{sec:intro}
To characterize and investigate the theoretical performance of \gls{FAS}, it is important to consider the specific implementation context. For instance, if the fluid antenna is integrated at the receiver, then one must account for the spatial correlation at the receiver. Similarly, if the transmitter is equipped with a fluid antenna, the spatial correlation at the transmitter end becomes paramount. Modeling the spatial correlation at the FAS is thus a crucial step in characterizing the performance, and can be done through geometric or mathematical approaches. 

Attention should be directed toward the FAS architecture. Options include \gls{1D}, \gls{2D}, or even \gls{3D} fluid antenna surfaces. Typically, the performance of \gls{FAS} tends to enhance as the dimension of the surface increases. The positioning, orientation, or length of the radiating element can be reconfigured discretely or continuously. Nevertheless, in the current stage of research, the emphasis of \gls{FAS} primarily centers around position reconfiguration. Moreover, the configuration may involve a single active radiating element or multiple elements within a given surface, necessitating consideration of isolation techniques, circuit settings, and mutual coupling effects. 

It is also vital to consider the carrier frequency and communication environment. The environment influences the number of scatterers, determining whether the area surrounding the transmitter and receiver is rich in scatterers or finite. Additionally, assessing whether the transmitter and receiver have \gls{LoS} or \gls{NLoS} is crucial. Furthermore, near-field and far-field propagation are determined not only by the fluid antenna size but also the operating frequency, as well as the distance between the transmitter and receiver. On the other hand, consideration should also be given to modeling accuracy versus mathematical tractability. For instance, while the classical Jakes' model is understood to accurately represent isotropic propagation effects, it is prohibitively challenging to analyze. The aforementioned factors shape the system models. In this section, we will discuss existing system models used for FAS, delving into their merits and drawbacks. This exploration aims to empower readers to make informed decisions when selecting or developing their system models. 

\begin{figure}[]
\centering{}\includegraphics[scale=0.45]{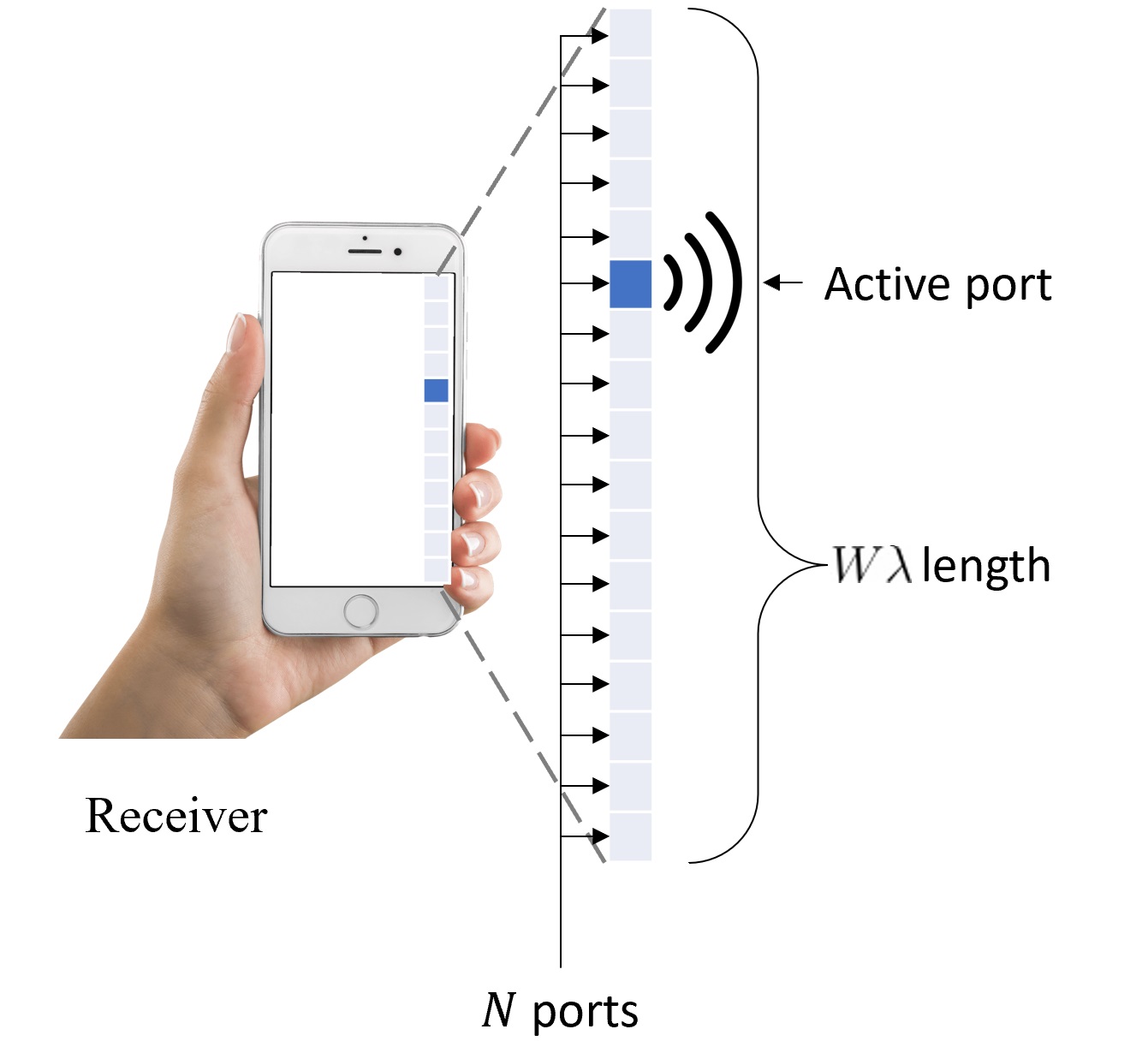}
\caption{A schematic of a \gls{1D} fluid antenna structure.}\label{1D_fluid_antenna_surface}
\end{figure}

\subsection{Simplified Channel Model}\label{ssec:scmodel}
We shall start by introducing one of the simplest channel models which was first used in \cite{9264694}. Without loss of generality, our focus will be on small-scale fading since the large-scale fading remains similar to \gls{TAS}. This assumption holds when the size of the fluid antenna is significantly smaller than the distance between the transmitter and receiver and the operating frequency is not tremendously high. In this model, a point-to-point channel is considered where the transmitter deploys a traditional fixed-position antenna and the receiver uses a linear fluid antenna with a single radiating element (also known as active port). The fluid antenna has $N$ preset locations (also known as ports) that are evenly distributed over a length of $W\lambda$, where $\lambda$ is the wavelength, as depicted in Fig. \ref{1D_fluid_antenna_surface}. The normalized channel vector between the transmitter and the $N$ ports of the \gls{FAS} receiver can be denoted as $\bm{h}=[h_{1},\dots,h_{N}]^{T}$, where $h_{n}\sim{\cal CN}(0,1)$ is the channel gain from the transmitter to the $n$-th port, with a standard complex Gaussian distribution. As the ports may be closely located, some entries of $\bm{h}$ are strongly correlated. Thus, the channel coefficients can be parameterized as in \cite{6094281}, where the first port is treated as a reference port. More concretely, the $n$-th entry of $\bm{h}$ can be computed as 
\begin{align} 
h_{n} & =\left(\sqrt{1-\mu_{n}^{2}}x_{n}+\mu_{n}x_{1}\right)\nonumber \\
 & +j\left(\sqrt{1-\mu_{n}^{2}}y_{n}+\mu_{n}y_{1}\right),~{\text{for}}~n=2,\dots,N,\label{eq:h_cls}
\end{align} 
where $x_{1},\dots,x_{N},y_{1},\dots,y_{N}$ are \gls{i.i.d.}~real Gaussian variables with zero-mean and variance of $\frac{1}{2}$, while $\mu_{n}$ in \eqref{eq:h_cls} is chosen as
\begin{equation}\label{mu_n}
\mu_{n}=J_{0}\left(2\pi\frac{\left|n-1\right|}{N-1}W\right),
\end{equation}
where $J_{0}\left(\cdot\right)$ is the zero-order Bessel function of the first kind. 

Since the $n$-th entry of $\bm{h}$ is coupled with the reference port only, this model greatly simplifies the performance analysis of \gls{FAS}. Nevertheless, it imposes a structure on the covariance of two different ports (e.g., $n$-th and $m$-th ports). Due to this structure, the spatial correlation between any two ports cannot be observed without the reference port. That is to say, if two ports are weakly correlated to the reference port, then they will have to be weakly correlated to each other even though they are close to each other but far from the reference port. To address this, \cite{wong2022closed} proposed a simple fix to characterize each channel coefficient using a common correlation parameter.
In particular, $\mu_{n}$ in (\ref{eq:h_cls}) can be replaced with
{\begin{equation}\label{mu}
\mu=\sqrt{2}\sqrt{_{1}F_{2}\left(\frac{1}{2};1,\frac{3}{2};-\pi^{2}W^{2}\right)-\frac{J_{1}(2\pi W)}{2\pi W}},
\end{equation} 
in which $\mu$ represents the common correlation parameter, $_{1}F_{2}(\cdot;\cdot;\cdot)$ denotes the generalized hypergeometric function, and $J_{1}(\cdot)$ is the first-order Bessel function of the first kind. This method links all the ports on the same fluid antenna together without a reference port. The choice of $\mu$ in (\ref{mu}) aims to mimic the average squared spatial correlation of an actual \gls{1D} fluid antenna structure, intending to improve the modelling accuracy while maintaining tractability.

Nevertheless, it was reported in \cite{10103838} that the channel model in (\ref{eq:h_cls}) with the spatial correlation parameter in (\ref{mu_n}) or even (\ref{mu}) may still not accurately capture the spatial correlation between different ports and result in an overly optimistic performance estimation. Most recently, an enhancement is made to improve the accuracy. Specifically, based on this model, \cite{ramirez2024new} introduced a new block spatial correlation model that strikes a balance between accuracy and analytical tractability. That model will be introduced and discussed in Section \ref{ssec:bscmodel}.
It is worth noting that compared to \gls{TAS} in a point-to-point scenario, where the transmitter and receiver deploy a traditional fixed-position antenna, the complex channel can be modeled as
\begin{equation}\label{TAS_p2p}
h_{1}=x_{1}+jy_{1},
\end{equation} which is equivalent to having one port in \gls{FAS}. Clearly, the fixed-position antenna of \gls{TAS} limits the channel dimensions.

\subsection{Fully Correlated Channel Model}\label{ssec:fcmodel}
To accurately capture the spatial correlation between different ports based on the Jakes' assumption in \cite{jakes1994microwave}, the work in \cite{10103838} introduced a generalized \gls{FAS} channel model. In this model, a similar point-to-point channel is considered where the transmitter adopts a traditional fixed-position antenna and the receiver uses a linear fluid antenna with a single radiating element, and $N$ ports uniformly distributed along a length of $W\lambda$. Denoting the covariance matrix of the channels by $\bm{J}$, the $(n,m)$-th element of $\bm{J}$ can be expressed as
\begin{align}
\bm{J}_{n,m} & ={\rm Cov}\left[h_{n},h_{m}\right]=J_{0}\left(2\pi\frac{\left|n-m\right|}{N-1}W\right).\label{eq:cov}
\end{align}
Each entry of $\bm{h}$ can be further modeled as a linear combination of $N$ \gls{i.i.d.}~\gls{CSCG} random variables by utilizing the covariance matrix $\bm{J}$ \cite{10103838}. Specifically, we can denote the eigenvalue decomposition of $\bm{J}$ as $\bm{Q}\bm{\varLambda}\bm{Q}^{H}$, where $\bm{Q}$ is the eigenvector matrix of $\bm{J}$ and $\bm{\varLambda}={\rm diag}\{\lambda_{1},\dots,\lambda_{N}\}$ is the diagonal eigenvalue matrix of $\bm{J}$ such that $\lambda_{1}\geq\dots\geq\lambda_{N}$. As such, we can write
\begin{equation}
\bm{h}=\bm{Q}\bm{\varLambda}^{\frac{1}{2}}\bm{g},\label{eq:h_jk}
\end{equation}
where $\bm{g}=[g_{1},\dots,g_{N}]^{T}$ and $g_{n}\sim{\cal CN}(0,1),\forall n$. Alternatively, the $n$-th element of $\bm{h}$ can be rewritten as
\begin{align}
h_{n} & =\sum_{m=1}^{N}q_{n,m}\sqrt{\lambda_{m}}g_{m}=\sum_{m=1}^{N}q_{n,m}\sqrt{\lambda_{m}}\left(x_{m}+jy_{m}\right),\label{eq:h_jkn}
\end{align}
where $q_{n,m}$ is the $(n,m)$-th element of $\bm{Q}$. Obviously, with (\ref{eq:h_jk}) or (\ref{eq:h_jkn}), $\bm{h}$ follows the Jakes' assumption, i.e., $\bm{h}\sim{\cal CN}(\bm{0},\bm{J})$. This model yields a more accurate performance estimation; however, it is very difficult to analyze the performance of \gls{FAS} using (\ref{eq:h_jk}) or (\ref{eq:h_jkn}) since the \gls{PDF} and \gls{CDF} usually result in expressions that involve $N$ nested integrals, which are non-computable and thus mathematically intractable \cite{10279640}. 

\begin{figure}
\centering \includegraphics[scale=0.45]{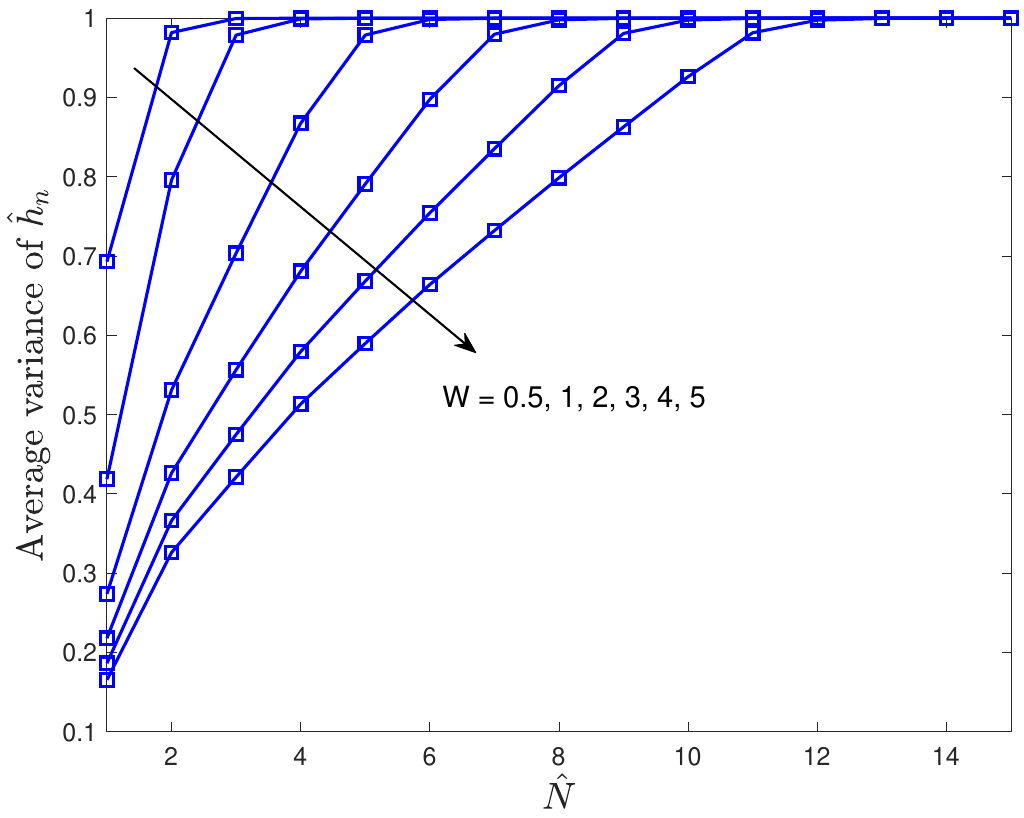} 
\caption{Average variance of $h_{n},\forall n\in\{1,\dots,N\}$ versus the approximation level $\hat{N}$ with $N=100$.}\label{ave_var_appro}
\end{figure}

Fortunately, \cite{10103838,10279640} have shown that the covariance matrix $\bm{J}$ of the channel vector $\bm{h}$ mainly focuses on a few largest eigenvalues if $N$ is sufficiently large or $W$ is small enough.\footnote{This property is also exploited in the block correlation model in \cite{ramirez2024new}.} This makes it possible to approximate each channel coefficient of $h_{n}$ by taking only $\hat{N}\ll N$ eigenvalues into account. More concretely, the entry $h_{n}$ in (\ref{eq:h_jkn}) can be approximated by 
\begin{align}
\hat{h}_{n}=\sum_{m=1}^{\hat{N}}q_{n,m}\sqrt{\lambda_{m}}\left(x_{m}+jy_{m}\right).\label{model3}
\end{align}
From (\ref{model3}), it can be verified that ${\hat{h}}_{n}\sim{\cal CN}(0,\sum_{m=1}^{\hat{N}}q_{n,m}^{2}\lambda_{m})$. Fig.~\ref{ave_var_appro} depicts the value of 
\begin{equation}\label{ave_var}
\frac{1}{N}\sum_{n=1}^{N}{\rm Cov}\left[\hat{h}_{n}\right]=\frac{1}{N}\sum_{n=1}^{N}\sum_{m=1}^{\hat{N}}q_{n,m}^{2}\lambda_{m}.
\end{equation}
Specifically, (\ref{ave_var}) computes the average variance of ${\hat{h}}_{n},\forall n\in\{1,$ $\dots,N\}$. In the extreme $\hat{N}=N$ case, the value of (\ref{ave_var}) is $1$ since $\sum_{m=1}^{N}q_{n,m}^{2}\lambda_{m}=\bm{J}{}_{n,n}=1$. As such, if $\hat{N}$ increases and $\hat{N}<N$, the value of (\ref{ave_var}) will gradually approach $1$. Fig.~\ref{ave_var_appro} shows that this is true and can be realized by a small $\hat{N}$ (in contrast to $N$). For example, when $W=0.5$ and $W=2$, the value of (\ref{ave_var}) is close to $1$ with $\hat{N}$, respectively, being $3$ and $6$. The exact channel model $h_{n}$ in (\ref{eq:h_jkn}) can thus be accurately approximated by taking into account only a few eigenvalues of $\bm{J}$. Using these properties, it is possible to approximate the \gls{PDF} and \gls{CDF} of $\bm{h}$ in closed-form expressions, and analyze the outage probability of single-user \gls{FAS} channels \cite{10103838,10130117}, and that for a two-user \gls{FAMA} system \cite{10279640}. 
Recall that in a point-to-point \gls{TAS} scenario, where the transmitter and receiver use a traditional fixed-position antenna, the complex channel is modeled as in \eqref{TAS_p2p}, which has limited channel dimension.

\subsection{Channel Model for \gls{2D} FAS}
The channel model in (\ref{eq:h_jk}) primarily focuses on a \gls{1D} fluid antenna surface implemented at the receiver. Nevertheless, we can extend the channel model to the scenario where both the transmitter and receiver are equipped with \gls{2D} fluid antenna surfaces by taking into account of the \gls{3D} environment. More concretely, we may assume that the \gls{2D} fluid antenna has $N_{i}^{s}$ ports uniformly distributed along a linear dimension of length $W_{i}^{s}\lambda$, where $i\in\left\{ 1,2\right\} $ and $s\in\left\{ {\rm rx},{\rm tx}\right\}$. Thus, the \gls{2D} fluid antenna has a size of $W_{s}=W_{1}^{s}\lambda\times W_{2}^{s}\lambda$ and $N_{s}=N_{1}^{s}\times N_{2}^{s}$ ports. To simplify the notation, we can refer the \gls{2D} indices of the ports from left to right and from top to bottom, and assign the resulting numbers as the new port indices. For instance, the $\left(n_{1},n_{2}\right)$-th port can be mapped to a new index
\begin{equation}\label{k_n1n2}
k_{\left(n_{1},n_{2}\right)}=\left(n_{2}-1\right)N_{1}+n_{1}.
\end{equation}

\begin{figure}[t]
\noindent \centering{}\includegraphics[scale=0.4]{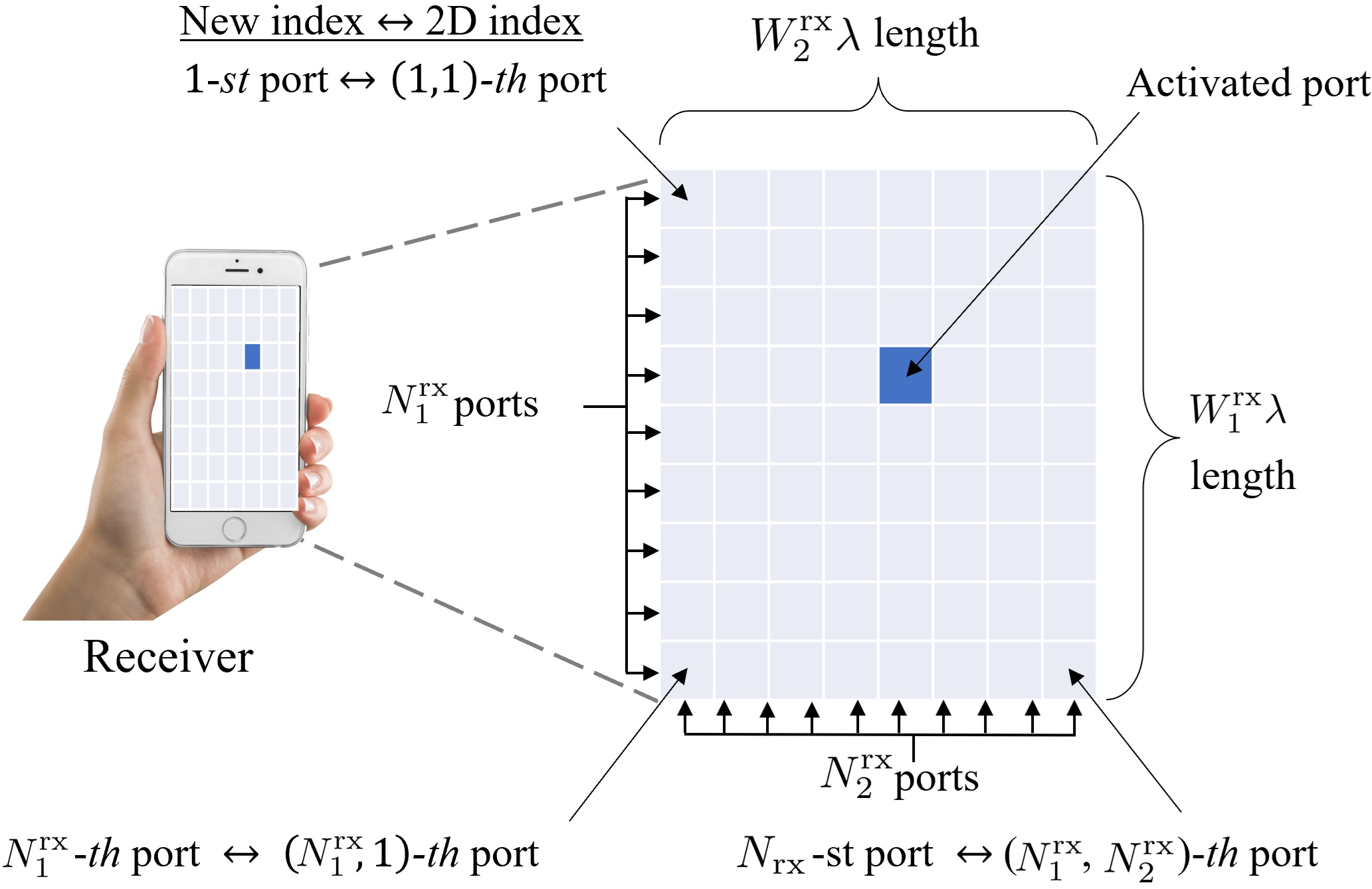}
\caption{An example of a \gls{2D} FAS receiver, illustrating the mapping between the \gls{2D} indices and \gls{1D} index.}\label{2D_fluid_antenna_surface}
\end{figure}

Fig.~\ref{2D_fluid_antenna_surface} illustrates an example of a \gls{2D} FAS surface at the receiver side. Similarly, we can use the covariance matrix $\boldsymbol{J}_{s}\in{\mathbb{C}}^{N_{s}\times N_{s}}$ to characterize the spatial correlation among all the ports at the $s$ side. For example, we can write the covariance matrix at the transmitter as
\begin{align}
 & \bm{J}_{\rm tx}=\nonumber \\
 & \begin{bmatrix}J_{1,1}^{\rm tx} & \cdots & J_{1,k_{\left(\tilde{n}_{1},\tilde{n}_{2}\right)}}^{\rm tx} & \cdots & J_{1,N_{\rm tx}}^{\rm tx}\\
\vdots & \ddots & \vdots & \ddots & \vdots\\
J_{k_{\left(n_{1},n_{2}\right)},1}^{\rm tx} & \cdots & J_{k_{\left(n_{1},n_{2}\right)},k_{\left(\tilde{n}_{1},\tilde{n}_{2}\right)}}^{\rm tx} & \cdots & J_{k_{\left(n_{1},n_{2}\right)},N_{\rm tx}}^{\rm tx}\\
\vdots & \ddots & \vdots & \ddots & \vdots\\
J_{N_{\rm tx},1}^{\rm tx} & \cdots & J_{N_{\rm tx},k_{\left(\tilde{n}_{1},\tilde{n}_{2}\right)}}^{\rm tx} & \cdots & J_{N_{\rm tx},N_{\rm tx}}^{\rm tx}
\end{bmatrix},
\end{align}
where $J_{k_{\left(n_{1},n_{2}\right)},k_{\left(\tilde{n}_{1},\tilde{n}_{2}\right)}}^{\rm tx}$ represents the spatial correlation between the $\left(n_{1},n_{2}\right)$-th and $\left(\tilde{n}_{1},\tilde{n}_{2}\right)$-th ports. By considering a \gls{3D} environment undergoing rich scattering (i.e., with a huge number of scatterers), the spatial correlation follows
\begin{multline}\label{J}
J_{k_{\left(n_{1},n_{2}\right)},k_{\left(\tilde{n}_{1},\tilde{n}_{2}\right)}}^{\rm tx}=\\
j_{0}\left(2\pi\sqrt{\left(\frac{\left|n_{1}-\tilde{n}_{1}\right|}{N_{1}^{{\rm tx}}-1}W_{1}^{{\rm tx}}\right)^{2}+\left(\frac{\left|n_{2}-\tilde{n}_{2}\right|}{N_{2}^{{\rm tx}}-1}W_{2}^{{\rm tx}}\right)^{2}}\right),
\end{multline}
where $j_{0}\left(\cdot\right)$ is the zero-order spherical Bessel function or the sinc function. Suppose we consider a similar \gls{2D} fluid antenna structure at the receiver, the covariance matrix at the receiver can be characterized in a similar fashion. Let us now denote the eigenvalue decomposition of $\bm{J}_{\rm tx}$ and $\bm{J}_{\rm rx}$ as $\bm{Q}_{\rm tx}\bm{\varLambda}_{\rm tx}\bm{Q}_{\rm tx}^{H}$ and $\bm{Q}_{\rm rx}\bm{\varLambda}_{\rm rx}\bm{Q}_{\rm rx}^{H}$, respectively. Then the complex channel between the transmitter and receiver with \gls{2D} fluid antenna surfaces can be modeled as \cite{10303274}
\begin{equation}\label{H}
\bm{H}=\bm{Q}_{\rm rx}\bm{\varLambda}_{\rm rx}^{\frac{1}{2}}\bm{G}\left(\bm{\varLambda}_{\rm tx}^{\frac{1}{2}}\right)^{H}\bm{Q}_{\rm tx}^{H},
\end{equation}
in which $\bm{G}\in{\mathbb{C}}^{N_{{\rm {rx}}}\times N_{{\rm tx}}}$ is a \gls{CSCG} random matrix, such that each entry is \gls{i.i.d.}~and they follow ${\cal CN}(0,1)$.\footnote{From (\ref{H}), it is easy to rewrite (\ref{eq:h_jk}) to accommodate the implementation of a fluid antenna solely on the transmitter.}

Overall, analyzing the performance of \gls{FAS} using this model is much more challenging than using (\ref{eq:h_jk}) because the rows and columns of (\ref{H}) are fully correlated. But this setup can significantly improve the performance of \gls{FAS}, given that both the transmitter and receiver are equipped with a single \gls{2D} fluid antenna surface and the scattering stems from a \gls{3D} environment. Note that considering multiple active radiating elements in \gls{FAS} further complicates the analysis \cite{6185738}.

\subsection{Block Spatial Correlation Model}\label{ssec:bscmodel}
The Jakes' model can accurately represent isotropic propagation effects \cite{jakes1994microwave}, although the correlation structure is generally imposed by specific propagation conditions and antenna radiation patterns (see, e.g., \cite{Aulin-1979}). Unfortunately, the analytical characterization of \gls{FAS} under Jakes' model is infeasible, and several efforts have been made to come up with simplified approximations, either by replacing Jakes' model with a constant correlation between ports (Section \ref{ssec:scmodel}) or by employing lower-rank approximations (Section \ref{ssec:fcmodel}). The former has a tendency to produce inaccurate results while the latter remains prohibitively complex (analytically intractable). 

An alternative approach has been recently proposed in \cite{ramirez2024new}, providing a smart way to approximate FAS performance not only under Jakes' model, but more generally under arbitrary correlation structures. Inspired by the coherence interval idea behind block-fading models, \cite{ramirez2024new} defines spatial blocks that are independent, but the spatial correlation remains constant within each spatial block, translating the concept of time coherence interval into space. Hence, the resulting (approximated) correlation matrix is block-diagonal of the form 
\begin{align}
\hat{\bf J} = \begin{pmatrix}
        \mathbf{A}_1 & \mathbf{0} & \cdots &\mathbf{0} \\
        \mathbf{0} & \mathbf{A}_2 & \cdots &\mathbf{0} \\
        \vdots & & \ddots & \cdots \\
        \mathbf{0} & \mathbf{0} & \cdots &\mathbf{A}_B
    \end{pmatrix}, \label{eq:Japprox}
\end{align}
where each submatrix $\mathbf{A}_b$ is a constant correlation matrix of size $L_b$ and correlation $\mu_b^2$, i.e., $\mathbf{A}_b$ for $b = 1,\dots,B$ has ones in the diagonal and the off-diagonal elements equal to $\mu^2_b$. Due to its block-diagonal structure, the approximation inherits the tractability of the constant model in \cite{wong2022closed}. 

The block sizes $L_b$ are chosen based on spectral analysis of the target (true) correlation matrix, aiming to approximate its eigenvalues. As the fluid antenna is densified and the number of ports or positions is increased for the same aperture, the resulting correlation matrix is dominated by a few eigenvalues, as predicted by statistical theory on large Toeplitz matrices (see Section \ref{ssec:fcmodel}), and therefore only a few eigenvalues need to be approximated. This agrees with sampling theory, i.e., increasing the number of ports is equivalent to oversampling the spatial correlation function, and thus many of the samples (ports) provide no extra information regarding the correlation structure. For simplicity in the approximation, one possible choice is to assume that all blocks have the same correlation parameter, i.e., $\mu_b = \mu, \forall b$. Under this choice, each block $\mathbf{A}_b$ yields a set of eigenvalues $\{\hat{\lambda}_{n'}\}_{n'=1}^{L_b}$ with \cite{ramirez2024new}
\begin{equation}
    \hat{\lambda}_{n'} = \begin{cases}
        (L_b-1)\mu^2 + 1 & \text{if } n' = 1\\
        1-\mu^2 & \text{if } n' = 2,\dots, L_b .
    \end{cases}
\end{equation}
Letting $\mu\rightarrow 1$ makes the multiple eigenvalues at $1-\mu^2$ conveniently close to $0$, and each block will produce a single dominant eigenvalue at $(L_b-1)\mu^2 + 1$. Therefore, each of the blocks in \eqref{eq:Japprox} can approximate a dominant eigenvalue of the true correlation matrix by setting 
\begin{equation}
    L_b = \left\lfloor\frac{\lambda_b - 1}{\mu^2}+1\right\rfloor, \label{eq:Lb}
\end{equation}
where $\{\lambda_b\}_{b=1}^B$ is the set of dominant eigenvalues of the true correlation matrix. To avoid getting an approximation matrix with more ports than the original one, the block sizes $L_b$ can be increased iteratively as proposed in \cite[Algorithm 1]{ramirez2024new}. This procedure leads to a block-diagonal matrix with as many blocks as dominant eigenvalues in the original correlation matrix, and approximately equal spectrum (set of eigenvalues), as illustrated in Fig.~\ref{fig:eig_comparison} using a linear FAS as an example.

\begin{figure}[t]
\centering
\includegraphics[trim = {0 7.5cm 0 7.5cm}, clip, width = \columnwidth]{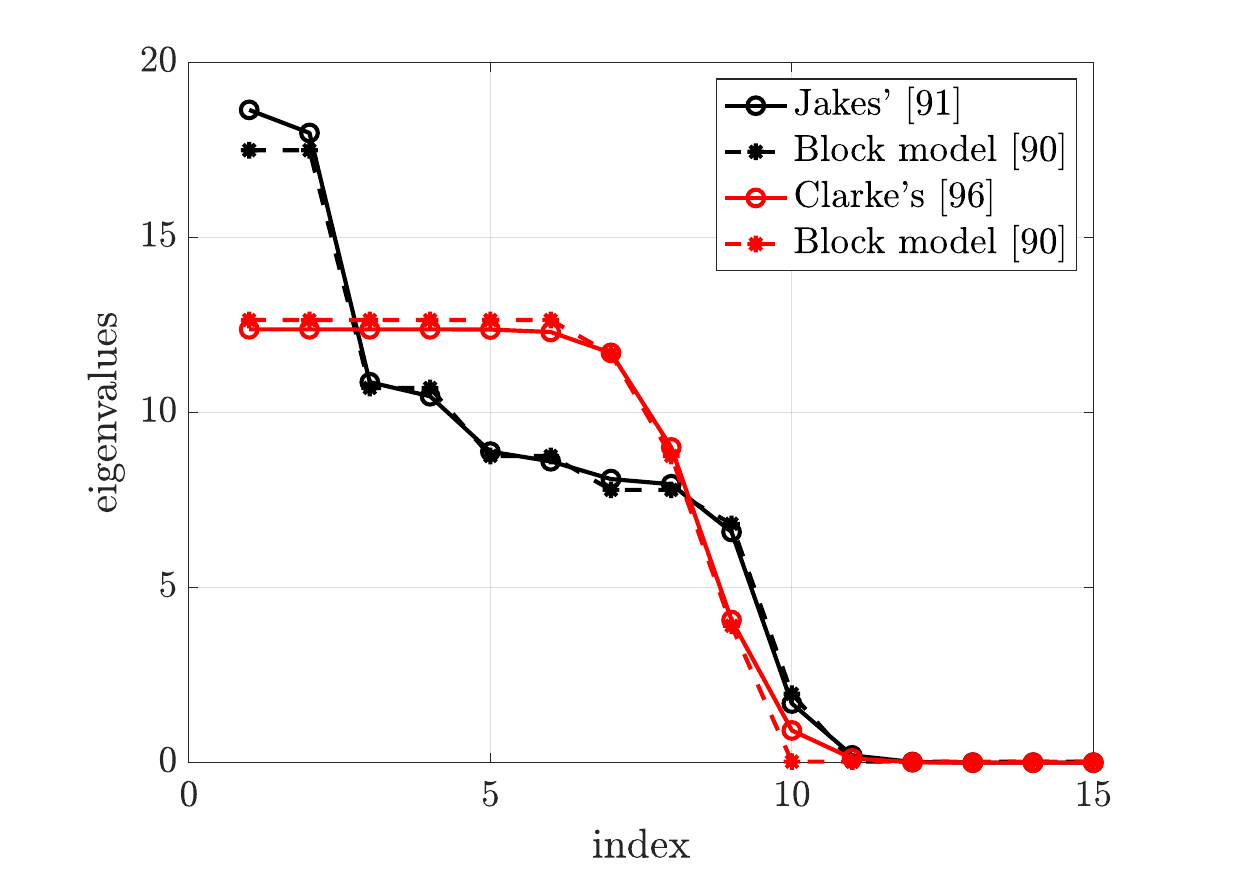}
\caption{Eigenvalues of correlation matrix for a linear fluid antenna with $W = 4$ and $N_{\rm rx} = 100$ under Jakes' model \cite{jakes1994microwave}, Clarke's model \cite{Aulin-1979} and the method in \cite{ramirez2024new} ($\mu^2 = 0.97$).}\label{fig:eig_comparison}
\end{figure}

Replacing the true correlation matrix (e.g., Jakes') by the above block-diagonal approximation seems to translate well into similar performance when analyzing FAS, as exemplified in Fig.~\ref{fig:outage}, where the outage probability of \gls{FAMA} is plotted under Jakes' correlation model, the block-diagonal approximation \cite{ramirez2024new} and the constant model \cite{wong2022closed}. As can be seen, while the block-diagonal model yields a tight approximation, the constant model considerably overestimates the performance.

In short, \cite{ramirez2024new} provides an alternative approach to analytically characterize FAS which (i) retains the tractability of \cite{wong2022closed} while yielding accurate results, tightly approximating the performance under realistic models such as Jakes', (ii) can be applied to any (arbitrary) correlation structure, as well as to linear (\gls{1D}) and planar (\gls{2D}) fluid antennas, and (iii) relieves the computational burden of simulating FAS since the block-diagonal correlation is much easier to generate than the cross-correlation inherent to conventional models like Jakes'.

\begin{figure}[t]
\centering
\includegraphics[trim = {0 7.5cm 0 7.5cm}, clip, width = \columnwidth]{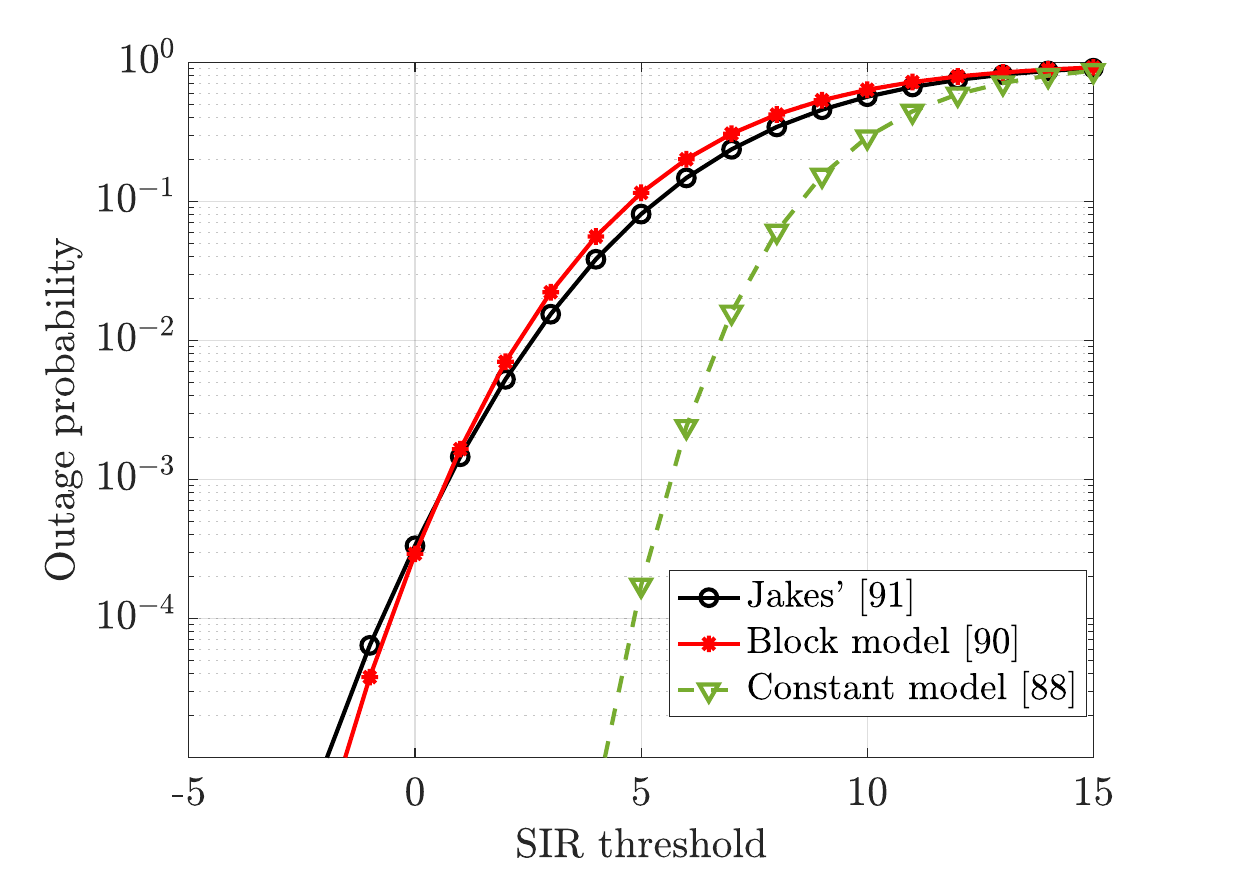}
\caption{Outage probability for 3-user slow \gls{FAMA} assuming a linear FAS at each user with $W = 7$ and $N_{\rm rx} = 150$  under different correlation models ($\mu^2 = 0.97$ for the block-diagonal approximation). Slow \gls{FAMA} will be discussed in Section \ref{ssec:sfama}.}\label{fig:outage}
\end{figure}

\subsection{Finite Scattering Channel Model}
Although the consideration of \gls{2D} fluid antenna at both ends can be represented by the channel model in (\ref{H}), it does not account for \gls{LoS}, \gls{3D} fluid antenna surface, or finite scatterers. These limitations can be addressed by using the planar-wave geometric model to characterize the channel \cite{6834753}. Specifically, the complex channel matrix $\boldsymbol{H}$ can be modeled as 

\begin{align}
\bm{H} & =\sqrt{\frac{K}{K+1}}e^{j\omega}\boldsymbol{a}_{r}\left(\theta_{0,r},\phi_{0,r}\right)\boldsymbol{a}_{t}\left(\theta_{0,t},\phi_{0,t}\right)^{H}\nonumber\\
 & +\sqrt{\frac{1}{L_p\left(K+1\right)}}\sum_{l=1}^{L_p}\kappa_{l}\boldsymbol{a}_{r}\left(\theta_{l,r},\phi_{l,r}\right)\boldsymbol{a}_{t}\left(\theta_{l,t},\phi_{l,t}\right)^{H},\label{eq:H_3D}
\end{align}
where $K$ is the Rice factor, $\omega$ denotes the phase of the \gls{LoS} component, $\kappa_{l}$ is the complex channel coefficient of the $l$-th scattered component, and $L_p$ is the total number of \gls{NLoS} paths. Also, $\boldsymbol{a}_{r}\left(\theta_{l,r},\phi_{l,r}\right)$ and $\boldsymbol{a}_{t}\left(\theta_{l,t},\phi_{l,t}\right)$ are the receive and transmit vector response functions, respectively. For instance, the receive steering vector can be expressed as
\begin{multline}\label{eq:steer}
\boldsymbol{a}_{r}\left(\theta_{r},\phi_{r}\right)=\\
\left[1~e^{-j\frac{2\pi}{\lambda}\boldsymbol{\psi}\left(\theta_{r},\phi_{r}\right)^{T}\boldsymbol{n}_{1}^{{\rm rx}}\lambda}~\cdots~e^{-j\frac{2\pi}{\lambda}\boldsymbol{\psi}\left(\theta_{r},\phi_{r}\right)^{T}\boldsymbol{n}_{N_{\rm rx}}^{{\rm rx}}\lambda}\right]^{T},
\end{multline}
where $\theta_{r}$ and $\phi_{r}$ denote the azimuth and elevation \gls{AoA}, respectively. Further, the $n$-th (vector) port of the receiver in the \gls{3D} coordinate system can be defined as}
\begin{equation}\label{steer_ports}
\boldsymbol{n}_{n}^{{\rm rx}}=\left[\frac{n_{3}^{{\rm rx}}-1}{N_{3}^{{\rm rx}}-1}W_{3}^{{\rm rx}},\frac{n_{2}^{{\rm rx}}-1}{N_{2}^{{\rm rx}}-1}W_{2}^{{\rm rx}},\frac{n_{1}^{{\rm rx}}-1}{N_{1}^{{\rm rx}}-1}W_{1}^{{\rm rx}}\right]
\end{equation}
and the wave vector can be defined as
\begin{equation}
\boldsymbol{\boldsymbol{\psi}}\left(\theta,\phi\right)=\left[\begin{array}{ccc}
\cos\phi\cos\theta & \cos\phi\sin\theta & \sin\phi\end{array}\right]^{T}. 
\end{equation}
The transmit steering vector can be written in a similar manner. It is noteworthy that (\ref{eq:H_3D}) reduces to an \gls{NLoS} environment if $K=0$. Furthermore, (\ref{eq:H_3D}) can be generalized to a rich scattering environment when $L_p\rightarrow\infty$. If there is only one port at the receiver or transmitter, we have $\boldsymbol{a}_{r}\left(\theta_{l,r},\phi_{l,r}\right)=1$ or $\boldsymbol{a}_{t}\left(\theta_{l,t},\phi_{l,t}\right)=1$, respectively. We can also reduce the fluid antenna structure to \gls{2D} or \gls{1D} by, respectively, setting 
\begin{equation}
\boldsymbol{n}_{n}^{{\rm rx}}=\left[0,\frac{n_{2}^{{\rm rx}}-1}{N_{2}^{{\rm rx}}-1}W_{2}^{{\rm rx}},\frac{n_{1}^{{\rm rx}}-1}{N_{1}^{{\rm rx}}-1}W_{1}^{{\rm rx}}\right]
\end{equation}
or 
\begin{equation}
\boldsymbol{n}_{n}^{{\rm rx}}=\left[0,0,\frac{n_{1}^{{\rm rx}}-1}{N_{1}^{{\rm rx}}-1}W_{1}^{{\rm rx}}\right]. 
\end{equation}
Although the model in (\ref{eq:H_3D}) is more general and accurate than the previously discussed models, it is much more difficult to analyze the performance of \gls{FAS} using this model because statistical tools are limited. Hence, this model is usually employed in simulations for evaluating the performance of \gls{FAS}. As $N_i^s\to\infty$ for $i\in\{1, 2, 3\}$, we have the asymptotic version of (\ref{eq:H_3D}), in which the positions of the radiating element are continuous. In this case, we may consider only the positions of the active radiating elements and simplify (\ref{eq:H_3D}) to the field response based channel model as proposed in \cite{zhu2024performance}.

\subsection{Copula-Based Channel Model}
The above channel models are mainly based on the system geometry, which sometimes may only establish linear correlations amongst the channel gains for any two or more ports. A linear correlation means that an increase or decrease in one random variable will cause the increase or decrease of another random variable by a proportional amount. While this seems logical in the channels over the FAS ports, this may not be the case under non-ideal conditions. For example, mutual coupling effects are multiplicative and non-linear. There could also be undesirable non-linear effects imposing on the channel ports specific to how the ports are actually connected to the antenna feed, perhaps via a shared medium (e.g., a surface integrated waveguide). On the other hand, if the FAS is electrically large, as would be the case in the application of RIS, then different ports could see channels coming from different paths, resulting in a totally different correlation structure. Furthermore, even under ideal situations, the interest would be on the eventual performance metric like \gls{SINR}, achievable rate, and so on, but not the channel. These performance metrics at different channel ports would correlate with each other in a more complicated, non-linear way.

In other words, geometric channel models might be inadequate to capture the dependence structure inherent in \gls{FAS} in real settings. Technically speaking, while the linear correlation coefficient is effective for elliptical multivariate distributions, it falls short to capture the correlation of non-elliptical multivariate distributions. For example, in the $N$-variate Nakagami-$m$ distribution (a widely used channel distribution), the linear correlation approximation falters in the tails, which is critical as bit errors or outages predominantly occur during deep fade. This deficiency might result in the loss of the diversity gain offered by \gls{FAS}. Furthermore, generating the joint distribution of $N$ correlated random variables can pose huge challenges in \gls{FAS}, especially when the system is highly complex.

To tackle these challenges, copula theory emerges as a flexible approach for modeling the fading channel and analyzing the performance of various wireless communication systems, e.g., \cite{gholizadeh2015capacity,ghadi2020copula,jorswieck2020copula,ghadi2022capacity,9762969,10415000}. Recent efforts have also seen copula theory applied to investigate the performance of FAS \cite{10253941,10364840,10319727,ghadi2023gaussian,FR1,FR2,FR3}. In general, the copula-based approach offers three advantages: 
\begin{itemize}
\item[(i)] It can characterize both linear and non-linear correlation, as well as positive and negative correlation among two or more arbitrary random variables; 
\item[(ii)] It has the capability to generate the multivariate distribution of two or more correlated random variables from different families; and
\item[(iii)] It significantly reduces the mathematical analysis complexity owing to its simple structures.
\end{itemize}
More specifically, an $N$-dimension copula $\mathcal{C}:\left[0,1\right]^{N}\rightarrow\left[0,1\right]$ is a joint \gls{CDF} of $N$ random vectors on the unit cube $\left[0,1\right]^{N}$ with uniform marginal distributions \cite{nelsen2006introduction}, i.e., 
\begin{align}
\mathcal{C}\left(b_{1},\dots,b_{N};\vartheta_{C}\right)=\Pr\left(B_{1}\leq b_{1},\dots,B_{N}\leq b_{N}\right),
\end{align}
in which $b_{n}=F_{\left|h_{n}\right|}\left(r_{n}\right)$, $F_{\left|h_{n}\right|}(\cdot)$ is the marginal \gls{CDF} of $|h_{n}|$, and $\vartheta_{C}$ is the copula parameter that measures the degree of dependency between the correlated random variables. The significance of the copula-based method stems from the Sklar's theorem which asserts that for any arbitrary $N$-dimension \gls{CDF} $F_{\left|h_{1}\right|,\dots,\left|h_{N}\right|}\left(r_{1},\dots,r_{N}\right)$ with univariate marginal distributions $F_{\left|h_{n}\right|}\left(r_{n}\right)$, there exists a copula function $\mathcal{C}$ such that for all $\left|h_{n}\right|$ in the extended real line domain $\mathbb{R}$, we have
\begin{align}
\hspace{-0.2cm}F_{\left|h_{1}\right|,\dots,\left|h_{N}\right|}\left(r_{1},\dots,r_{N}\right)=\mathcal{C}\left(F_{\left|h_{1}\right|}\left(r_{1}\right),\dots,F_{\left|h_{N}\right|}\left(r_{N}\right)\right).\label{eq-cdf-copula}
\end{align}
By applying the chain rule, the corresponding joint \gls{PDF} can be obtained as 
\begin{align}
 & f_{\left|h_{1}\right|,\dots,\left|h_{N}\right|}\left(r_{1},\dots,r_{N}\right)\nonumber \\
 & =\frac{\partial^{N}C\left(F_{\left|h_{1}\right|}\left(r_{1}\right),\dots,F_{\left|h_{N}\right|}\left(r_{N}\right);\vartheta_{C}\right)}{\partial F_{\left|h_{1}\right|}\left(r_{1}\right)\dots\partial F_{\left|h_{N}\right|}\left(r_{N}\right)}\prod_{n=1}^{N}f_{\left|h_{n}\right|}\left(r_{n}\right).\label{eq-pdf-copula}
\end{align}
Interestingly, (\ref{eq-cdf-copula}) and (\ref{eq-pdf-copula}) are valid for any choice of arbitrary fading distribution. This highlights the benefits of using the copula-based channel model. Nonetheless, it is worth pointing out that determining an appropriate copula function $\mathcal{C}$ can be challenging in certain cases \cite{ghadi2023gaussian}.


\subsection{Basic Circuit and Antenna Models}
In general, more than one ports, say $n_s$, can be activated at the $s$ side for processing, if there are $n_s$ RF chains. In this case, we can express the equivalent channel as
\begin{equation}\label{eq: H_bar}
\bar{\boldsymbol{H}}=\boldsymbol{\mathcal{A}}_{{\rm rx}}\boldsymbol{H}\boldsymbol{\mathcal{A}}_{{\rm tx}},
\end{equation}
where $\boldsymbol{\mathcal{A}}_{{\rm tx}}=\left[\boldsymbol{\alpha}_{1}^{\rm tx},\ldots,\boldsymbol{\alpha}_{n_{\rm tx}}^{\rm tx}\right]$ and $\boldsymbol{\mathcal{A}}_{{\rm rx}}=\left[\boldsymbol{\alpha}_{1}^{\rm rx},\ldots,\boldsymbol{\alpha}_{n_{\rm rx}}^{\rm rx}\right]^{T}$ are the activation port matrices at the transmitter and receiver, respectively, such that $\boldsymbol{\alpha}_{l}^{\rm tx}$ and $\boldsymbol{\alpha}_{l}^{\rm rx}$ are standard basis vector, i.e., $\boldsymbol{\alpha}_{m}^{s}\in\left\{ \boldsymbol{e}_{1},\ldots,\boldsymbol{e}_{N_{s}}\right\}$ and $\boldsymbol{e}_{m}$ is an all-zero vector except the $m$-th entry being unity.

As multiple ports are activated, it is necessary to account for any mutual coupling between them, a phenomenon influenced by circuit and antenna theories. Circuit and antenna designs play pivotal roles in determining the mutual coupling effect. To consider this effect accurately, the mutual coupling matrix should be added to $\bm{h}$ or $\boldsymbol{H}$, which usually involves computing the scattering parameter matrix or mutual impedance matrix \cite{1310320} tailored to the specific hardware design and circuit configuration. For example, if liquid-based or mechanical movable antennas are used, the channel $\bar{\boldsymbol{H}}$ with mutual coupling effect can be modeled as \cite{8058474}
\begin{equation}\label{eq:C1}
\boldsymbol{\bar{H}}_{{\rm mc}}=\boldsymbol{Z}_{{\rm mc}}^{\rm rx}\boldsymbol{\bar{H}}\boldsymbol{Z}_{{\rm mc}}^{\rm tx},
\end{equation}
where $\boldsymbol{Z}_{{\rm mc}}^{\rm rx}$ and $\boldsymbol{Z}_{{\rm mc}}^{\rm tx}$ are the mutual coupling matrices which can be pre-computed offline if the number of ports at side $s$ is finite. The mutual coupling matrix can computed as
\begin{equation}\label{eq:C2}
\boldsymbol{Z}_{{\rm mc}}^{s}=\left(Z_{A}^{s}+Z_{L}^{s}\right)\left(\boldsymbol{Z}_{s}+Z_{L}^{s}\boldsymbol{I}\right)^{-1},
\end{equation}
where $Z_{A}^{s}$, $Z_{L}^{s}$ and $\boldsymbol{Z}_{s}$ denote the antenna impedance, load impedance and mutual impedance matrix of the active ports at side $s$, respectively. To compute them, it is necessary to consider the structures and positions of the radiating elements. This can be done by using the antenna toolbox in MATLAB\textsuperscript{\textregistered} considering different lengths/sizes, types, and array structures of the radiating elements. Alternatively, it is also possible to mathematically model these structures \cite{balanis2016antenna}. Note that with efficient hardware designs, the mutual coupling effect can sometimes be made trivial even when few active ports are closed, i.e., $\boldsymbol{Z}_{{\rm mc}}^{s}\approx\boldsymbol{I}$. However, if pixel-based antennas or any similar designs are used, the mutual coupling is usually non-trivial when the number of ports is large. Hence, in practice, efforts are required to improve the scattering parameter matrix via circuit setting such as multiport conjugate matching, or antenna design such as isolation techniques. 

Furthermore, the scattering parameter matrix and mutual impedance matrix are related by \cite{8081282}
\begin{equation}\label{eq:C3}
\boldsymbol{Z}_{{\rm mc}}^{s}=Z_{0}\left(\boldsymbol{I}-\boldsymbol{S}_{s}\right)^{-1}\left(\boldsymbol{I}+\boldsymbol{S}_{s}\right),
\end{equation}
where $Z_{0}$ is the reference impedance and $\boldsymbol{S}_{s}$ is the scattering parameter matrix at side $s$. The scattering parameter matrix can be affected by factors such as the operating frequency, material of the radiating elements, impedance mismatch, transmission line effects, circuit elements and components, isolation techniques, and many more. Apart from these, optimizing (\ref{eq:C1}) can be extremely difficult due to the presence of $\boldsymbol{Z}_{{\rm mc}}^{\rm rx}$ and $\boldsymbol{Z}_{{\rm mc}}^{\rm tx}$, not to mention that they are also influenced by the circuit setting and antenna design. Nevertheless, it is important to highlight that mutual coupling is not always harmful. In some cases, it can be used to improve the performance of \gls{FAS} \cite{10303274}.

Compared to \gls{TAS} with $M_{s}=M_{1}^{s} \times M_{2}^{s}$ multiple fixed-position antennas that are separated by a distance $d_{i}^{s}$ on the $s$ side, where $W_{i}^{s} \geq d_{i}^{s} \geq 0.5$, the spatial correlation between the $\left(m_{1},m_{2}\right)$-th and $\left(\tilde{m}_{1},\tilde{m}_{2}\right)$-th antennas can be modeled as
\begin{multline}
J_{k_{\left(m_{1},m_{2}\right)},k_{\left(\tilde{m}_{1},\tilde{m}_{2}\right)}}^{\rm s}=\\
j_{0}\left(2\pi\sqrt{\left(\left|m_{1}-\tilde{m}_{1}\right| d_{1}^{s}\right)^{2}+
\left(\left|m_{2}-\tilde{m}_{2}\right|d_{2}^{s}\right)^{2}}\right),
\end{multline} where $m_{i}\in\left\{ 1,\dots,\lfloor\frac{W_{i}^{s}}{d_{i}^{s}}\rfloor+1\right\} ,\forall i$. The complex channel of TAS in a MIMO setup can then be generated using similar steps as in \eqref{J}-\eqref{H}. Alternatively, in a finite scattering model, we can consider $d_{i}^{s}, \forall i,$ in \eqref{steer_ports}. Unlike \gls{FAS}, notice that $M_{i}^{s}$ in \gls{TAS} is restricted by $\lfloor \frac{W_{i}^{s}}{d_{i}^{s}}\rfloor+1$, yielding limited spatial resolution. Furthermore, for a fair comparison, only $n_{s}$ antennas should be selected for transmission. This is also referred to as \gls{MIMO} with antenna selection.

\subsection{Other Existing Models}
\begin{figure}[t]
\noindent \centering{}\includegraphics[scale=0.55]{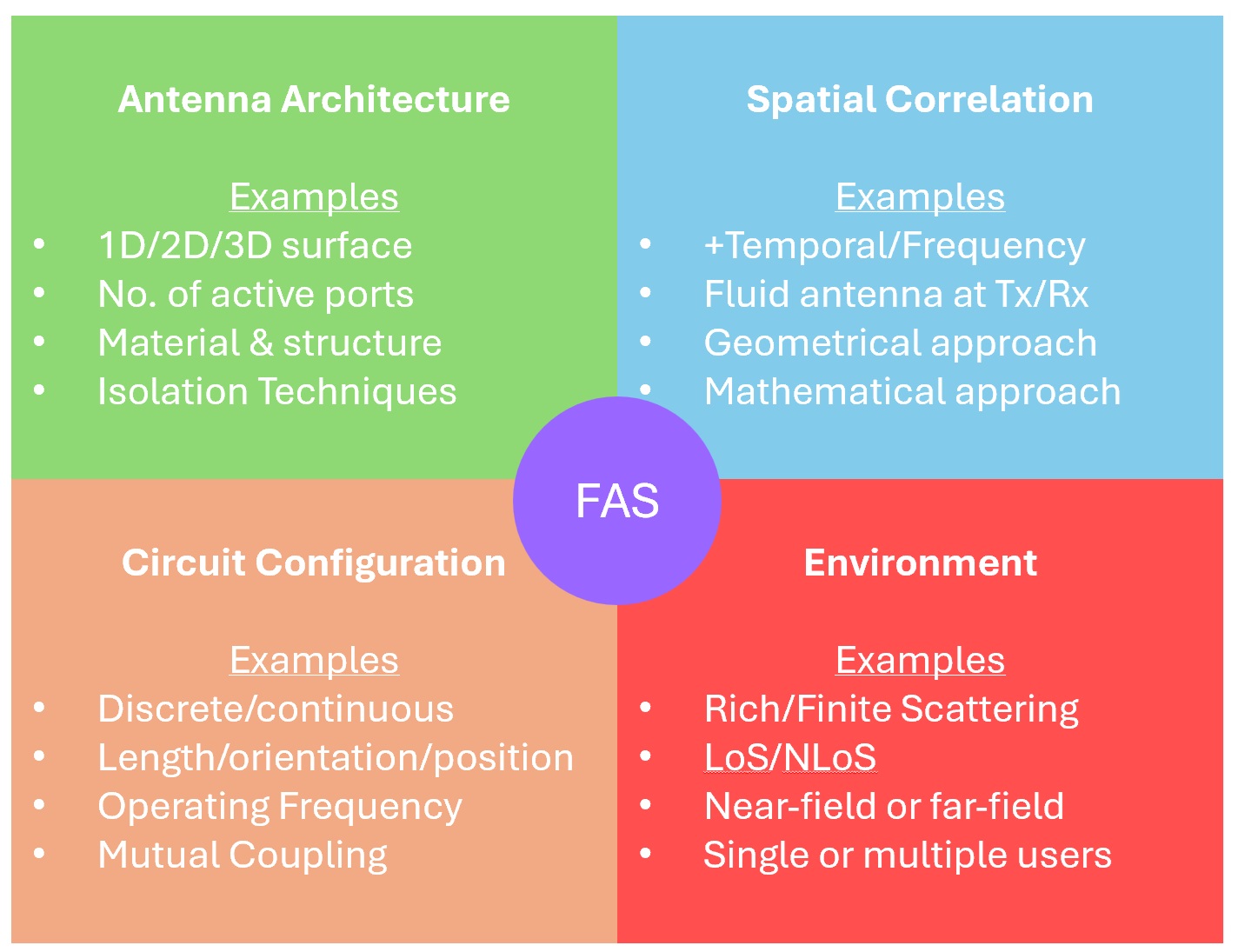}
\caption{Some key considerations in \gls{FAS} models.}\label{Considerations}
\end{figure}

It is essential to acknowledge that various system models exist in the field of \gls{FAS}. As a matter of fact, the repositioning of the radiating elements can be optimized as a continuous function rather than a discrete one, as investigated in \cite{10309171,10243545}. The fluid antenna structure may appear in the form of a uniform linear, planar, or circular array \cite{10188603}. Reconfiguring the lengths or heights of the radiating elements is also possible to best serve different operating frequencies \cite{8105795,8762117}, and adjusting the \gls{3D} orientations of the radiating elements can improve the performance as shown in \cite{Shao-2024-1,Shao-2024-2}. Moreover, correlation can actually take place in the time and frequency domains, in addition to the spatial domain \cite{zhu2024performance,10207934}. These diverse channel models, along with the previously discussed ones, can be extended to scenarios involving multiple \gls{FAS} users, as evidenced in \cite{FAS_NOMA,xu2023capacity,ISAC_FAS,HKFAMA}. Stochastic geometry can also be seamlessly integrated into \gls{FAS} \cite{9992289,9838484,10184308}, where the locations of multiple transmitters are distributed randomly. Rather than going through all existing channel models in literature, the materials above aim to equip readers with the essential knowledge regarding the factors to consider and extensions that can be made. This foundation enables researchers to adopt existing channel models or develop new ones tailored to their applications. On the other hand, circuit and antenna theories can help develop a more physics- and electromagnetic-compliant modeling of \gls{FAS} and enhance its performance. In a nutshell, some key considerations include antenna architecture, circuit configuration, spatial correlation, and environment, as highlighted in Fig.~\ref{Considerations}. 

\section{Channel Estimation}\label{sec:ce}
In this section, we discuss the process of channel estimation for \gls{FAS}. Unlike \gls{TAS} where each port or preset location of the antennas requires estimation, \gls{FAS} simplifies the process by necessitating the channel knowledge from only a few locations for complete recovery of the full \gls{CSI}. This efficiency is achieved through leveraging the strong spatial correlation or channel sparsity inherent in \gls{FAS}. Moreover, we will elucidate how \gls{ML} and mathematical approaches can be effectively used for channel state acquisition. Specifically, we provide concrete examples of how \gls{ML} can be employed for channel estimation in rich scattering environments and how mathematical approaches can be applied for channel estimation in a finite scattering environment. Nevertheless, it is worth noting that both techniques are not limited by specific environments.

\begin{figure*}[htbp]
\centering \includegraphics[scale=0.9]{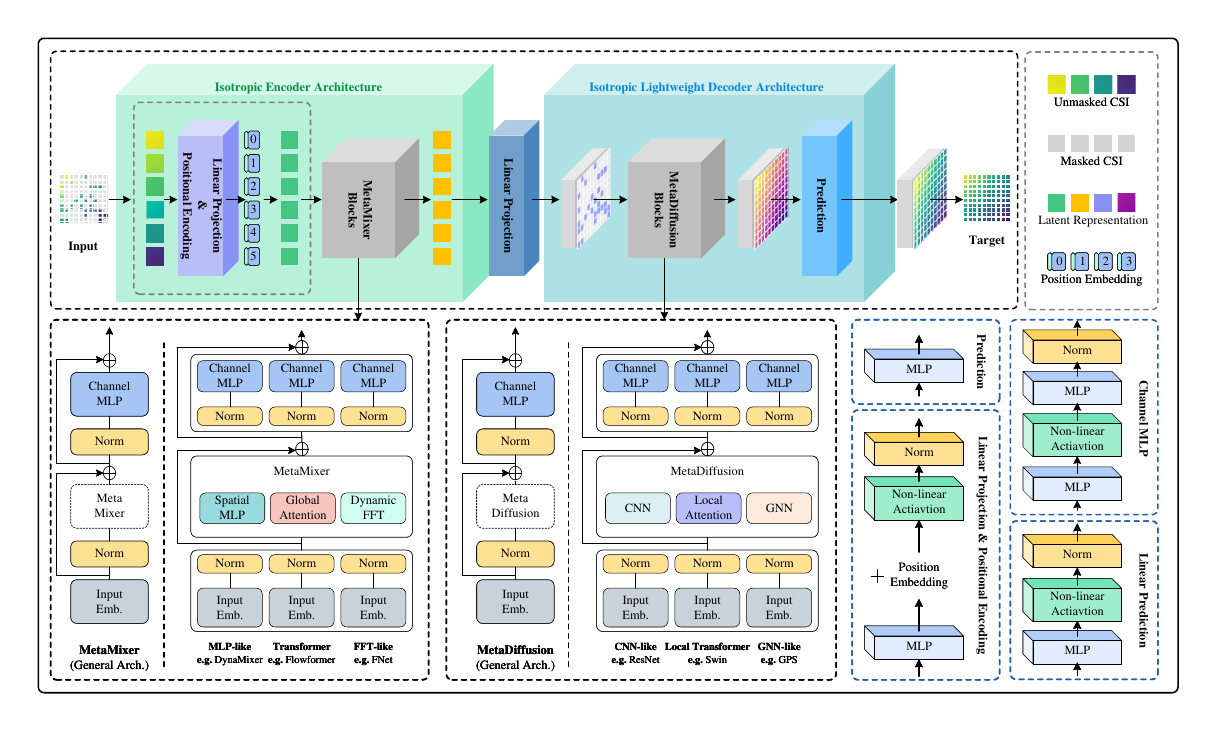} \caption{A UAMA architecture for CSI extraplation.}\label{UAMA_model}
\end{figure*}

\subsection{Rich Scattering Environment}
To obtain an efficient performance in \gls{FAS}, it is important to obtain reliable \gls{CSI} of all the ports. Nevertheless, estimating each of the ports can lead to substantial hardware switching and system overhead. To tackle this, we can apply state-of-the-art techniques such as deep learning to extrapolate the \gls{CSI} of the unknown ports from that of a few known ports (also referred to as the observable ports). In particular, it is possible to exploit the fact that a channel-to-channel mapping exists if the position-to-channel mapping is bijective \cite{9048929}, which is the case in a static \gls{FAS} communication environment (including geometry, materials, positions, etc). Therefore, there exists a channel mapping $\mathbf{\Phi}_{\mathbf{h}}:\left\{ \boldsymbol{h}_{\mathcal{O}}\right\} \rightarrow\left\{ \boldsymbol{h}_{\mathcal{U}}\right\} $, where $\boldsymbol{h}_{\mathcal{O}}$ and $\boldsymbol{h}_{\mathcal{U}}$ are the \gls{CSI} of observable and unknown ports, respectively. 

The function of channel extrapolation can be mathematically represented as
\begin{equation}\label{extrapolation}
\boldsymbol{h}_{\mathcal{U}}=f_{{\rm net}}\left(\boldsymbol{h}_{\mathcal{O}},\mathbf{\Theta}\right),
\end{equation}
where $f_{{\rm net}}\left(\cdot\right)$ serves as the neural network operation and $\mathbf{\Theta}=\left\{ \theta_{\mathrm{P}},\theta_{\mathrm{E}},\theta_{\mathrm{M}},\theta_{\mathrm{D}},\theta_{\mathrm{R}}\right\}$ is the set of learnable parameters. Given the aforementioned channel extrapolation problem, a \gls{UAMA} architecture can be employed to address this issue, as illustrated in Fig.~\ref{UAMA_model}. The overall operation of \gls{UAMA} can be defined as
\begin{align}
f_{{\rm net}}^{\left(\mathbf{\Theta}\right)}=&\operatorname{Pre-mapper}^{\left(\theta_{\mathrm{P}}\right)}\notag\\
\circ&\operatorname{Encoder}^{\left(\theta_{\mathrm{E}}\right)}\notag\\
\circ&\operatorname{Mid-mapper}^{\left(\theta_{\mathrm{M}}\right)}\notag\\
\circ&\operatorname{Decoder}^{\left(\theta_{\mathrm{D}}\right)}\notag\\
\circ&\operatorname{Post-mapper}^{\left(\theta_{\mathrm{R}}\right)},\label{UAMA_equation}
\end{align}
which lists the compositions of $5$ trainable modules, denoted by the `$\circ$' notation. Note that the learnable parameters of the neural networks are placed in superscripts for clarity in order to differentiate from standard functions.

Multiple \gls{UAMA} modules are combined to create a \gls{UAMA} model as illustrated in (\ref{extrapolation}). In the following, we summarize the role of each module.
\begin{itemize}
\item \textbf{Input:} The input to \gls{UAMA} consists of the real and imaginary parts of \gls{CSI} from the observable ports, $\boldsymbol{h}_{\mathcal{O}}\triangleq\left[\Re\left(\boldsymbol{h}_{\mathcal{O}}\right),\Im\left(\boldsymbol{h}_{\mathcal{O}}\right)\right]$. But it is also possible to extrapolate the channel using complex-valued inputs \cite{9209095}.

\item \textbf{Pre-mapper} (with parameters $\theta_{\mathrm{P}}$): This module is employed for non-linear projection and position encoding. First, it projects $\boldsymbol{h}_{\mathcal{O}}$ into a high-dimensional space, yielding a latent representation $\boldsymbol{v}$, thus enabling the exploration of the channel-to-channel mapping relationships within a larger solution space. Afterwards a position encoding is performed with respect to the port locations to obtain positional embeddings, which facilitates the exploration of spatial coupling relationships amongst different ports. Note that the non-linear mapping is achieved through the \gls{MLP} and non-linear activation functions (e.g., Gaussian error linear unit---GELU \cite{hendrycks2016gaussian}, rectified linear unit---RELU \cite{pmlr-v15-glorot11a}, etc.). Conventional positional encoding techniques are primarily categorized into three classes: (i) absolute positional encoding, (ii) relative positional encoding, and (iii) learnable positional encoding. The selection of an appropriate technique depends on the specific channel characteristics.

\item \textbf{Encoder} (with parameters $\theta_{\mathrm{E}}$): This module is utilized to construct basis vectors based on $\boldsymbol{v}$, which are yielded from the observable ports. Various adaptive mechanisms are utilized to construct basis vectors in response to the \gls{CSI} of observable ports, which are subsequently used for the linear representation of the \gls{CSI} of unknown ports. For instance, attention mechanisms can be employed, wherein the basis vectors are constructed based on the similarity of \gls{CSI} across different observable ports. Additionally, there exist spatial and frequency domain interaction mechanisms, such as the spatial \gls{MLP} (e.g., DynaMixer \cite{pmlr-v162-wang22i}) and the dynamic \gls{FFT} (e.g., FNet \cite{DBLP:journals/corr/abs-2105-03824}), as shown in Fig.~\ref{UAMA_model}. We collectively refer to these various mechanism modules as MetaMixer. Note that all these mechanisms possess a global receptive field.

\item \textbf{Mid-mapper} (with parameters $\theta_{\mathrm{M}}$): This module is employed to reduce the dimensionality of the basis vectors. This is very important to help alleviate the computational complexity during the decoding process.

\item \textbf{Decoder} (with parameters $\theta_{\mathrm{D}}$): This module is utilized to recover the \gls{CSI} of unknown ports. On a planar array with a certain resolution, the channels often exhibit local correlations and smoothness. Therefore, we can linearly represent the \gls{CSI} of the remaining unknown ports in a local diffusion using the basis vectors generated by the Encoder, which essentially involves learning the coefficients of these basis vectors. As depicted in Fig.~\ref{UAMA_model}, the modules with inductive bias for capturing local correlations and smoothness include \gls{CNN} (e.g., ResNet \cite{he2016deep}), Local Attention (e.g., Swin \cite{Liu_2021_ICCV}), \gls{GNN} (e.g., GPS \cite{NEURIPS2022_5d4834a1}), and etc. We collectively refer to these local diffusion architectures as MetaDiffusion. Note that the introduction of the aforementioned inductive bias significantly reduces the complexity of channel extrapolation.

\item \textbf{Post-mapper} (with parameters $\theta_{\mathrm{R}}$): This module reduces the dimensionality of the high-dimensional \gls{CSI} to the dimensions of the original space. Specifically, it maps the latent representation $\boldsymbol{v}$ to the predicted output $\hat{\boldsymbol{h}}_{\mathcal{U}}$. 
\end{itemize}

Besides the aforementioned key modules, each module incorporates normalization techniques such as LayerNorm \cite{ba2016layer}, BatchNorm \cite{pmlr-v37-ioffe15}, InstanceNorm \cite{ulyanov2017instance}, etc. Their purposes are to expedite the convergence speed and enhance the gradient stability of backpropagation. Finally, \gls{MSE} is typically used as the objective function.

Note that recent study of channel estimation/extrapolation using deep learning has primarily focused on mechanisms such as \gls{MLP}-like \cite{9430899,8400482,8693948,8949757,8052521}, attention \cite{9526282}, \gls{CNN}-like \cite{8353153}, \gls{LSTM}-like \cite{9715064}, and etc. These methods all fall within the \gls{UAMA} framework. For instance, they can be divided into two steps: (i) constructing basis vectors from the observable \gls{CSI} and (ii) extrapolating the remaining unknown \gls{CSI} based on the former.

\begin{figure}[tp]
\noindent \centering \includegraphics[width=0.53\textwidth]{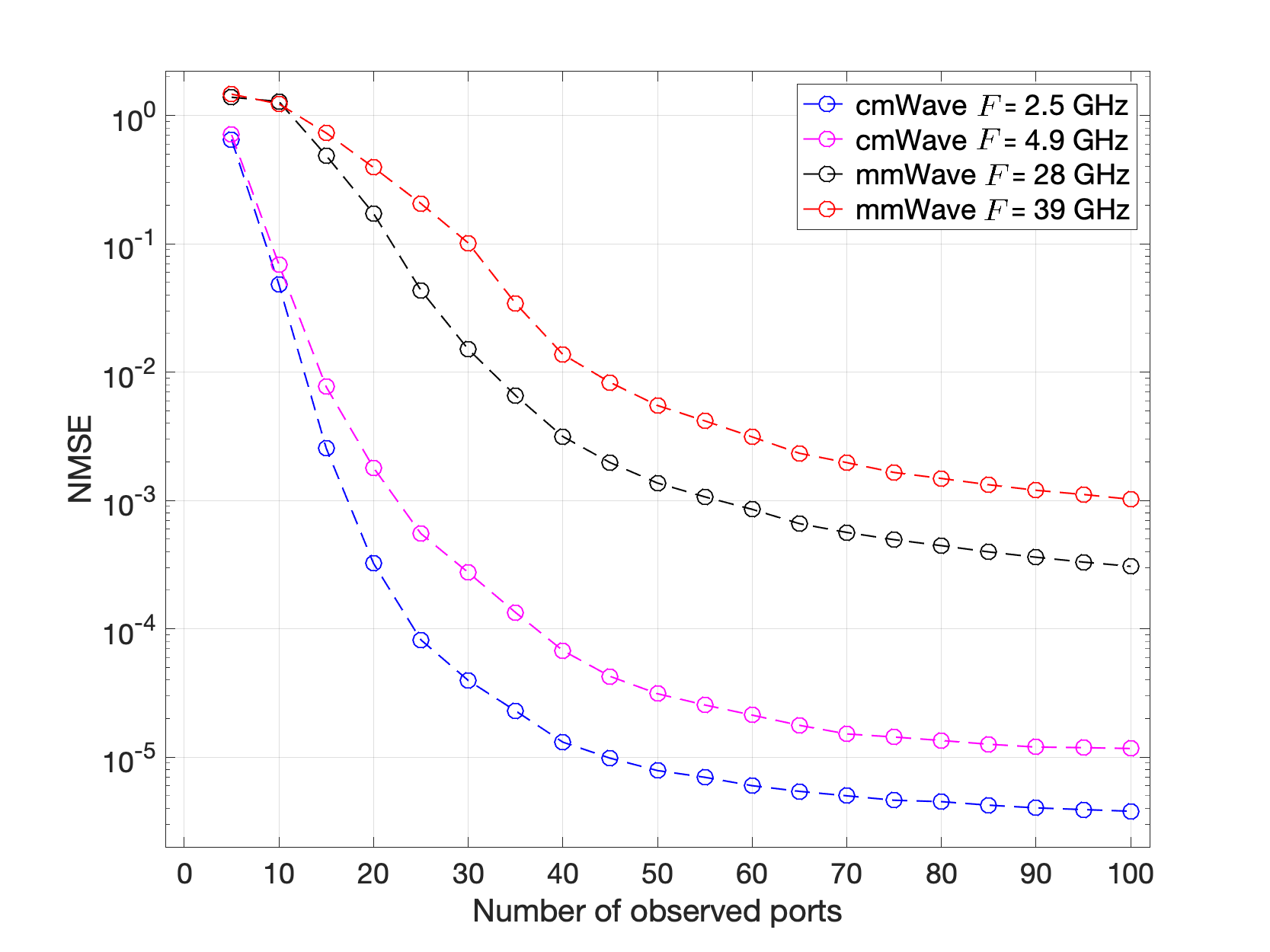}
\caption{The NMSE for \gls{CSI} extrapolation versus the number of observed ports at different frequencies $F$ with number of ports $(N_{1},N_{2})=(20,40)$, number of users $U=10$, and the planar \gls{FAS} size $(W_{1},W_{2})=(2{\rm cm},4{\rm cm})$.}\label{NMSE}
\end{figure}

In Fig.~\ref{NMSE}, we employ \gls{UAMA} and examine the relationship between the number of observable ports and the \gls{NMSE} at different frequencies $F$ while the physical size of the FAS is fixed.\footnote{This means that the electrical size of the FAS increases with the carrier frequency. This also implies that weaker correlation between the observable ports is anticipated at higher frequency.} The \gls{NMSE} metric is defined as ${\rm NMSE}_{\boldsymbol{h}_{\mathcal{U}},\hat{\boldsymbol{h}}_{\mathcal{U}}}=\frac{\sum_{t=1}^{S_{\text{test }}}\|\boldsymbol{h}_{\mathcal{U}}^{(t)}-\hat{\boldsymbol{h}}_{\mathcal{U}}^{(t)}\|^{2}}{\sum_{t=1}^{S_{\text{test }}}\|\boldsymbol{h}_{\mathcal{U}}^{(t)}\|^{2}}$, where $S_{\text{test}}$ is the number of test samples and the superscript $(t)$ specifies the $t$-th sample. In this simulation, we employ a transformer \cite{vaswani2017attention} architecture as the encoder, while the decoder utilizes a \gls{CNN}-like architecture similar to ResNet \cite{he2016deep}. The computational complexity of the attention mechanism is $\mathcal{O}(N_{O}^{2}l_{1}+N_{O}l_{1}^{2})$, in which $N_{O}$ represents the number of observable ports and $l_{1}$ represents the hidden layer dimensions. The computational complexity of the \gls{CNN}-like mechanism is $\mathcal{O}(Nc_{o}^{2}l_{2}^{2}+Nl_{2}^{2})$, where $N=N_{1}\times N_{2}$ is the total number of ports, $c_{o}$ is the convolution kernel size, and $l_{{\rm 2}}$ represents the hidden layer dimensions. Hence, the overall complexity is $\mathcal{O}(N_{O}^{2}l_{{\rm 1}}+N_{O}l_{{\rm 1}}^{2}+Nc_{o}^{2}l_{2}^{2}+Nl_{2}^{2})$ where typically $N_{O},c_{o}$ $\ll N$.

As can be observed from the results in Fig.~\ref{NMSE}, the \gls{NMSE} for CSI extrapolation at a FAS receiver decreases with the number of observable ports. This is because when the number of observable ports increases, the constructed basis vectors in the channel space become more precise, and thus resulting in smaller errors in linearly approximating the \gls{CSI} of unknown ports. Furthermore, only a small percentage of observable ports is required to accurately estimate the \gls{CSI} of all the ports. For instance, we can achieve an \gls{NMSE} of $10^{-3}$ or lower for the frequency range from $2.5~{\rm GHz}$ to $39~{\rm GHz}$ by using only $100$ observable ports out of $800$ ports. In the following subsection, we will investigate different channel estimation methods for finite scattering environments via a mathematical approach. 

\subsection{Finite Scattering Environment}

\begin{table*}[ht]
\caption{A brief summary of papers on channel estimation in \gls{FAS}.}
\label{summ_CE}
\centering{}%
\begin{tabular}{|r|>{\raggedright}p{14cm}|}
\hline 
\centering Ref. & \centering Summary\tabularnewline
\hline 
\hline 
\centering \cite{9992289} & a) A multi-cell homogeneous network was considered based on stochastic geometry. There were a \gls{BS} and a user equipment (UE) in each cell.\\
b) The BS used a single omnidirectional fixed-position antenna, while each UE used multiple linear FASs that form a ring.\\
c) A skipped-enabled linear minimum mean-squared error (LMMSE)-based channel estimation (SeCE) technique was proposed to estimate the channel at only some selected ports.
d) This paper proposed to estimate the CSI at only a few observable ports using the LMMSE-based method, and then simply take the CSI of an observable port as that of other ports in its neighborhood. Although this scheme can be applied in both the rich and finite scattering environments, it introduces not only estimation error but also approximation error.\tabularnewline
\hline 
\centering \cite{10375559} & a) A multiuser uplink millimeter-wave (mmWave) system was considered.\\
b) Each transmitter was equipped with a linear FAS while the receiver used multiple fixed-position antennas.\\
c) A \gls{L3SCR} method was proposed to estimate and reconstruct the channel, and its performance in terms of the estimation accuracy and complexity was compared with the \gls{OMP} method.\tabularnewline
\hline 
\centering \cite{10233765} & a) A point-to-point mmWave system was considered.\\
b) The transmitter used a fixed-position antenna while the receiver used a linear FAS.\\
c) The channel parameters were estimated based on the least squares regression scheme.\tabularnewline
\hline 
\centering \cite{10236898} & a) A point-to-point system was considered.\\
b) Both the transmitter and the receiver were equipped with a 2D FAS.\\
c) A successive transmitter-receiver compressed sensing (STRCS) method was proposed to estimate the channel.\tabularnewline
\hline 
\centering \cite{xiao2023channel} & a) The same system as in \cite{10236898} was considered. The aim was to solve the estimation error propagation problem in \cite{10236898} and reduce the estimation overhead.\\
b) The channel parameters were jointly estimated by the \gls{OMP} algorithm.\\
c) Different measurement position setups were studied.\tabularnewline
\hline 
\centering \cite{zhang2023successive} & a) A point-to-point system was considered.\\
b) The transmitter used a fixed-position antenna while the receiver used a linear FAS that has multiple antennas.\\
c) The successive Bayesian reconstructor (S-BAR) was proposed to estimate the channel.\tabularnewline
\hline 
\end{tabular}
\end{table*}

As summarized in Table \ref{summ_CE}, there have been efforts tackling the CSI estimation problem in FAS. In \cite{9992289}, it was proposed to estimate the CSI at a few ports using the LMMSE-based method, and then simply take the estimated CSI of an observable port as that of other ports in its neighborhood. Although this scheme can be applied in both rich and finite scattering environments, it introduces not only estimation error but also approximation error. Under finite scattering, a more efficient and common practice is to estimate the sparse parameters of the channel and then reconstruct the CSI at all ports based on these parameters \cite{10375559,10233765,10236898,xiao2023channel,zhang2023successive}.

\begin{figure}
\centering \includegraphics[scale=0.4]{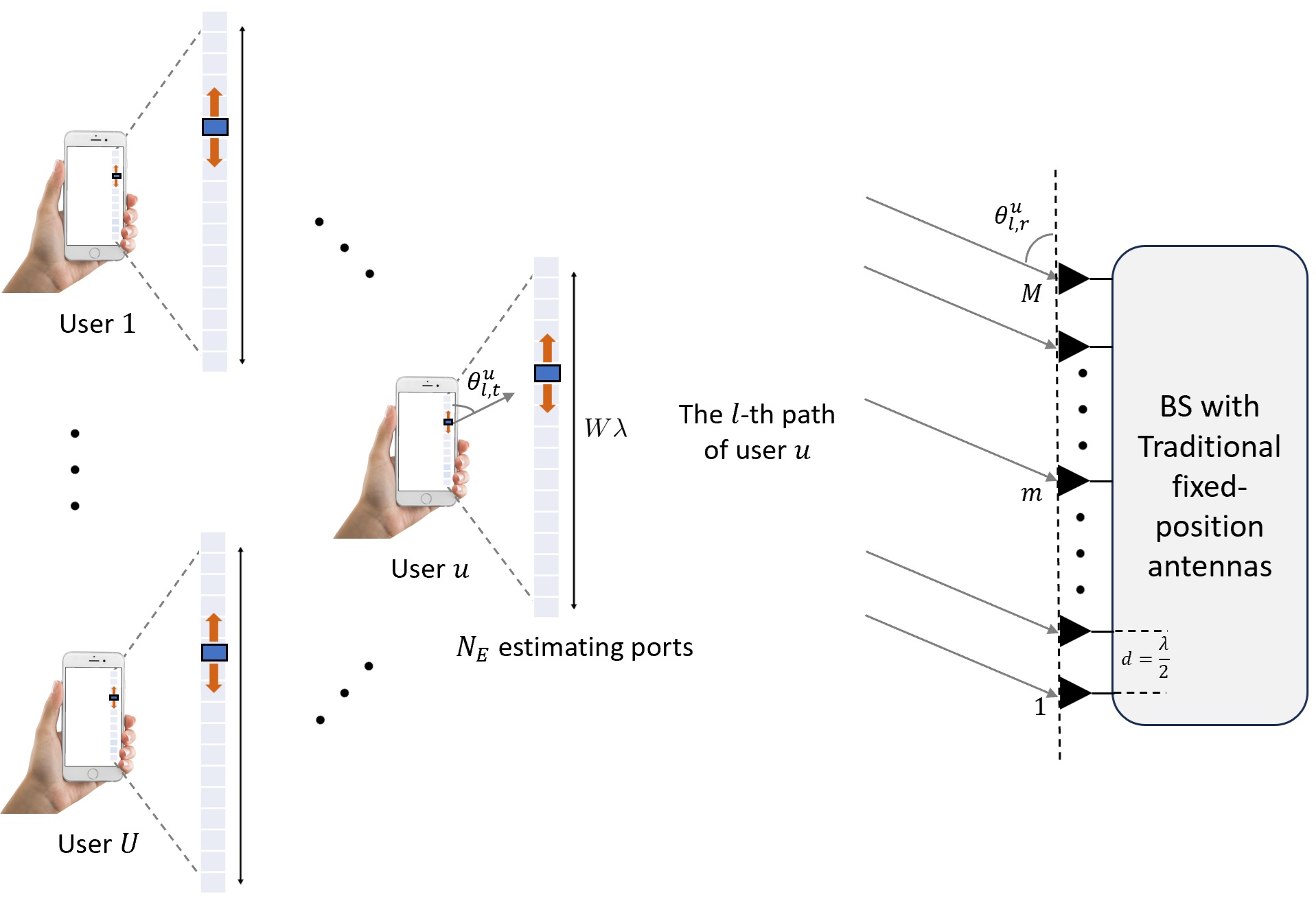}
\caption{Channel estimation for a multiuser uplink system, where each user is equipped with a linear \gls{FAS} while the \gls{BS} has multiple traditional fixed-position antennas.}\label{system_channel_est}
\end{figure}

In this subsection, we use the multiuser uplink system in \cite{10375559} as an example. Consider a simple setup as depicted in Fig.~\ref{system_channel_est}, where the \gls{BS} has $M$ fixed-position antennas that are separated by $\Delta=\frac{\lambda}{2}$ and each user has a \gls{1D} fluid antenna, with $N$ selectable ports uniformly distributed along a linear dimension of length $W\lambda$. In the multiuser uplink, the channel vector from the $n$-th port of user $u$ to the \gls{BS} can be denoted by $\bm{h}_{u,n}\in{\cal C}^{M\times1}$. By stacking $\bm{h}_{u,n}$ for all $n\in\{1,\dots,N\}$, we can obtain the channel matrix between the \gls{BS} and all the ports of user $u$ as
\begin{equation}\label{H_u}
\bm{H}_{u}=[\bm{h}_{u,1},\dots,\bm{h}_{u,N}].
\end{equation}

Channel estimation in \gls{FAS} aims to estimate $\bm{H}_{u}, \forall u$. In the following, we discuss several methods that can do it.

\subsubsection{Least square method}
A direct, straightforward approach to estimate $\bm{H}_{u}$ involves all the users to transmit orthogonal pilot sequences while switching the radiating elements synchronously across all $N$ ports of their FASs. In each port, a pilot sequence is transmitted $T$ times, where $T$ subframes are dedicated to CSI estimation and each subframe contains $T_{s}$ time slots. Upon receiving the pilot sequences at the \gls{BS}, the least square estimate of $\bm{h}_{u,n}$ for user $u$ can be obtained  by post-multiplying the normalized received signal matrix by the corresponding pilot sequence, resulting in the estimate
\begin{align}
\hat{\bm{h}}_{u,n}^{\rm LS}=\bm{h}_{u,n}+\hat{\bm{\sigma}}_{u,n},\label{y_nt_qu}
\end{align}
where $\hat{\bm{\sigma}}_{u,n}$ is the estimation error row vector caused by the noise. By stacking (\ref{y_nt_qu}) for all $n\in\{1,\dots,N\}$, we can obtain the least square estimate of $\bm{H}_{u}$ as
\begin{align}
\hat{\bm{H}}_{u,\text{LS}}=\bm{H}_{u}+\hat{\bm{\Sigma}}_{u},\label{Gu_LS}
\end{align}
where $\hat{\bm{\Sigma}}_{u}=\left[\hat{\bm{\sigma}}_{u,1},\dots,\hat{\bm{\sigma}}_{u,N}\right]^{T}$ denotes the noise matrix. The performance of the least square method is mainly determined by the number of times a pilot sequence is repeated, i.e., $T$, and the transmit power. Nonetheless, due to the large number of ports, i.e., $N$, requiring the radiating element of each user to switch across all ports for channel measurement involves high hardware switching and pilot overhead. Hence, the least square method is not particularly practical in real-world scenarios and is only provided here as an obvious benchmark.

\subsubsection{\gls{L3SCR} method}
Let us consider a finite scattering environment where the channel has sparsity, i.e., the channel can be characterized using some sparse parameters by considering only a few main propagation paths. In this case, \gls{L3SCR} can be applied to estimate the channel \cite{10375559}. In contrast to (\ref{Gu_LS}), this method proves effective in reducing hardware switching and pilot overhead greatly. Suppose that during the estimation period, each fluid antenna switches to only $N_{O}\ll N$ observable ports, which are uniformly distributed with adjacent spacing $\tau$, to acquire $\bm{H}_{o,u}=[\bm{h}_{u,1},\dots,\bm{h}_{u,N_{O}}]\in{\mathbb{C}}^{M\times N_{O}}$. By using the plane-wave geometric model, $\bm{H}_{o,u}$ can be rewritten as\footnote{For ease of expositions, we here consider a \gls{1D} fluid antenna surface with $K=0$ in (\ref{eq:H_3D}). Without loss of generality, the constant factor $\sqrt{\frac{1}{L_p}}$ in (\ref{eq:H_3D}) is normalized for brevity.} 
\begin{equation}\label{H1}
\bm{H}_{o,u}=\sqrt{MN_{O}}\sum_{l=1}^{L_{u}}\kappa_{l}^{u}\bm{a}_{u,r}\left(\theta_{l,r}^{u}\right)\bm{a}_{u,t}\left(\theta_{l,t}^{u}\right)^{H},
\end{equation}
where $L_{u}$ is the number of propagation paths between the \gls{BS} and user $u$, and $\kappa_{l}^{u}$ is the complex channel gain. Furthermore, $\bm{a}_{u,r}(\theta_{l,r}^{u})$ and $\bm{a}_{u,t}(\theta_{l,t}^{u})$ are, respectively, the steering vectors at the \gls{BS} and user $u$, which are given as 
\begin{multline}\label{a_RT}
\bm{a}_{u,r}\left(\theta_{l,r}^{u}\right)=\\
\frac{1}{\sqrt{M}}\left[1,e^{-j\frac{2\pi}{\lambda}\Delta\cos\theta_{l,r}^{u}},\dots,e^{-j\frac{2\pi}{\lambda}\left(M-1\right)\Delta\cos\theta_{l,r}^{u}}\right]^{T},
\end{multline}
and
\begin{multline}
\bm{a}_{u,t}\left(\theta_{l,t}^{u}\right)=\\
\frac{1}{\sqrt{N_{O}}}\left[1,e^{-j\frac{2\pi}{\lambda}\tau\cos\theta_{l,t}^{u}},\dots,e^{-j\frac{2\pi}{\lambda}\left(N_{O}-1\right)\tau\cos\theta_{l,t}^{u}}\right]^{T},
\end{multline}
where $\theta_{l,r}^{u}$ and $\theta_{l,t}^{u}$ are, respectively, the \gls{AoA} and \gls{AoD} of the $l$-th path. Moreover, let us denote
\begin{equation}\label{A_RT}
\left\{\begin{aligned}\bm{K}_{u} & ={\rm diag}\{\kappa_{1}^{u},\dots,\kappa_{L_{u}}^{u}\}\in{\mathbb{C}}^{L_{u}\times L_{u}},\\
\bm{A}_{u,r} & =\left[\bm{a}_{u,r}\left(\theta_{1,r}^{u}\right),\dots,\bm{a}_{u,r}\left(\theta_{L_{u},r}^{u}\right)\right]\in{\mathbb{C}}^{M\times L_{u}},\\
\bm{A}_{u,t} & =\left[\bm{a}_{u,t}\left(\theta_{1,t}^{u}\right),\dots,\bm{a}_{u,t}\left(\theta_{L_{u},t}^{u}\right)\right]\in{\mathbb{C}}^{N_{O}\times L_{u}}.
\end{aligned}\right.
\end{equation}
Then $\bm{H}_{o,u}$ can be expressed in matrix form as 
\begin{equation}
\bm{H}_{o,u}=\sqrt{MN_{O}}\bm{A}_{u,r}\bm{K}_{u}\bm{A}_{u,t}^{H}.\label{H2}
\end{equation}
In \gls{L3SCR}, the main idea is to obtain the least square estimate, $\hat{\bm{H}}_{u,{\rm LS}}$, of (\ref{H2}) utilizing the least square method, and then estimate the sparse parameters of the channel based on $\hat{\bm{H}}_{u,{\rm LS}}$. Below we give details on how to estimate the number of paths, \gls{AoA}, \gls{AoD}, channel gain, and reconstruct the channel.
\begin{itemize}
\item \textbf{Estimation Number of Paths and \gls{AoA}}: Denote $\bm{\varOmega}\in\mathbb{C}^{M\times M}$ as the normalized \gls{DFT} matrix, and multiply its conjugate transpose to $\frac{1}{\sqrt{MN_{O}}}\hat{\bm{H}}_{u,{\rm LS}}$. Then we have
\begin{align}\label{Y_DFT}
\hat{\bm{H}}_{u,{{\rm LS}}}^{\rm DFT}=\bm{\varOmega}^{H}\bm{A}_{u,r}\bm{K}_{u}\bm{A}_{u,t}^{H}+\frac{1}{\sqrt{MN_{O}}}\bm{\varOmega}^{H}\hat{\bm{\Sigma}}_{o,u}.
\end{align}
When the number of fixed-position antennas at the BS, $M$, is large, $\hat{\bm{H}}_{u,{\rm LS}}^{\rm DFT}$ is a row sparse matrix with full column rank. As such, we can calculate the total power of each row of $\hat{\bm{H}}_{u,{\rm LS}}^{\rm DFT}$ and consider the count of distinct power peaks $\hat{L}_{u}$ as an estimate of the number of propagation paths, $L_{u}$. In addition, we can estimate the \gls{AoA} from the indices of these power peaks, denoted as $\{\hat{m}_{1},\dots,\hat{m}_{\hat{L}_{u}}\}$. It is essential to note that the accuracy of the \gls{DFT}-based method is limited by the resolution $\frac{1}{M}$. By solely relying on this method, the estimates for the number of paths and \gls{AoA} may be not quite reliable. To improve the estimation accuracy, the angular rotation operation can be applied to compensate the angular mismatch \cite{8493600}. Let $\bm{\varPsi}\in{\mathbb{C}}^{M\times M}$ denote the diagonal angular rotation matrix with rotation parameter $\psi\in\left[-\frac{1}{2M},\frac{1}{2M}\right]$. Applying both the \gls{DFT} and angular rotation matrices to $\frac{1}{\sqrt{MN_{O}}}\hat{\bm{H}}_{u,{\rm LS}}$, we obtain
\begin{multline}\label{Y_DFT_rota}
\hat{\bm{H}}_{u,{\rm LS}}^{{\rm DFT},{\rm ro}}=\bm{\varOmega}^{H}\bm{\varPsi}^{H}\bm{A}_{u,r}\bm{K}_{u}\bm{A}_{u,t}^{H}\\
+\frac{1}{\sqrt{MN_{O}}}\bm{\varOmega}^{H}\bm{\varPsi}^{H}\hat{\bm{\Sigma}}_{o,u}.
\end{multline}
With the angular rotation matrix, the power beam at index $m$ is rotated to $m-M\psi$, which can vary continuously within $[m-\frac{1}{2},m+\frac{1}{2}]$ as $\psi\in[-\frac{1}{2M},\frac{1}{2M}]$ corresponds to a new angle for each $\psi$. The remaining challenge is to determine the optimal $\psi$ for a given power peak $\hat{m}_{l}\in\{\hat{m}_{1},\dots,\hat{m}_{\hat{L}_{u}}\}$, ensuring that the estimated angle after compensation closely aligns with the actual angle. This can be achieved by discretizing the interval $[-\frac{1}{2M},\frac{1}{2M}]$ and utilizing a \gls{1D} search method. After that, the estimates of \gls{AoA}, i.e., $\hat{\theta}_{l,r}^{u}$ and $\hat{\bm{A}}_{u,r}$, can then be obtained from the power beam indices and $\psi$.
\item \textbf{Estimation of \gls{AoD} and Channel Gains:} After obtaining $\hat{\bm{A}}_{u,r}$, we can multiply it to $\frac{1}{\sqrt{MN_{O}}}\hat{\bm{H}}_{u,{\rm LS}}^{H}$. It is noted that each column of $\frac{1}{\sqrt{MN_{O}}}\hat{\bm{H}}_{u,{\rm LS}}^{H}\hat{\bm{A}}_{u,r}$ contains the \gls{AoD} and channel gain information of a specific path. As a result, this presents a $1$-sparse reconstruction problem, and hence we can apply low-complexity matched filters to estimate the \gls{AoD} and channel gains. Specifically, let $\bm{D}\in{\mathbb{C}}^{N_{O}\times D}$ be the dictionary matrix, where $D$ is the dictionary size, and each column of $\bm{D}$ is the steering vector with angle $\left(\frac{d_{\theta}-1}{D}\right)\pi$. By pre-multiplying $\bm{D}$ and post-multiplying $\hat{\bm{A}}_{u,r}$ to $\frac{1}{\sqrt{MN_{O}}}\hat{\bm{H}}_{u,{\rm LS}}^{H}$, we can obtain
\begin{multline}\label{approx2}
\frac{1}{\sqrt{MN_{O}}}\bm{D}^{H}\hat{\bm{H}}_{u,{\rm LS}}^{H}\hat{\bm{A}}_{u,r}\\
\approx \Big[\kappa_{1}^{u*}\bm{D}^{H}\bm{a}_{u,t}\left(\theta_{1,t}^{u}\right)+\bm{D}^{H}\bm{v}_{u,1},\dots,\\
\kappa_{L_{u}}^{u}\bm{D}^{H}\bm{a}_{u,t}\left(\theta_{L_{u},t}^{u}\right)+\bm{D}^{H}\bm{v}_{u,L_{u}}\Big],
\end{multline}
where $\bm{v}_{u,l}$ denotes the $l$-th column of $\boldsymbol{V}_{u}$ and $\boldsymbol{V}_{u}=\frac{1}{\sqrt{MN_{O}}}\bm{D}^{H}\hat{\bm{\Sigma}}_{o,u}$. When $D$ is large, each column vector in (\ref{approx2}) exhibits a power peak, which corresponds to an \gls{AoD} and channel gain pair. These give us the estimates of the \gls{AoD} $\hat{\theta}_{l,t}$ and channel gain $\hat{\kappa}_{l}^{u}$. 
\item \textbf{Channel Reconstructions: }Once the number of paths, \gls{AoA}, \gls{AoD}, and channel gains are estimated, the complete channel matrix $\bm{H}_{u}$ can be reconstructed based on the plane-wave geometric model. In particular, the channel matrix can be reconstructed as
\begin{equation}\label{G_hat}
\hat{\bm{H}}_{u,{\rm L3SCR}}=\sqrt{MN}\sum_{l=1}^{\hat{L}_{u}}\hat{\kappa}_{l}^{u}\bm{a}_{u,r}(\hat{\theta}_{l,r}^{u})\hat{\bm{a}}_{u,t}(\hat{\theta}_{l,t}^{u})^{H},
\end{equation}
where $\bm{a}_{u,r}(\hat{\theta}_{l,r}^{u})$ can be obtained based on (\ref{a_RT}) and 
\begin{multline}\label{a_T_hat}
\hat{\bm{a}}_{u,t}\left(\hat{\theta}_{l,t}^{u}\right)=\\
\frac{1}{\sqrt{N}}\left[1,e^{-j2\pi\frac{W}{N-1}\cos\hat{\theta}_{l,t}^{u}},\dots,e^{-j2\pi W\cos\hat{\theta}_{l,t}^{u}}\right]^{T}.
\end{multline}
\end{itemize}

\subsubsection{\gls{OMP} method}
In the \gls{L3SCR} method, we can see that the sparse parameters of the channel are estimated in two steps, with the first step handling the joint estimation problem of the number of paths and \gls{AoA}s utilizing the \gls{DFT}-based method and angle rotation. After that, in the second step, the \gls{AoD}s and channel gains are estimated using matched filters. Note that the accuracy of the first step depends significantly on $M$, i.e., the number of the \gls{BS} antennas. When $M$ is sufficiently large, the estimates of the number of paths and \gls{AoA}s are accurate; otherwise, they are not. Clearly, the estimation errors from the first step affect the overall estimation performance. To improve the accuracy, \gls{OMP} in \cite{7458188} can be adopted to jointly estimate the \gls{AoA}-\gls{AoD} pair in an alternative manner, at the expense of higher computational complexity. The performance of \gls{OMP} in channel estimation for \gls{FAS} has been verified in \cite{10236898,xiao2023channel} in a point-to-point communication system, where both the transmitter and receiver are equipped with a \gls{FAS}. In the following, we explain how to apply this method to estimate the channel for the considered system.

The \gls{OMP} method operates through several steps. First of all, similar to the \gls{L3SCR} method, all \gls{FAS} users sequentially transmit orthogonal pilot sequences at $N_{O}$ observable ports, and the \gls{BS} obtains the least square estimate $\hat{\bm{H}}_{u,{\rm LS}}$ of (\ref{H2}). Next, a set of quantized angle grids is selected, and based on this selection, the sensing matrix is computed. Note that the number of angle grids should exceed the number of observable ports at the transmitters, i.e. $N_{O}$, and the number of antennas at the receiver, i.e. $M$. Furthermore, the residual vector is initialized to be the vectorization of $\hat{\bm{H}}_{u,{\rm LS}}$. Following this, the \gls{OMP} method iteratively chooses the column of the sensing matrix that is most strongly correlated with the residual vector. The column index corresponds to a pair of grids, from which an \gls{AoA}-\gls{AoD} pair can be estimated. Subsequently, the channel gain associated with the corresponding \gls{AoA}-\gls{AoD} pair can be estimated by solving the least square problem. The residual vector is then updated by subtracting the contributions of the chosen column vectors from the vectorization of $\hat{\bm{H}}_{u,{\rm LS}}$. These steps continue iteratively until the difference between the residual vector in the current iteration and that of the previous iteration falls below a predetermined threshold. Upon estimating the sparse parameters, the final step involves reconstructing the complete channel matrix to obtain ${\hat{\bm{H}}}_{u,{\rm OMP}}$, mirroring the concluding step in the \gls{L3SCR} method.

\begin{figure}
\centering \includegraphics[scale=0.46]{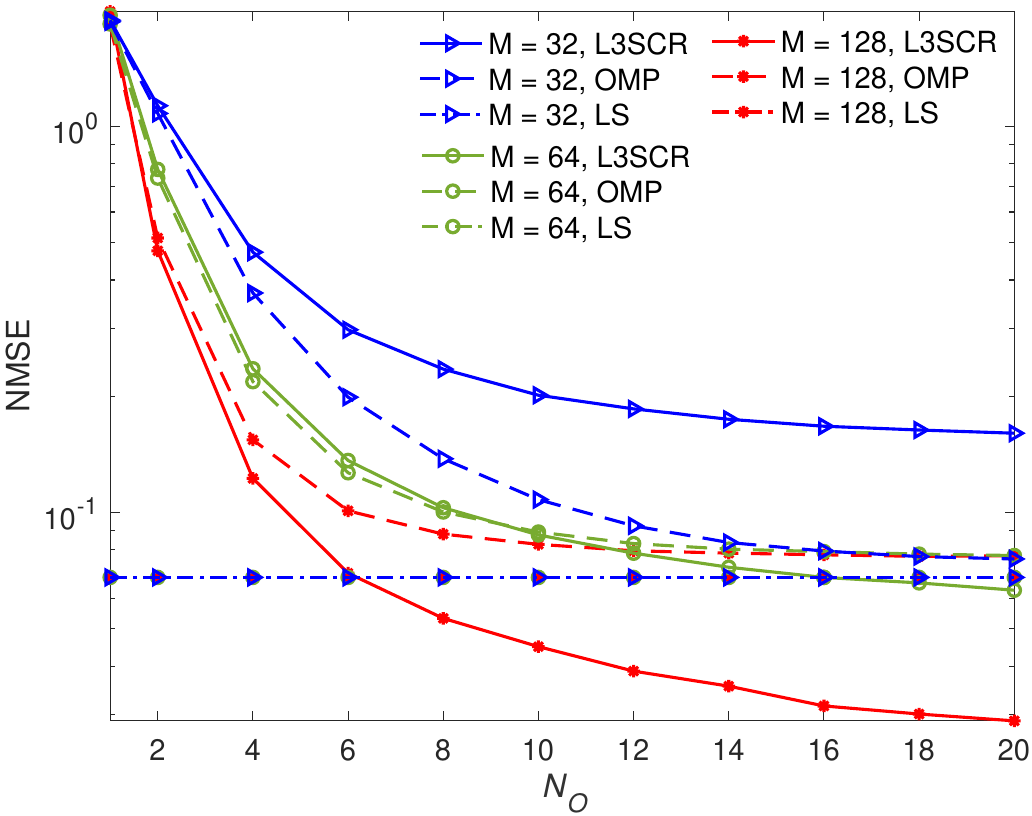} \caption{\gls{NMSE} versus $N_{O}$.}\label{NMSE_VS_KM}
\end{figure}

\begin{figure}
\centering \includegraphics[scale=0.46]{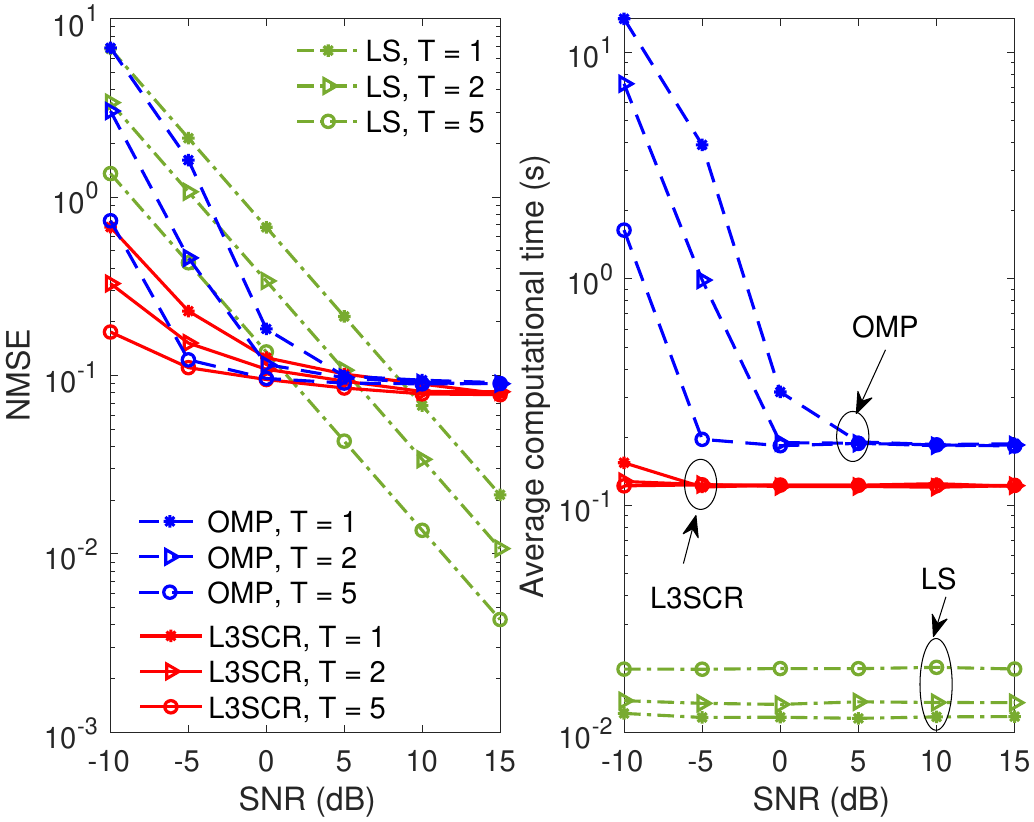}
\caption{The \gls{NMSE} and computational time for different estimation methods with $M=64$ and $N_{O}=10$.}\label{NMSE_time_VS_PT}
\end{figure}

In Figs.~\ref{NMSE_VS_KM} and \ref{NMSE_time_VS_PT}, we show the \gls{NMSE} obtained by various estimation methods and the average required computational time for each channel realization across different parameters. Several observations can be made from the results in these figures. Firstly, when $M$ is small, the \gls{OMP} method outperforms the \gls{L3SCR} method in terms of the \gls{NMSE}, whereas the situation reverses if $M$ is large. Secondly, in numerous configurations, the least square method outperforms the \gls{L3SCR} and \gls{OMP} methods in terms of the \gls{NMSE}. However, it is important to recognize that the least square method requires significantly higher hardware switching and pilot overhead compared to the other two methods. This is owing to the requirement for the antennas of all users to switch and transmit pilot sequences at all $N$ ports, which may lead to lower spectral efficiency in practice. In addition, it is evident that the \gls{OMP} method incurs much higher computational time than the \gls{L3SCR} method. This is attributed to the need for matrix inversions in the \gls{OMP} method when handling least square problems, contributing to high computational intensity.

\subsubsection{Other schemes}
So far, we have shown that in the finite-scattering environment, the geometric model can be adopted to characterize the channel of an \gls{FAS}-assisted uplink system. Based on this model, some standard tools can be employed to estimate the sparse channel parameters at some prescribed observable ports, based on which the \gls{CSI} at all $N$ ports can then be reconstructed. Besides the \gls{L3SCR} and \gls{OMP} methods, there are still some other schemes that can be applied for channel estimation in \gls{FAS}-assisted systems, such as the S-BAR method \cite{zhang2023successive}, the multiple signal classification (MUSIC) method \cite{guo2017millimeter}, estimation of signal parameter via the rotational invariance technique (ESPRIT) \cite{roy1989esprit}, the unitary ESPRIT algorithm \cite{haardt1995unitary}, the space-alternating generalized expectation-maximization (SAGE) scheme \cite{fleury1996wideband}, etc. These methods differ mainly in terms of the estimation accuracy they achieve and the computational complexity levels they demand.

With different estimation methods familiar to the readers, we will discuss the fundamentals of \gls{FAS} in the next section.

\section{Fundamentals of FAS}\label{sec:funda}
In this section, we discuss the benefits and unique features of \gls{FAS} in comparison to \gls{TAS} across various fundamental setups. Following the conventional approach in point-to-point communication channels, these setups are basically divided into four main categories: \gls{SISO}-\gls{FAS}, \gls{SIMO}-\gls{FAS}, \gls{MISO}-\gls{FAS} and \gls{MIMO}-\gls{FAS}. For fair comparison, each of these categories is, respectively, compared to \gls{SISO}, \gls{SIMO}, \gls{MISO} and \gls{MIMO} setups in \gls{TAS} in which the number of active radiating elements is the same. Unlike \gls{TAS}, there are several more cases in \gls{FAS} because the transmitters and receivers may be equipped with either fluid antennas or traditional antennas. Therefore, it will be more convenient to have specific terminologies for different setups and cases. The terminologies are introduced in Table \ref{terminologies_setups}.

We will begin with the simplest setup, i.e., \gls{SISO}-\gls{FAS} and then build to more advanced setups, such as, \gls{SIMO}-\gls{FAS} and \gls{MISO}-\gls{FAS}. We will show how these setups are closely related to the downlink or uplink communications in which multiple users are being served by an access point or \gls{BS}. After that, a more complicated setup, i.e., \gls{MIMO}-\gls{FAS} is then discussed. With the terminologies, we can easily identify the superiority of \gls{SISO}-\gls{FAS}, \gls{SIMO}-\gls{FAS}, \gls{MISO}-\gls{FAS} and \gls{MIMO}-\gls{FAS} as compared to the conventional \gls{SISO}, \gls{SIMO}, \gls{MISO} and \gls{MIMO} fixed-position antenna systems, respectively.\footnote{Here, our results are presented using (\ref{H}), where $N_{1}^{s}=N_{2}^{s}$ and $W_{1}^{s}=W_{2}^{s}$, unless stated otherwise. Note that (\ref{H}) is an extension of (\ref{eq:h_jk}) and (\ref{eq:h_jkn}).} This section is concluded by extending \gls{MIMO}-\gls{FAS} to multiuser \gls{MIMO}-\gls{FAS}, which is highly relevant to cellular networks and WiFi. 

\vspace{.2cm}
\subsection{\gls{TAS}: Benchmarking Scheme}\label{subsec:TAS}
Before discussing the benefits and uniqueness of \gls{FAS}, we find it useful to first discuss the technical details of the existing \gls{TAS}. In general, any wireless communication system with a fixed antenna configuration, where the system parameters such as position, shape, and other characteristics remain unchanged, can be interpreted as \gls{TAS}.  Thus, for benchmarking, we can compare \gls{FAS} with position-reconfigurable antennas  to \gls{TAS} with fixed-position antennas. In this comparison, it is sufficient to evaluate both systems with the same number of RF chains, as this allows for comparable signal processing capabilities. For example, \gls{MRT} can only be performed if multiple active ports/antennas are considered at the transmitter side, while \gls{MRC} requires multiple active ports/antennas at the receiver side. Likewise, parallel transmission can only be achieved if multiple active ports/antennas are employed at both ends.

In certain cases, it is useful to consider the same antenna structure, such as position-reconfiguration dimensions and identical antenna sizes. In this context, multiple fixed-position antennas can be used in \gls{TAS} but the key distinction between \gls{FAS} and \gls{TAS} is that the antennas in \gls{TAS} require at least half a wavelength of spacing. Besides, since \gls{FAS} is not limited to antenna-position reconfiguration alone, \gls{TAS} can similarly be extended to exclude other reconfigurable characteristics, such as antenna orientation and shape, for benchmarking purposes.

\begin{table*}[t]
\noindent \begin{centering}
\caption{Terminologies for different setups and cases.}\label{terminologies_setups}
\par\end{centering}
\noindent \centering{}%
\begin{tabular}{|>{\centering}p{2.5cm}|>{\centering}p{2.5cm}|>{\centering}p{2.5cm}|>{\centering}p{2.5cm}|>{\centering}p{2.5cm}|>{\centering}p{2.5cm}|}
\hline 
Transmitter & Receiver & Terminologies & Transmitter & Receiver & Setup\tabularnewline
\hline 
\hline 
Single\\
traditional antenna${}^\sharp$ & Single\\
traditional antenna & SISO & Multiple \\
traditional antennas & Single\\
traditional antenna & MISO\tabularnewline
\hline 
Single\\
traditional antenna & Single\\
fluid antenna & Rx-SISO-FAS & Multiple \\
traditional antennas & Single\\
fluid antenna & Rx-MISO-FAS\tabularnewline
\hline 
Single\\
fluid antenna & Single\\
traditional antenna & Tx-SISO-FAS & Multiple\\
fluid antennas & Single\\
traditional antenna & Tx-MISO-FAS\tabularnewline
\hline 
Single\\
fluid antenna & Single\\
fluid antenna & Dual-SISO-FAS & Multiple\\
fluid antennas & Single\\
fluid antenna & Dual-MISO-FAS\tabularnewline
\hline 
Single\\
traditional antenna & Multiple \\
traditional antennas & SIMO & Multiple \\
traditional antennas & Multiple \\
traditional antennas & MIMO\tabularnewline
\hline 
Single\\
traditional antenna & Multiple\\
fluid antennas & Rx-SIMO-FAS & Multiple \\
traditional antennas & Multiple\\
fluid antennas & Rx-MIMO-FAS\tabularnewline
\hline 
Single\\
fluid antenna & Multiple\\
traditional antennas & Tx-SIMO-FAS & Multiple\\
fluid antennas & Multiple \\
traditional antennas & Tx-MIMO-FAS\tabularnewline
\hline 
Single\\
fluid antenna & Multiple\\
fluid antennas & Dual-SIMO-FAS & Multiple\\
fluid antennas & Multiple\\
fluid antennas & Dual-MIMO-FAS\tabularnewline
\hline 
\end{tabular}
\begin{tablenotes}
\item ${}^\sharp$A `traditional' antenna corresponds to a fixed-position antenna in conventional communication systems.
\end{tablenotes}
\end{table*}

\subsection{SISO-FAS: The Basic Principles}\label{subsec:SISO-FAS}
\begin{figure}[t]
\centering{}\includegraphics[scale=0.4]{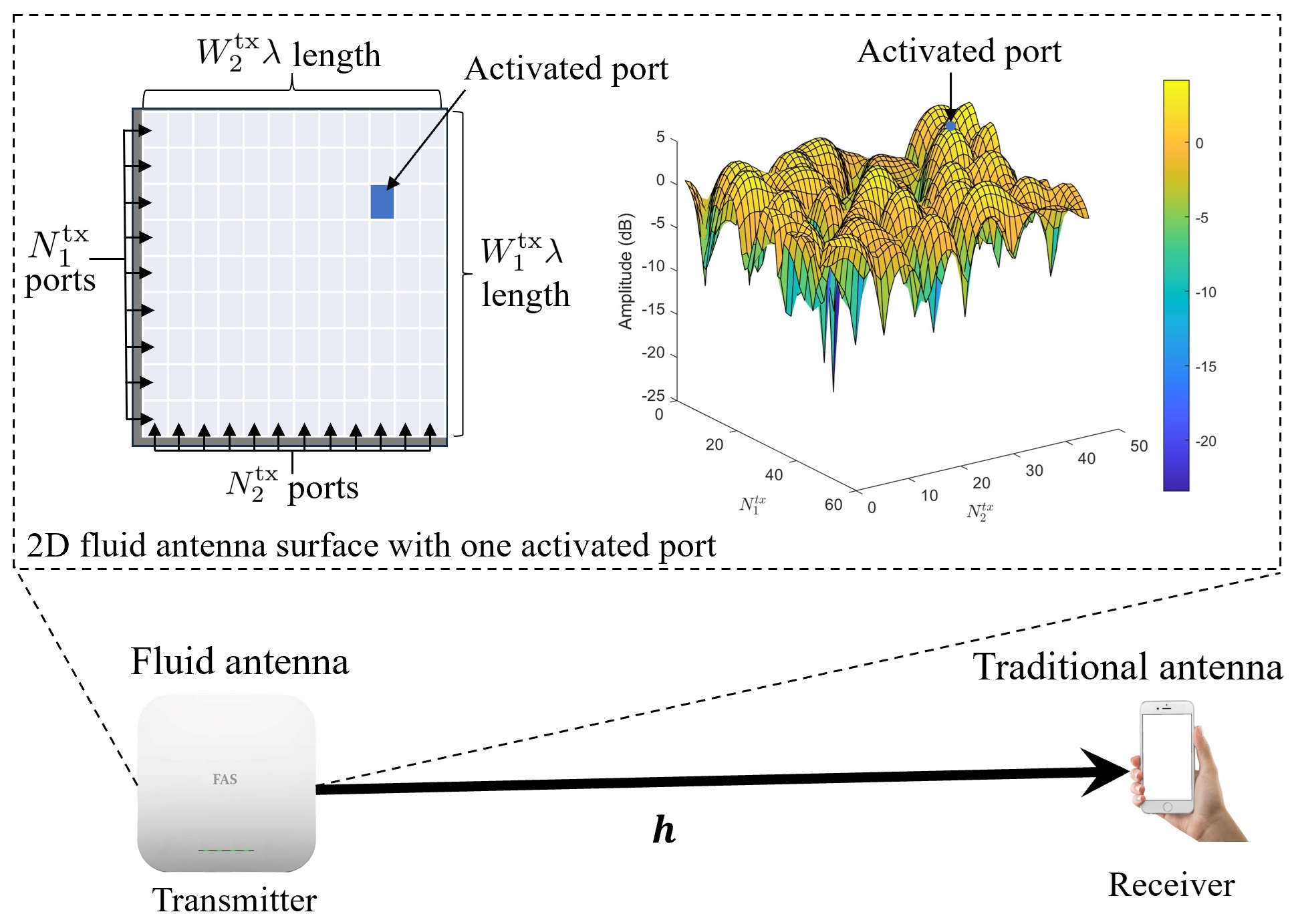}
\caption{A schematic of Tx-SISO-FAS with \gls{2D} FAS.}\label{Tx_SISO_FAS}
\end{figure}

Let us now consider a Tx-\gls{SISO}-\gls{FAS} as shown in Fig.~\ref{Tx_SISO_FAS}. Since the same analogy can be applied to Rx-\gls{SISO}-\gls{FAS}, our discussion here will be sufficient to understand both cases. However, interested readers may refer to \cite{10103838,10130117} for a more comprehensive treatment on the case of Rx-\gls{SISO}-\gls{FAS}. At the end of this subsection, we will look into Dual-\gls{SISO}-\gls{FAS} and highlight its outstanding performance, followed by Tx/Rx-\gls{SISO}-\gls{FAS}, as compared to the traditional \gls{SISO} system.

Let us consider a \gls{2D} fluid antenna surface that consists of one \gls{RF} chain and $N_{{\rm tx}}=N_{1}^{{\rm tx}}\times N_{2}^{{\rm tx}}$ ports. We assume that the $N_{i}^{{\rm tx}}$ ports are uniformly distributed along a linear dimension of length $W_{i}^{{\rm tx}} \lambda$, where $i\in\left\{ 1,2\right\}$ and $W_{{\rm tx}}=W_{1}^{{\rm tx}} \lambda  \times W_{2}^{{\rm tx}} \lambda$. Furthermore, we assume that only one port can be activated at a time. The complex channel vector between the transmitter and receiver can be denoted as $\boldsymbol{h}$. As discussed previously, the complex channel between the $n$-th port of the transmitter and the receiver can similarly be expressed as (\ref{eq:h_jkn}).\footnote{Since $\boldsymbol{Q}\boldsymbol{Q}^{H}=\boldsymbol{I}$, it is apparent that Tx-\gls{SISO}-\gls{FAS} and Rx-\gls{SISO}-\gls{FAS} are fundamentally similar. Thus, there is no significant difference if the fluid antenna is implemented at the transmitter or the receiver in \gls{SISO}-\gls{FAS}.} Since $\bm{h}\sim{\cal CN}(\bm{0},\bm{J})$, $\left|h_{1}\right|,\ldots,\left|h_{N_{{\rm tx}}}\right|$ are correlated Rayleigh random variables. Moreover, since a \gls{CSCG} random variable is preserved by linearity, $h_{n}$ is still a \gls{CSCG} random variable and thus $\left|h_{n}\right|$ is a Rayleigh random variable for a fixed $n$. This implies that \gls{FAS} is equivalent to \gls{TAS} if the same port is always activated regardless of the other factors. Intuitively, there is no difference between a fluid antenna and a traditional antenna if the position of the radiating element remains fixed.

To obtain the best performance, the port with the maximum amplitude should be activated \cite{9264694}, i.e., 
\begin{equation}
\left|h_{{\rm FAS}}\right|=\max\left\{ \left|h_{1}\right|,\ldots,\left|h_{N_{{\rm tx}}}\right|\right\} .
\end{equation}
By choosing the optimal port, the rate of Tx-\gls{SISO}-\gls{FAS} can be computed as
\begin{equation}
R_{{\rm FAS}}=\log\left(1+{\rm SNR}\left|h_{{\rm FAS}}\right|^{2}\right),
\end{equation}
where ${\rm SNR}$ is the \gls{SNR} and the outage probability of Tx-\gls{SISO}-\gls{FAS} can be expressed as
\begin{equation}
\mathbb{P}\left(R_{{\rm FAS}}<R_{{\rm min}}\right),
\end{equation}
where $R_{{\rm min}}$ is the minimum rate requirement. 

Following this, the diversity gain of Tx-\gls{SISO}-\gls{FAS} can be evaluated as
\begin{equation}\label{eq:D_gain}
D_{\text{FAS}}=\min\left\{ N_{{\rm tx}},N'_{{\rm tx}}\left(W_{{\rm tx}}\right)\right\},
\end{equation}
in which $N'_{{\rm tx}}\left(W_{{\rm tx}}\right)$ denotes the maximum diversity that can be obtained for a fixed $W_{{\rm tx}}$ as $N_{{\rm tx}}\rightarrow\infty$. Since the ports are spatially correlated, it would be inconceivable to have infinite diversity just by increasing $N_{{\rm tx}}$ for a fixed $W_{{\rm tx}}$. Thus, (\ref{eq:D_gain}) suggests that, even if $N_{{\rm tx}}\rightarrow\infty$, there is only finite spatial diversity available in \gls{FAS} due to a finite size $W_{{\rm tx}}$.

\begin{figure}
\noindent \centering{}\includegraphics[scale=0.6]{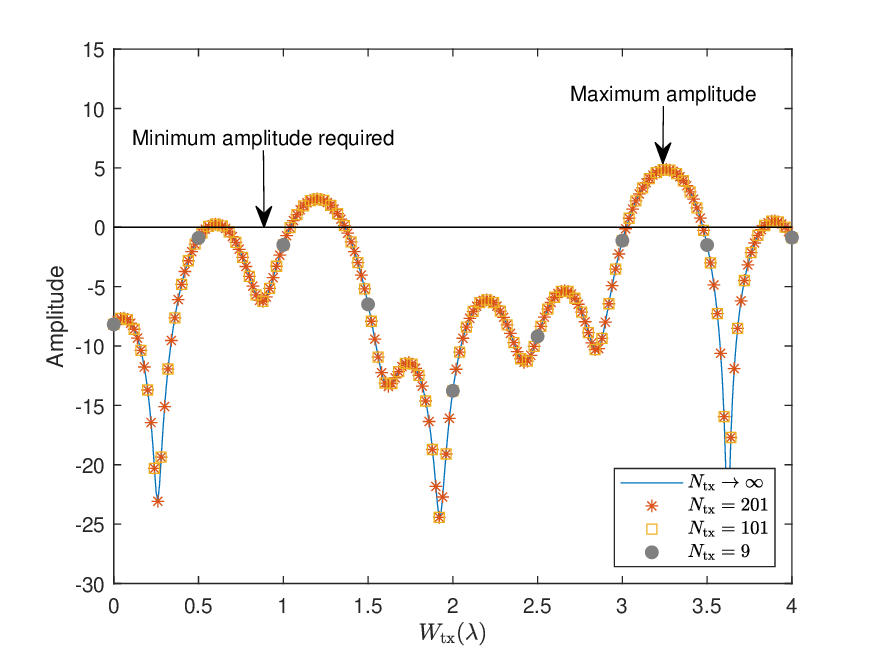}
\caption{Amplitude of FAS versus $W_{{\rm tx}}$.}\label{amp_vs_w}
\end{figure}

To understand this at a more intuitive level, we illustrate the amplitude of \gls{FAS} versus $W_{{\rm tx}}$ in a \gls{1D} FAS and consider different values of $N_{{\rm tx}}$ in Fig.~\ref{amp_vs_w}. For any finite $N_{{\rm tx}}$, each marker represents the amplitude that is accessible by \gls{FAS}. For practical relevance, we also assume that a minimum amplitude is required by the \gls{FAS} receiver which corresponds to the minimum rate requirement. As it is seen, if $N_{{\rm tx}}$ is small (e.g., $N_{{\rm tx}}=9$), \gls{FAS} is unable to satisfy the minimum amplitude and thus outage occurs more easily. A much larger $N_{{\rm tx}}$ is required in \gls{FAS} to fully exploit the fine resolutions of the spatial diversity in a given space, which \emph{makes \gls{FAS} different from antenna selection in \gls{TAS}}. But when $N_{{\rm tx}}$ is sufficiently large (e.g., $N_{{\rm tx}}=201$ or $101$), the spatial diversity cannot further be improved because the total peaks and valleys are statistically similar for a fixed $W_{{\rm tx}}$ due to the spatial correlation. Moreover, there is also a maximum amplitude that can be obtained by \gls{FAS} as $W_{{\rm tx}}$ increases. This means that the amplitude cannot be improved indefinitely just by increasing $W_{{\rm tx}}$, and thus the rate remains limited unless other techniques are employed. However, if $W_{{\rm tx}}$ is increased, the total peaks and valleys may also increase and thus more spatial diversity can be harnessed. The same principles can be applied to Rx-\gls{SISO}-\gls{FAS} or higher dimensional fluid antenna surface. 

\begin{figure}[t]
\noindent \begin{centering}
\subfloat[]{\noindent \includegraphics[scale=0.6]{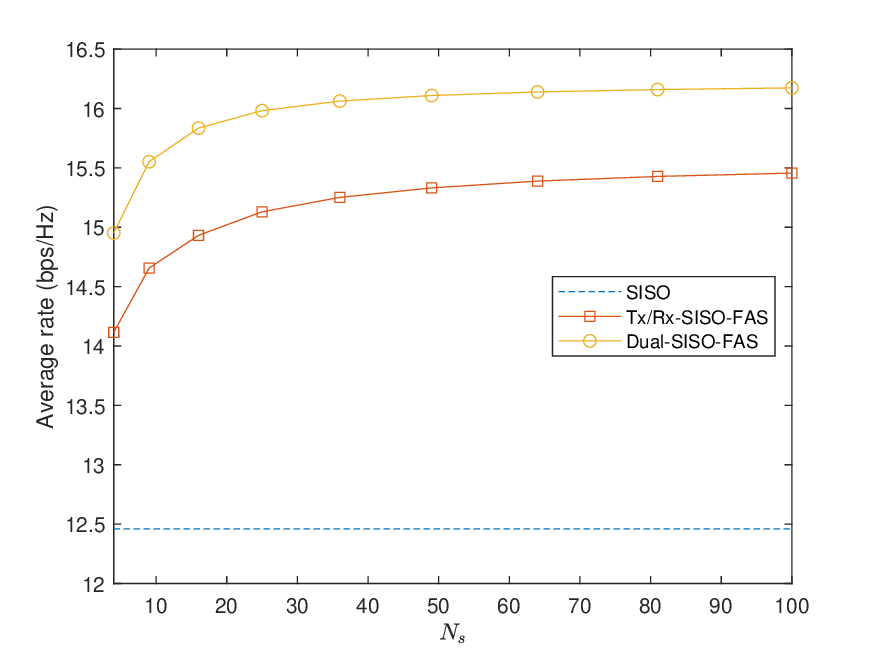}}
\par\end{centering}
\centering{}\subfloat[]{\centering{}\includegraphics[scale=0.6]{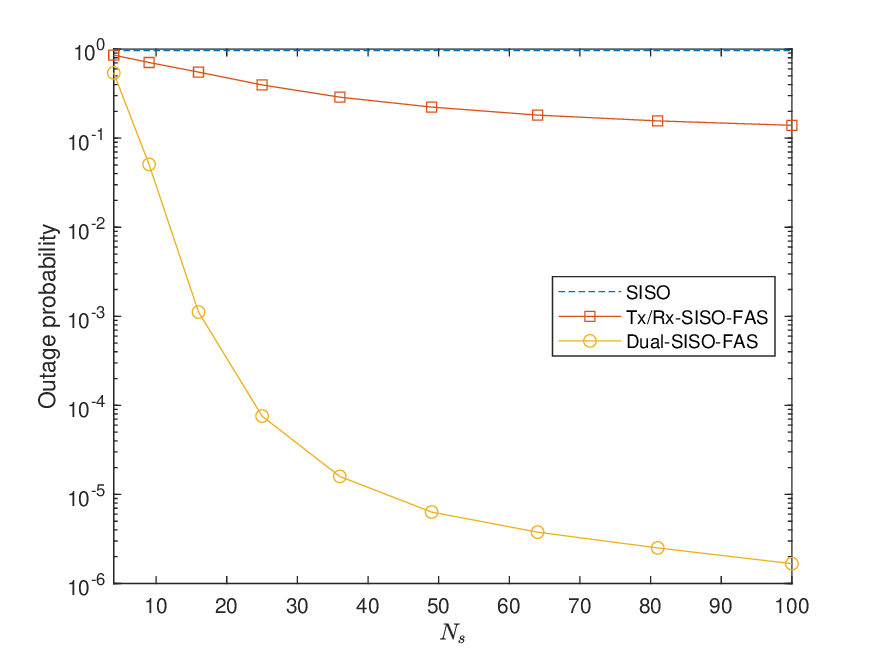}}
\caption{The performance of different SISO-FAS cases versus $N_{s}$, where ${\rm SNR}=40~{\rm dB}$, $R_{{\rm min}}=15~{\rm bps/Hz}$, and $W_{s}=2\lambda\times 2\lambda$: a) average rate; and b) outage probability.}\label{SISO_FAS_results}
\end{figure}

Obviously, we can implement a fluid antenna at the transmitter and receiver to obtain better performance, i.e. Dual-\gls{SISO}-\gls{FAS}. This changes the channel vector $\boldsymbol{h}$ to the channel matrix $\boldsymbol{H}$. For optimal performance, the transmitter and receiver can activate the port that provides the overall highest amplitude. Using these optimal strategies, we can investigate the performance of a traditional \gls{SISO} system, Tx/Rx-\gls{SISO}-\gls{FAS} and Dual-\gls{SISO}-\gls{FAS} over different $N_{s},s\in\left\{ {\rm rx},{\rm tx}\right\}$.

As illustrated in Fig.~\ref{SISO_FAS_results}, the performance of Dual-\gls{SISO}-\gls{FAS} outperforms Tx/Rx-\gls{SISO}-\gls{FAS} followed by the traditional \gls{SISO} system. The rationale behind the superiority of \gls{FAS} is that Dual-\gls{SISO}-\gls{FAS} can be interpreted as a fully correlated \gls{MIMO} system where only one input and one output are used. Consequently, the performance is understandably much better than the traditional \gls{SISO} system. Likewise, Tx-\gls{SISO}-\gls{FAS} and Rx-\gls{SISO}-\gls{FAS} correspond to the correlated \gls{MISO} and \gls{SIMO} systems but only one input or output can be accessed by the transmitter or receiver, respectively. Since the \gls{SISO} system is a subset of the correlated \gls{MIMO}/\gls{MISO}/\gls{SIMO} system, it is straightforward that \gls{FAS} is more superior than \gls{TAS}. Thus, despite having the same number of radiating elements, \gls{FAS} outperforms \gls{TAS} because an extreme number of correlated channels within a given space can be exploited.

\begin{figure}[t]
\noindent \includegraphics[scale=0.6]{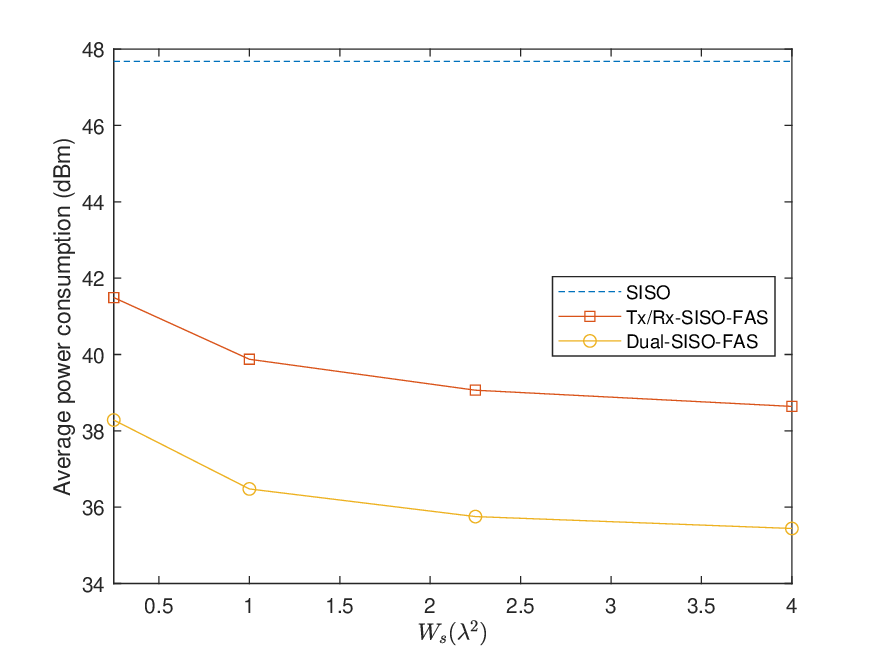}
\centering{}\caption{The power consumption of different SISO-FAS cases versus $W_{s}$, where $R_{{\rm min}}=15~{\rm bps/Hz}$ and $N_{s}=100$.}\label{SISO_FAS_energy}
\end{figure}

It is worth pointing out that \gls{FAS} can also yield higher energy efficiency than \gls{TAS} due to the diversity gain. Specifically, \gls{FAS} requires less transmit power than \gls{TAS} to achieve a specific rate. This can be verified by computing the average power consumption to satisfy a fixed rate based on the above optimal strategies. Fig.~\ref{SISO_FAS_energy} shows the average power consumption of different \gls{SISO}-\gls{FAS} cases versus $W_{s}$. As observed, the average power consumption of Dual-\gls{SISO}-\gls{FAS} and Tx/Rx-\gls{SISO}-\gls{FAS} is significantly lower that of the traditional \gls{SISO} system. This superiority follows the same principle in diversity gain. In other words, the extreme diversity gain of \gls{FAS} can be used to reduce power consumption or improve energy efficiency.

\subsection{SIMO-FAS and MISO-FAS: The Connection with Broadcast Channel and Medium Access Channel}\label{ssec:nomafas}

\begin{figure}
\noindent \begin{centering}
\subfloat[]{\includegraphics[scale=0.39]{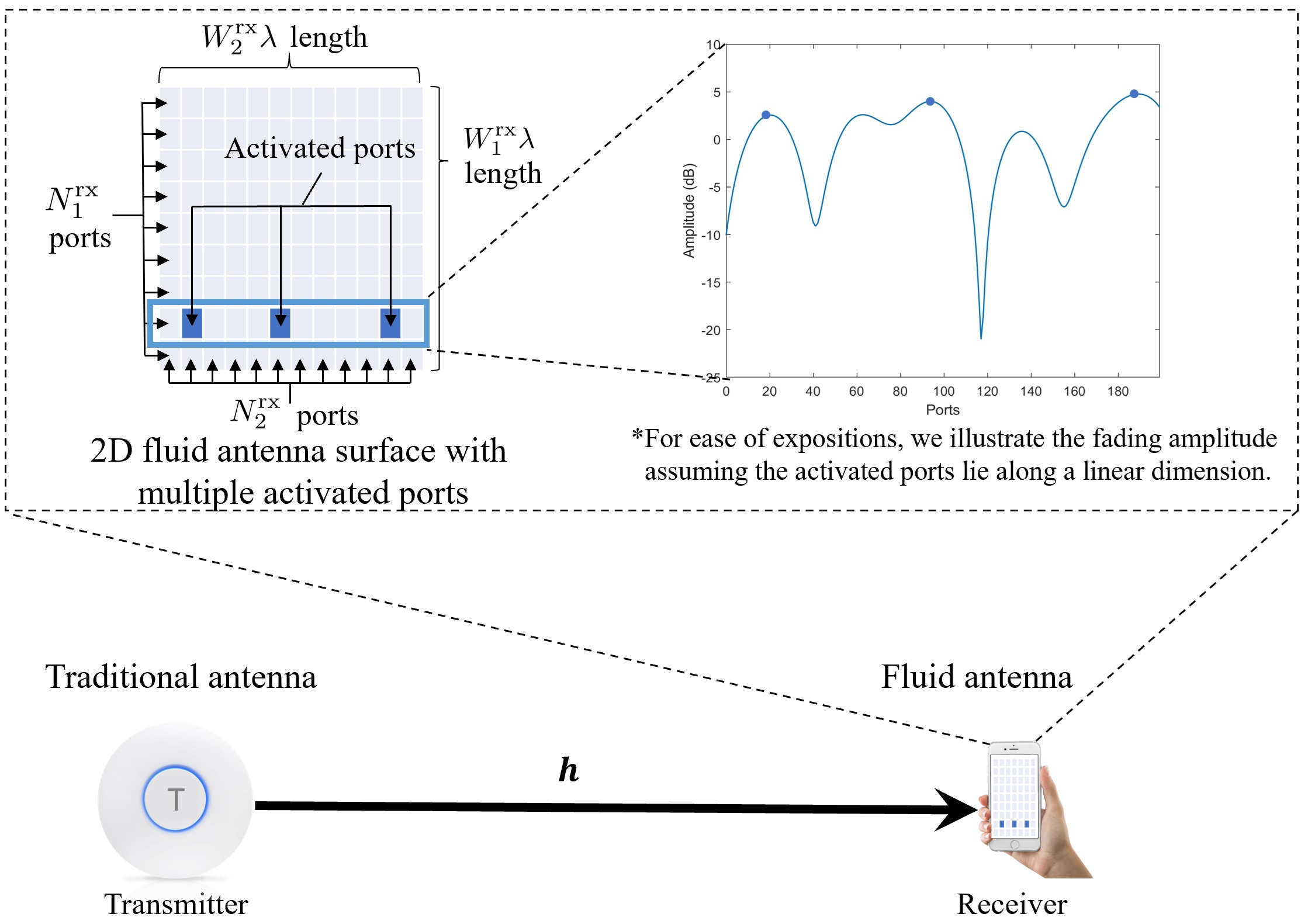}}
\par\end{centering}
\noindent \centering{}\subfloat[]{\includegraphics[scale=0.4]{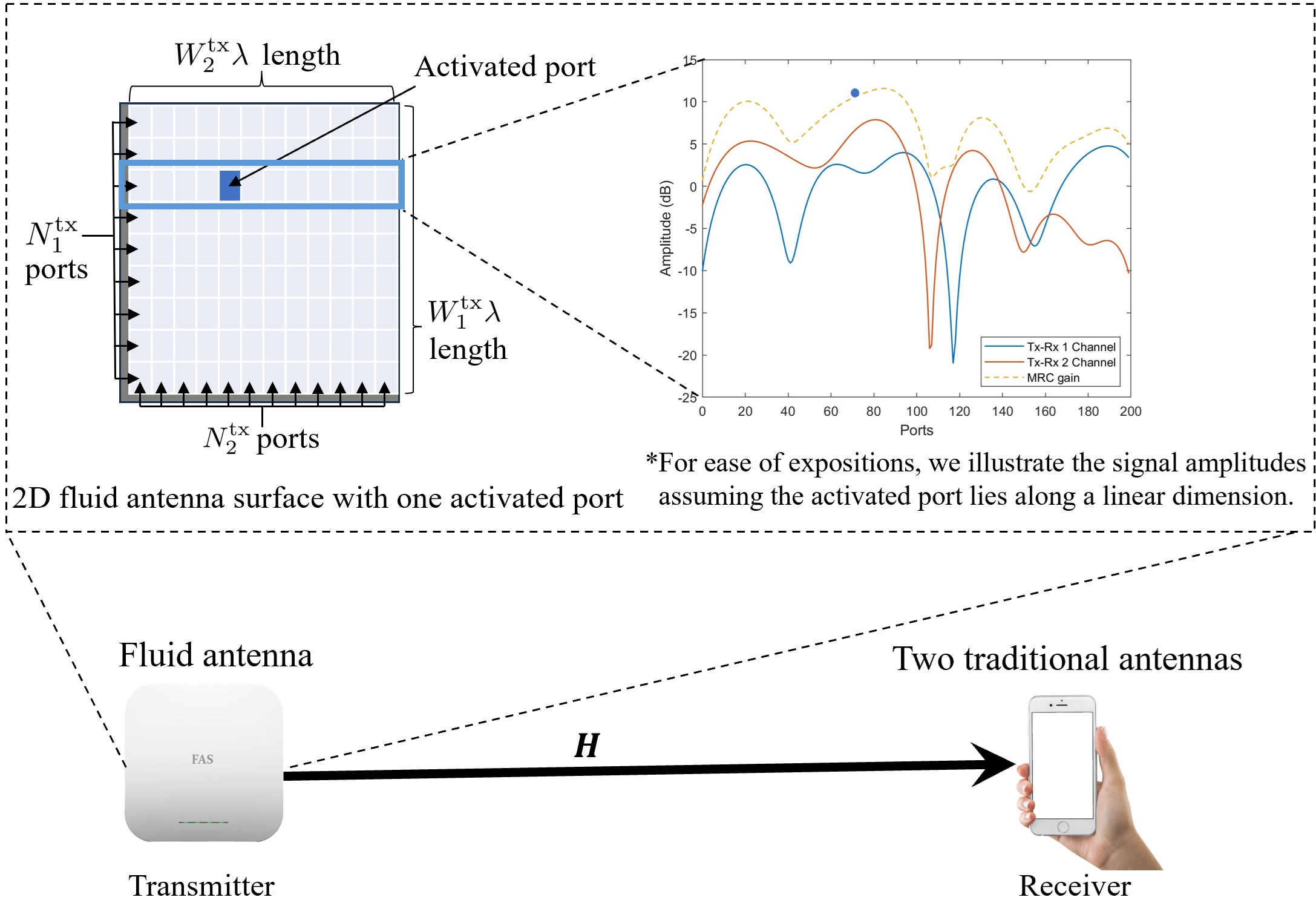}}
\caption{Special cases of SIMO-FAS: a) Rx-SIMO-FAS; and b) Tx-SIMO-FAS.}\label{SIMO_FAS}
\end{figure}

\begin{figure}
\noindent \begin{centering}
\subfloat[]{\includegraphics[scale=0.6]{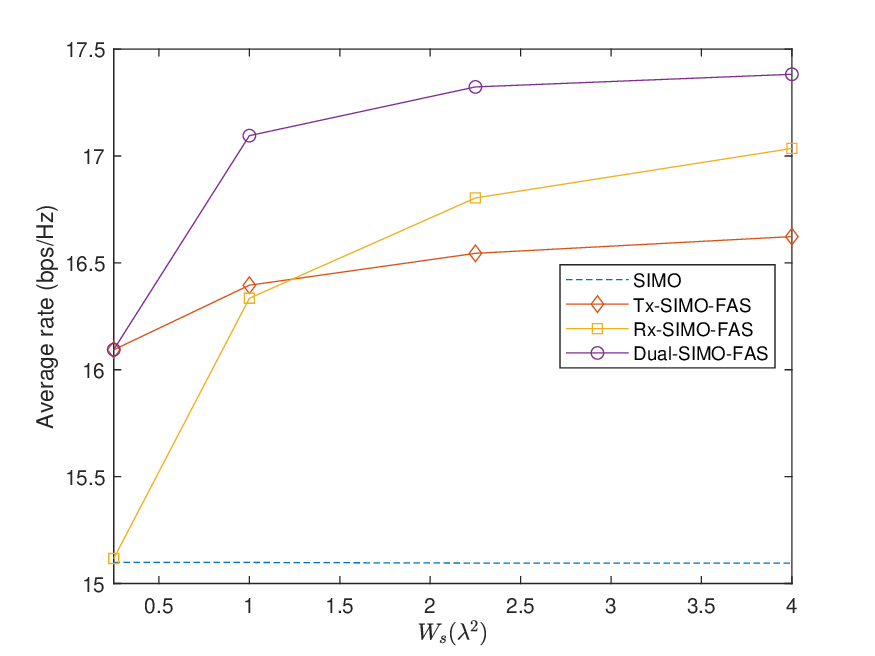}}
\par\end{centering}
\noindent \centering{}\subfloat[]{\includegraphics[scale=0.6]{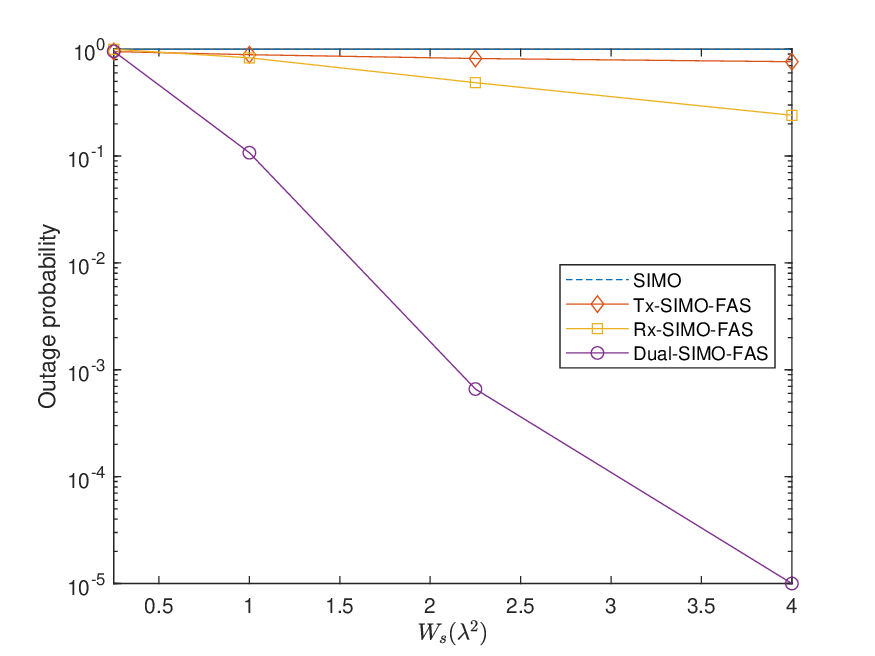}}
\caption{The performance of different SIMO-FAS cases versus $W$, where ${\rm SNR}=40~{\rm dB}$, $R_{{\rm min}}=16.8~{\rm bps/Hz}$, and $N_{s}=100$: a) average rate; and b) outage probability.}\label{SIMO_FAS_results}
\end{figure}

A natural extension of \gls{SISO}-\gls{FAS} is \gls{SIMO}-\gls{FAS} and \gls{MISO}-\gls{FAS}. For brevity, we focus on \gls{SIMO}-\gls{FAS} as similar principles can be applied to \gls{MISO}-\gls{FAS}. If the receive active ports are co-located, then Rx-\gls{SIMO}-\gls{FAS} is similar to a correlated \gls{SIMO} system as seen in Fig.~\ref{SIMO_FAS}(a), resemblance to antenna selection with an extreme number of compactly-placed fixed-position antennas in a given space. To obtain the optimal performance, the receiver in Rx-\gls{SIMO}-\gls{FAS} can activate ports that provide the highest \gls{MRC} gain in the presence of mutual coupling. Matching networks can also be employed to improve the performance. Alternatively, the receiver may suboptimally select ports that provide the highest \gls{MRC} gain while ensuring a minimum distance between the active ones to prevent mutual coupling.\footnote{This approach works for liquid-based fluid antennas. However, depending on how FAS is implemented, unselected ports may still cause mutual coupling and in this case, keeping distance between active ports are not useful.} Compared to traditional \gls{SIMO} systems, Rx-\gls{SIMO}-\gls{FAS} \emph{is capable of achieving additional gain by finely reconfiguring the positions of the active radiating elements in response to the channel conditions}. The gain is more obvious when the size of FAS increases. 

In contrast, Tx-\gls{SIMO}-\gls{FAS} is more similar to a correlated \gls{MIMO} system where an antenna is selected by the transmitter and the receiver performs \gls{MRC}. The main distinction between traditional \gls{MIMO} with antenna selection at the transmitter and Tx-\gls{SIMO}-\gls{FAS} is that in \gls{FAS}, only a single fluid antenna is required at the transmitter as opposed to having an extreme number of fixed-position antennas within the predefined space. Due to position flexibility, the gain of Tx-\gls{SIMO}-\gls{FAS} is more apparent when the FAS size is small. To obtain the optimal performance, the transmitter in Tx-\gls{SIMO}-\gls{FAS} can activate the port with the largest \gls{MRC} gain. Similar to \gls{SISO}-\gls{FAS}, we can implement the fluid antennas at both ends and optimize their ports using the same principle in Dual-\gls{SIMO}-\gls{FAS}. 

In Fig.~\ref{SIMO_FAS_results}, we present the performance of different \gls{SIMO}-\gls{FAS} cases against the FAS size, $W_{s}$. It is assumed that the number of active ports is $4$. As anticipated, Dual-\gls{SIMO}-\gls{FAS} outperforms Tx/Rx-only \gls{SIMO}-\gls{FAS} significantly, followed by the fixed-position \gls{SIMO} antenna system. When $W_{s}$ is small, Tx-\gls{SIMO}-\gls{FAS} outperforms Rx-\gls{SIMO}-\gls{FAS} while the opposite occurs when $W_{s}$ is large. The main reason is that in the \gls{SIMO} case, there are multiple fluid antennas in the Rx-only case but only one fluid antenna in the Tx-only case, which means that correlation hurts the Rx-only case more than the Tx-only case.

\begin{figure}
\centering{}\includegraphics[scale=0.32]{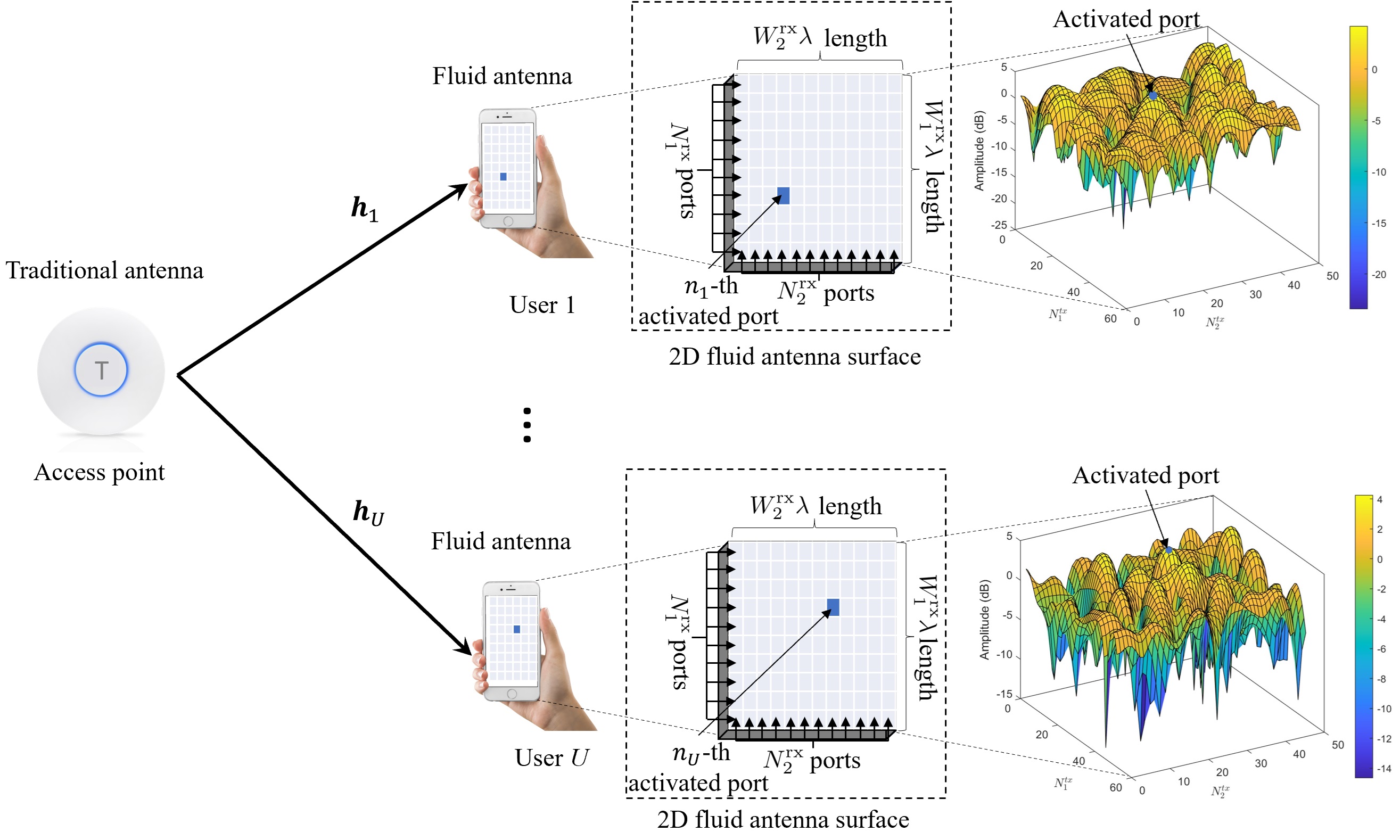}
\caption{A schematic of an access point or \gls{BS} serving multiple FAS users with 2D FAS in the downlink.}\label{DL}
\end{figure}

The above mainly focuses on co-located receive active ports. However, what would happen if the receive active ports are distributed? In this case, Rx-\gls{SIMO}-\gls{FAS} is no longer similar to a correlated \gls{SIMO} system because the channel from the transmitter to each receive active port can then be regarded as \gls{i.i.d.} Due to higher spatial diversity, its performance can be improved using cooperative \gls{MRC} as compared to the co-located case. A prevalent scenario is the multiuser assumption. In \gls{TAS}, this assumption converts the \gls{SIMO} system into a broadcast channel, which is also referred to as the downlink communications. To understand the performance of \gls{FAS} in a broadcast channel, let us consider a scenario where multiple \gls{FAS} users are being served by an access point with a traditional antenna in the downlink as shown in Fig.~\ref{DL}. 

\begin{figure}
\noindent \centering{}\includegraphics[scale=0.63]{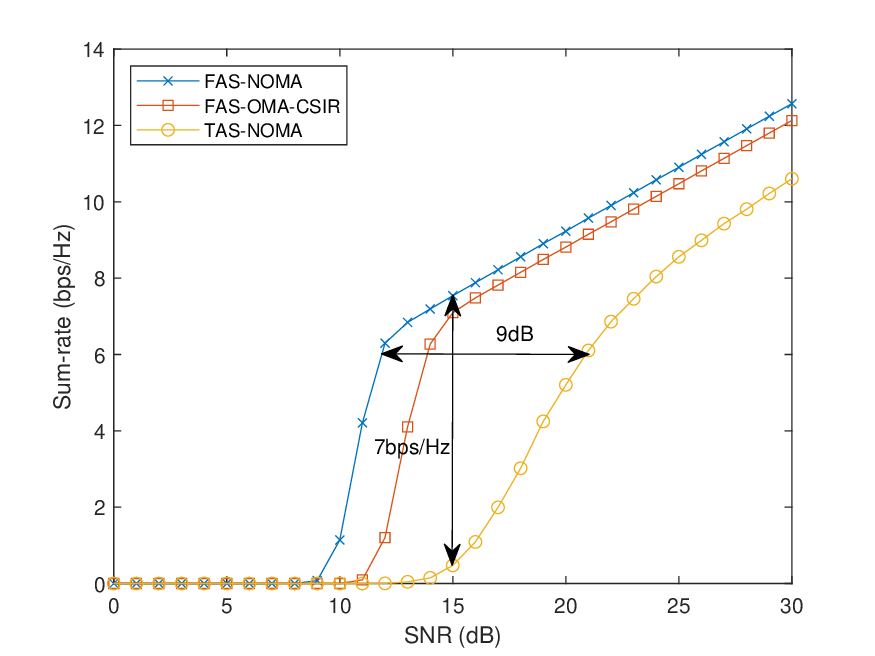}
\caption{The performance of FAS and TAS in the downlink with $4$ users, where $N_{{\rm rx}}=100$ and $W_{{\rm rx}}=4\lambda^{2}$.}\label{DL_results}
\end{figure}

Since the input signals originate from the same antenna, it is possible to use multiple access schemes that can prevent multiuser interference. It is well known that superposition coding and \gls{SIC} (e.g., power-domain \gls{NOMA}) are capacity-achieving techniques in degraded broadcast channels \cite{9693417}.\footnote{Since NOMA is capacity-achieving in this setup, \gls{RSMA} is not considered. However, it is worth noting that \gls{RSMA} can be more effective in situations when for example, \gls{CSI} is imperfect, the channel is non-quasi-degraded, or only one layer of \gls{SIC} is employed. In these scenarios, \gls{RSMA} might offer performance gains compared to \gls{NOMA}.} Therefore, it makes sense to continue employing power-domain \gls{NOMA} when the receivers are equipped with fluid antennas and the number of users is not large. For ease of exposition, we refer the use of traditional antenna in \gls{NOMA} and fluid antenna as \gls{TAS}-\gls{NOMA} and \gls{FAS}-\gls{NOMA}, respectively. We also consider the use of \gls{OMA} with only \gls{CSI} at the receiver and refer to this suboptimal scheme as \gls{FAS}-\gls{OMA}-CSIR. The sum-rate results with $4$ users are given in Fig.~\ref{DL_results}. The results reveal that \gls{FAS}-\gls{NOMA} provides an enormous rate improvement of $7~{\rm bps/Hz}$ as compared to \gls{TAS}-\gls{NOMA} when the \gls{SNR} is $15~{\rm dB}$. From another perspective, \gls{FAS}-\gls{NOMA} requires $9~{\rm dB}$ less than \gls{TAS}-\gls{NOMA} to reach $6~{\rm bps/Hz}$. The performance of \gls{FAS}-\gls{NOMA} clearly outperforms \gls{TAS}-\gls{NOMA}, especially at medium \gls{SNR}. Remarkably, it is observed that even \gls{FAS}-\gls{OMA}-CSIR outperforms \gls{TAS}-\gls{NOMA} at all \gls{SNR}. In other words, a suboptimal scheme in \gls{FAS} can outperform the capacity-achieving scheme in \gls{TAS}, \emph{establishing a new possibility for performance leap}. To further understand the impact of different multiple access schemes in \gls{FAS} as compared to \gls{TAS}, see \cite{FAS_NOMA}.

The above discussion can be generalized to \gls{MISO}-\gls{FAS}. For example, if the transmit active ports are co-located within a given space, Tx-\gls{MISO}-\gls{FAS} is similar to an existing correlated \gls{MISO} system with antenna selection in which a huge number of fixed-position antennas is deployed within a given space. Besides, Rx-\gls{MISO}-\gls{FAS} resembles a correlated \gls{MIMO} system where an output is selected by the receiver. Instead of \gls{MRC}, \gls{MRT} is used at the transmitter. It is also expected that Dual-\gls{MISO}-\gls{FAS} outperforms Tx/Rx-\gls{MISO}-\gls{FAS}, followed by the fixed-position \gls{MISO} system. On the other hand, Tx-\gls{MISO}-\gls{FAS} with distributed transmit active ports is the reverse setup of Rx-\gls{SIMO}-\gls{FAS} with distributed receive active ports. Thus, we can relate them to the broadcast channel (downlink) and the medium access channel (uplink) by considering the multiuser setup, according to where the user signals come from and where they are being sent to.

\subsection{MIMO-FAS: Optimization and \gls{DMT} Overview}

\begin{figure}
\noindent \centering{}\includegraphics[scale=0.4]{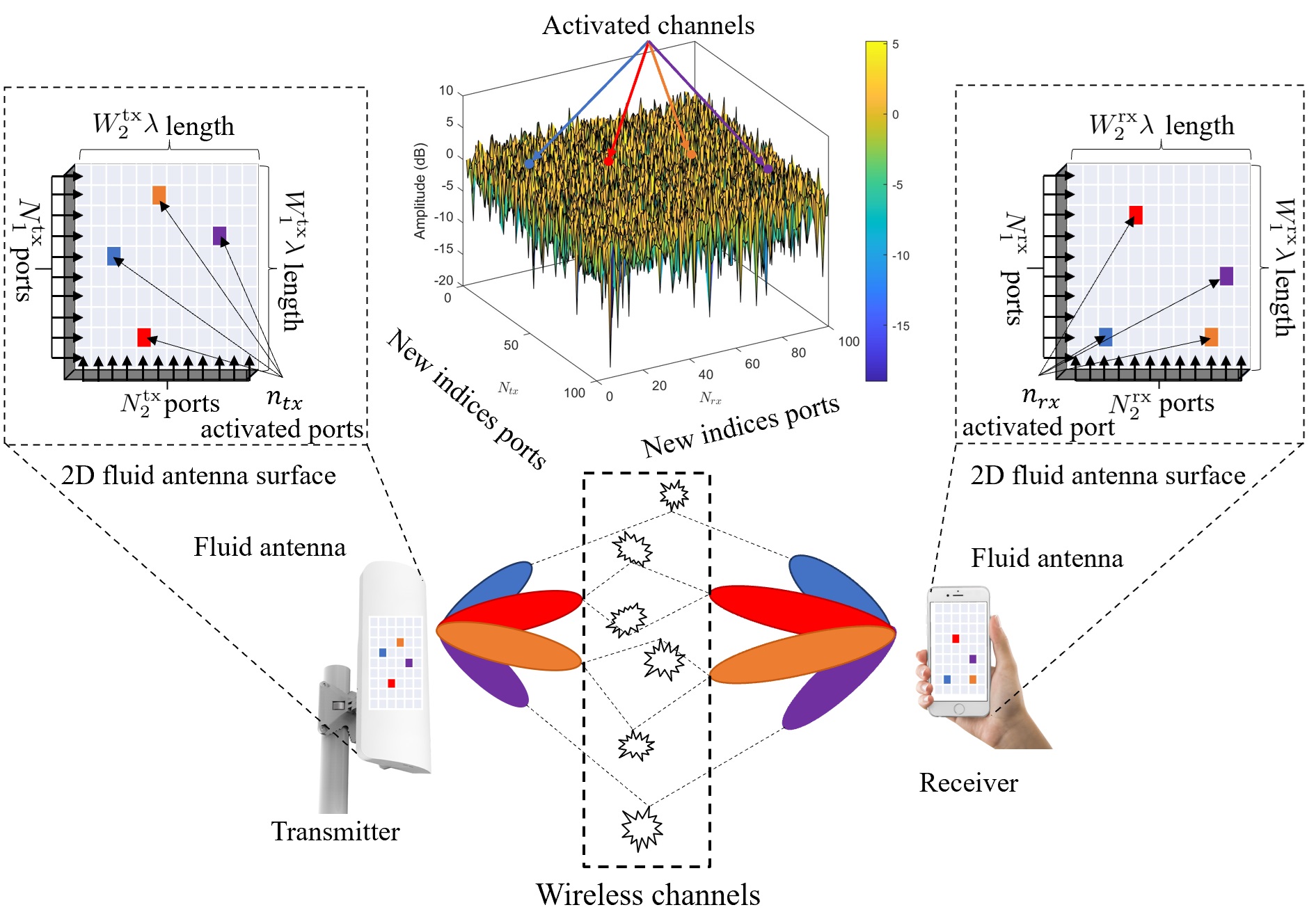}
\caption{A schemetic of Dual-MIMO-FAS with 2D FAS.}\label{MIMO_FAS}
\end{figure}

\gls{MIMO}-\gls{FAS} is closely related to a traditional fixed-position \gls{MIMO} antenna system. The main distinctive feature of \gls{FAS} is the very fine resolution in the spatial domain that can harness additional diversity. Conceptually, \gls{MIMO}-\gls{FAS} is equivalent to MIMO with an extreme number of antennas being deployed in a given space while only having a subset being activated. As a result, \gls{MIMO}-\gls{FAS} \emph{can jointly optimize the active ports, beamforming and power allocation for a specific metric}. 

One important problem is to maximize the achievable rate for Dual-\gls{MIMO}-\gls{FAS}, which is unfortunately recognized as an NP-hard problem. However, a simple yet near-optimal solution can be obtained in the high \gls{SNR} regime. Specifically, the optimization can be split into two sub-problems: (i) optimal port selection and (ii) optimal beamforming and power allocation. To solve the port selection subproblem, it is possible to exploit a property related to the achievable sum-rate and formulate a relaxed problem for which a near-optimal solution at high \gls{SNR} can be obtained utilizing strong \gls{RRQR} factorization \cite{doi:10.1137/0917055}. Given that the port selection is known, the optimal beamforming and power allocation subproblem can then be solved using \gls{SVD} and waterfilling power allocation \cite{1203154}. For other \gls{SNR}s or more complicated problems, various approaches such as alternating optimization, generalized Bender decomposition, genetic algorithms, penalty methods, and other relaxation techniques can be used to address the challenges \cite{cheng2023sumrate,pi2023multiuser,xiao2023multiuser,10354003,qin2023antenna,wu2023movable}. Interestingly, the activated ports, beamforming and power allocation of \gls{MIMO}-\gls{FAS} can also be optimized jointly with only statistical \gls{CSI} \cite{10328751,chen2023joint}.

To better understand the distinction between \gls{FAS} and \gls{TAS}, consider Dual-\gls{MIMO}-\gls{FAS}, as shown in Fig.~\ref{MIMO_FAS}, and compare it with the traditional \gls{MIMO} system. For any \gls{MIMO} system, there is fundamental tradeoff between diversity and multiplexing gains. Specifically, a \gls{MIMO} system is said to achieve a multiplexing gain of $r$ and a diversity gain of $d$ if \cite{1197843}
\begin{equation}
\lim_{{\rm SNR}\rightarrow\infty}\frac{R_{{\rm sys}}\left({\rm SNR}\right)}{\log{\rm SNR}}=r,
\end{equation}
and the outage probability satisfies
\begin{equation}
\lim_{{\rm SNR}\rightarrow\infty}\frac{\log\left(\mathbb{P}_{{\rm sys}}^{{\rm out}}\left({\rm SNR},r\right)\right)}{\log{\rm SNR}}=-d\left(r\right),
\end{equation}
in which $R_{{\rm sys}}\left({\rm SNR}\right)$ and $\mathbb{P}_{{\rm sys}}^{{\rm out}}\left({\rm SNR},r\right)$ are, respectively, the rate and outage probability of the \gls{MIMO} system. Since Dual-\gls{MIMO}-\gls{FAS} is also a \gls{MIMO} system, it is natural to have this tradeoff as well. If $W_{\rm tx}$ and $W_{\rm rx}$ are sufficiently large, then the \gls{DMT} of a traditional \gls{MIMO} system is a piece-wise curve connecting the points $\left(r,d\left(r\right)\right)$ where \cite{4137906}
\begin{equation}\label{eq:d_MIMO}
d\left(r\right)=\left(n_{{\rm tx}}-r\right)\left(n_{{\rm rx}}-r\right),
\end{equation}
and $r\in\left[0,n_{\min}\right]$, in which $n_{\min}=\min\left\{ n_{{\rm tx}},n_{\rm rx}\right\}$, $n_{{\rm tx}}$ and $n_{{\rm rx}}$ are the number of traditional antennas at the transmitter and receiver, respectively. In contrast, according to \cite{10303274}, the \gls{DMT} of Dual-\gls{MIMO}-\gls{FAS} is a piece-wise curve connecting the points $\left(n_{\min},0\right)$ and 
\begin{equation}\label{eq:d_MIMO_FAS}
\left\{ r,\left(N_{\rm tx}^{'}\left(W_{\rm tx}\right)-r\right)\left(N_{\rm rx}^{'}\left(W_{\rm rx}\right)-r\right)\right\},~r=0,\dots,\tilde{N},
\end{equation}
where $N_{\rm tx}^{'}\left(W_{\rm tx}\right)$ is the maximum transmit diversity that can be obtained for a fixed $W_{\rm tx}$ as $N_{\rm tx}\rightarrow\infty$, $N_{\rm rx}^{'}\left(W_{\rm rx}\right)$ is defined in a similar fashion, and
\begin{equation}
\tilde{N}=\arg\min_{{\eta\in\mathbb{Z}\atop 0\leq\eta\leq n_{\min}-1}}\frac{\left(N_{{\rm tx}}^{'}\left(W_{{\rm tx}}\right)-\eta\right)\left(N_{{\rm rx}}^{'}\left(W_{{\rm rx}}\right)-\eta\right)}{n_{\min}-\eta}.
\end{equation}

In \gls{FAS}, we may interpret that $n_{{\rm tx}}$ and $n_{{\rm rx}}$ are the number of active transmit and receive ports. Under the assumption where $W_{\rm tx}$ and $W_{\rm rx}$ are sufficiently large, we have $n_{{\rm tx}}\leq N_{\rm tx}^{'}\left(W\right)$ and $n_{{\rm rx}}\leq N_{\rm rx}^{'}\left(W\right)$. After some manipulations, we can similarly obtain the \gls{DMT} of Tx/Rx-\gls{MIMO}-\gls{FAS} by connecting the points $\left(n_{\min},0\right)$ and 
\begin{equation}
\left\{ r,\left(N_{s}^{'}\left(W_{s}\right)-r\right)\left(n_{\bar{s}}-r\right)\right\},~r=0,\dots,\tilde{N_{s}},
\end{equation}
where $\bar{s}$ is the complement of $s$, and 
\begin{equation}
\tilde{N_{s}}=\arg\min_{{\eta\in\mathbb{Z}\atop 0\leq\eta\leq n_{\min}-1}}\frac{\left(N_{s}^{'}\left(W_{s}\right)-\eta\right)\left(n_{\bar{s}}-\eta\right)}{n_{\min}-\eta}.
\end{equation}

\begin{figure}
\noindent \centering{}\includegraphics[scale=0.6]{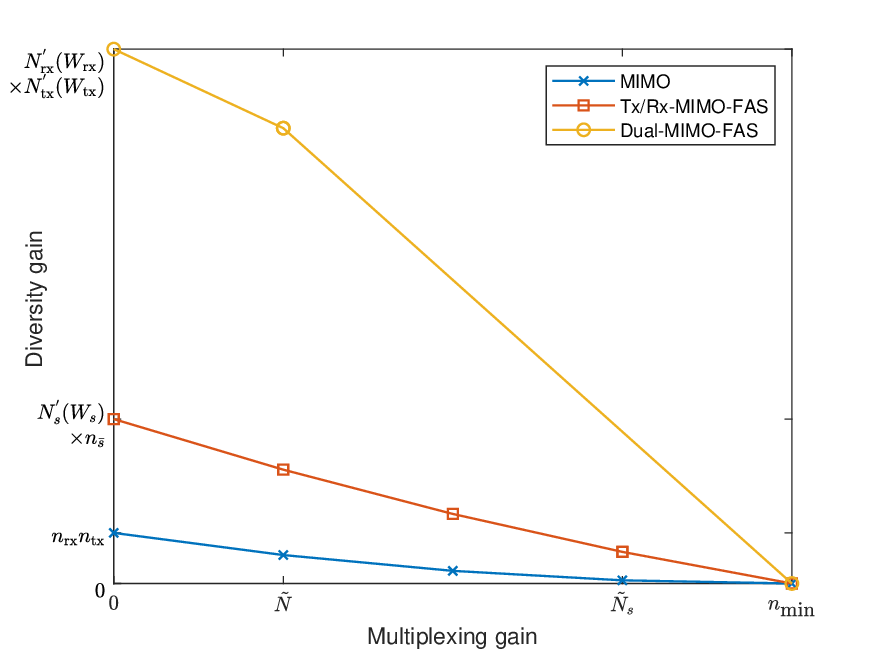}
\caption{The DMT of different MIMO-FAS cases.}\label{DMT_result}
\end{figure}

In Fig.~\ref{DMT_result}, we present the \gls{DMT} results of different \gls{MIMO}-\gls{FAS} cases. The DMT of Dual-\gls{MIMO}-\gls{FAS} \emph{is an outer bound} of Tx/Rx-\gls{MIMO}-\gls{FAS}, followed by \gls{MIMO}. To obtain a good sense of how much diversity gain can be obtained in \gls{MIMO}-\gls{FAS}, a simple method was proposed in \cite{10303274} to approximate the values of $N_{{\rm tx}}^{'}\left(W_{{\rm tx}}\right)$ and $N_{{\rm rx}}^{'}\left(W_{{\rm rx}}\right)$. When $W_{{\rm tx}}=W_{{\rm rx}}=0.25\lambda^{2}$, the maximum diversity gain of \gls{MIMO}-\gls{FAS} with \gls{2D} fluid antenna surface at both sides is $169$ while the maximum diversity of the traditional \gls{MIMO} system with \gls{2D} antenna surface at both sides is $16$. To help readers better understand the performance gain, the estimated maximum diversity of Dual-\gls{MIMO}-\gls{FAS} and traditional \gls{MIMO} are given in Table \ref{Reduced_rank}. The rationale behind the superiority of \gls{MIMO}-\gls{FAS} is that all ports must experience deep fading in order for an outage to occur. From the \gls{DMT}, it is seen that the multiplexing gains cannot be improved in \gls{MIMO}-\gls{FAS} but it is worth highlighting that \gls{MIMO}-\gls{FAS} provides some rate gain when compared to traditional \gls{MIMO}. To further understand the performance gain of \gls{MIMO}-\gls{FAS}, the readers may refer to \cite{10303274,10243545}. Similar to previous setups, the extreme diversity gain in \gls{MIMO}-\gls{FAS} can be exploited to improve the energy efficiency. 

\begin{table}
\caption{Maximum diversity based on different $W_{s}$.}\label{Reduced_rank}
\begin{centering}
\begin{tabular}{|c|c|c|}
\hline 
$W_{s}\left(\lambda^{2}\right)$ & MIMO & Dual-MIMO-FAS\tabularnewline
\hline 
\hline 
$0.5\times0.5$ & $4\times4=16$ & $13\times13=169$\tabularnewline
\hline 
$1\times1$ & $9\times9=81$ & $23\times23=529$\tabularnewline
\hline 
$1.5\times1.5$ & $16\times16=256$ & $34\times34=1156$\tabularnewline
\hline 
$2\times2$ & $25\times25=625$ & $48\times48=2304$\tabularnewline
\hline 
$2.5\times2.5$ & $36\times36=1296$ & $60\times60=3600$\tabularnewline
\hline 
$3\times3$ & $49\times49=2401$ & $73\times73=5329$\tabularnewline
\hline 
\end{tabular}
\begin{tablenotes}
\item The above results were obtained using the method proposed in \cite{10303274} with the assumption that $\xi=10^{-3}$, and $N_{s}=100$.
\end{tablenotes}
\par\end{centering}
\end{table}

\subsection{Multiuser \gls{MIMO}-\gls{FAS}: The \gls{ML} Approach}\label{ssec:mmimofas}
\begin{figure}
\noindent \centering{}\includegraphics[scale=0.3]{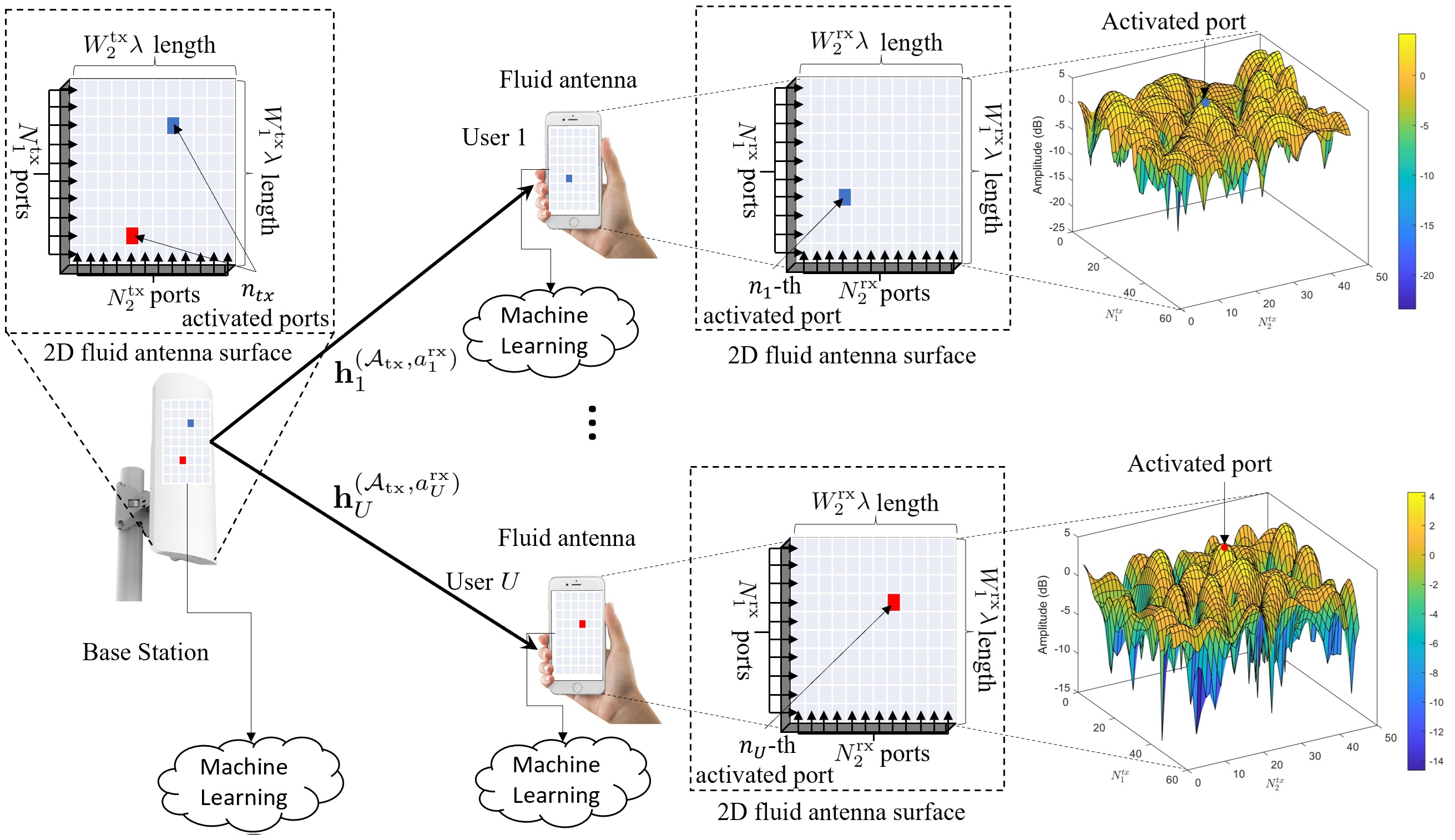}
\caption{A schematic of \gls{ML}-based multiuser dual-\gls{MIMO}-\gls{FAS} with 2D \gls{FAS}.}\label{MU_MIMO_FAS}
\end{figure}

In this subsection, we extend the concept of \gls{MIMO}-\gls{FAS} to multiuser \gls{MIMO}-\gls{FAS}, a setup applicable to cellular networks or even WiFi, as depicted in Fig.~\ref{MU_MIMO_FAS}. In contrast to the above, we shift our focus to the optimization complexity and discuss state-of-the-art methods to overcome this problem. For brevity, we assume that the \gls{BS} is equipped with a \gls{2D} fluid antenna surface with $M_{{\rm tx}}$ \gls{RF}-chains, where the dimension of the fluid antenna surface is $W_{{\rm tx}}=W_{1}^{{\rm tx}}\lambda\times W_{2}^{{\rm tx}}\lambda$ and there are a total of $N_{{\rm tx}}=N_{1}^{{\rm tx}}\times N_{2}^{{\rm tx}}$ ports. On the user side, we consider two cases: (i) a single traditional antenna with one \gls{RF}-chain and (ii) a \gls{2D} fluid antenna surface with one \gls{RF}-chain, a dimension of $W_{{\rm rx}}=W_{1}^{{\rm rx}}\lambda\times W_{2}^{{\rm rx}}\lambda$ and $N_{{\rm rx}}=N_{1}^{{\rm rx}}\times N_{2}^{{\rm rx}}$ ports. Note that the earlier represents multiuser Tx-\gls{MIMO}-\gls{FAS} and the latter represents multiuser Dual-\gls{MIMO}-\gls{FAS}. Let $\mathcal{A}=\left\{ 1,\dots,N_{{\rm tx}}\right\}$ denote the set of ports. Then the subset $\mathcal{A}_{{\rm tx}}=\left\{ a_{1}^{{\rm tx}},\dots,a_{U}^{{\rm tx}}\right\} \subset\mathcal{A}$, with a cardinality of $U$, represents the indices of the activated ports of the \gls{BS} to serve $U$ users, such that $U\le M_{{\rm tx}}$.\footnote{$\{a_{n}^{\rm tx}\}$ represent the entries where the column vectors of $\boldsymbol{\mathcal{A}}_{\rm tx}$ are unity.} Given the set $\mathcal{A}_{{\rm tx}}$, precoding is then performed to eliminate the multiuser interference based on the \gls{CSI} information of $\mathcal{A}_{{\rm tx}}$. In the following, we discuss the models of multiuser Tx-\gls{MIMO}-\gls{FAS} and multiuser Dual-\gls{MIMO}-\gls{FAS} and formulate their corresponding optimization problems that maximize their achievable sum-rates. 

\subsubsection{Multiuser Tx-\gls{MIMO}-\gls{FAS}}
Given the set $\mathcal{A}_{{\rm tx}}$, the received signal of the $u$-th user with traditional antennas is 
\begin{align}
y_{u}&=\boldsymbol{h}_{u}^{\left(\mathcal{A}_{{\rm tx}}\right)}\boldsymbol{W}^{\left(\mathcal{A}_{{\rm tx}}\right)}\boldsymbol{s}+\zeta_{u},\\
\boldsymbol{h}_{u}^{\left(\mathcal{A}_{{\rm tx}}\right)}&=\left[h_{u}\left(a_{1}^{{\rm tx}}\right),\dots,h_{u}\left(a_{U}^{{\rm tx}}\right)\right],
\end{align}
where $\boldsymbol{h}_{u}^{\left(\mathcal{A}_{{\rm tx}}\right)}$ represents the channel between user $u$ with the activated ports of the \gls{BS}, $h_{u}\left(a_{m}^{{\rm tx}}\right)$ denotes the complex channel coefficient between user $u$ and the $a_{m}^{{\rm tx}}$-th port of the \gls{BS}, and $\boldsymbol{W}^{\left(\mathcal{A}_{{\rm tx}}\right)}=\left[\boldsymbol{w}_{1}^{\left(\mathcal{A}_{{\rm tx}}\right)},\dots,\boldsymbol{w}_{U}^{\left(\mathcal{A}_{{\rm tx}}\right)}\right]\in\mathbb{C}^{U\times U}$ denotes the precoding matrix with transmit power of $P_{{\rm tx}}$. Moreover, $\boldsymbol{s}=\left[s_{1},\dots,s_{U}\right]^{T}\in\mathbb{C}^{U\times1}$ is the vector of the transmitted symbols of all the users with $\mathbb{E}\left\{ |s_{u}|^{2}\right\} =1$, and $\zeta_{u}$ is the \gls{AWGN} with $N_{0}$ being the noise power, i.e., $\zeta_{u}\sim\mathcal{CN}\left(0,N_{0}\right)$. We use the superscript $\mathcal{A}_{{\rm tx}}$ to indicate that $\boldsymbol{W}^{\left(\mathcal{A}_{{\rm tx}}\right)}$ is a function of the set $\mathcal{A}_{{\rm tx}}$.

Consequently, the \gls{SINR} of the $u$-th user is given by
\begin{equation}\label{Eq:SINR:1}
{\rm SINR}_{u}^{\left(\mathcal{A}_{{\rm tx}},\boldsymbol{W}^{(\mathcal{A}_{\rm tx})}\right)}=\frac{\left|\boldsymbol{h}_{u}^{\left(\mathcal{A}_{{\rm tx}}\right)}\boldsymbol{w}_{u}^{\left(\mathcal{A}_{{\rm tx}}\right)}\right|^{2}}{\sum_{\bar{u}\ne u}\left|\boldsymbol{h}_{u}^{\left(\mathcal{A}_{{\rm tx}}\right)}\boldsymbol{w}_{\bar{u}}^{\left(\mathcal{A}_{{\rm tx}}\right)}\right|^{2}+N_{0}}.
\end{equation}
To maximize the sum-rate of multiuser Tx-\gls{MIMO}-\gls{FAS}, we optimize the set $\mathcal{A}_{{\rm tx}}$ and the precoding matrix $\boldsymbol{W}^{\left(\mathcal{A}_{{\rm tx}}\right)}$. More concretely, the optimization problem can be formulated as 
\begin{equation}\label{eq:MIMO:1}
\max_{\mathcal{A}_{{\rm tx}},\boldsymbol{W}^{(\mathcal{A}_{\rm tx})}}\sum_{u=1}^{U}\log_{2}\left(1+{\rm SINR}_{u}^{\left(\mathcal{A}_{\rm tx},\boldsymbol{W}^{(\mathcal{A}_{\rm tx})}\right)}\right),
\end{equation}
where additional constraints can be added if needed.

\subsubsection{Multiuser Dual-\gls{MIMO}-\gls{FAS}}
In this case, we let $\mathcal{A}_{{\rm rx}}=\left\{ a_{1}^{{\rm rx}},\dots,a_{U}^{{\rm rx}}|a_{u}^{{\rm rx}}\in\left[1,N_{{\rm rx}}\right]\right\}$ represent the set of the active ports of the users, where $a_{u}^{{\rm rx}}$ denotes the index of the activated port of the $u$-th user. Moreover, let $\boldsymbol{W}^{\left(\mathcal{A}_{{\rm tx}},\mathcal{A}_{{\rm rx}}\right)}=[\boldsymbol{w}_{1}^{\left(\mathcal{A}_{{\rm tx}},\mathcal{A}_{{\rm rx}}\right)},\dots,\boldsymbol{w}_{U}^{\left(\mathcal{A}_{{\rm tx}},\mathcal{A}_{{\rm rx}}\right)}]$ be the precoding matrix that is related to both of the active ports of the \gls{BS} and users. Then the received signal of the $a_{u}^{{\rm rx}}$-th port of user $u$ is 
\begin{align}
y_u(a_{u}^{\rm rx})&=\boldsymbol{h}_{u}^{\left(\mathcal{A}_{\rm tx},a_{u}^{\rm rx}\right)}\boldsymbol{w}_{u}^{\left(\mathcal{A}_{{\rm tx}},\mathcal{A}_{{\rm rx}}\right)}\boldsymbol{s}+\zeta_u^{(a_{u}^{\rm rx})},\label{eq:yu}\\
\boldsymbol{h}_{u}^{\left(\mathcal{A}_{\rm tx},a_{u}^{\rm rx}\right)}&=\left[h_{u}\left(a_{1}^{{\rm tx}},a_{u}^{\rm rx}\right),\dots,h_{u}\left(a_{U}^{\rm tx},a_{u}^{\rm rx}\right)\right],
\end{align}
where $h_{u}\left(a_{m}^{{\rm tx}},a_{u}^{{\rm rx}}\right)$ denotes the complex channel coefficient between the $a_{u}^{{\rm rx}}$-th port of user $u$ and the $a_{m}^{{\rm tx}}$-th port of the \gls{BS}, while $\zeta_{u}^{(a_{u}^{\rm rx})}\sim\mathcal{CN}\left(0,N_{0}\right)$ is the \gls{AWGN} at the $a_{u}^{{\rm rx}}$-th port of user $u$. The \gls{SINR} of the $a_{u}^{{\rm rx}}$-th port of user $u$ is
\begin{multline}
{\rm SINR}_{u}^{\left(\mathcal{A}_{{\rm tx}},a_{u}^{{\rm rx}},\boldsymbol{W}^{\left(\mathcal{A}_{{\rm tx}},\mathcal{A}_{{\rm rx}}\right)}\right)}\\
=\frac{\left|\boldsymbol{h}_{u}^{\left(\mathcal{A}_{{\rm tx}},a_{u}^{\rm rx}\right)}\boldsymbol{w}_{u}^{\left(\mathcal{A}_{{\rm tx}},\mathcal{A}_{{\rm rx}}\right)}\right|^{2}}{\sum_{\bar{u}\ne u}\left|\boldsymbol{h}_{u}^{\left(\mathcal{A}_{{\rm tx}},a_{u}^{\rm rx}\right)}\boldsymbol{w}_{\bar{u}}^{\left(\mathcal{A}_{{\rm tx}},\mathcal{A}_{{\rm rx}}\right)}\right|^{2}+N_{0}}.
\end{multline}
Then we can maximize the achievable sum-rate of multiuser Dual-\gls{MIMO}-\gls{FAS}:
\begin{align}\label{Eq:MIMO:2}
\max_{\mathcal{A}_{{\rm tx}},\mathcal{A}_{rx},\atop \boldsymbol{W}^{\left(\mathcal{A}_{{\rm tx}},\mathcal{A}_{{\rm rx}}\right)}}\sum_{u=1}^{U}\log_{2}\left(1+{\rm SINR}_{u}^{\left(\mathcal{A}_{{\rm tx}},a_{u}^{{\rm rx}},\boldsymbol{W}^{\left(\mathcal{A}_{{\rm tx}},\mathcal{A}_{{\rm rx}}\right)}\right)}\right).
\end{align}
As expected, both (\ref{eq:MIMO:1}) and (\ref{Eq:MIMO:2}) pose optimization challenges because selecting the subset $\mathcal{A}_{{\rm tx}}$ from $\mathcal{A}$ involves a combinatorial optimization problem, and the design of the precoding matrix is intricately coupled with the activated ports, i.e., $\mathcal{A}_{{\rm tx}}$.

Besides mathematical optimization approaches, we can use deep learning to solve more complicated problems \cite{AI_FAS}. For instance, \cite{9715064} used a mixture of \gls{ML} techniques, including the \gls{SPO} framework \cite{doi:10.1287/mnsc.2020.3922} to create algorithms that can select the best port with minimal observations. These methods significantly reduce outage probability, even with just one port observation. Meanwhile, \cite{10299674} proposed a novel online learning framework to address port selection challenges in time-varying channel conditions. Leveraging bandit learning, the algorithm dynamically learns the optimal port selection without relying on full instantaneous \gls{CSI}. Exploring the feasibility of reinforcement learning for addressing combination optimization challenges in this context is of great interest. One viable approach involves treating the \gls{BS} as an agent within a reinforcement learning framework \cite{bello2017neural}, utilizing metrics such as outage probability as rewards for non-gradient-based optimization of the port selection model. In scenarios where both users and \gls{BS} are equipped with fluid antennas, the application of multi-agent reinforcement learning \cite{9839183} represents a potential avenue. More concretely, it offers the prospect of enabling individual users to make informed port selection decisions without requiring knowledge of the \gls{CSI} of other users. Recently, a successful application appears in \cite{Waqar-2024} where deep reinforcement learning has been shown to be effective in tackling the distributed optimization problem for opportunistic \gls{FAMA}. Furthermore, \cite{Waqar-2024} also managed to incorporate game theory into deep learning for improving the learning efficiency for self-optimizing \gls{FAS} users.

\begin{figure*}[t]
\centering \includegraphics[width=1\linewidth]{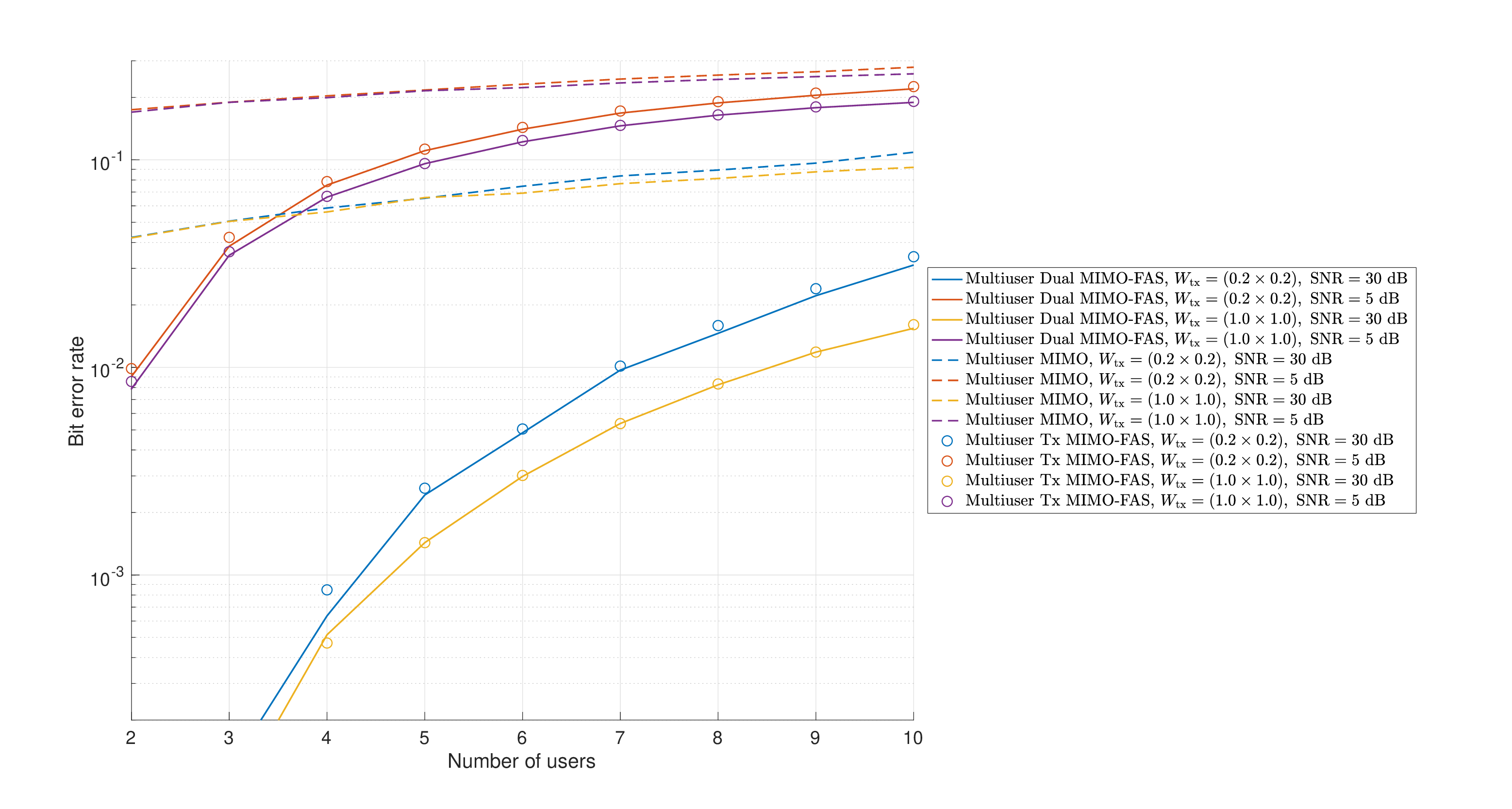}
\caption{Bit error rate versus the number of users with different antenna configurations.}\label{fig:BER}
\end{figure*}

To gain valuable insights into the new flexibility introduced by multiuser \gls{MIMO}-\gls{FAS} as well as its performance improvements compared to the traditional multiuser \gls{MIMO} system, we simplify (\ref{eq:MIMO:1}) and (\ref{Eq:MIMO:2}) by employing \gls{RZF} precoding. In particular, we consider a scenario with a carrier frequency of $4.5~{\rm GHz}$ and the \gls{BS} is equipped with \gls{2D} fluid antenna surface with $30\times30$ ports and a dimension of $W_{{\rm tx}}=1{\rm m}\times1{\rm m}$ or $W_{{\rm tx}}=0.2{\rm m}\times0.2{\rm m}$. The dimension of the user \gls{2D}'s fluid antenna surface is $0.1{\rm m}\times0.1{\rm m}$ and the number of ports is $3\times3$. To optimize the performance, we use multi-agent reinforcement learning in \cite{ISAC_FAS}. Here, we consider quaternary phase shift keying (QPSK) modulation. Fig.~\ref{fig:BER} shows the average bit error rate versus the number of users for different $W_{{\rm tx}}$ and \gls{SNR}. As it is seen, the performance of \gls{FAS} is related to $W_{{\rm tx}}$, and a larger $W_{{\rm tx}}$ can offer higher diversity gain as discussed previously. Furthermore, the gain is more significant at high \gls{SNR}. For a more comprehensive discussion and coverage, readers may refer to \cite{ISAC_FAS}. 

\begin{figure}[t]
\begin{centering}
\subfloat[]{\centering{}\includegraphics[scale=0.6]{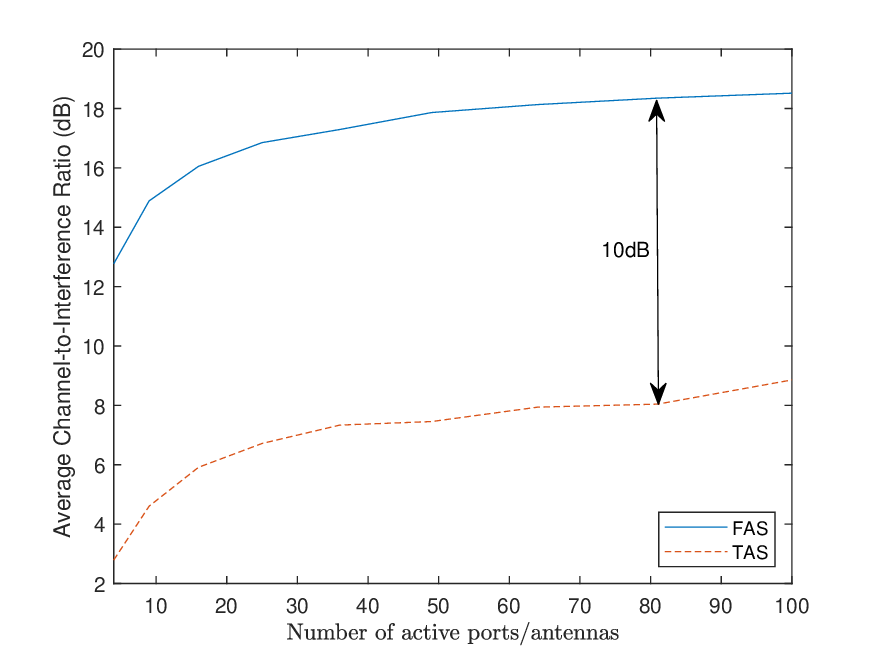}}
\par\end{centering}
\centering{}\subfloat[]{\noindent \includegraphics[scale=0.6]{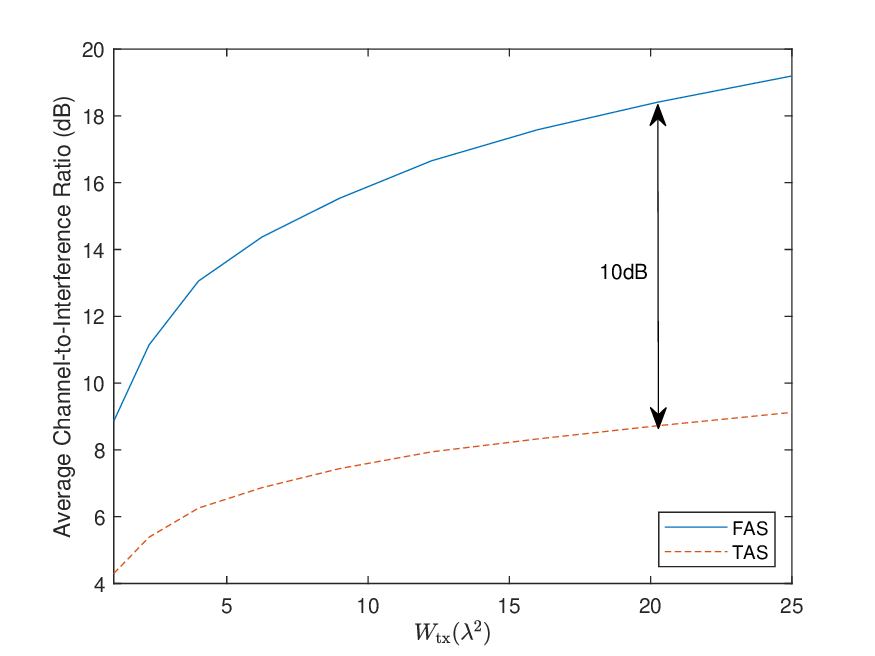}}
\caption{The average channel-to-interference ratio comparison of multiuser \gls{MIMO}-\gls{FAS} and multiuser \gls{MIMO}-\gls{TAS}: a) the effect of active ports/antennas, where $N_{{\rm tx}}=50\times 50$ and $W_{{\rm tx}}=4.5\lambda\times 4.5\lambda$; and b) the effect of $W_{{\rm tx}}$, where $N_{{\rm tx}}=10n_{\rm tx}\times 10n_{\rm tx}$ and $n_{{\rm tx}}={\scriptstyle \left(\left\lfloor \frac{2\sqrt{W_{{\rm tx}}}}{\lambda}\right\rfloor +1\right)}^{2}$.}\label{Multiuser_MIMO}
\end{figure}

The unique advantages of multiuser \gls{MIMO}-\gls{FAS} over traditional multiuser \gls{MIMO} also extend to the newfound capabilities in \emph{flexible beamforming} and \emph{rapid channel hardening}. In Fig.~\ref{Multiuser_MIMO}, we compare the average channel-to-interference ratio between \gls{FAS} and \gls{TAS} in rich scattering environments. Given the ability to finely reconfigure the positions of radiating elements within a given space, it is evident that the average channel-to-interference ratio of \gls{FAS} can be significantly enhanced as compared to \gls{TAS} when considering an equal number of active ports or antennas. Interestingly, the gain of \gls{FAS} increases with the size of the fluid antenna surface.

\begin{figure}
\centering{}\includegraphics[scale=0.6]{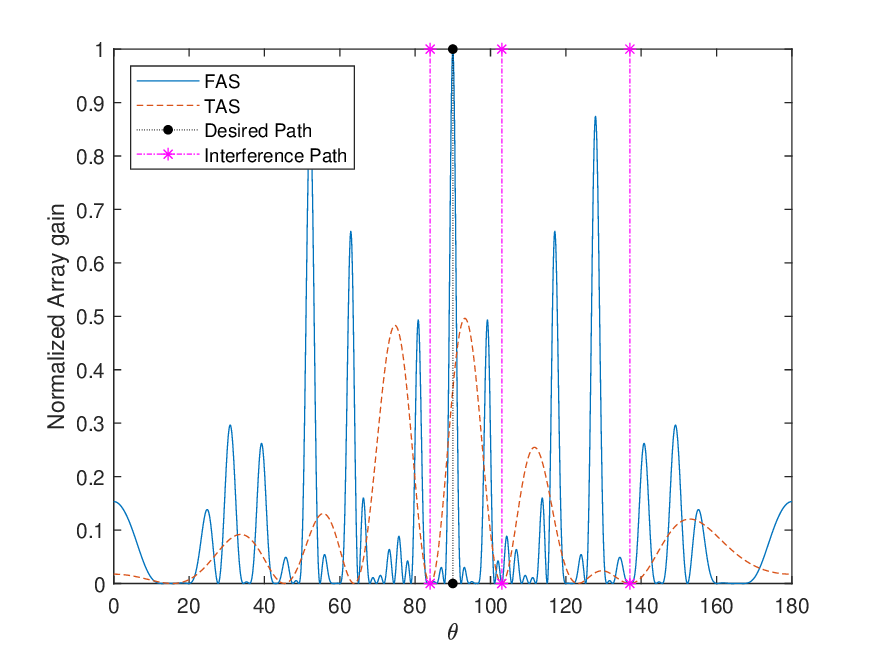}
\caption{Normalized array gain of \gls{FAS} and \gls{TAS} in different directions, with $n_{\rm tx}=8$, $\theta_{0,{\rm des}}=90^\circ$ (\gls{AoA} of the desired user signal), $\theta_{{\rm 1,int}}=84^\circ$ (\gls{AoA} of the first interfering signal), $\theta_{2,{\rm int}}=103^\circ$ (\gls{AoA} of the second interfering signal), and $\theta_{{\rm 3,int}}=107^\circ$ (\gls{AoA} of the third interfering signal).}\label{fig:Array_gain}
\end{figure}

To understand the underlying principle, let us examine their behaviors more closely in a finite scattering environment. As studied in \cite{10278220}, \gls{TAS} in some cases may lose array gain over the desired signal when nulling the interfering directions due to the fixed positions of radiating elements. In contrast, \gls{FAS} can achieve the full array gain in these cases, giving rise to superior performance. Fig.~\ref{fig:Array_gain} illustrates an example of such cases, in which the positions of radiating elements in TAS are separated by half a wavelength, and the positions in \gls{FAS} are optimally reconfigured using the closed-form expressions from \cite{10278220}. Note that this capability can be extended to multi-beam forming at a cost of some array gains, as studied in \cite{10382559}. 

\begin{figure}
\centering{}\includegraphics[scale=0.6]{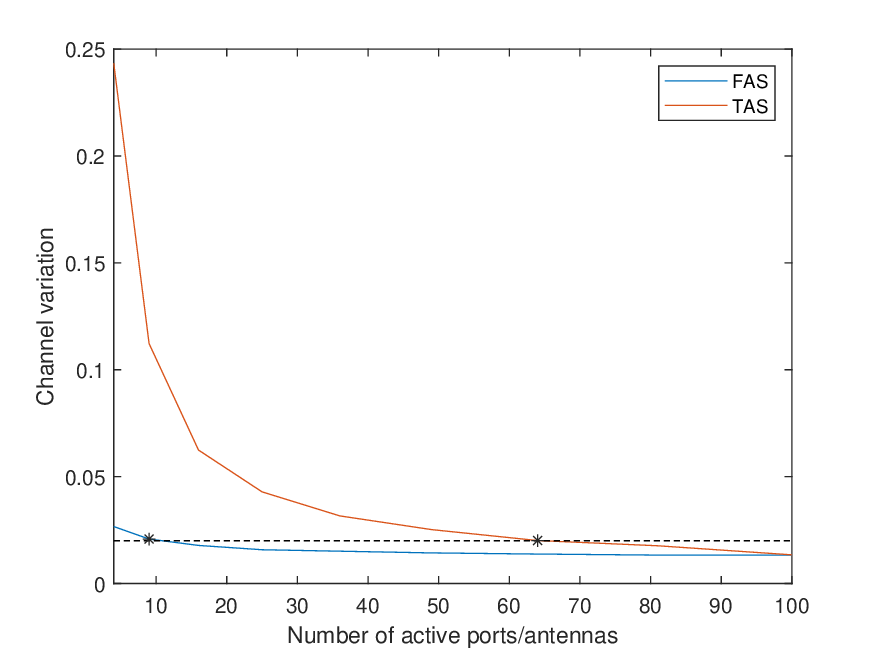}
\caption{Channel variation comparison for \gls{FAS} and \gls{TAS} versus the number of active ports/antennas, where $N_{\rm tx}=50\times 50$ and $W_{\rm tx}=4.5\lambda\times 4.5\lambda$.}\label{fig:Ch_hardening}
\end{figure}

Moreover, \gls{FAS} enjoys rapid channel hardening with significantly fewer radiating elements as compared to \gls{TAS}. Unlike \gls{TAS}, which relies on the law of large number, \gls{FAS} exploits the extreme value theorem to realize strong channel hardening \cite{zheng2023flexibleposition}. Fig.~\ref{fig:Ch_hardening} provides an example of the channel hardening effect in \gls{FAS} and \gls{TAS}. To achieve a $0.02$ channel variation, \gls{FAS} only requires $9$ active radiating elements, compared to $64$ fixed-position antennas in \gls{TAS}. Hence, by reconfiguring radiating elements, \gls{FAS} improves the performance of cellular networks in new ways. In the subsequent section, we explore multiuser \gls{FAS} communications more extensively. 

\section{New Methods for Multiple Access}\label{sec:fama}
With position flexibility, \gls{FAS} can innovate multiple access and change the way in which interference is mitigated. This new technique is referred to as \gls{FAMA} \cite{9650760,10066316}. Specifically, unlike existing techniques, \gls{FAMA} does not require any \gls{CSI} at the transmitter nor \gls{SIC} at the receiver. The idea behind \gls{FAMA} is to enable receiver to access a desirable spatial moment for communications where the interference suffers from deep fade. This is in contrast to usual spatial multiplexing schemes where signals are carefully mixed to create artificial interference null, as in the case of multiuser or massive MIMO.

\gls{FAMA} can be categorized into two types: slow \gls{FAMA} \cite{10066316,Waqar-2024,10279640,hao2024revisit,10018377,10354059,10122674} and fast \gls{FAMA} \cite{9650760,9953084,https://doi.org/10.1049/cmu2.12592}. In slow \gls{FAMA}, the receiver's antenna position changes whenever the channel changes while in fast \gls{FAMA}, the receiver changes its position on a symbol-by-symbol basis. In both schemes, the interference is treated as noise and thus no advanced signal processing is required. To understand the working principle, in Section \ref{ssec:HK}, we first consider a simple scenario where there are only two pairs of transmitters and receivers. By leveraging information theory, we will show that slow \gls{FAMA} can be capacity-near-optimal. A general setup with any number of users will be considered in Section \ref{ssec:sfama} via some approximation techniques. In Section \ref{ssec:ffama}, our attention then turns to the fast \gls{FAMA} approach for symbol-level port switching for extreme massive connectivity. It is worth noting that fast \gls{FAMA} has the capability of accommodating hundreds of users on the same channel while slow \gls{FAMA} is typically not expected to handle $10$ users or more. Finally, we introduce \gls{CUMA}, a variant of slow \gls{FAMA}, that improves the multiple access capability with slow port switching. We conclude this section by commenting on how \gls{FAMA} may inspire a scalable new multiple access scheme for future mobile networks. 

\subsection{\gls{HK} Comparison}\label{ssec:HK}
\begin{figure}
\centering{}\includegraphics[scale=0.32]{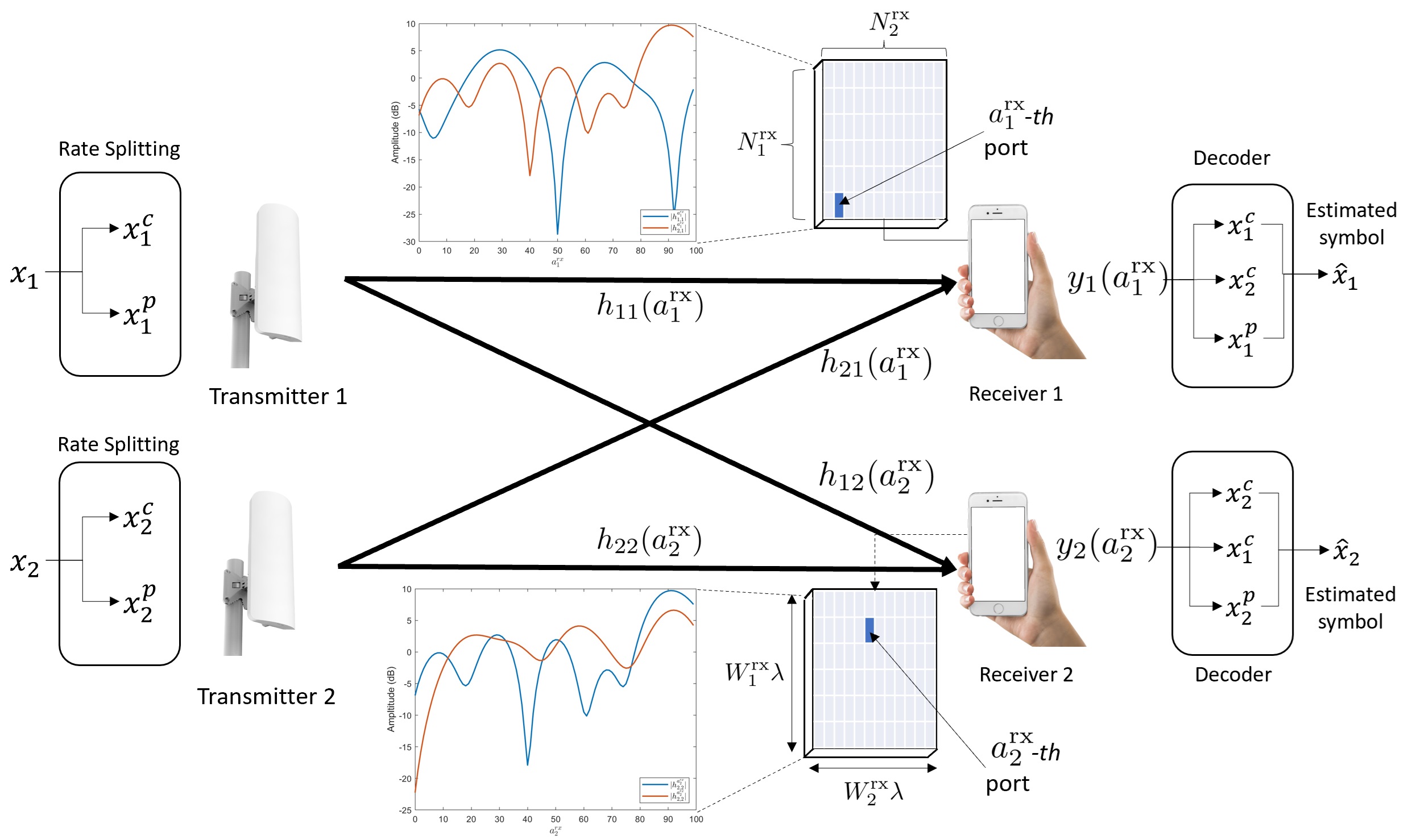}
\caption{A schematic of \gls{HK}-\gls{FAMA} for an interference channel with two transmitter-receiver pairs.}\label{fig:HK_FAMA}
\end{figure}

Consider an interference channel in Fig.~\ref{fig:HK_FAMA}, in which there are two pairs of transmitters and receivers. Each transmitter aims to send the intended information signal to its respective receiver but it interferes with the other receiver. To simplify discussion, we assume that each transmitter is equipped with a traditional fixed-position antenna while each receiver has a \gls{2D} \gls{FAS} with $N_{{\rm rx}}$ ports uniformly distributed in a grid structure over an area of $W_{{\rm rx}}$. This setup can be viewed as the classical two-user Gaussian interference channel if each receiver uses a fixed port. In information theory, it is widely known that the best achievable scheme of a two-user Gaussian interference channel is the \gls{HK} scheme which uses rate-splitting and joint-decoding to perform all possible strategies \cite{4675741,9314200}. It is capable of approaching the capacity to within one bit for all values of channel configurations by setting the interference to noise level \cite{4675741}. Intuitively, given the availability of global \gls{CSI} at the transmitters and receivers, it is possible to improve the rate performance when optimal rate-splitting and power-splitting are computed for each port and the optimal port is selected for communications. We refer to this scheme as \gls{HK}-\gls{FAMA}, which is mainly used for benchmarking \cite{HKFAMA}. 

To understand the working principle of \gls{FAMA}, we introduce the concept of \gls{gdof} in non-symmetric reconfigurable channel for \gls{FAS} with finite \gls{SNR}.\footnote{Note that channel reconfiguration is only useful here when the SNR is finite and not asymptotically high.} Specifically, the \gls{gdof} of a scheme is defined as 
\begin{equation}
{\rm {gdof}}\triangleq\frac{R_{{\rm {sys}}}\left(\mathcal{A}_{{\rm rx}}\right)}{C^{*}},
\end{equation}
where 
\begin{multline}
C^{*}=\max_{\mathcal{A}_{\rm rx}}\log\left(1+\frac{P_{{\rm tx}}\left|h_{11}(a_{1}^{\rm rx})\right|^{2}}{N_{0}}\right)\\
+\log\left(1+\frac{P_{\rm tx}\left|h_{22}(a_{2}^{\rm rx})\right|^{2}}{N_{0}}\right),
\end{multline}
$\left|h_{uu}(a_{u}^{\rm rx})\right|^{2}$ is the channel gain between the transmitter and the $a_{u}^{{\rm rx}}$-th port of the desired receiver, and $R_{\rm sys}\left(\mathcal{A}_{\rm rx}\right)$ is the maximum system sum-rate as a function of the optimal port. The \gls{gdof} can be interpreted as the ratio of the maximum sum-rate of a system to the maximum sum-rate without interference. Thus, we say that the system has full dof if $R_{{\rm {sys}}}\left(\mathcal{A}_{{\rm rx}}\right)=C^{*}$, i.e., the interference has no effect on the receivers. In contrast, the system has zero dof if $R_{{\rm {sys}}}\left(\mathcal{A}_{rx}\right)=0$, i.e., the interference affects the system to the extent that no communication is possible. In addition, we find it useful to define $\alpha$ and $\beta$ as the ratio of interference-to-noise ratio to \gls{SNR}, in decibels. These variables represent the interference level. If $\alpha=\beta$, we have a symmetric interference channel. If $\alpha$ or $\beta\geq0$, the system is operating in the interference-limited regime. Otherwise, it is operating in the noise-limited regime where noise dominates. 

\begin{figure}[t]
\begin{centering}
\subfloat[]{\centering{}\includegraphics[scale=0.6]{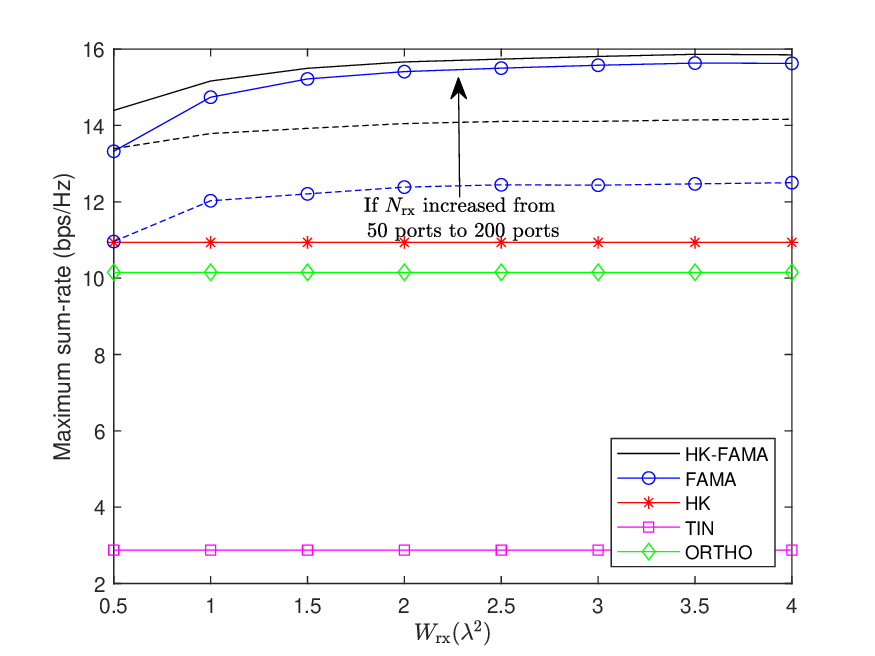}}
\par\end{centering}
\begin{centering}
\subfloat[]{\noindent \includegraphics[scale=0.6]{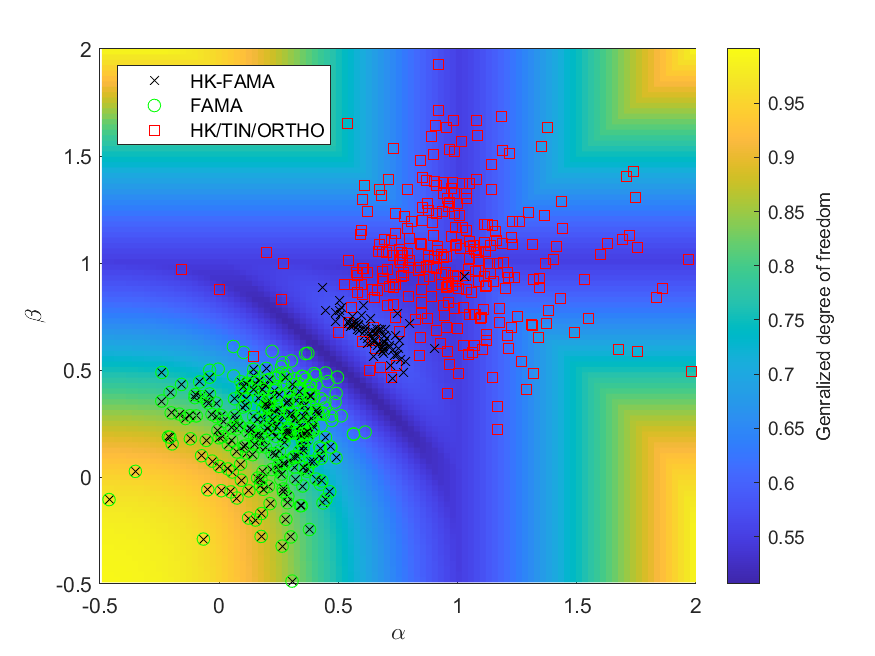}}
\caption{The performance of \gls{HK}-\gls{FAMA}, \gls{FAMA} and other existing schemes: a) the effect of $N_{{\rm rx}}$ and $W_{{\rm rx}}$ on the sum-rate, where $W_{2}^{{\rm rx}}=\lambda$ and $N_{2}^{{\rm rx}}=10$; b) the maximum gdof over $300$ independent channel realizations.}\label{HK_FAMA}
\par\end{centering}
\vspace{-.5cm}
\end{figure}

We illustrate the sum-rate performance of \gls{HK}-\gls{FAMA} and \gls{FAMA} as well as other existing schemes such as \gls{HK}, \gls{ORTHO} and \gls{TIN} for the two-user interference channel in Fig.~\ref{HK_FAMA}. Note that the performance of \gls{HK}, \gls{ORTHO} and \gls{TIN} can also be obtained in \gls{FAS} by always activating the same port or by setting $N_{{\rm rx}}=1$. As shown in Fig.~\ref{HK_FAMA}(a), the performance of \gls{FAMA} approaches to that \gls{HK}-\gls{FAMA} as $N_{{\rm rx}}$ and $W_{{\rm rx}}$ increase, meaning that \gls{HK} is unnecessary and \gls{FAMA} can be near-optimal if $N_{\rm rx}$ and $W_{\rm rx}$ are sufficiently large. To help understand this, we investigate the maximum \gls{gdof} that can be achieved by these schemes over different realizations as demonstrated in Fig.~\ref{HK_FAMA}(b). Here, \gls{HK}-\gls{FAMA} and \gls{FAMA} have the ability to reconfigure the channel in each channel realization (i.e., they can adjust the values of $\alpha$ and $\beta$ by selecting a different port while \gls{HK}, \gls{ORTHO} and \gls{TIN} cannot). It can be observed that \gls{HK}-\gls{FAMA} and \gls{FAMA} tend to make the values of $\alpha$ and $\beta$ small to obtain a higher \gls{gdof}, implying that the noise-limited regime is operationally more desirable if such channel reconfiguration is possible. In contrast, \gls{HK}, \gls{ORTHO} and \gls{TIN} are unable to reconfigure the channel because the active port is fixed or $N_{\rm rx}=1$. In other words, the performance of those schemes are limited by the randomness of the channel configuration in each realization while \gls{FAS} provides receivers the unique ability to reconfigure the channel to operate in the noise-limited regime. 

\subsection{Slow FAMA: Approximation Techniques}\label{ssec:sfama}

\begin{figure}
\centering \includegraphics[scale=0.4]{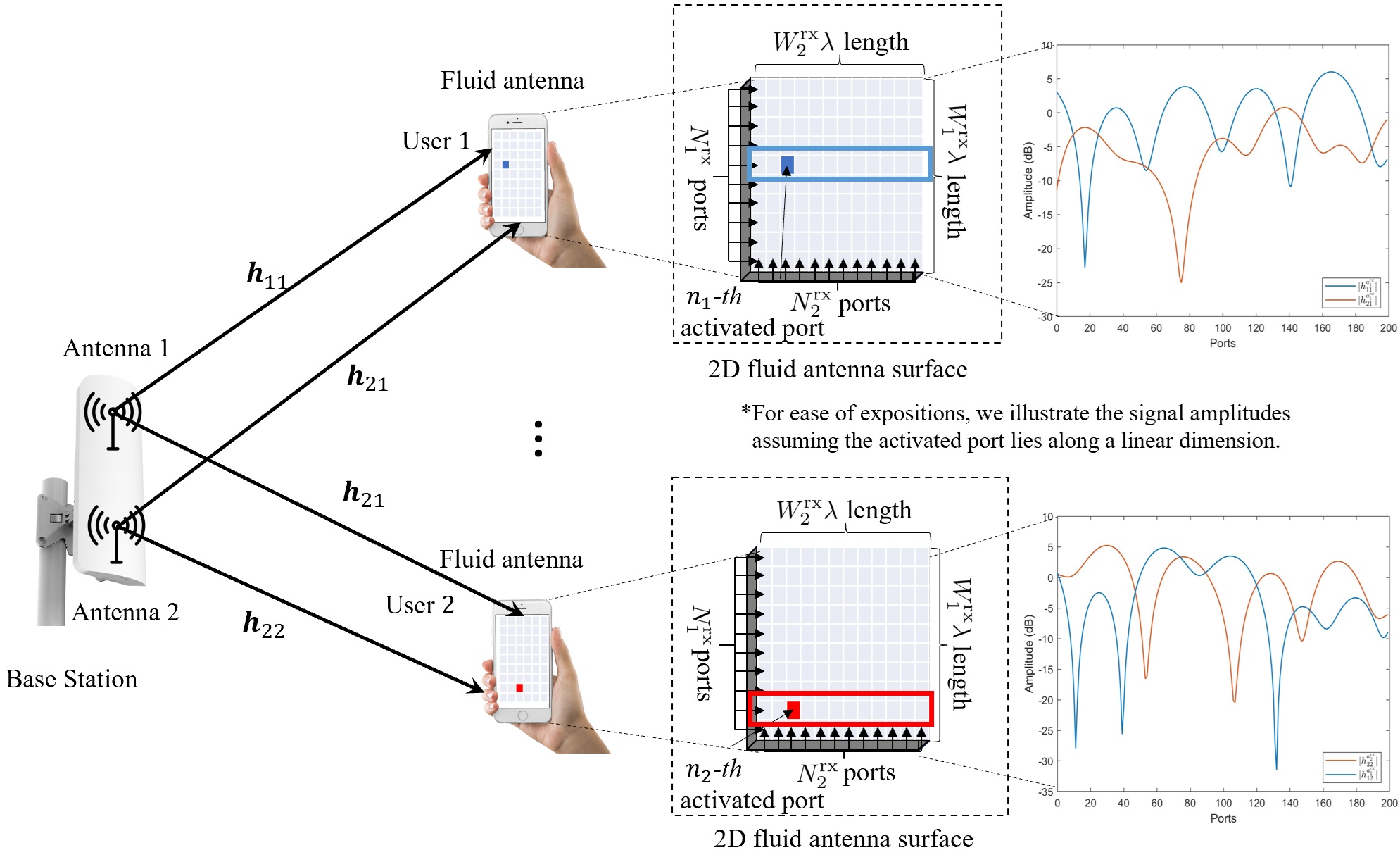} 
\caption{A schematic of downlink slow \gls{FAMA}, where a \gls{BS} is equipped with two traditional fixed-position antennas and the two users are equipped with fluid antennas.}\label{CM_slow_FAMA}
\end{figure}

Here, we continue our discussion of \gls{FAMA} in the case of two users but deviate from the capacity-centric \gls{HK} comparison. The setup depicted in Fig.~\ref{CM_slow_FAMA} is considered, in which each transmitter corresponds to a traditional fixed-position antenna of a \gls{BS} and each user has a \gls{2D} \gls{FAS}. A virtue of \gls{FAMA} is that \gls{CSI} is no longer necessary at the transmitter side and the receiver ends also do not rely on \gls{SIC}.\footnote{Evidently, \gls{FAS} can also be used as a new dof to improve existing multiple access schemes, as discussed in the case of NOMA in Section \ref{ssec:nomafas} and multiuser \gls{MIMO} in Section \ref{ssec:mmimofas}.} However, in \gls{FAMA}, interference is not completely eliminated. The interest would be to understand how much the overall performance is affected by the interference. In this model, the $u$-th user is exclusively served by antenna $u\in\{1,2\}$ with a transmit power of $P_{u}$. In the interference-limited regime, where the interference level is much greater than the noise power level, the \gls{SINR} of the $u$-th user at the $a_{u}^{\rm rx}$-th port can be approximated as\footnote{We refer to this as the signal-to-interference ratio approximation.} 
\begin{equation}\label{SINR_kn}
{\rm SINR}_u(a_u^{\rm rx})\approx\frac{P_{u}\left|h_{uu}(a_{u}^{\rm rx})\right|^{2}}{P_{\bar{u}}\left|h_{\bar{u}u}(a_{u}^{\rm rx})\right|{}^{2}},
\end{equation}
where ${\bar{u}}$ is the complement of $u$ and $h_{\bar{u}u}(a_{u}^{\rm rx})$ is the complex channel coefficient from the $\bar{u}$-th antenna to the $a_{u}^{{\rm rx}}$-th port of user~$u$. The performance of user $u$ can be generally evaluated using the outage probability, approximately given as 
\begin{align}
 & \mathbb{P}_{{\rm out},u}^{\rm FAMA}\left(R_{u}\right)\notag\\
\approx & \mathbb{P}\left\{ \max\left\{ \frac{\left|h_{uu}(1)\right|^{2}}{\left|h_{\bar{u}u}(1)\right|{}^{2}},\dots,\frac{\left|h_{uu}(N_{{\rm rx}})\right|^{2}}{\left|h_{\bar{u}u}(N_{{\rm rx}})\right|{}^{2}}\right\} <\frac{\gamma_{\rm th}P_{\bar{u}}}{P_{u}}\right\} \notag\\
= & \mathbb{P}\left\{ \frac{\left|h_{uu}(1)\right|}{\left|h_{\bar{u}u}(1)\right|}<q_{u},\dots,\frac{\left|h_{uu}(N_{{\rm rx}})\right|^{2}}{\left|h_{\bar{u}u}(N_{{\rm rx}})\right|{}^{2}}<R_{u}\right\},
\end{align}
where $\gamma_{\rm th}$ represents the minimum \gls{SINR} threshold without outage and $R_{u}\triangleq\sqrt{\gamma_{\rm th}P_{\bar{u}}/P_{u}}$. Similar to (\ref{eq:h_jk}), we can define $\bm{h}_{\bar{u}u}=[h_{\bar{u}u}(1),\dots,h_{\bar{u}u}(N_{{\rm rx}})]^{T}$ and $\bm{K}_{\bar{u}u}=\xi_{\bar{u}u}^{2}\bm{J}_{u}$ as the covariance matrix of $\bm{h}_{\bar{u}u}$, where $\xi_{\bar{u}u}^{2}$ denotes the large-scale fading from the $\bar{u}$-th antenna to user $u$. According to \cite{10103838,hao2024revisit}, $\bm{h}_{\bar{u}u}$ can be generated through the eigenvalue decomposition on $\bm{J}_{u}$ and the introduction of $2N_{{\rm rx}}$ \gls{i.i.d.} Gaussian random variables. We refer to this as the ``exact channel model''. Under this model, $\mathbb{P}_{{\rm out},u}^{\rm FAMA}\left(R_{u}\right)$ can be analyzed analytically but the obtained expression involves $N_{\rm rx}$ nested integrals, which are computationally intractable \cite{hao2024revisit}. To overcome this, approximation techniques are required.

Utilizing the strategies in \cite{10103838}, the exact channel model is approximated in two stages, leading to the approximations of outage probability in \cite{hao2024revisit}. In the first stage, it is recognized that the exact channel model is mainly determined by a few largest eigenvalues. This allows the approximation of each channel coefficient by considering only $\hat{N}_{\rm rx}$ dominant eigenvalues, where $\hat{N}_{\rm rx}$ is considerably smaller than the number of ports $N_{\rm rx}$. This is known as the first-stage approximation, providing a closed-form expression for the outage probability. However, despite the significant simplification achieved in this stage, the approximated outage probability remains challenging to compute since it involves a $4\hat{N}_{\rm rx}$-fold integral.

To further simplify the analysis in the second stage, we begin by defining a random matrix ${\hat{\bm{H}}}_{\bar{u}u}$ of size $N_{{\rm rx}}\times N_{1}$. This matrix serves as as an $N_{1}$-dimensional extension of ${\hat{\bm{h}}}_{\bar{u}u}$, where each column of ${\hat{\bm{H}}}_{\bar{u}u}$ shares the same distribution as ${\hat{\bm{h}}}_{\bar{u}u}$, and different columns are statistically independent. The parameter $N_{1}$ plays a key role in influencing the accuracy of the approximation and requires careful design. Then, another random matrix ${\bar{\bm{H}}}_{\bar{u}u}$ of the same size, featuring independent rows and dependent columns, is introduced. The similarity between ${\bar{\bm{H}}}_{\bar{u}u}$ and ${\hat{\bm{H}}}_{\bar{u}u}$ is quantified by measuring the distance between their covariance matrices, a metric minimized through the appropriate design of $N_{1}$. This new model allows for an approximation of the outage probability expressed as a $2$-fold integral in closed form, which is easy to compute.

\begin{figure}
\centering \includegraphics[scale=0.6]{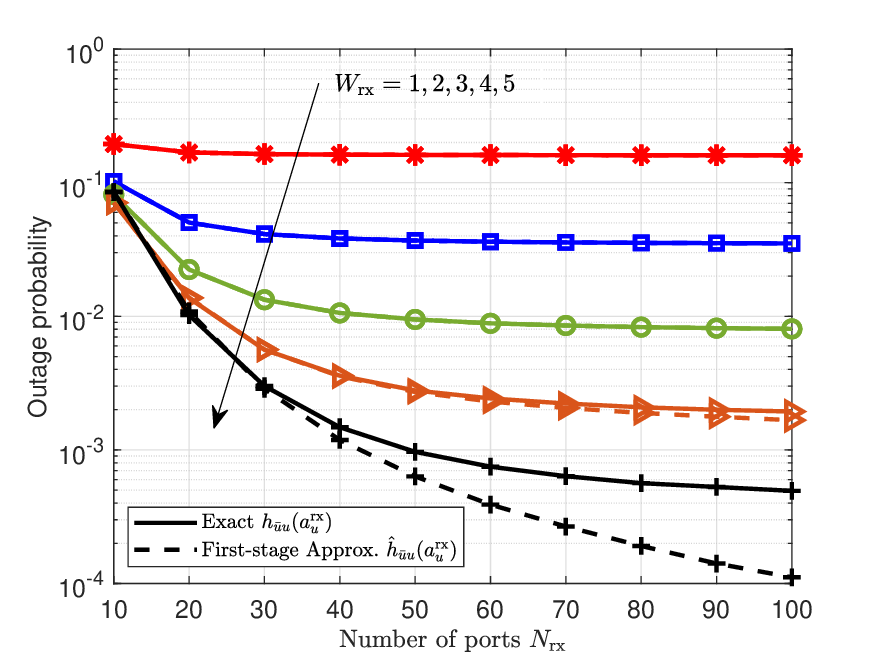} 
\caption{Outage probability and the first-stage approximation, where $\gamma_{\rm th}=5~{\rm dB}$, $P_{\bar{u}}=P_{u}$, and $\hat{N}_{\rm rx}=10$.}\label{Pr_out_VS_N}
\end{figure}

\begin{figure}
\centering \includegraphics[scale=0.6]{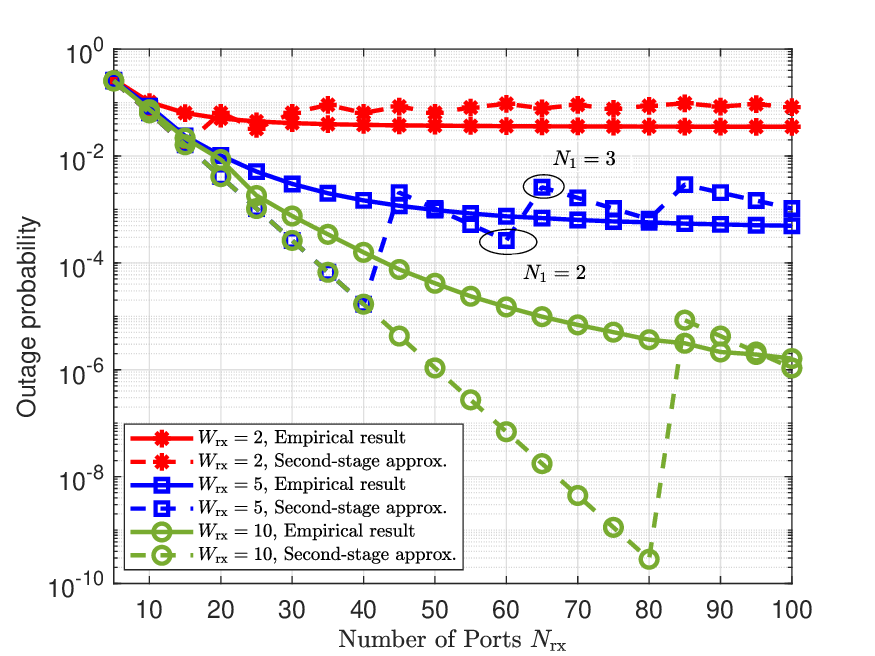} 
\caption{Outage probability and the second-stage approximation, where $\gamma_{\rm th}=5~{\rm dB}$, $P_{\bar{u}}=P_{u}$ and $\hat{N}_{{\rm rx}}=10$.}\label{Empi_VS_seco_N}
\end{figure}

In Figs.~\ref{Pr_out_VS_N} and \ref{Empi_VS_seco_N}, the approximations are evaluated. In the simulation results, we consider a \gls{1D} \gls{FAS} with $N_{{\rm rx}}$ ports and a length of $W_{{\rm rx}}\lambda$ at each user. As observed, when $W_{{\rm rx}}$ is small (e.g., $W_{{\rm rx}}=1$), the outage probability remains almost constant as $N_{{\rm rx}}$ increases. In contrast, when $W_{{\rm rx}}$ is sufficiently large, the outage probability initially decreases greatly with $N_{{\rm rx}}$ before it gradually saturates. This observation reveals that an excessive increase in $N_{{\rm rx}}$ does not yield additional gains when $W_{{\rm rx}}$ is fixed, which resembles the behavior observed in a point-to-point \gls{FAS} where only a maximum diversity can be achieved for a fixed $W_{{\rm rx}}$. Evidently, the results illustrate that the outage probability significantly decreases as $W_{{\rm rx}}$ increases. This indicates that increasing the size of \gls{FAS} can significantly enhance the performance of \gls{FAMA}, especially when $W_{{\rm rx}}$ is small. The comparison between the results in the two figures also discovers distinct performance characteristics of the two approximation strategies. With a well-designed $N_{1}$, the curves obtained by the first-stage approximation almost coincide with those obtained by the exact channel model. As for the second-stage approximation, despite being not as accurate as the first scheme, it exhibits a relatively good performance when the outage probability is above $10^{-6}$. Readers are referred to \cite{hao2024revisit} for more technical details and mathematical expressions. 
 
Certainly, \gls{FAMA} can handle more than two users, and as considered above, this belongs to the slow version of \gls{FAMA} where the port is selected based on maximizing the received \gls{SINR} at each user. That is, at the $u$-th user, it aims to find
\begin{equation}\label{eq:sfama-select}
(a_u^{\rm rx})^*=\arg\max_{a_u^{\rm rx}\in\{1,\dots,N_{\rm rx}\}}\frac{P_u|h_{uu}(a_u^{\rm rx})|^2}{\sum_{\tilde{u}\ne u}P_{\tilde{u}}|h_{\tilde{u}u}(a_u^{\rm rx})|^2+N_0}.
\end{equation}

In the general case with any number of users, the outage probability for slow \gls{FAMA} has been analyzed in \cite{10066316} and the performance analysis of \gls{FAMA} in conjuction with opportunistic scheduling was also further given in \cite{10078147} but under the simplified channel model in Section \ref{ssec:scmodel}. Most recently, an accurate performance evaluation for the general slow FAMA was accomplished using the new block-correlation model in \cite{ramirez2024new}. \gls{ML}-based approaches that perform joint optimization of port selection and scheduling can also be found in \cite{Waqar-2024}. 

In short, \emph{the huge diversity gain of \gls{FAS} can be exploited and translated into an uncanny ability to mitigate interference}, under the concept of \gls{FAMA} without requiring \gls{SIC} at the users nor \gls{CSI} at the transmitter for precoding optimization.

\subsection{Fast \gls{FAMA}: Symbol-Level Switching}\label{ssec:ffama}
In fact, the first version of \gls{FAMA} in \cite{9650760} implies fast port switching on a per-symbol basis. This is in contrast to the slow \gls{FAMA} approach \cite{10066316} that chooses the best port to adapt to the instantaneous channel conditions for the maximum received \gls{SINR}, i.e., (\ref{eq:sfama-select}). If the \gls{CSI} remains unchanged, slow \gls{FAMA} keeps the same port selection, which is understandably more practical. This is not the case for fast \gls{FAMA}, in which each \gls{FAS}-enabled user chooses the port that maximizes the ratio between the instantaneous desired user's signal energy and the energy of the instantaneous sum-interference plus noise signal for a much approved multiple access capability, i.e.,
\begin{equation}\label{eq:ffama-select}
(a_u^{\rm rx})^*=\arg\max_{a_u^{\rm rx}\in\{1,\dots,N_{\rm rx}\}}\frac{|h_{uu}(a_u^{\rm rx})|^2}{\left|\sum_{\tilde{u}\ne u}h_{\tilde{u}u}(a_u^{\rm rx})s_{\tilde{u}}+\zeta_u(a_u^{\rm rx})\right|^2}.
\end{equation}

A major difference from (\ref{eq:sfama-select}) is that here in fast FAMA, the solution to (\ref{eq:ffama-select}) is data-dependent and specifically a function of the data from the interfering users and even the noise sample. In terms of interference mitigation, fast FAMA exceeds slow FAMA by a huge margin. The reason is that for slow FAMA, it relies on the existence of some port where the sum of the interference power is weak, which is not likely as the number of interfering users becomes large. On the contrary, the sum-interference plus noise signal is complex Gaussian distributed, which is particularly true when $U$ is large. This implies that the magnitude of the sum-signal will be Rayleigh distributed and deep fade does occur. In other words, there will be natural phenomenon that on a per-symbol scale, the sum-interference plus noise signal will vanish at some port. Fast FAMA (\ref{eq:ffama-select}) is designed to exploit this phenomenon which is impossible for slow FAMA and that explains its increased capability.

\begin{figure}
\begin{centering}
\subfloat[]{\centering{}\includegraphics[scale=0.2]{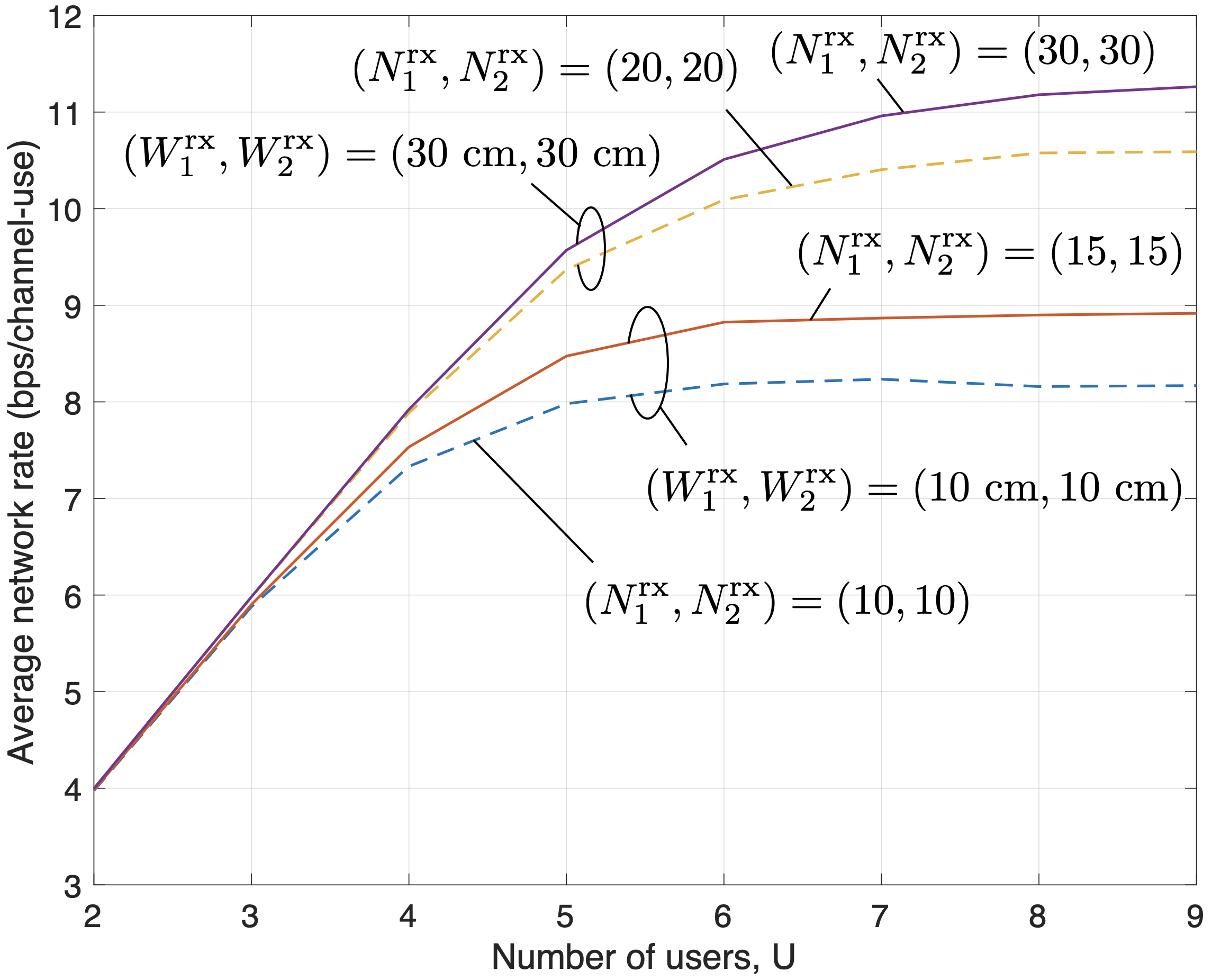}}
\par\end{centering}
\begin{centering}
\subfloat[]{\noindent \includegraphics[scale=0.2]{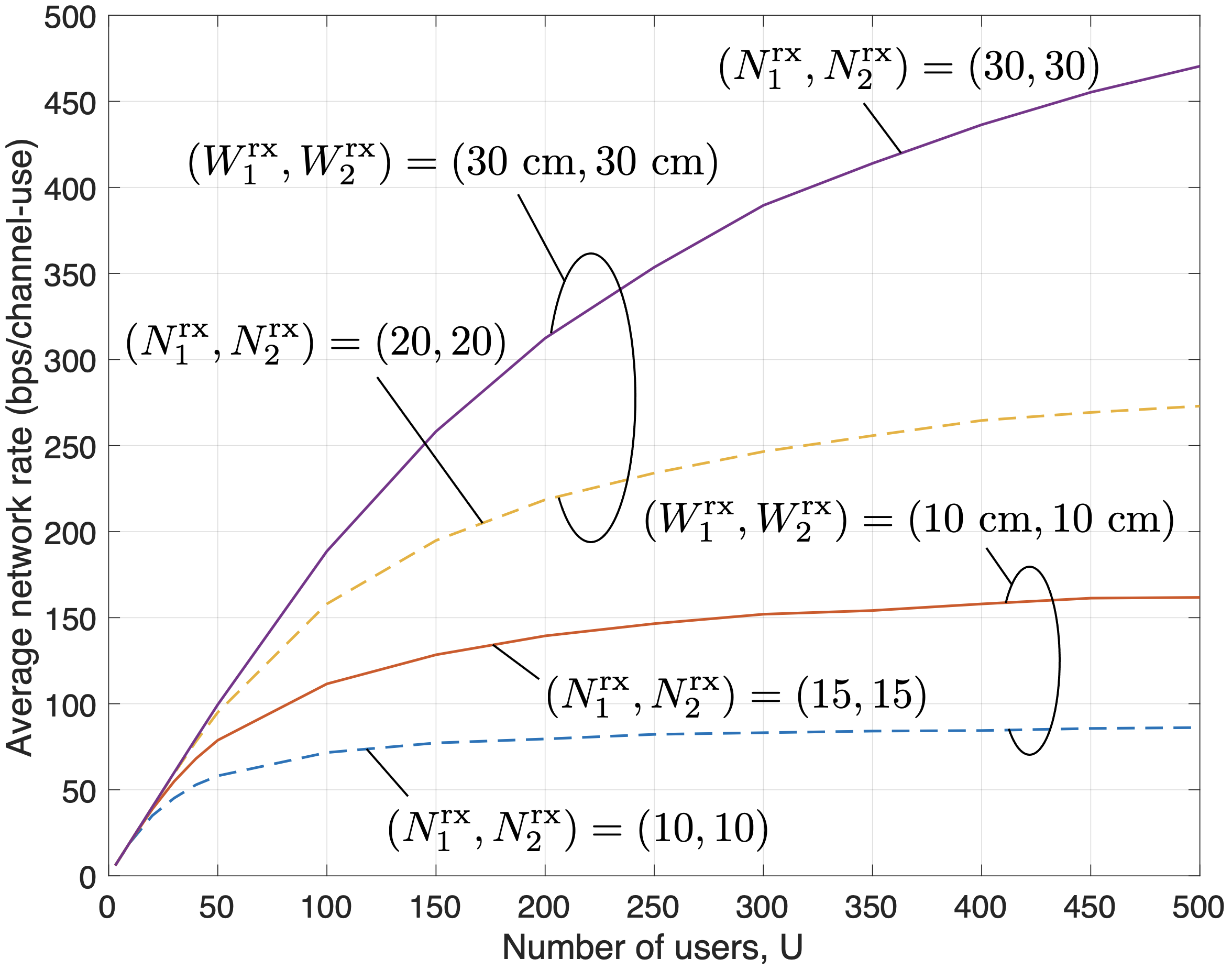}}
\par\end{centering}
\caption{Data rates of a) slow \gls{FAMA} and b) fast \gls{FAMA} against the number of UEs with varying sizes and resolutions of \gls{FAS} at each UE. Each UE is equipped with a \gls{2D} \gls{FAS} offering $N_{\rm rx}=N_1^{\rm rx}\times N_2^{\rm rx}$ flexible positions or ports. The data rate is computed assuming binary symmetric channels, employing uncoded QPSK transmissions. The system operates at a frequency of $39~{\rm GHz}$ and the channel has a Rice factor of $7$ with two scattered paths, i.e, $K=7$ and $L_p=2$.}\label{fig:slow-fast-fama}
\end{figure}

The results in Figs.~\ref{fig:slow-fast-fama}(a) and \ref{fig:slow-fast-fama}(b) investigate the average network rate performance of both slow and fast FAMA under finite scattering channels with Rice factor $K=7$ and $2$ scattered paths and considering QPSK transmission. The channel conditions reflect what would typically feature in the mmWave band. In the simulations, we considered $39~{\rm GHz}$. The network rate was obtained by estimating the bit error rate, then computing the channel capacity of a binary symmetric channel and averaging over a lot of independent channel realizations. Two \gls{FAS} sizes are considered, i.e., $(W^{\rm rx}_1,W_2^{\rm rx})=(10~{\rm cm},10~{\rm cm})$ or $(30~{\rm cm},30~{\rm cm})$. The former represents the size of a typical handheld device while the latter may be interpreted as the size of a laptop. As can be seen, for both slow and fast \gls{FAMA}, the network rate will increase, if the \gls{FAS} size increases and/or its port resolution increases. It is worth pointing out that if we continue to increase the port resolution, then the rate will not increase without bound, as already reported in single-user \cite{10103838,10130117} and multiuser systems \cite{ramirez2024new,hao2024revisit}. On the other hand, it is observed that the network rate generally increases with the number of users but will eventually plateau if there are too many users, exceeding the capability of \gls{FAMA}. The results indicate that slow \gls{FAMA} can handle $6$ co-channel users if the size is $(W^{\rm rx}_1,W_2^{\rm rx})=(10~{\rm cm},10~{\rm cm})$. This number can be increased to $9$ users if the size becomes $(30~{\rm cm},30~{\rm cm})$. The results for fast \gls{FAMA} are even more impressive and hundreds of users can be accommodated. The results demonstrate that fast \gls{FAMA} can deal with $300$ users (with a smaller \gls{FAS}) and serving $500$ users is possible if a larger \gls{FAS} is allowed.

Despite the unbelievable multiple access capability without CSI at the transmitter side, fast FAMA is not practically ready because it is indeed a challenge for each UE to estimate the energy ratio in (\ref{eq:ffama-select}) for the maximization and then switch to the optimal port instantly. Recent attempts have addressed this to some extent if the CSI of the desired user's channel and the received signals at all the ports are perfectly known \cite{9953084}. On the other hand, the concept of virtual \gls{FAS} in \cite{Wong-vfas2024} is expected to be useful to estimate the received signals at the unobserved ports if only a subset of ports are observable. Overall, however, it is fair to say that this remains largely an open problem.

\subsection{\gls{CUMA}: An Enhancement to Slow \gls{FAMA}}\label{ssec:cuma}
As discussed above, slow FAMA is practical but has limited multiple access capability while fast FAMA is not known to be practically realizable despite its extraordinary connectivity. Therefore, there is desire to keep the practicality but improve the performance of slow FAMA, which has led to the \gls{CUMA} technique in \cite{10318083}. In \gls{CUMA}, rather than focusing on one best port for reception, a large number of ports are activated and their signals are summed in the analogue domain to produce the output signal for detection. The key is to select the correct ports where the desired user's channels would be aligned to give user a boost in the desired signal and hence an advantage over the aggregate interference signal. Specifically, the signal alignment may be achieved by ensuring the in-phase (or real part) components of the desired user's channel at the selected ports are all positive, or all negative. Similarly, this can be done focusing on the quadrature (or imaginary part) components of the channel at the selected ports. The output signals at different selected port groups can then be combined for detection.

Mathematically, assuming that the \gls{BS} is equipped with only fixed-position transmit antennas and each antenna is assigned to send the information symbol to a designated user, the signals that would have been received at the \gls{FAS} ports of the $u$-th user can be written in vector form as
\begin{equation}\label{eq:yu-vect}
\boldsymbol{y}_u=\boldsymbol{h}_{uu}s_u+\sum_{\tilde{u}\ne u}\boldsymbol{h}_{\tilde{u}u}s_{\tilde{u}}+\boldsymbol{\zeta}_u,
\end{equation}
where the notations above follow from (\ref{eq:yu}) with slight modification. Following (\ref{eq:H_3D}), we can write
\begin{multline}
\boldsymbol{h}_{\tilde{u},u}=\sqrt{\frac{K}{K+1}}e^{j\omega_{\tilde{u},u}}\boldsymbol{a}_{r}\left(\theta_{0,r}^{(\tilde{u},u)},\phi_{0,r}^{(\tilde{u},u)}\right)\\
+\sqrt{\frac{1}{L_p\left(K+1\right)}}\sum_{l=1}^{L_p}\kappa_{l}^{(\tilde{u},u)}\boldsymbol{a}_{r}\left(\theta_{l,r}^{(\tilde{u},u)},\phi_{l,r}^{(\tilde{u},u)}\right),
\end{multline}
where $\boldsymbol{a}_{r}(\theta,\phi)$ has been defined by (\ref{eq:steer}).

\begin{figure*}
\centering \includegraphics[scale=0.23]{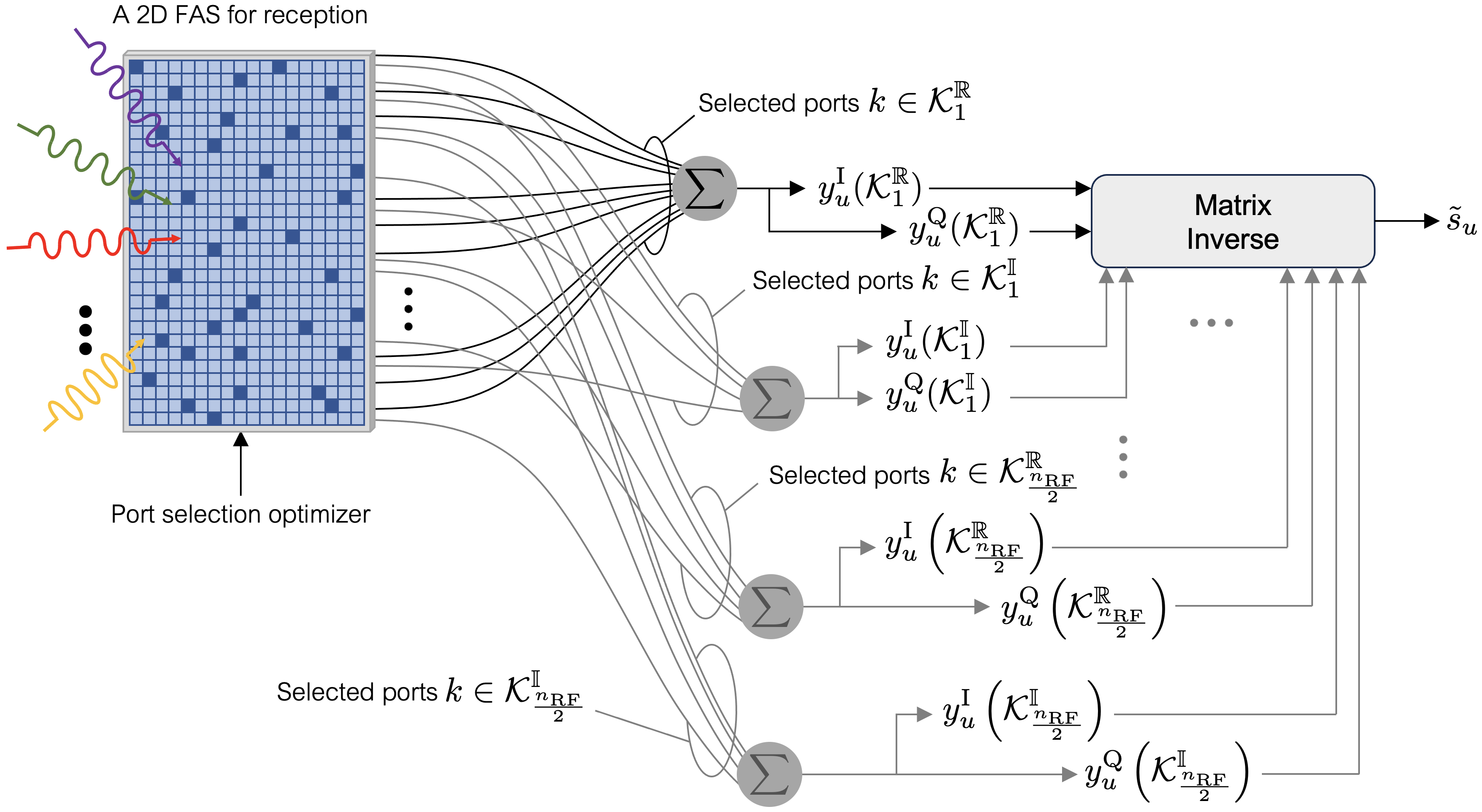} 
\caption{The receiver architecture for \gls{CUMA} with $n_{\rm RF}$ \gls{RF} chains.}\label{fig:cuma}
\end{figure*}

An important step in \gls{CUMA} is that user $u$ forms the set, say ${\cal K}$, of the selected ports according to some criteria. Note that ${\cal K}$ here replaces the parameters $a_u^{\rm rx}$ in (\ref{eq:yu}), and this vector version (\ref{eq:yu-vect}) collects the received signals at all the ports. With ${\cal K}$, \gls{CUMA} proceeds to aggregate the received signals of the selected ports in the analogue domain to get
\begin{equation}\label{eq:yIQ}
\left\{\begin{aligned}
y_u^{\rm I}({\cal K})&=\sum_{k\in{\cal K}}{\rm real}\left([\boldsymbol{y}_u]_k\right),\\
y_u^{\rm Q}({\cal K})&=\sum_{k\in{\cal K}}{\rm imag}\left([\boldsymbol{y}_u]_k\right),
\end{aligned}\right.
\end{equation}
where $[\cdot]_k$ returns the $k$-th entry of the input vector. The real-valued signals in (\ref{eq:yIQ}) correspond to, respectively, the in-phase and quadrature of the aggregated complex signals from the set of ports ${\cal K}$. Denoting the complex information symbol as $s_u=s_u^{\rm I}+js_u^{\rm Q}$, the aggregate received signals are given by
\begin{multline}\label{eq:yI}
y_u^{\rm I}({\cal K})=\left[\sum_{k\in{\cal K}}{\rm real}\left([\boldsymbol{h}_{uu}]_k\right)\right]s_u^{\rm I}\\
+\left[-\sum_{k\in{\cal K}}{\rm imag}\left([\boldsymbol{h}_{uu}]_k\right)\right]s_u^{\rm Q}\\
+\sum_{k\in{\cal K}}{\rm real}\left(\left[\sum_{\tilde{u}=1\atop\tilde{u}\ne u}^U\boldsymbol{h}_{\tilde{u}u}s_{\tilde{u}}+\boldsymbol{\zeta}_u\right]_k\right)
\end{multline}
and
\begin{multline}\label{eq:yQ}
y_u^{\rm Q}({\cal K})=\left[\sum_{k\in{\cal K}}{\rm imag}\left([\boldsymbol{h}_{uu}]_k\right)\right]s_u^{\rm I}\\
+\left[\sum_{k\in{\cal K}}{\rm real}\left([\boldsymbol{h}_{uu}]_k\right)\right]s_u^{\rm Q}\\
+\sum_{k\in{\cal K}}{\rm imag}\left(\left[\sum_{\tilde{u}=1\atop\tilde{u}\ne u}^U\boldsymbol{h}_{\tilde{u}u}s_{\tilde{u}}+\boldsymbol{\zeta}_u\right]_k\right).
\end{multline}
As the aggregation of the signals at the selected ports is done without scaling nor phase shifting, only one \gls{RF} chain suffices to obtain the two real-valued output signals (\ref{eq:yIQ}).

In \cite{10318083}, two \gls{RF} chains were assumed, one concentrating on aligning the in-phase component of the desired user's channel and another performing the same for the quadrature component of the desired user's channel. Specifically, the first set of the selected ports, ${\cal K}^{\mathbb{R}}_1$, is obtained based on several criteria. First is to shortlist the ports based on the parameter $\rho$ such that 
\begin{equation}
{\cal K}^+=\left\{k: {\rm real}\left([\boldsymbol{h}_{uu}]_k\right)\ge \rho\max_\ell{\rm real}\left([\boldsymbol{h}_{uu}]_\ell\right)\right\}.
\end{equation}
A similar criterion can also be used to obtain another set
\begin{equation}
{\cal K}^-=\left\{k: {\rm real}\left([\boldsymbol{h}_{uu}]_k\right)\le \rho\min_\ell{\rm real}\left([\boldsymbol{h}_{uu}]_\ell\right)\right\}.
\end{equation}
Then a second criterion is applied to limit the number of ports in the set (${\cal K}^+$ or ${\cal K}^-$) to be at most $N_{\rm max}$. To do so, we can randomly select up to $N_{\rm max}$ entries to form a new set, i.e.,
\begin{equation}\label{eq:rand-K+}
\left\{\bar{\cal K}^+\subseteq{\cal K}^+||\bar{\cal K}^+|\le N_{\rm max}\right\}.
\end{equation}
Similarly, for ${\cal K}^-$, we also have
\begin{equation}\label{eq:rand-K-}
\left\{\bar{\cal K}^-\subseteq{\cal K}^-||\bar{\cal K}^-|\le N_{\rm max}\right\}.
\end{equation}
Note that in the above, the selection of each entry is equally probable without replacement. Finally, the set of selected ports, ${\cal K}^{\mathbb{R}}_1$, can be chosen between $\bar{\cal K}^+$ and $\bar{\cal K}^-$ by
\begin{equation}
{\cal K}^{\mathbb{R}}_1=\arg \left|\sum_{k\in\bar{\cal K}^+}{\rm real}\left([\boldsymbol{h}_{uu}]_k\right)\right|
\mathop{\gtrless}_{\bar{\cal K}^-}^{\bar{\cal K}^+}
\left|\sum_{k\in\bar{\cal K}^-}{\rm real}\left([\boldsymbol{h}_{uu}]_k\right)\right|.
\end{equation}
The same procedure can be adopted focusing on the quadrature component of the desired user's channel to give ${\cal K}^{\mathbb{I}}_1$.

Due to the random sampling in (\ref{eq:rand-K+}) and (\ref{eq:rand-K-}), if we repeat the above procedures, then different sets, ${\cal K}^{\mathbb{R}}_2$ and ${\cal K}^{\mathbb{I}}_2$, will be produced. As mentioned above, the aggregation of the selected signals from a set requires a dedicated \gls{RF} chain. Thus, if the \gls{FAS} receiver has $n_{\rm RF}$ \gls{RF} chains, then we can repeat the above procedures $n_{\rm RF}/2$ times, as shown in Fig.~\ref{fig:cuma}. Recently, \cite{Wong-spawc2024} considered the case of $n_{\rm RF}=4$ for \gls{CUMA} and revealed a substantial performance gain over \gls{CUMA} with $n_{\rm RF}=2$.

In general, with $n_{\rm RF}\ge 2$ (assuming an even number), the information symbol for the $u$-th UE can be estimated by
\begin{equation}
\tilde{\boldsymbol{s}}_u=\left[\begin{array}{c}
\tilde{s}_u^{\rm I}\\
\tilde{s}_u^{\rm Q}
\end{array}\right]=\left[\begin{array}{c}
{\bf\Psi}({\cal K}^{\mathbb{R}}_1)\\
----\\
{\bf\Psi}({\cal K}^{\mathbb{I}}_1)\\
----\\
\vdots\\
----\\
{\bf\Psi}\left({\cal K}^{\mathbb{R}}_{\frac{n_{\rm RF}}{2}}\right)\\
----\\
{\bf\Psi}\left({\cal K}^{\mathbb{I}}_{\frac{n_{\rm RF}}{2}}\right)
\end{array}\right]^{-1}\left[\begin{array}{c}
y^{\rm I}_u({\cal K}^{\mathbb{R}}_1)\\
y^{\rm Q}_u({\cal K}^{\mathbb{R}}_1)\\
----\\
y^{\rm I}_u({\cal K}^{\mathbb{I}}_1)\\
y^{\rm Q}_u({\cal K}^{\mathbb{I}}_1)\\
----\\
\vdots\\
----\\
y^{\rm I}_u\left({\cal K}^{\mathbb{R}}_{\frac{n_{\rm RF}}{2}}\right)\\
y^{\rm Q}_u\left({\cal K}^{\mathbb{R}}_{\frac{n_{\rm RF}}{2}}\right)\\
----\\
y^{\rm I}_u\left({\cal K}^{\mathbb{I}}_{\frac{n_{\rm RF}}{2}}\right)\\
y^{\rm Q}_u\left({\cal K}^{\mathbb{I}}_{\frac{n_{\rm RF}}{2}}\right)
\end{array}\right],
\end{equation}
where
\begin{equation}
{\bf\Psi}({\cal K})\triangleq\left[\begin{array}{lc}
\sum_{k\in{\cal K}}{\rm real}\left([\boldsymbol{h}_{uu}]_k\right) & -\sum_{k\in{\cal K}}{\rm imag}\left([\boldsymbol{h}_{uu}]_k\right)\\
\sum_{k\in{\cal K}}{\rm imag}\left([\boldsymbol{h}_{uu}]_k\right) & ~\sum_{k\in{\cal K}}{\rm real}\left([\boldsymbol{h}_{uu}]_k\right)
\end{array}\right].
\end{equation}
Note that the port resolution of \gls{FAS} in \gls{CUMA} should be high to be meaningful. Otherwise, the different port selection sets would not result in different communication experience. Also, it is noteworthy that given the finite size of \gls{FAS} at each user, the performance of \gls{CUMA} would not keep on improving even if $n_{\rm RF}$ increases without bound. That said, the most important cases are $n_{\rm RF}=1, 2$ or $4$, with $n_{\rm RF}=1$ being the simplest version, $n_{\rm RF}=2$ being the original version of \gls{CUMA} in \cite{10318083} and $n_{\rm RF}=4$ perhaps being the expected version, knowing that \gls{5G} requires mobile devices to have $4$ \gls{RF} chains to operate. Finally, \gls{CUMA} can be viewed as an upgraded version of slow \gls{FAMA} as it maintains the port switching to a per-block basis but deviates only in terms of the receiver architecture.

\begin{figure}
\begin{centering}
\subfloat[]{\centering{}\includegraphics[scale=0.2]{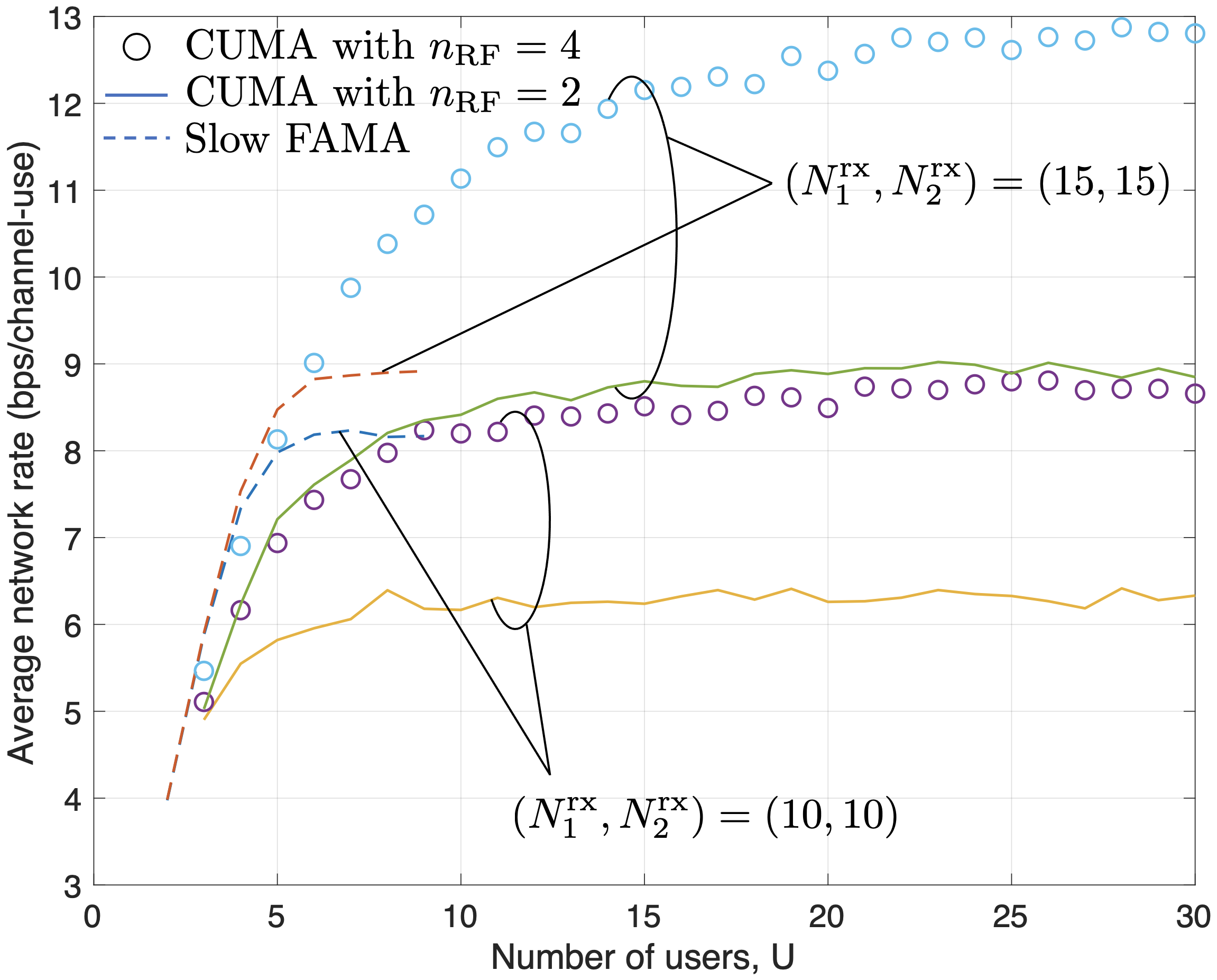}}
\par\end{centering}
\begin{centering}
\subfloat[]{\noindent \includegraphics[scale=0.2]{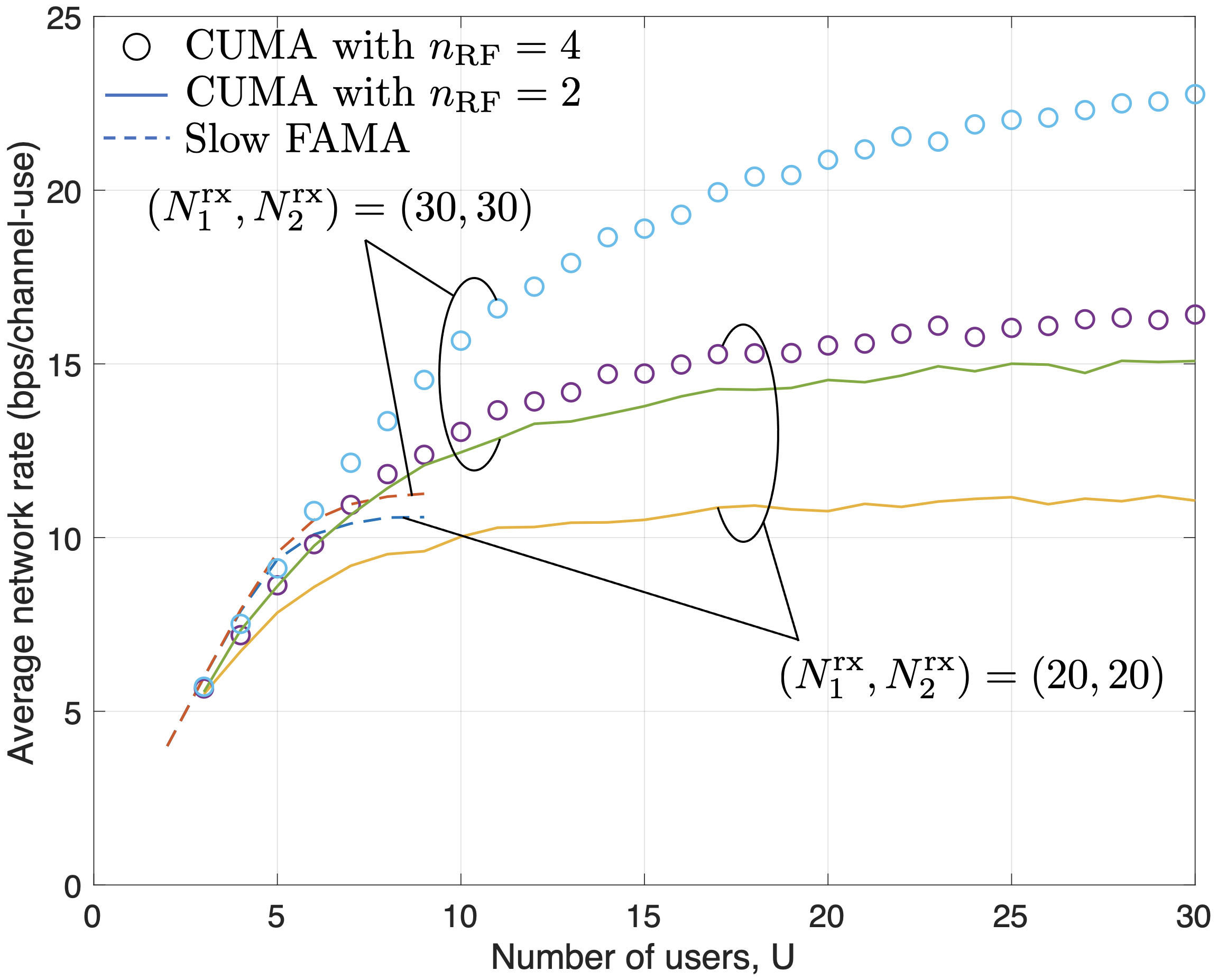}}
\par\end{centering}
\caption{Average data rates of \gls{CUMA} with different values of $n_{\rm RF}$ versus slow \gls{FAMA} with \gls{FAS} size of a) $(W_1^{\rm rx},W_2^{\rm rx})=(10~{\rm cm},10~{\rm cm})$ and b) $(W_1^{\rm rx},W_2^{\rm rx})=(30~{\rm cm},30~{\rm cm})$. Other parameters are the same as that in Fig.~\ref{fig:slow-fast-fama}.}\label{fig:sfamaVScuma}
\end{figure}

We here provide the average rate results comparing \gls{CUMA} and slow \gls{FAMA} in Fig.~\ref{fig:sfamaVScuma}. To be more precise, it would be necessary to take into account the potential mutual coupling effects among the ports in the system model for performance evaluation. Nonetheless, as shown in \cite{Wong-ojcom2024}, the effects can be ignored if the return loss and isolation at each port are kept at $-15~{\rm dB}$ and $-20~{\rm dB}$, respectively, which are practically achievable \cite{4685877,4232615}. For simplicity, we therefore did not consider mutual coupling effects in the simulations. In Fig.~\ref{fig:sfamaVScuma}, the same parameters as in Fig.~\ref{fig:slow-fast-fama} were assumed. As before, we will focus on two cases, $(W^{\rm rx}_1,W_2^{\rm rx})=(10~{\rm cm},10~{\rm cm})$ and $(30~{\rm cm},30~{\rm cm})$. The results of the former case are given in Fig.~\ref{fig:sfamaVScuma}(a) while Fig.~\ref{fig:sfamaVScuma}(b) deals with the latter.

With a smaller size and low resolution of \gls{FAS}, as illustrated in Fig.~\ref{fig:sfamaVScuma}(a), the benefits of \gls{CUMA} are not obvious. In fact, slow \gls{FAMA} can perform much better than \gls{CUMA} with $n_{\rm RF}=2$. With sufficient port resolution, i.e., $(N_1^{\rm rx},N_2^{\rm rx})=(15,15)$, \gls{CUMA} starts to catch up with slow \gls{FAMA} as the number of users to serve increases. The great performance of \gls{CUMA} begins to show when $n_{\rm RF}=4$ and the port resolution of \gls{FAS} is sufficiently high. As we can see, in this case, \gls{CUMA} greatly outperforms slow \gls{FAMA} and is capable of supporting more than $30$ users, as compared to only $6$ using slow \gls{FAMA}. The results are even more encouraging if the \gls{FAS} size is increased to $(30~{\rm cm},30~{\rm cm})$ in Fig.~\ref{fig:sfamaVScuma}(b). It can be observed that with sufficient size and port resolution, \gls{CUMA} with $n_{\rm RF}=2$ can even achieve much higher rate than slow \gls{FAMA} and serve many more users. In summary, \gls{CUMA} can support more than $30$ users if the \gls{FAS} has sufficient size and resolution.

\subsection{\gls{CSI}-Less Extreme Massive Connectivity}\label{ssec:fama4ss}
As discussed, \gls{FAMA} can offer a much simpler alternative to \gls{CSI}-based massive \gls{MIMO} for massive multiple access, the one that permits transmitter \gls{CSI}-free multiple access. But in practice, how important it is to avoid the need of \gls{CSI} at the transmitter side? To answer this question, we may check out how much \gls{CSI} a \gls{BS} is practically possible to have in order to preform precoding. In \gls{5G}, we are able to have a $64$-antenna \gls{BS} performing precoding to support a maximum of $12$ UEs. That is $12\times 64=768$ instantaneous complex channel coefficients to be estimated and fed back to the \gls{BS}. Moreover, this needs to be conducted once every few milliseconds. 

Now, if we contemplate the idea of delivering $1~{\rm kbps/Hz}$ in \gls{6G} \cite{10183795}, this may mean that we are using a $1000$-antenna 
\gls{BS} to serve $1000$ UEs on the same physical channel by precoding. This demands an estimation of $1000\times 1000=10^6$ channels to be available at the \gls{BS} every few milliseconds. This does not seem remotely possible. However, if the channel is \gls{LoS}-dominant, precoding may be effectively designed based on the \gls{LoS} link which typically varies only very slowly, or not at all. This will present a much more practical setup for precoding to operate in massive user access scenarios. 

In particular, recalling from the finite scattering channel model (\ref{eq:H_3D}), we have the user channel
\begin{multline}\label{eq:H_mmimo}
\bm{H}=\sqrt{\frac{K}{K+1}}e^{j\omega}\boldsymbol{a}_{t}\left(\theta_{0,t},\phi_{0,t}\right)^{H}\\
+\sqrt{\frac{1}{L_p\left(K+1\right)}}\sum_{l=1}^{L_p}\kappa_{l}\boldsymbol{a}_{t}\left(\theta_{l,t},\phi_{l,t}\right)^{H},
\end{multline}
where the index for UE is omitted for conciseness and the UE is assumed to have a single fixed receive antenna. For MRT precoding, the full channel vector $\bm{H}$ for every UE is required at the \gls{BS} so that the precoding vector can be set as
\begin{equation}\label{eq:MRT}
\boldsymbol{\omega}_{\rm MRT}=\frac{\bm{H}^H}{\|\bm{H}\|}.
\end{equation}

Alternatively, we can perform \gls{LoS}-only precoding, i.e.,
\begin{equation}\label{eq:LoSprecoding}
\boldsymbol{\omega}_{\rm LoS}=\frac{\boldsymbol{a}_{t}\left(\theta_{0,t},\phi_{0,t}\right)}{\|\boldsymbol{a}_{t}\left(\theta_{0,t},\phi_{0,t}\right)\|},
\end{equation}
which only requires the knowledge of \gls{AoD}s of the \gls{LoS}, $\theta_{0,t}$ and $\phi_{0,t}$ which can be considered as deterministic channel parameters for low-mobility users. In other words, it is much more practical to have the \gls{AoD}s of the \gls{LoS} of the users ready at the \gls{BS} and perform \gls{LoS}-only precoding (\ref{eq:LoSprecoding}).

\begin{figure}
\centering \includegraphics[scale=0.2]{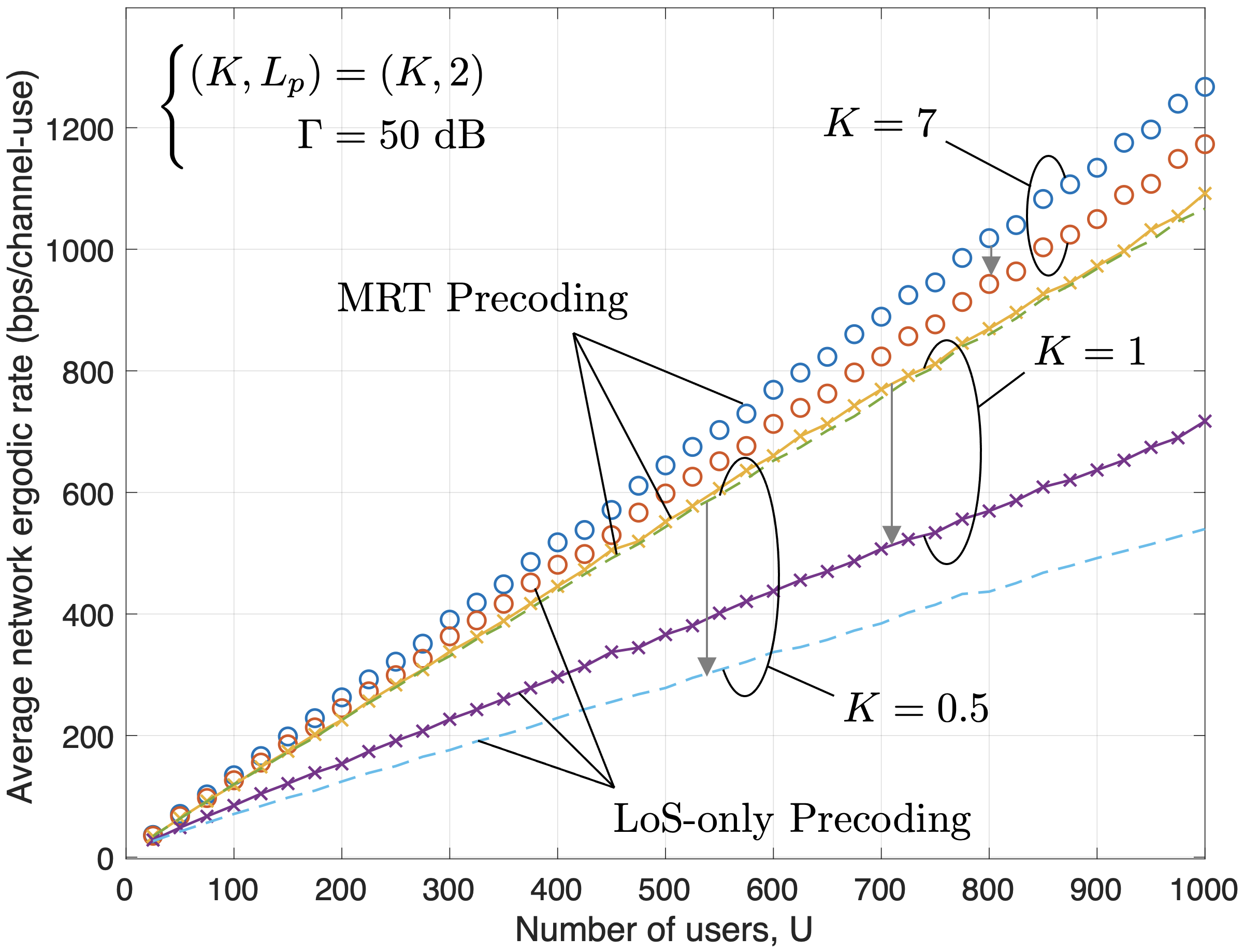} 
\caption{Average ergodic rates for massive \gls{MIMO}.}\label{fig:mmimo_erate}
\end{figure}

Fig.~\ref{fig:mmimo_erate} illustrates the rate results for massive \gls{MIMO} with the two precoding schemes above for different values of Rice factor, $K$. The wireless channel is assumed to only have two scattered paths, i.e., $L_p=2$ and the average \gls{SNR} is set to $\Gamma=50~{\rm dB}$. The results indicate that \gls{LoS}-only precoding can be effective and performs close to the full \gls{CSI}-based MRT precoding but this only happens if $K$ is large, meaning that there is a very strong \gls{LoS} link. When $K$ decreases, e.g., $K=1$ or even $K=0.5$, \gls{LoS}-only precoding loses its grip and huge performance gaps from MRT precoding begin to appear, reaffirming the importance of full \gls{CSI}. The fact that recent developments of \gls{6G} place emphasis on \gls{RIS}, suggests that the operations of \gls{6G} be likely in situations where the \gls{LoS} is not strong and the use of \gls{RIS} is necessary to repair the link.

The results in Fig.~\ref{fig:mmimo_erate} have revealed the main problem of massive \gls{MIMO} and its reliance on full \gls{CSI} to perform well. We are therefore interested in whether the situations would be different if \gls{FAS} is used at the UEs. In this case, we have the channel matrix given in (\ref{eq:H_3D}). Given the full $\bm{H}$, one can use \gls{SVD} to decompose the channel and choose the principal eigenvector as the precoding vector for the given user. This is referred to as \gls{SVD} precoding in this article. 

Similarly, we can focus on the core \gls{LoS} channel
\begin{equation}\label{eq:H_3D_LoS}
\bm{H}_{\rm LoS}=\boldsymbol{a}_{r}\left(\theta_{0,r},\phi_{0,r}\right)\boldsymbol{a}_{t}\left(\theta_{0,t},\phi_{0,t}\right)^{H},
\end{equation}
which can easily be constructed by knowing the \gls{AoD}s, $\theta_{0,t}$ and $\phi_{0,t}$, and \gls{AoA}s, $\theta_{0,r}$ and $\phi_{0,r}$, of the \gls{LoS} link. The \gls{LoS}-only precoding for a \gls{FAS} user therefore can be chosen to be the principal eigenvector of $\bm{H}_{\rm LoS}$. Here, we assume that each \gls{FAS} UE is using the \gls{CUMA} architecture with $n_{\rm RF}=2$.

\begin{figure}
\begin{centering}
\subfloat[]{\centering{}\includegraphics[scale=0.2]{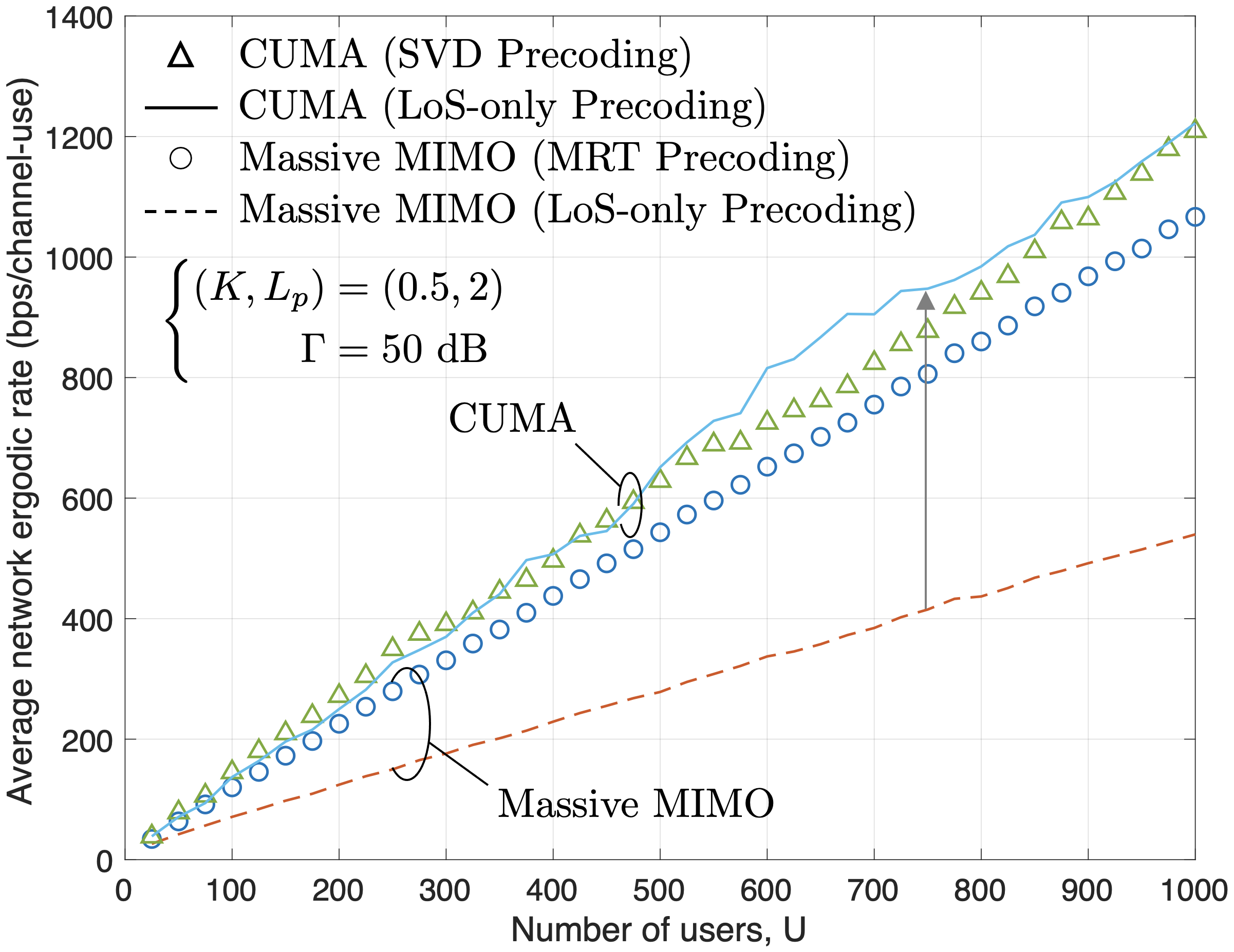}}
\par\end{centering}
\begin{centering}
\subfloat[]{\noindent \includegraphics[scale=0.2]{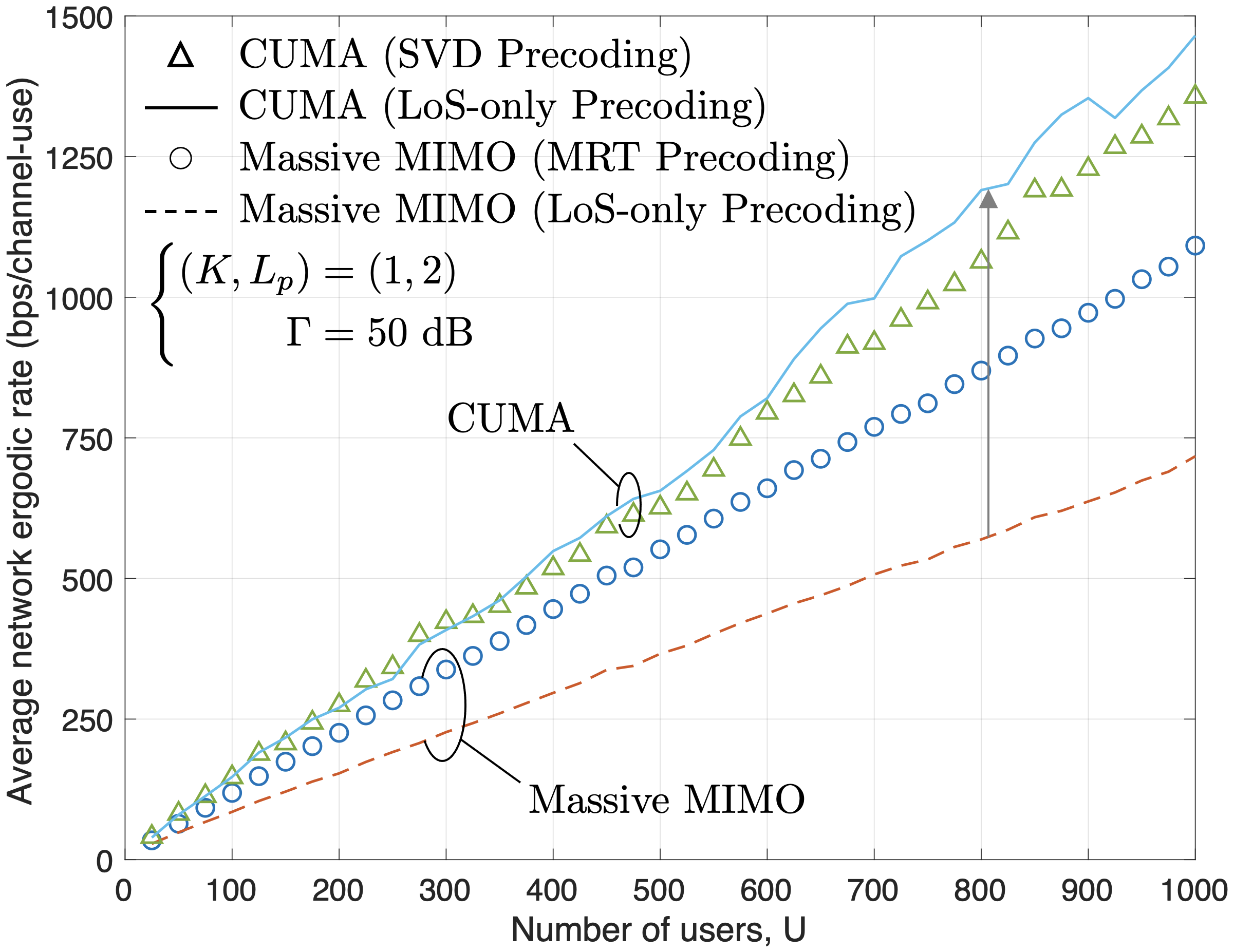}}
\par\end{centering}
\begin{centering}
\subfloat[]{\noindent \includegraphics[scale=0.2]{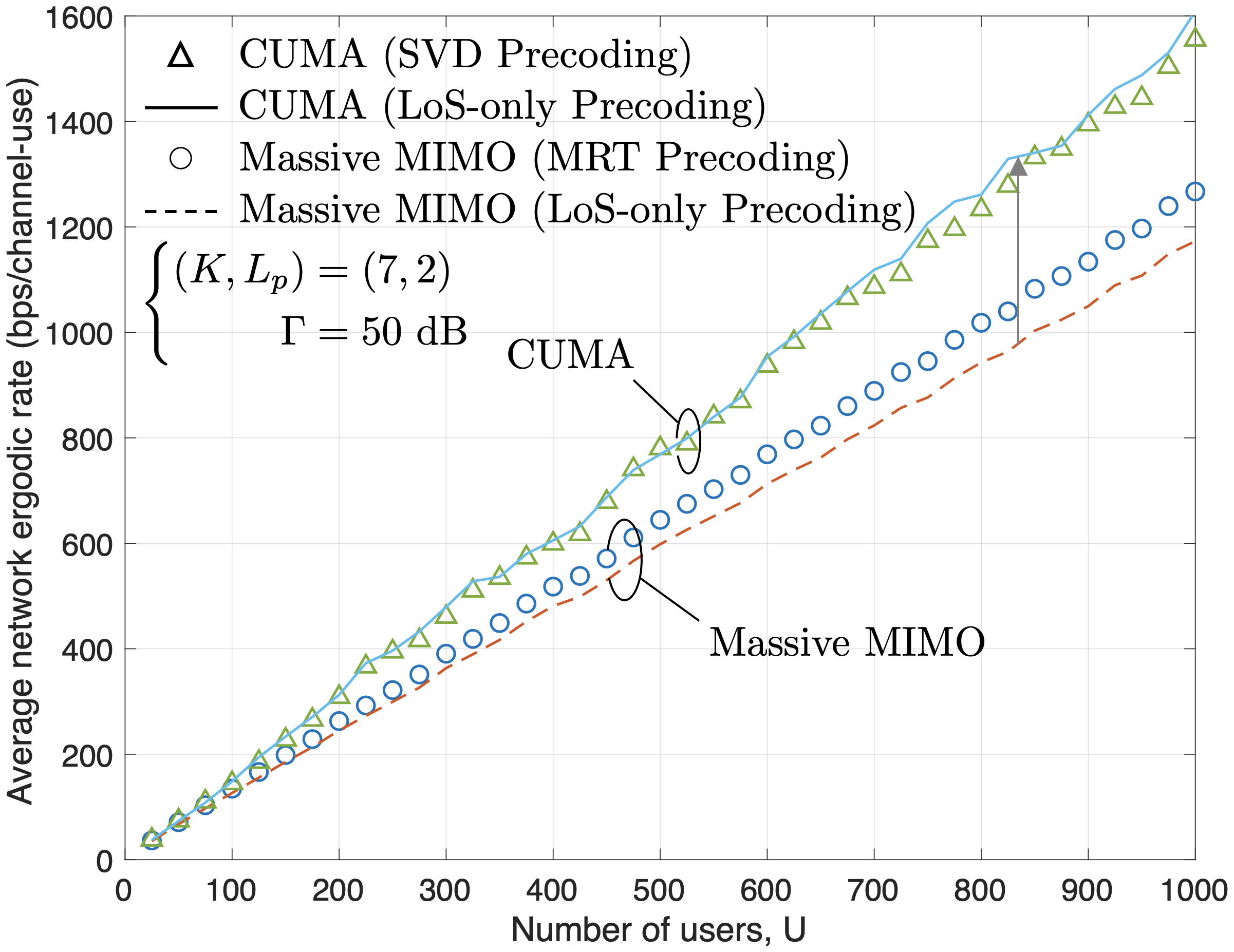}}
\par\end{centering}
\caption{Comparison of the average ergodic rates of \gls{CUMA} with precoding and massive \gls{MIMO} when a) $K=0.5$ (weak \gls{LoS} scenarios), b) $K=1$ (equal power of \gls{LoS} and \gls{NLoS} scenarios) and c) $K=7$ (strong \gls{LoS} scenarios). For \gls{CUMA}, each UE has a \gls{2D}-\gls{FAS} receiver with size of $(W_1^{\rm rx},W_2^{\rm rx})=(13\lambda,7\lambda)$ and $(N_1^{\rm rx},N_2^{\rm rx})=(27,14)$, i.e., $378$ ports.}\label{fig:cumaVSmmimo}
\end{figure}

Fig.~\ref{fig:cumaVSmmimo} provides the rate results of \gls{CUMA} with different precoding schemes in comparison with that of massive \gls{MIMO} under the settings similar to that in Fig.~\ref{fig:mmimo_erate}. A few important observations can be made. First, \gls{CUMA} outperforms massive \gls{MIMO} with MRT precoding greatly, regardless of whether \gls{LoS}-only or \gls{SVD} precoding is adopted. This means that \gls{FAS} can be a key ingredient that improves massive \gls{MIMO} without adding burden to the \gls{BS} in terms of \gls{CSI} requirements. Secondly and perhaps surprisingly, \gls{LoS}-only precoding performs better than \gls{SVD} precoding in \gls{CUMA} systems. This is possible because \gls{SVD} precoding is not optimal, especially when the combining at each UE receiver is only performing the \gls{CUMA} signal alignment but not matching to the eigenvector. Thirdly, the results demonstrate that \gls{CUMA} with \gls{LoS}-only precoding performs well regardless of $K$, which means that the \gls{AoD}s and \gls{AoA}s of the \gls{LoS} link are sufficient to enable massive spatial multiplexing if \gls{CUMA} is in place. Additionally, though in the case of large $K$, the performance difference between massive \gls{MIMO} and \gls{CUMA} both with \gls{LoS}-only precoding is smaller, \gls{CUMA} still performs substantially better. The results further discover that hitting $>1000~{\rm bps/Hz}$ is now a realistic target and more than $1000$ users can be served by simple \gls{LoS}-only precoding and each UE using a $2$-\gls{RF}-chain \gls{FAS}. The results also project that a larger performance gap over massive \gls{MIMO} is anticipated with a larger number of users. On the other hand, it is seen that the rate of \gls{CUMA} improves as $K$ increases.

Real environments tend to have mixed channel conditions, some users with strong \gls{LoS} while others with weak \gls{LoS}. The above results for \gls{CUMA} therefore are particularly encouraging. The fact that \gls{CUMA} with \gls{LoS}-only precoding performs best for a wide range of $K$ values, confirms that \gls{CUMA} is robust and will perform well with users of mixed conditions. Nonetheless, the above results have assumed that all the users are in the far-field but with such a large array at the \gls{BS} and so many UEs, some users are bound to be in the near-field. A more rigorous study will be needed in the future.

\section{Overview of \gls{FAS} Hardware Designs}\label{sec:hardware}
In the concept of \gls{FAS}, the hardware designs and implementation methods are deliberately left open, affording engineers the flexibility to determine the optimal configuration. The key concept is to empower antennas with the capability to flexibly reconfigure the position, orientation, shape, or dimension(s) of the radiator. Consequently, \gls{FAS} encompasses a broad spectrum of reconfigurable flexible antennas. Four primary designs can realize the concept of \gls{FAS}: i) mechanical movable antennas, ii) liquid-based antennas, iii) pixel-based antennas, and iv) hybrid antennas. In the following, we will discuss these state-of-the-art designs and highlight their advantages and disadvantages of facilitating the hardware development of \gls{FAS}.

\subsection{Mechanical Movable Antennas}

\begin{figure}
\noindent \centering{}\includegraphics[scale=0.35]{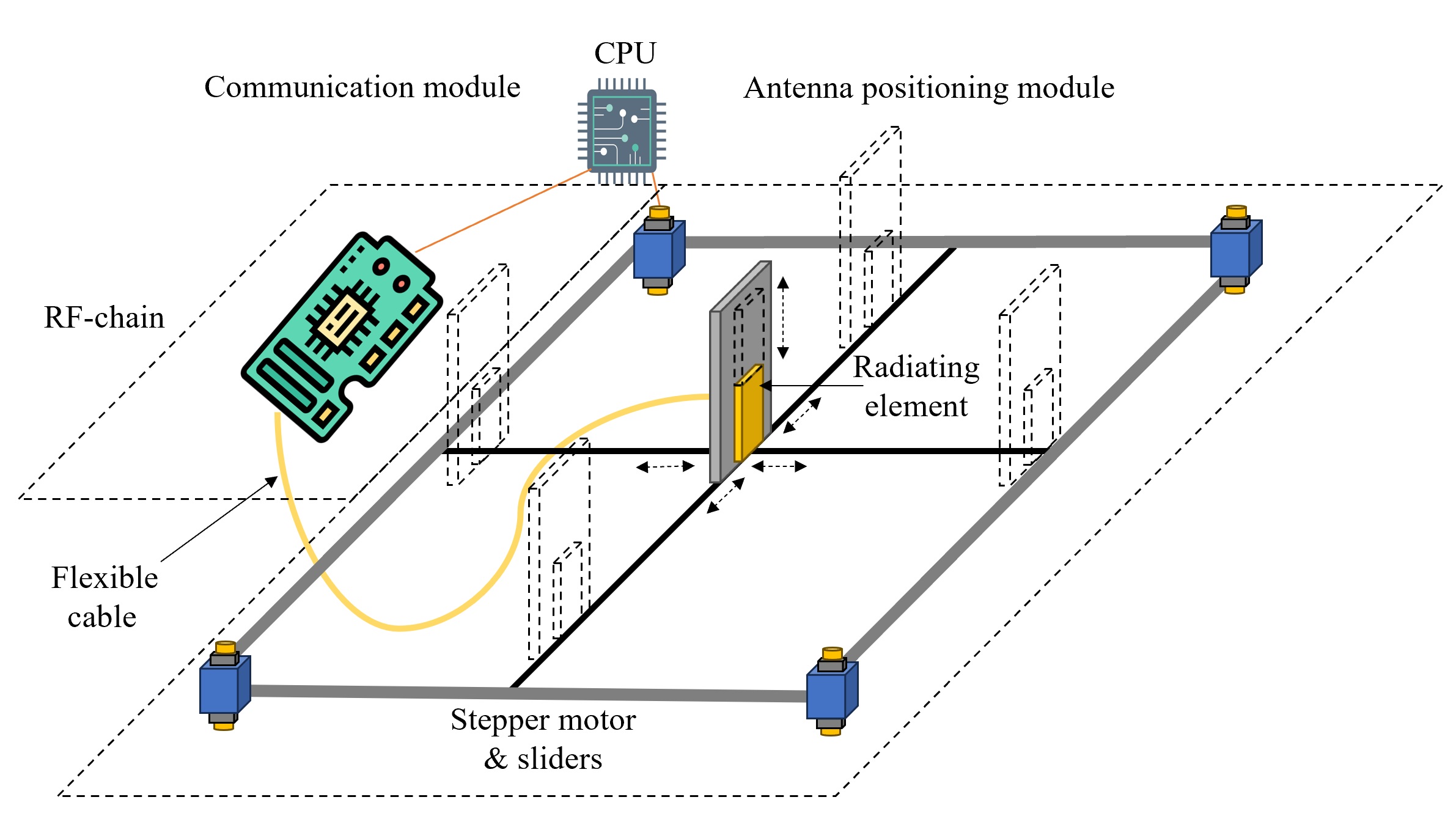}
\caption{A schematic of mechanical movable FAS \cite{10286328}.}\label{MAS}
\end{figure}

One of the most obvious methods to realize the concept of \gls{FAS} is the use of mechanical movable antennas, as shown in Fig.~\ref{MAS}. This design can be separated into two parts \cite{10286328}: the communication module and the antenna positioning module. In the communication module, the antennas are connected to the \gls{RF}-chains. Flexible cables are required instead to facilitate the antenna positioning module. In the antenna positioning module, the antennas are installed on a mechanical slide, which is driven by stepper motors to reconfigure the antenna positions in a \gls{3D} Cartesian coordinate \cite{7360379,9770353}. Besides, the orientation of the antennas at a given position aided by a servo motor can offer additional dof for the antenna movement \cite{8060521}. To balance durability, speed, weight considerations, and conductivity, the materials used in this design should be chosen carefully. The communication module and antenna positioning module are interconnected via a \gls{CPU} for digital signal processing and antenna positioning. Specifically, the motors can cooperatively relocate the antennas to their target positions and perform beamforming upon receiving the control signal from the \gls{CPU}. It is crucial to note that the size of the antenna framework may far surpass the available movement area, and the speed at which the antennas can be relocated is restricted. Additionally, the dependence on the mechanical structure makes this design prone to considerable wear and tear, even over a short period of time.

\subsection{Liquid-based Antennas}
Another idea to realize \gls{FAS} is to exploit the flexible nature of liquid or fluid as the radiating element of an antenna. Both metallic and non-metallic liquids can be used in designing the radiating element. One of them is gallium-based alloys, such as eutectic gallium indium (eGaIn) or Galinstan. These non-toxic and non-flammable alloys exhibit high electrical conductivity and show great promise for thermal properties relevant to \gls{RF} applications. Interestingly, their melting points are typically close to or lower than room temperature \cite{liquid_metals}. Hence, gallium-based alloys have become one of the main subjects of extensive exploration in the realm of liquid-based antennas \cite{paracha2019liquid,9388928}.

Encouraged by the flexibility and versatility of liquid-based antennas, researchers have explored various reconfigurability. This includes reconfigurability in frequency \cite{8762117,Murray-2014,Konca-2015,Singh-2019}, radiation pattern \cite{Qian-2017,Chen-tap2017,8889301,8822465,Ren-tap2020,Hao-2021}, and polarization \cite{Wang-awpl2019,Zhen-tap2019,Xu-tap2020,Chen-tap2018}. Additionally, embedding fluid switches within the antenna structure is another attractive method. This allows for dynamic control over antenna functionality in response to the dynamic communication environments, e.g., \cite{8105795,Batra-2014,Alkaraki-2021}.

\begin{figure}
\noindent \begin{centering}
\subfloat[]{\includegraphics[scale=0.2]{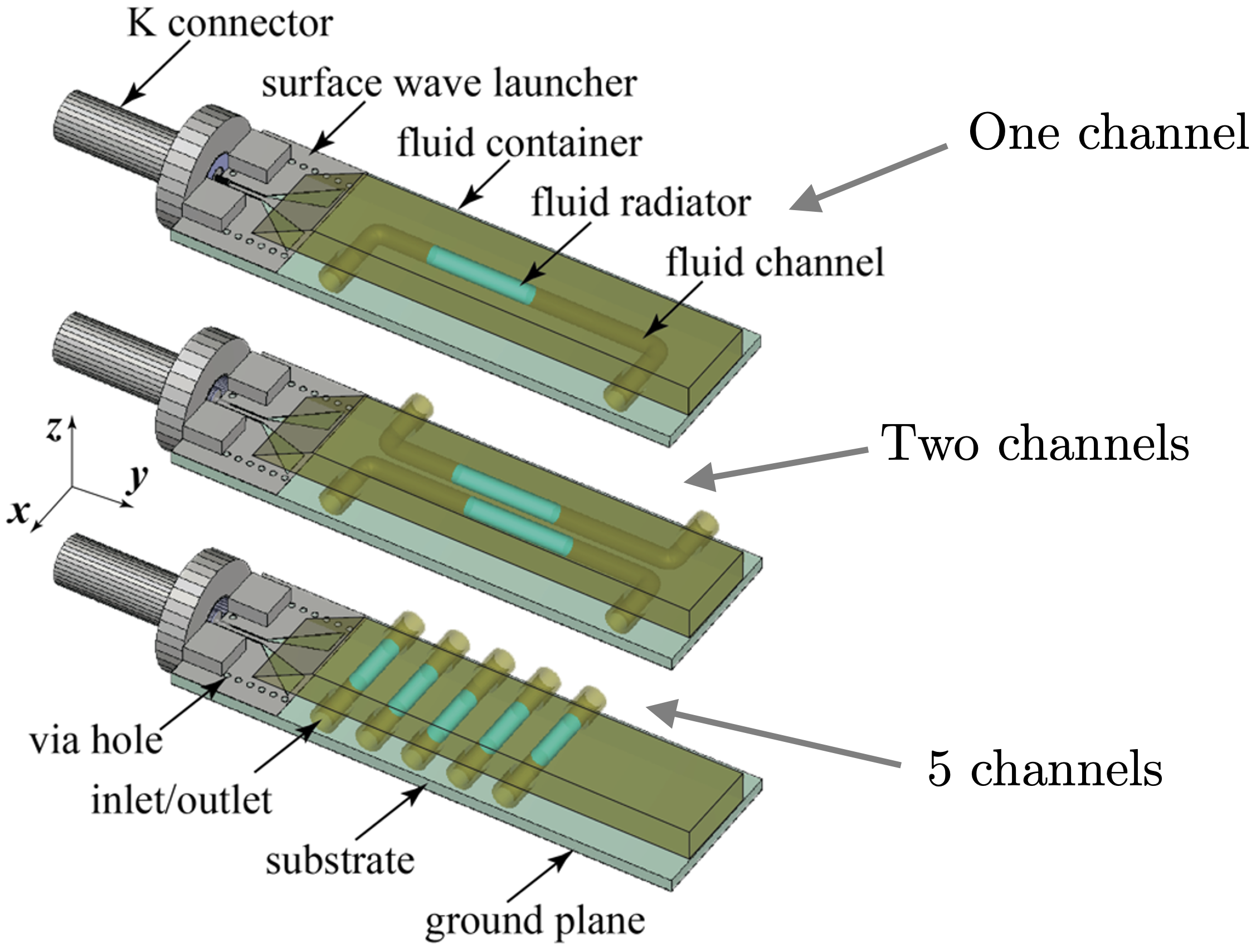}}
\par\end{centering}
\noindent \centering{}\subfloat[]{\includegraphics[scale=0.7]{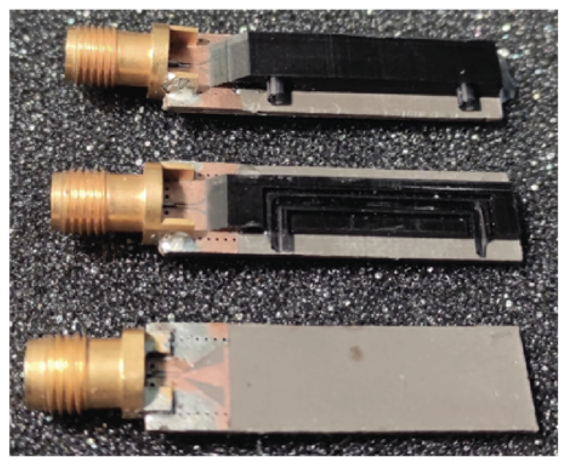}}
\caption{A fluid-channel-based FAS antenna design \cite{Shen-fas2021,Shen-fas2022a,Shen-fas2022b}: a) schematic; and b) prototype.}\label{Liquid}
\end{figure}

Despite these initial achievements, position-flexible liquid-based \gls{FAS} designs only emerged recently \cite{Shen-fas2021,Shen-fas2022a,Shen-fas2022b}. Specifically, the designs in \cite{Shen-fas2021,Shen-fas2022a,Shen-fas2022b} allow for the controlled movement of radiating elements within preset fluid channels using a micro-pump, as illustrated in Fig.~\ref{Liquid}. Upon receiving a signal from the \gls{RF}-chain, a surface wave launcher converts the input signal into a surface wave, propagating to the radiating element. As the surface wave interacts with the radiating elements, it is scattered into the wireless medium (if transmitting). The reverse process occurs for the reception of wireless signals. In \cite{9977471}, a pump-less design employing the \gls{EWOD} method in \cite{s10404-006-0085-8,1454032} was obtained, achieving a motion speed of up to $10~{\rm mm/s}$.

\begin{figure}
\noindent \begin{centering}
\subfloat[]{\includegraphics[scale=0.22]{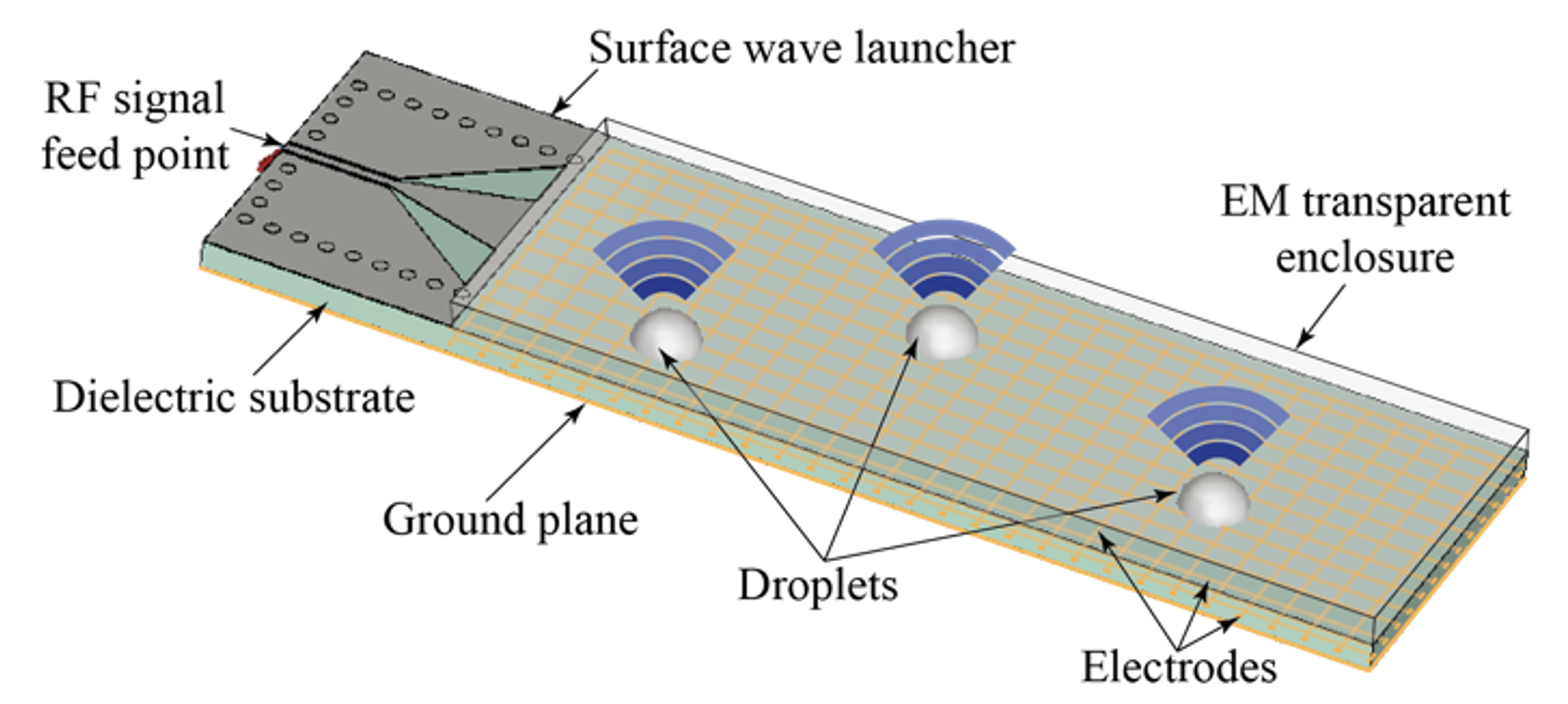}}
\par\end{centering}
\vspace{0.2cm}
\noindent \begin{centering}
\subfloat[]{\includegraphics[scale=0.22]{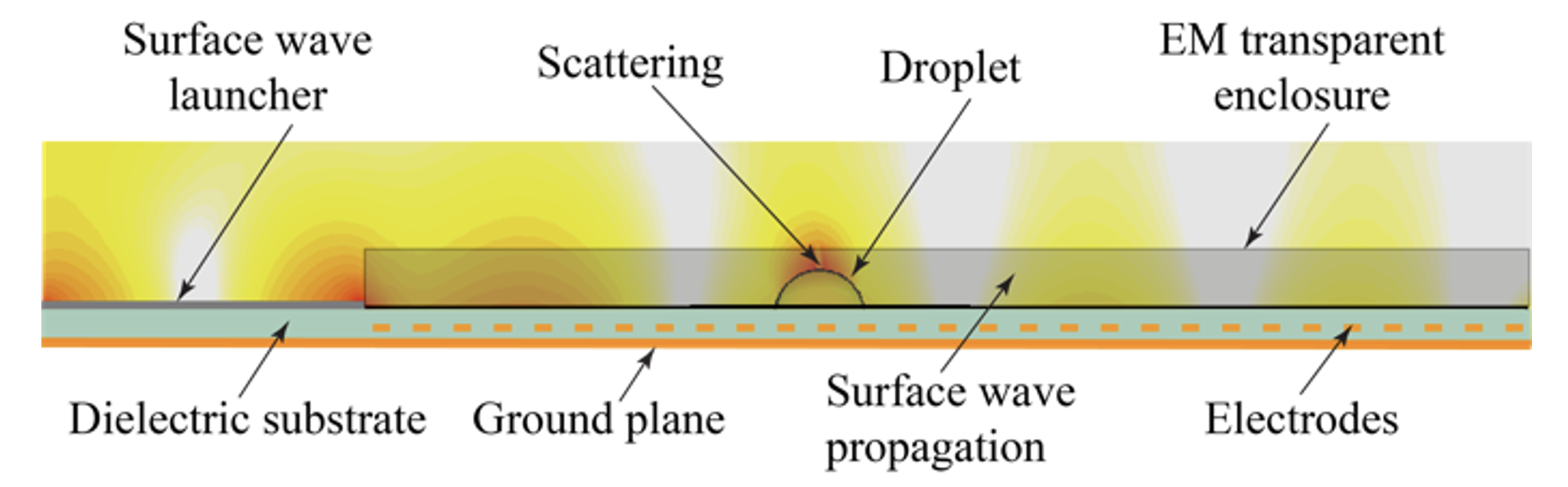}}
\par\end{centering}
\noindent \centering{}
\caption{A liquid-based FAS antenna design using EWOD \cite{9977471}: a) 3D view without enclosure; and b) side view.}\label{EWOD}
\end{figure}

In the \gls{EWOD} design, as shown in Fig.~\ref{EWOD}(a) and (b), the positions of multiple droplets can be independently adjusted by manipulating the electric field through electrodes placed in the dielectric layer. The surface wave launcher \cite{Barlow-1962,Wan-apr2019,Chu-Jan2024} enable propagation of the signal between the \gls{RF}-chain and the radiating elements on the dielectric layer. This design offers not only independent adjustment of droplet sizes but also enables the splitting or combining of droplets. Such flexibility in droplet manipulation has the potential to optimize antenna efficiency and operating frequency. Reconfiguring droplet sizes contributes to enhanced antenna performance, while the ability to split and combine droplets can improve energy efficiency. 

Recently, it was reported based on experimental results that the liquid-based \gls{FAS} in Fig.~\ref{Liquid} could greatly improve outage probability and multiplexing gain for mmWave communications \cite{Shen-tap_submit2024}. Similar experimental results on \gls{FAS} using \gls{EWOD} are expected in the future and would validate the concept of \gls{FAS} further. It is fair to say that liquid-based \gls{FAS} designs are gathering momentum though such designs might be sensitive to fluctuations in environmental conditions, 

\subsection{Reconfigurable Pixel Antennas}\label{ssec:pixel}
To obtain the full potential of \gls{FAS}, the speed at which we can reconfigure the antenna will need to be in the millisecond range or less, in order to respond to the change in \gls{CSI}. Physical limits on the acceleration and velocity of the structures inside mechanical movable and liquid-based antennas could prevent their use at these speeds. To overcome this switching speed challenge, it would be necessary to consider antennas that can be reconfigured using electronic switches to realize \gls{FAS}. 

Approaches to the design of antennas that can be reconfigured using electronic switches have been well explored 
\cite{4267912,2009-TAP-GA,2010-TMTT-S_matrix_IMPM,2010-AWPL-CP_L_slot,2012-AWPL-CP_loop_slot,9531465} and they are usually referred to as reconfigurable antennas. The basic principle of reconfigurable antennas is to change the geometry of the antenna element by electronically switching in or out supplementary metallic structures. This subsequently alters the distribution of the \gls{RF} currents over the antenna surface and hence changes the antenna's radiation characteristics. A defining characteristic is that the reconfiguration of the antenna is performed internally within the antenna element and not externally through the use of phase shifters or additional antennas. 

The electronic switching elements utilized in reconfigurable antennas are typically based on PIN diodes, varactors, MEMS switches, transistors, etc. \cite{4267912}. The total number of possible antenna geometries that can be formed is referred to as the number of states. If there are $Q_p$ switches, then an upper bound to the number of states will be $S_p=2^{Q_p}$. In practice, only a subset of the $S_p$ states will be used since many of the states will not function well as an antenna. To apply reconfigurable antennas to \gls{FAS}, one possible way is to equate each state to a possible \gls{FAS} port. That is, the $N_s$ FAS ports can be thought of as $N_s$ possible states in a reconfigurable antenna. To realize \gls{FAS}, one possible approach is to exploit pixel reconfigurable antennas (PRA) \cite{1303859,2012-TAP-Pixel_slot,2014-TAP-Song_pixel,2014-TAP-Pixel_all_parameters}. In \cite{2017-TAP-Parisa_pixel}, there are $49$ switches in the antenna, potentially providing $2^{49}$ states. Below we provide an example design of PRA \gls{FAS}.  

\begin{figure}
\noindent \centering{}\includegraphics[scale=0.35]{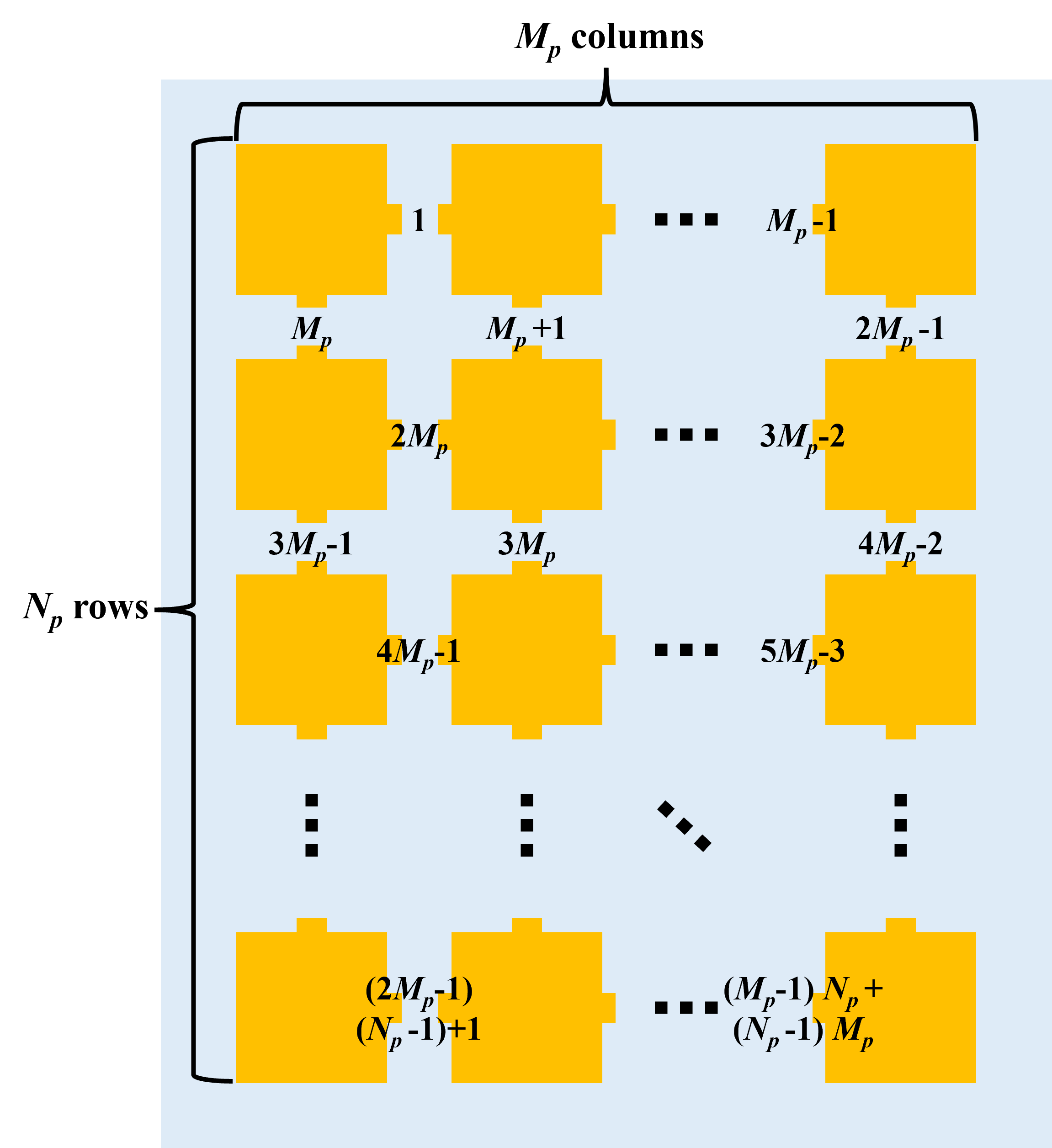}
\caption{Planar view of a pixel surface with $M_p \times N_p$ pixels. The positions for the switches between adjacent pixels are labelled with numbers and in total, there are $Q=M_p(N_p-1) + N_p(M_p-1)$ switches.}\label{geometry}
\end{figure}

The concept of PRA can be most easily described using Fig.~\ref{geometry} where a surface is divided into $N_p \times M_p$ elements. The size of each element can be arbitrary but when they are less than around $\lambda/10 \times \lambda/10$, we refer to them as pixels. In total, there are  $Q_p=M_p(N_p-1) + N_p(M_p-1)$ possible connections between adjacent pixels in the surface and therefore $2^{Q_p}$ possible states. An individual pixel in the PRA cannot radiate alone because it is significantly less than a wavelength in size. Adjacent pixels must be connected together, either with fixed wires or switches, to form a radiating surface. 

\begin{figure}
\noindent \centering{}\includegraphics[scale=0.12]{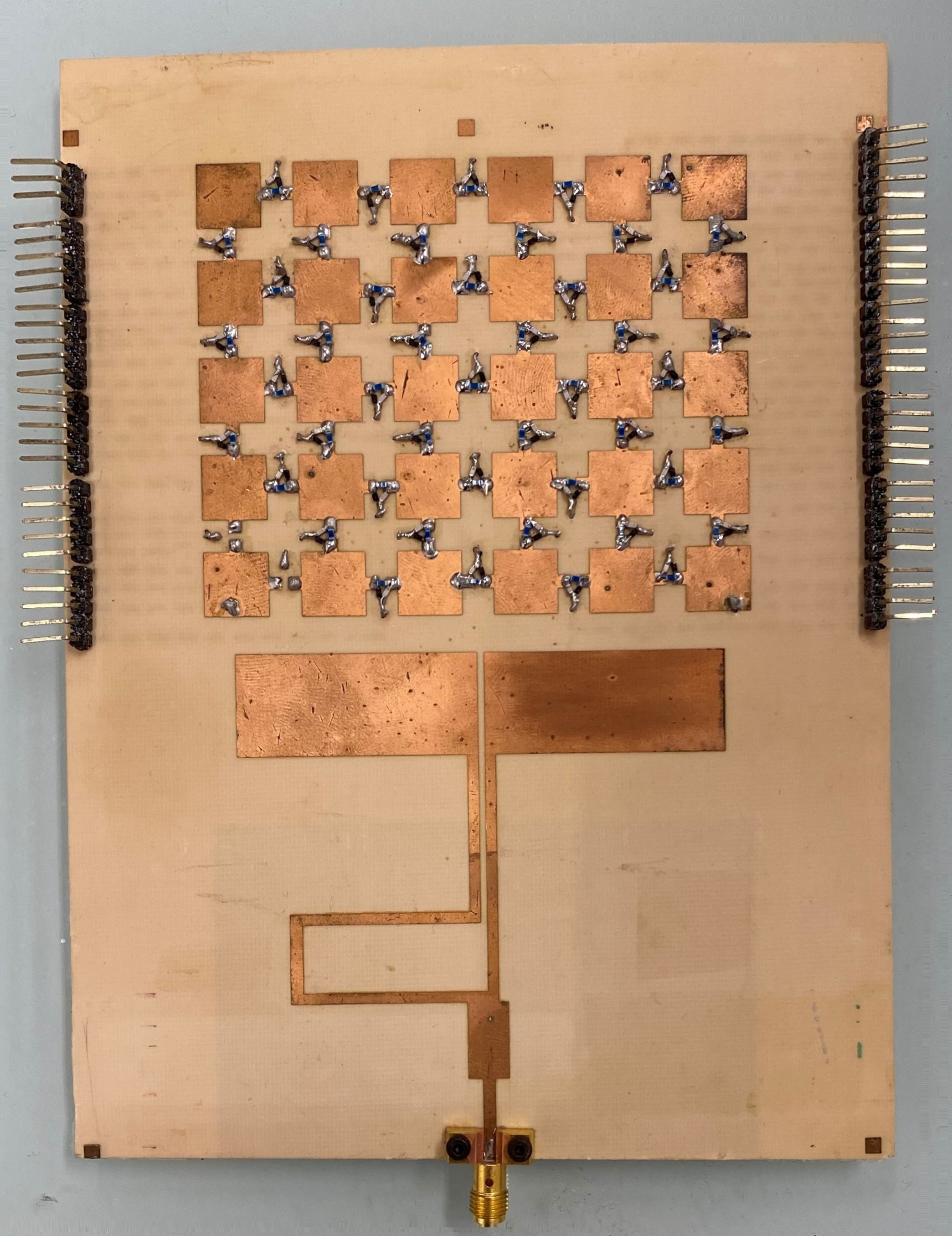}
\caption{Photo of a PRA prototype  that operates at $2.5~{\rm GHz}$ and radiates in the endfire direction (up the page) \cite{2017-TAP-Parisa_pixel}. In this prototype, $30$ pixels (with size of $10 \times 10~{\rm mm}^2$) can be seen and there are $49$ PIN diode switches between the pixels. The pixel surface is excited by a dipole located on its bottom side (the large middle element).}\label{Parisa}
\end{figure}

For the pixel surface to radiate, it must be excited by an \gls{RF} source such as a radiator (or direct \gls{RF} feed) from the side or underneath the surface. A photo of a complete PRA prototype is provided in Fig.~\ref{Parisa} which operates at $2.5~{\rm GHz}$ with endfire radiation \cite{2017-TAP-Parisa_pixel}. In the prototype, $30$ pixels can be seen and there are $49$ switches between the pixels. In the design, the pixel surface is excited by a dipole radiator located on its bottom side (the large middle element). In particular, the single \gls{RF} SMA antenna feed is located at the bottom of the photo and this feeds the dipole element (middle large element) which radiates onto the pixel surface. The connections to the $49$ digital control lines for the switches are located vertically on the left and right sides of the antenna. The routing to the PIN diodes for the digital control lines are located on the backside. By appropriately turning ``on" or ``off" the switches, using the digital control lines, a wide variety of antenna patterns can be formed. In \cite{2017-TAP-Parisa_pixel}, it was shown that a directive beam can be steered in $12$ different directions using $12$ of the $2^{49}$ possible states.  A large literature of PRA designs have been published previously \cite{1303859,2012-TAP-Pixel_slot,2014-TAP-Song_pixel,2014-TAP-Pixel_all_parameters,8952871,9769906,9743796} where multiport designs can also be considered \cite{2018-Saber-pixel}. 

\begin{figure}
\noindent \centering{}\includegraphics[scale=0.4]{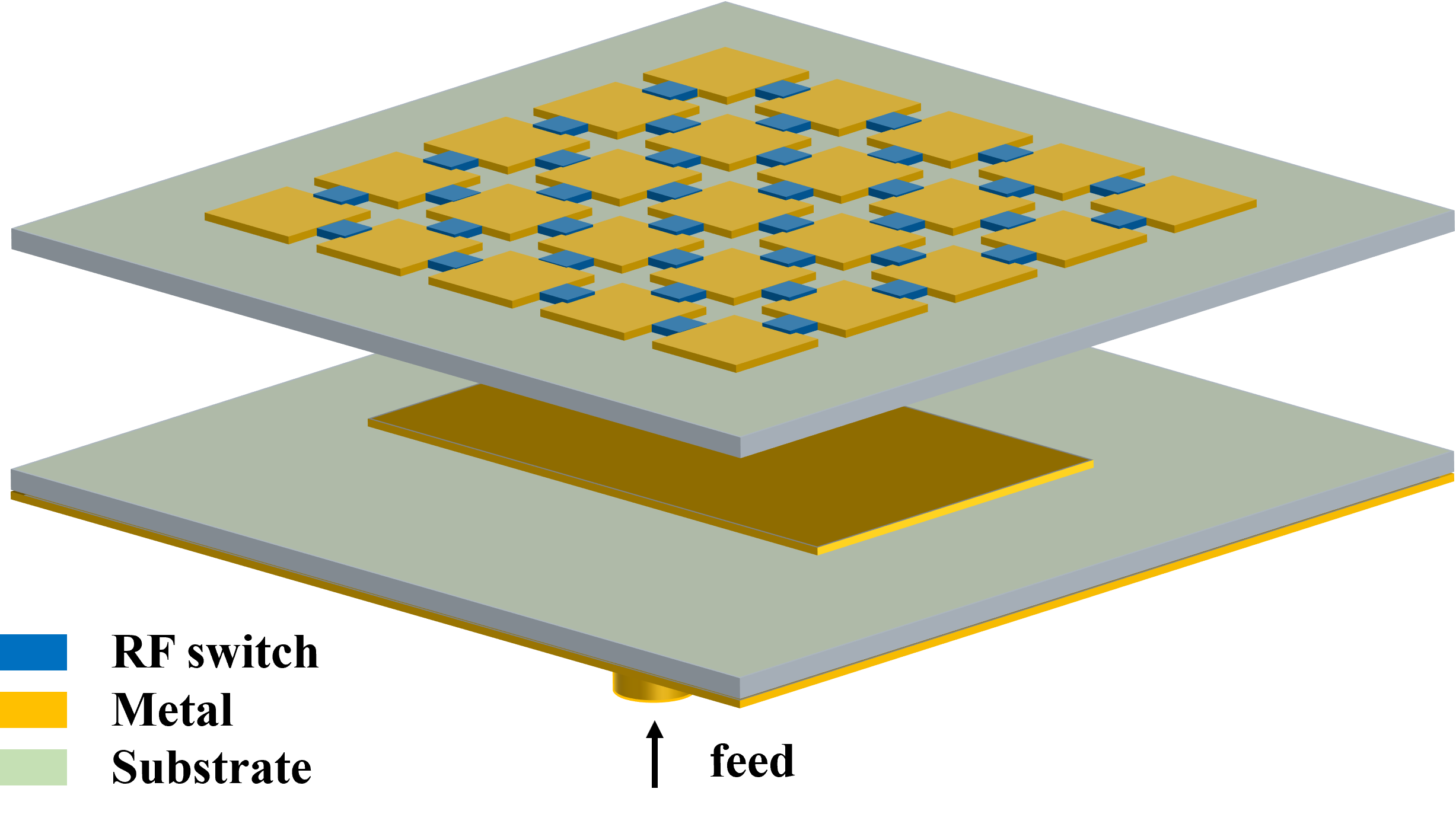}
\caption{The PRA design based on a patch antenna structure with $5 \times 5$ pixels. The pixel surface forms the top patch of the antenna and is excited by a radiator from underneath.}\label{Patch}
\end{figure}

When a PRA with broadside radiation is required instead of endfire, the structure shown in Fig.~\ref{Patch} can be used. The design is based on a patch antenna structure and in this example, there are $5 \times 5$ pixels and it is excited by a radiator from underneath. The planar ground plane of the patch antenna is underneath the bottom substrate and the SMA \gls{RF} feed connects through the bottom substrate to a parasitic radiator on top of the substrate that excites the pixel surface above it. 

To understand how to design PRA, we define its state as a vector that captures the state of each connection between adjacent PRA pixels, given by
\begin{equation}\label{Eq_pixels}
\mathbf{x} = [x_1, x_2, x_3, \dots ,x_{Q_p}],
\end{equation}
where $x_q \in \{0,1\}$ refers to whether the $q$-th connection is open (``0'') or closed (``1''). We find it useful to define $\mathcal{U}$ as the set of all possible PRA states and this set has $2^{Q_p}$ elements.

For a given PRA state $\bf{x}$, the PRA will exhibit a certain radiation pattern and input impedance. These antenna parameters can be linked to $\bf{x}$ through non-linear functions which we denote as $\mathbf{E}(\bm{\Omega},\mathbf{x})$ and $z(\mathbf{x})$, respectively, where $\bm{\Omega}$ is the three-dimensional pattern angle and the vector $\bf{E}$ represents the two far-field components of the electric field. The expressions for the non-linear functions can be determined straightforwardly and this has been well documented in previous work \cite{2014-TAP-Song_pixel,9769906}. A feature of these approaches is that a single full electromagnetic simulation can be utilized offline to find the impedance and patterns for all PRA states. 

In FAS, the number of PRA states we wish to use is set to $N_s$ which corresponds to the number of FAS ports. Thus, we can define $\mathcal{V}_l$ as a particular subset of $\mathcal{U}$ which contains $N_s$ unique vectors or states. Furthermore, we can denote the elements or states in the subset $\mathcal{V}_l$ as $(\mathbf{x}_1,\dots,\mathbf{x}_n,\dots, \mathbf{x}_{N_s})$. Theoretically, there are totally ${2^{Q_p} \choose N_s}$ different subsets $\mathcal{V}_l$ that can be constructed from  $\mathcal{U}$. When used in FAS, the $N_s$ states in $\mathcal{V}_l$ cannot be arbitrary. In fact, not all the pixel configurations will have good impedance matching. Consequently, the search space can be significantly reduced by first selecting only those subsets $\mathcal{V}_l$ that achieve a good match for each state $\mathbf{x}_n \in \mathcal{V}_l$. This matching constraint can be written as
\begin{equation}\label{Eq_match}
S_{\mathrm{feed}}\left( \mathbf{x}_n \right) < -10~{\rm dB},~\mbox{for } n=1,2,\dots,N_{s}.
\end{equation}
where
\begin{equation}
S_{\mathrm{feed}}\left( \mathbf{x}_n \right)\triangleq 20\log \left|\frac{z\left( \mathbf{x}_n \right)-Z_0}{z\left( \mathbf{x}_n \right)+Z_0} \right|~{\rm dB}
\end{equation}
is known as the return loss of the PRA and $Z_0\approx 377~\Omega$ is the impedance for free space. Thus, in the optimization process, we use only subsets $\mathcal{V}_l$ that satisfy (\ref{Eq_match}). We can then search through the selected $\mathcal{V}_l$ to find the optimum PRA design by
\begin{equation}\label{Eq_objective}
\mathop{\min} \limits_{\mathcal{V}^*_l } \sum_{n} \sum_{n'} \left|  \left|\mu(\mathbf{x}_n,\mathbf{x}_{n'}) \right| - \left| \mu^*(\mathbf{x}_n,\mathbf{x}_{n'})\right| \right|,
\end{equation}
where 
\begin{equation}\label{Eq_patterns}
\mu({\bf x}_n,{\bf x}_{n'})=\int_0^{2\pi} \int_0^{\pi} \mathbf{E} \left(\bm{\Omega},\mathbf{x}_n \right) \cdot \mathbf{E} \left(\bm{\Omega},\mathbf{x}_{n'} \right) \mathrm{d} \mathbf{\Omega}
\end{equation}
evaluates the correlation between state $n$ and state $n'$ in terms of their associated patterns, $\mu^*$ represents some desirable correlation structures, and $\mathcal{V}^*_l$ denotes the set of states that best satisfies the required FAS correlation characteristics. 

The optimization of this problem is difficult because there is an enormous number of possible subsets $\mathcal{V}_l$ and the objective function is non-linear. To overcome this, heuristic optimization techniques can be used \cite{8952871,7762757,9491941}. A recent work on how to use PRA for realizing \gls{FAS} can be found in \cite{Zhang-pFAS2024}.

\subsection{Hybrid Antennas}
It is possible to combine the above techniques for a better design specific to an application. For example, in pixel-based antennas, one can leverage liquid-based materials to adjust the conductivity of the pixels or address mutual coupling issues, such as injecting a highly conductive solution into on-pixels or a low-conductive solution into off-pixels. Moreover, with a limited codebook, one can use a liquid solution to refine the beam towards a particular direction. While the hybrid antenna design offers superior flexibility for reconfiguration, it also introduces a higher level of design complexity. Furthermore, interactions between different antenna designs could potentially give rise to unforeseen challenges.

\section{Standardization Discussion}\label{sec:standard}

Incorporating \gls{FAS} into either the \gls{BS} or UE, or both, in wireless networks has implications for standardization. It is evident that channel estimation involving \gls{FAS} is more complex than in conventional systems, as ports need to be switched successively to observe the pilots. If \gls{FAS} is combined with traditional beamforming at the \gls{BS}, more feedback capacity will be required to send this high-definition \gls{CSI} from the UEs to the \gls{BS}. This will necessitate changes in how the \gls{CSI} is quantized and fed back to the \gls{BS}. Evidently, in this case, the optimization at the \gls{BS} will involve both beamforming and antenna positions, impacting resource allocation to the UEs. The same is expected in the uplink. If \gls{FAS} is adopted at both the transmitter and receiver sides, it will be necessary to ensure that there are sufficient control messages dedicated to passing the essential parameters in the optimization procedure.


With that said, the introduction of \gls{FAS} into the UEs could actually reduce the reliance upon standardization activities for improving wireless network performance. To see this, recalling from the results in Section \ref{ssec:fama4ss}, it is known that using \gls{FAS} in particular the \gls{CUMA} architecture is highly beneficial because the \gls{BS} is simplified to perform \gls{LoS}-only precoding, greatly reducing the burden for \gls{CSI} feedback while still maintaining extraordinary multiple access capability. Since \gls{LoS}-only precoding is used, the \gls{CSI} estimation and feedback process does not change even if the number of \gls{BS} antennas is increased and/or if a UE switches to a better or worse \gls{FAS} handset. A \gls{FAS} terminal is just a plug-and-play UE which, when present, helps lift the network performance. A more advanced \gls{FAS} UE should also automatically get better performance. Evidently, a \gls{FAS} terminal should only come in several standard sizes, with appropriate channel estimation processes in place. Note that due to spatial correlation, the estimation processes depend only on size $W_s$ but not the resolution $N_s$ of the \gls{FAS}. They also get easier in higher bands as scattering becomes fewer.

To summarize, standardization efforts are required for the inclusion of \gls{FAS}. In particular, resources should be in place to facilitate the channel acquisition process. If \gls{FAS} is utilized at the transmitter side to bring in additional dof, then modifications are required on the feedback channel. However, once the \gls{CSI} can be reliably estimated for a \gls{FAS}-equipped terminal, it will always deliver better reception performance. Specifically, a more advanced \gls{FAS} handset will not only improve its own quality-of-experience (QoE), but also automatically elevate the overall network performance without requiring the \gls{BS} to do anything differently. Additionally, standards can be future-proof if \gls{FAS} terminals only come in standard sizes.

From a standardization timeline perspective, studies for 6G in 3GPP are expected to commence with Release 20, which will consider advanced MIMO such as XL-MIMO~\cite{6gstandard}. The first TSG-wide 6G workshop is anticipated in March 2025, just before the planned Release 19 RAN1 functional freeze in June 2025. The RAN Working Groups will continue their efforts over a 21-month period, from the third quarter of 2025 until the first quarter of 2027. We expect to see active discussions on fluid antenna and/or meta-antenna-based XL-MIMO.

\vspace{.5cm}
\section{Promising Research Directions in \gls{FAS}}\label{sec:Promising-Research-Directions}
This section is dedicated to the discussion of new challenges as well as the promising research directions in \gls{FAS}. We will begin by considering the new challenges of \gls{FAS}, spanning from channel models and estimations to theoretical foundations and performance limits. Then, we will deliberate on various promising research directions, focusing on the interaction between \gls{FAS} and other state-of-the-art technologies such as AI, \gls{FD} communications, green communications, ISAC, near-field communications, \gls{NGMA}, \gls{NTN}, \gls{PLS}, \gls{RIS}, \gls{THz} communications and \gls{XL}-\gls{MIMO} or \gls{CAP}-\gls{MIMO}.

\vspace{.2in}
\subsection{New Challenges within \gls{FAS}}
\subsubsection{Hardware Development and System Models}
This paper mainly focuses on theoretical channel models, as empirical channel models are currently lacking within the field of \gls{FAS}. Developing an empirical channel model requires conducting numerous channel measurements, a process contingent on the maturity of \gls{FAS} devices. However, the current state of \gls{FAS} devices remains in its infancy, with existing prototypes not yet ready for practical applications, although dedicated efforts have been given in this matter \cite{8105795,8762117,Shen-fas2022a,Shen-fas2022b,9977471,dash2023selection}. Addressing this challenge requires collaborative efforts from physicists, electromagnetic specialists, and antenna experts to create reliable \gls{FAS} prototypes and conduct comprehensive channel measurements across various environments. The measurement data can then be employed to construct empirical channel models tailored to different settings.

Also, our tutorial places primary emphasis on the spatial correlation of the ports but it is imperative to recognize that correlation exists not only in the spatial domain but also in the time and frequency domains. Recent developments have led to a temporal-spatial correlation model \cite{10207934} and frequency-spatial correlation model \cite{zhu2024performance}. But there is a need for a comprehensive joint frequency-temporal-spatial correlation model, which has the potential to further enhance the performance of \gls{FAS}. The development of theoretical channel models should also account for near-field spherical wave effects, atmospheric conditions, weather variations, and other environmental factors, especially when dealing with \gls{THz} communications \cite{tlebaldiyeva2023outage,9882323}. In addition, mutual coupling emerges as a significant challenge, particularly for pixel-based fluid antennas. Strategies such as leveraging circuit and antenna theories, which involve matching networks and employing isolation techniques in antenna design, can help mitigate mutual coupling effects \cite{4685877,4232615,10373407}. Furthermore, the response time or switching time of the fluid antenna plays a crucial role in determining the performance of \gls{FAS} \cite{10278619}. While studies like \cite{10188603,evans2023latency} indicated that the switching time issue could be addressed in liquid-based antenna designs, an alternative direction may emphasize on delay-free fluid antenna designs.

\subsubsection{System Management and Optimization}
To ensure seamless operation of \gls{FAS} within \gls{6G} networks, it is imperative to design an optimal system management. As discussed earlier, \gls{FAS} needs to estimate the channels of a subset of ports to reconstruct the \gls{CSI} of all ports. In this context, advanced techniques such as \gls{ML} and various mathematical approaches, have demonstrated highly efficient capabilities in reducing the required channel acquisition to only $10-20\%$ of all the ports \cite{10207934,ISAC_FAS,9715064,10018377}. However, the minimum number of observable ports necessary to satisfactorily recover the \gls{CSI} for different configurations of the fluid antenna surface remains unknown, including different size and dimensions. 

One possible approach to answer this question is to leverage the Nyquist-Shannon sampling theorem, which outlines the requirements for accurately reconstructing a continuous signal from its sampled, discrete version without loss of information. According to this theorem, to accurately reconstruct a continuous signal, the sampling rate must be at least twice the highest frequency present in the signal \cite{proakis2007digital}. This implies that an efficient \gls{FAS} needs to obtain information about the frequency of the fading in each channel realization to estimate the channel with the minimum number of ports efficiently. However, achieving this remains a non-trivial problem.

Another approach is to explore the functional dof from electromagnetic information theory. In electromagnetic information theory, the functional dof refers to the minimum number of required samples to reconstruct a given electromagnetic field \cite{zhu2022electromagnetic,9154219}. Contemporary studies postulate that half a wavelength is sufficient to reconstruct a continuous field \cite{9798854,di2023electromagnetic}. Nevertheless, a preliminary investigation in \gls{FAS} reveals otherwise, as the signal is space-limited and the spatial frequency spans over an infinite range. Hence, sampling less than half a wavelength can significantly reduce the \gls{NMSE}. This suggests that achieving perfect channel reconstruction based on half-wavelength sampling is still an open issue.

While the above knowledge may fill an important gap, it is crucial to recognize that the number of channels in \gls{FAS} is significantly larger than that in \gls{TAS}. As a matter of fact, this key distinction serves as the primary factor contributing to the superior performance of \gls{FAS}, even when both systems have an identical number of radiating elements. However, managing and optimizing such an extreme number of channels introduces complexity. While \gls{ML} methods offer scalability benefits, conventional management and mathematical optimization methods should not be overlooked for the sake of tractability. In other words, striking a balance between advanced techniques like \gls{ML} for scalability and conventional or mathematical approaches for tractability is vital in addressing the management and optimization complexities associated with the increased number of channels in \gls{FAS}, especially within the evolving landscape of \gls{6G} networks. Without exacerbating matters, the handover process might also become more challenging due to the channel reconfiguration capability that can complicate decision-making \cite{9539048}. Therefore, extensive efforts are still required in the management and optimization of \gls{FAS}.

\subsubsection{Theoretical Foundations and Performance Limits}
In this tutorial paper, we have delved into various technical aspects of \gls{FAS}, exploring topics such as extreme diversity gain, higher energy efficiency, and innovative techniques enabling scalable multiple access without \gls{CSI} at the transmitter and \gls{SIC} at the receiver. Nevertheless, fundamental investigations still remain in certain areas. For example, obtaining $N'\left(W\right)$ (see (\ref{eq:D_gain})) is a non-trivial task, although efforts have been given in this state-of-affair \cite{10130117,10303274}. It is important to note that $N'\left(W\right)$ is not the number of independent channels or functional dof, rather it is the number of significant eigenvalues that can enhance diversity. The current approach relies on approximating the non-negligible eigenvalues of the \gls{FAS} channel under the assumption that the number of ports is extremely large, approaching infinity. However, this approach is often hindered by numerical issues, and numerical analysis techniques can be employed to improve the approximation \cite{burden2011numerical}. Another alternative is to directly employ functional analysis.

Furthermore, the analysis of \gls{FAS} performance proves to be extremely difficult in many cases \cite{lopez2023novel,10308603,alvim2023performance,9771633,9833952,Leila2023B5G,10375698}. Making this matter worse, even minor adjustments to the system model can render the problem intractable. Therefore, novel approaches are crucial for effective performance analysis. Moreover, it is observed that the rate does not show drastic improvement simply by increasing the number of ports or the size of the fluid antenna surface. Consequently, innovative solutions are still required to enhance the network multiplexing gain, meeting the demands of massive communication. Additionally, the superior performance of \gls{FAS}-NOMA is applicable only when the access point has a single antenna. In cases where the access point is equipped with multiple antennas, RSMA may be more efficient in some cases \cite{mao2018rate}. Thus, exploring the integration of \gls{FAS} and RSMA, and other multiple access schemes remains of great interest.

Under the assumption that \gls{CSI} and \gls{SIC} are unavailable at the transmitter and receivers, respectively, \gls{FAMA} opens up new possibilities for multiple access techniques. But relying on straightforward concepts like slow \gls{FAMA} or fast \gls{FAMA} solely may not be sufficient to achieve TK$\mu$ extreme connectivity as these schemes are still interference-limited. In fact, the pursuit of TK$\mu$ extreme connectivity calls for the development of more advanced schemes. In this context, \gls{CUMA} or other variations of \gls{FAMA} could play more pivotal roles. For instance, \gls{CUMA} presents the potential to achieve $1~{\rm Kbps/Hz}$ by serving one thousand users per a frequency-time resource. Nevertheless, analyzing the performance of \gls{CUMA} in the presence of mutual coupling or computing the performance at an extreme scale poses considerable challenges. Moreover, the development of more innovative solutions would be encouraging in advancing the quest for TK$\mu$ extreme connectivity. Lastly, developing stochastic geometry frameworks that consider accurate spatial correlation, mutual coupling, and optimization effects can be immensely valuable since it allows researchers to evaluate the impacts of large-scale deployment in future \gls{6G} networks.

\subsection{Synergy of \gls{FAS} and Other Technologies}
\subsubsection{AI}
\gls{ML} techniques can play essential roles in tackling the complexity and scalability challenges when serving multiple \gls{FAS} users. Techniques such as \gls{CNN}, \gls{GNN}, \gls{LSTM} and reinforcement learning, among others, play crucial roles in tasks such as channel estimation, resource allocation optimization, handover procedures, task-oriented management, and more \cite{ISAC_FAS,10299674,10018377,9830377}. For instance, unsupervised learning can autonomously discover joint frequency-temporal-spatial correlations, optimizing the performance of \gls{FAS} \cite{10146286}. Deep learning can classify index patterns for \gls{FAS} based on index modulation, a system proposed in \cite{Elio2023,zhu2023index,Xu-2024,Faddoul-2024}, or explore optimal ports for beamforming \cite{ISAC_FAS}. \gls{ML} has also proven valuable for handover and task-oriented management, particularly in systems with massive channels \cite{9896861}. Conversely, \gls{FAS} can improve the performance of AI applications such as computation accuracy, as evidenced in \cite{zhang2023fluid,cheng2023movable}. Interestingly, \gls{FAS} applications in autonomous vehicles and unmanned aerial vehicles (UAVs) show tremendous potential, given that the positions of both the radiating elements and the communication objects themselves can be controlled \cite{zheng2023flexibleposition}.

\subsubsection{\gls{FD} Communications}
Compared to conventional half-duplex communications, \gls{FD} communications is a promising technology that can potentially double the spectral efficiency. This capability has garnered significant interest from both the research community and the industry \cite{10158724}. The combination of \gls{FAS} and \gls{FD} communications holds the potential for mutual benefits. Although \gls{FD} communications is anticipated to boost the network spectral efficiency by overlapping the uplink and downlink signals if self-interference can be mitigated, this introduces both intra- and inter-cell interference, posing a threat to the performance of large-scale multi-cell networks. The extra dof by \gls{FAS} can thus be important to address multiuser interference by selecting the port that either maximizes the \gls{SINR} or experiences minimal interference. For instance, \cite{10167904} adopted a multi-port fluid antenna technology to propose a cooperative communication strategy to elevate the performance of a multiuser \gls{FD} NOMA network. Moreover, an analytical framework utilizing stochastic geometry has been derived to evaluate the outage and average sum-rate performance of fluid antenna-assisted \gls{FD} cellular networks \cite{10184308}.

\subsubsection{Green Communications}
As global warming and climate change pose serious concerns, researchers are actively exploring techniques to make wireless communication systems more sustainable. In this context, \gls{FAS} has emerged as a valuable technology. On the one hand, \gls{FAS} can minimize transmission power while meeting user rate requirements \cite{10354003,qin2023antenna,wu2023movable}. On the other hand, \gls{FAS} can maximize energy efficiency in various scenarios \cite{10354059,10208068}, thanks to the diversity gains it offers. Furthermore, a paradigm shift from traditional information transmission to power transfer has occurred, influenced by groundbreaking works such as \cite{4595260} and \cite{6489506}. Nevertheless, wireless information transfer and power transfer exhibit distinct behaviors. For instance, interference can be utilized to enhance received power for energy harvesting, as suggested by \cite{7024146}, but high received power does not necessarily translate to maximum information transfer. Conventional approaches typically focus on maximizing either wireless information or power transfer, using techniques like power allocation, beamforming, and time switching or power splitting. In contrast, \gls{FAS} introduces a novel approach to address this tradeoff \cite{lin2023performance,Zhang-Hu-faswpt}. Specifically, the \gls{SINR} and received power over the ports in \gls{FAS} exhibit different trends. This unique capability of \gls{FAS} suggests that specific ports can be dedicated to maximizing the \gls{SINR} for information reception, while others can be assigned to maximize the harvested energy \cite{10146286}. 

\subsubsection{ISAC}
Cellular networks have undergone a transformation, shifting from their initial purpose of providing communication services to becoming ISAC networks, also referred to as dual-functional radar-communication networks. The ultimate goal of ISAC is to create perceptive mobile networks where both communication and radar sensing functions are seamlessly combined within a single system that shares the same frequency band and hardware. The sensing operation of ISAC involves extracting essential information about targets and their surroundings. Notably, \gls{MIMO} schemes have gained widespread recognition for their pivotal role in this configuration because they offer precoding capabilities for achieving spatial beamforming and waveform shaping \cite{9585321}. 

Presumably, \gls{FAS} could provide an additional dimension for enhancing the performance of ISAC \cite{ISAC_FAS}. Specifically, for ISAC transmitter, we can employ \gls{FAS} to optimize the ports and precoder jointly to establish favorable communication and sensing channels for performing sensing and communications simultaneously. On the other hand, for ISAC receiver, we can employ slow or fast \gls{FAMA} to leverage the variations in the channel envelope to eliminate interference due to radar sensing \cite{10066316,9953084}. Also, \gls{FAS} enables a flexible coexisting approach to use the spatial resources for realizing radar sensing and communications simultaneously. In fact, the greatest impediment for the ISAC network is to balance communication and sensing performance of the dual-functional waveform \cite{9296833}. 

\gls{FAS} provides a novel approach to unify the communication and sensing functionality by creating an ISAC channel that can make them reinforce each other. In particular, the spatial steering vectors of the sensing and communication channels can be aligned using \gls{FAS}. In such cases, the functionalities of communication and sensing can be achieved with a common signal waveform, simplifying the network design significantly. Furthermore, since the communication waveform can be utilized for sensing directly, removing the need for dedicated signals for sensing, the interference due to the sensing requirement can be reduced greatly. A crucial challenge in a \gls{FAS}-assisted ISAC system is however the need to optimize the port and beamforming jointly, which is an NP-hard optimization problem. In \cite{ISAC_FAS}, a deep reinforcement learning and pointer network was proposed to build an end-to-end learning network for addressing the intractable joint optimization problem.

\begin{figure}
\centering \includegraphics[scale=0.3]{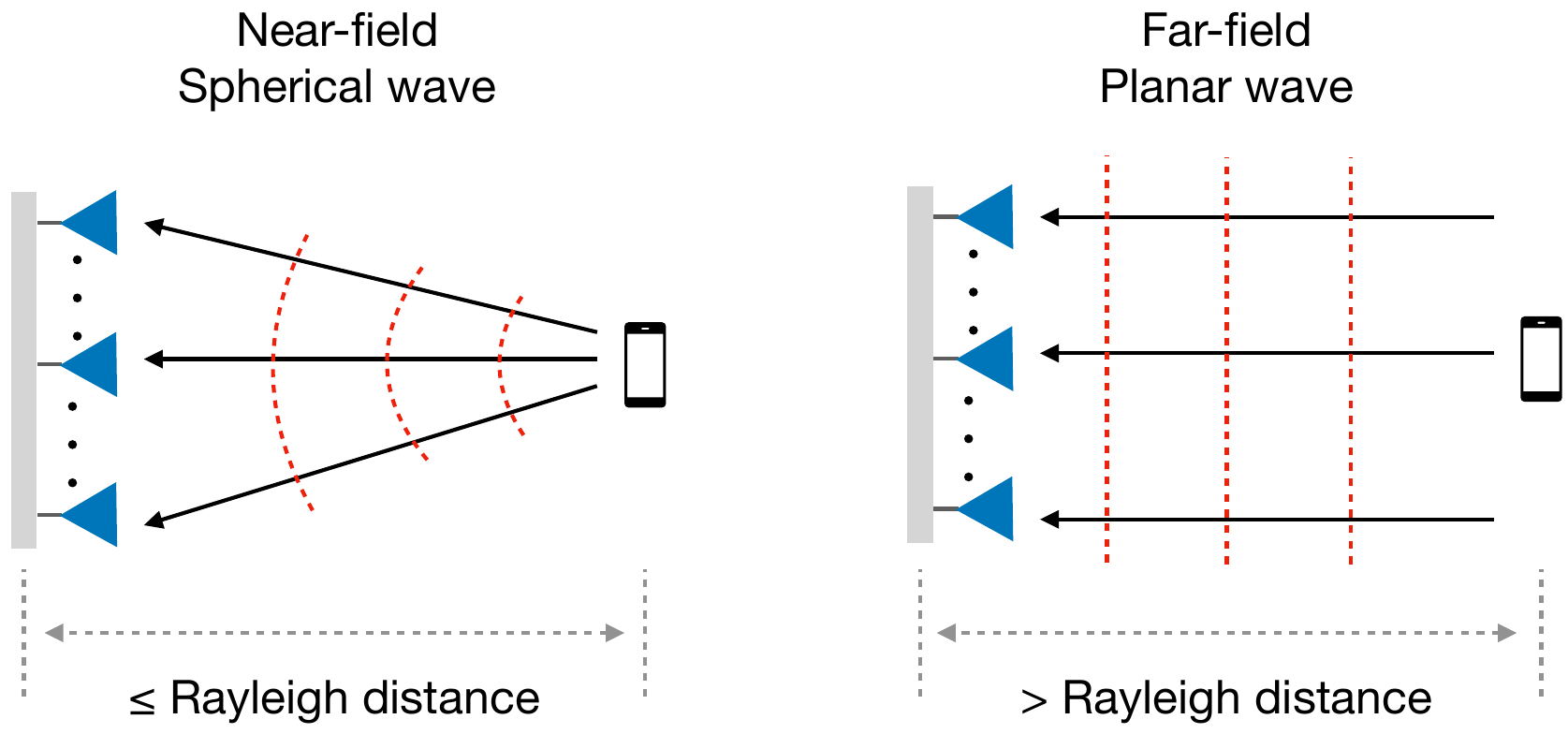} \caption{Near-field and far-field communications.}\label{near_field}
\end{figure}

\subsubsection{Near-Field Communications}
Traditionally, wireless systems are assumed to operate in the far field, where the wavefront can be approximated using planar waves. However, with increasing operating frequencies and antenna array sizes, this approximation may no longer hold accurate, necessitating the adoption of a near-field model. A fundamental premise of near-field communications is the Rayleigh distance, which defines the boundary between the near-field and far-field regions. The Rayleigh distance is proportional to the product of the square of the array aperture and the operating frequency \cite{7942128,9184098}. As shown in Fig.~\ref{near_field}, if the distance between the transmitter and the receiver exceeds the Rayleigh distance, the region is considered far-field, in which the electromagnetic field can be approximately modeled by planar waves. If the distance is less than the Rayleigh distance, then the region is considered near-field, and the electromagnetic field should be modeled using spherical waves \cite{9598863,9903389}. Unlike the far-field case where signal beams can focus energy at a certain angle due to different electromagnetic wave propagation models, near-field signal beams can concentrate energy at a specific location with a given angle and distance \cite{9536436}. This property can be exploited to mitigate inter-user interference and enhance spectrum efficiency. Therefore, near-field beams provide additional resolution in the distance domain, enabling possibilities like location division multiple access \cite{10123941}. Since the near-field effect becomes more pronounced as the operating frequency and antenna array size increase, it is crucial to investigate the performance of \gls{FAS} in the context of near-field communications, as a large surface area is desired in \gls{FAS}. In addition, exploring how more users can be accommodated using the concept of \gls{FAMA} in a near-field environment is intriguing.

\subsubsection{\gls{NGMA}}
To facilitate massive communications in \gls{6G}, the adoption of \gls{NGMA} is evidently crucial \cite{9693417,9806417}. One avenue of exploration involves examining the interplay between existing multiple access methods such as \gls{OMA}, NOMA, RSMA, grant-free access and \gls{FAMA}. A notable advantage in this research direction lies in the availability of \gls{CSI} at the transmitter, enabling effective interference management and enhancing network connectivity and capacity \cite{zheng2023fas}. Specifically, \gls{FAS}-\gls{NGMA} should adeptly support multiple users across allocated resource blocks, encompassing spatial moments, time slots, frequency bands, spreading codes, and power levels. This approach has the potential to further enhance the connectivity, spectral efficiency, energy efficiency, and latency of wireless communication systems. On an alternative trajectory, one can delve into the exploration of novel multiple access schemes, such as slow or fast \gls{FAMA}, opportunisitic \gls{FAMA}, \gls{CUMA} or other variations of \gls{FAMA} schemes where \gls{CSI} and \gls{SIC} are not mandated at the transmitter and receiver, respectively. The advantage in this research direction lies in the simplification of the multiple access protocol, as \gls{CSI} acquisition, beamforming, and power allocation are not required at the transmitter, and \gls{SIC} is not necessitated at the receiver. However, under certain situations, \gls{CSI} and \gls{SIC} are feasible and can bring additional performance gains. It is important to look into the synergy and see how \gls{FAS} can benefit from the presence of \gls{CSI} and \gls{SIC}. Given the emergence of novel multiple access schemes, it is also crucial to analyze their performance to obtain deeper insights. Consequently, some efforts have been given in this matter, see e.g., \cite{10078147,10122674,wong2023rate}, but consideration of more practical assumptions in the analysis needs to be sought.

\subsubsection{\gls{NTN}}
In order to realize ubiquitous connectivity, \gls{6G} is anticipated to extend beyond terrestrial boundaries, incorporating technologies such as drones, balloons, satellites, and more \cite{9609875}. Compared to terrestrial devices, non-terrestrial devices may experience \gls{LoS} links, introducing challenges related to strong intra-cell and inter-cell interference. To address these issues, researchers have actively explored multiple access and interference management schemes \cite{9860935,10122514,9070200,9459605,new2021application,9750898}. Interestingly, \gls{FAS} presents promising solutions for \gls{NTN}. It was shown in \cite{10278220} that the full array gain in \gls{FAS} can be harnessed over the desired direction while null steering can be realized over all undesired directions. Building upon the concepts from \cite{ibrahim2021intelligent,9482610,9906810}, the position reconfiguration of fluid antennas can also be strategically employed to increase the rank-deficiency of the \gls{LoS} channels. To a certain degree, fluid antennas can also be utilized with optimal trajectories or positioning to jointly configure the small-scale and large-scale fading effects for \gls{NTN}. Conversely, trajectory optimization, which is widely adopted for UAV communications, may also be applied in \gls{FAS} \cite{zheng2023flexibleposition}. Moreover, handover issues may be addressed via the use of fluid antennas and \gls{ML} methods.

\subsubsection{\gls{PLS}}
Since \gls{FAS} provides an additional dof and significant communication gains over \gls{TAS}, some researchers have investigated the potential of \gls{FAS} in the context of \gls{PLS} \cite{10092780,PLS_FAS,cheng2023enabling,hu2023secure,JD24}. Specifically, \cite{10092780} considered a \gls{FAS}-assisted wiretap channel consisting of a legitimate transmitter (Alice), a legitimate receiver (Bob), and an eavesdropper (Eve). In this setup, Alice transmitted not only the information-bearing signal but also a jamming signal to disrupt Eve. The secrecy rate of the system was maximized by port selection and power control. The study revealed that employing \gls{FAS} yielded secrecy performance comparable to the scenario where Bob utilized multiple traditional antennas and \gls{MRC}. Nevertheless, in \cite{10092780}, Alice adopted Gaussian noise for jamming, causing interference not only to Eve but also to Bob as well. To further improve this, \cite{PLS_FAS} revisited the same system in which Alice adopted a different jamming mode to transmit encoded codewords instead of Gaussian noise. This allowed Bob to successfully decode the jamming signal and cancel the interference, while Eve could not. With such an approach, the secrecy performance of the system was greatly enhanced. Different from \cite{10092780} and \cite{PLS_FAS}, where the fluid antenna was deployed at Bob, \cite{cheng2023enabling} and \cite{hu2023secure} investigated scenarios with multiple fluid antennas at the transmitter, i.e., Alice. In \cite{cheng2023enabling}, a single Eve was considered while \cite{hu2023secure} examined scenarios with multiple Eves. Algorithms were proposed to optimize the beamforming vector and the positions of the fluid antennas, demonstrating that \gls{FAS} can significantly enhance the secrecy performance. Furthermore, \cite{JD24} provided a thorough secrecy analysis for \gls{FAS} in spatially correlated Nakagami-$m$ fading channels. \gls{FAS} has also recently been considered to maximize the average monitoring rate in proactive monitoring systems that oversee suspicious communication \cite{Yao-2024cl}.

Despite the work above, there are still many open questions that remain unanswered. For example, if both Alice and Bob are equipped with multiple fluid antennas, how should the system be designed to approach or achieve the optimal secrecy performance? Can Alice's or Bob's fluid antennas still improve the system secrecy if Eve also uses fluid antennas to enhance its wiretapping capability? These questions point to the need for further investigation into the intricate interactions and dynamics in scenarios where all parties employ \gls{FAS}. Future research efforts could delve into the design and optimization strategies for secure communication systems involving multiple \gls{FAS}-equipped entities, addressing potential challenges and uncovering new insights in the realm of \gls{PLS}.

\subsubsection{\gls{RIS}}
With the shortening of communication distance due to the use of higher operating frequencies in \gls{5G} and beyond systems, \gls{RIS} has emerged as an important fix to repair broken links between the \gls{BS} and UEs, reducing the need for network densification which is extremely costly. However, the doubly fading of the cascaded channels in \gls{RIS} means that the received signal power is usually weak compared to noise, despite the beam-focusing of \gls{RIS}. \gls{FAS} here can thus provide the diversity desperately needed at the UE to mitigate the noise effects. In \cite{Ghadi-2024wcl}, it was shown that \gls{FAS} can significantly improve the outage probability and delay outage rate for the \gls{RIS} setup. In addition, recently, the joint optimization of the \gls{BS} precoding, the \gls{RIS} phase shifting matrix and the \gls{FAS} positions in the multiuser setup was investigated in \cite{Boyi-2024fasris}. The results showed that the required transmit power of the \gls{BS} could be reduced substantially while meeting the users' \gls{SINR} constraints.

While \gls{RIS} and \gls{FAS} clearly can serve as additional dof and benefit from each other, they can also synergize each other from a different perspective. In particular, slow \gls{FAMA} works by finding a favorable position in \gls{FAS} for each UE such that the aggregate interference power is minimal. This is likely only if the channels have sufficient fluctuation in space to exploit, such as in rich scattering scenarios. Otherwise, the interference mitigation capability of slow \gls{FAMA} degrades significantly, as would be in the mmWave band \cite{wong2022extra}. In this case, it is possible to utilize \gls{RIS} as artificial scatterers that recreate rich scattering conditions for \gls{FAMA} to function well \cite{wong2022bruce,9770295}. This idea is easy to realize as it involves no optimization of the \gls{RIS} and the beam-focusing is performed by nature as the multiple paths combine at the \gls{FAS} of each UE. This technique was explored in \cite{10146262} where it was demonstrated that artificial scatterers via random \gls{RIS}s effectively restored the performance of \gls{FAMA}.

Apparently, it is still not well understood how to bring the best of both when they combine. Future work is needed to explore proper situations and applications where the performance of \gls{RIS}-\gls{FAS} justifies the increased complexity. Evidently, it is also of great interest to study its performance limits.

\subsubsection{\gls{THz} Communications}
While mmWave communications ($30-100~{\rm GHz}$) is anticipated to unlock the capacity in \gls{5G}, achieving data rates in the terabits-per-second (Tbps) range remains an impossible feat. Accordingly, \gls{THz} communications ($0.1-10~{\rm THz}$) have emerged as a highly promising solution for future \gls{6G} networks to address spectrum scarcity and capacity limitations \cite{9782674,9794668}. In contrast to mmWave- and centimeter-wave communications, \gls{THz} communications offers advantages such as broad bandwidth, narrow beamwidth, and high directivity. However, \gls{THz} communications also comes with the drawbacks of high propagation loss, susceptibility to \gls{LoS} blockage, and rank-deficient channel. Recognizing that \gls{FAS} provides an additional dof by dynamically reconfiguring radiating elements, integrating these two technologies becomes crucial. For example, in \cite{tlebaldiyeva2023outage}, the outage performance of a point-to-point \gls{THz} communication exploiting \gls{FAS} over correlated $\alpha$-$\mu$ fading channels was studied. The results showed that \gls{FAS} can address key challenges such as high path loss, blockages, and molecular absorption effects in the \gls{THz} band.

\subsubsection{\gls{XL}- and \gls{CAP}-\gls{MIMO}}
Another emerging technology that has attracted significant interest is \gls{XL}-\gls{MIMO} or \gls{CAP}-\gls{MIMO}. As discussed earlier, the antenna aperture of \gls{XL}-\gls{MIMO} can be discrete or continuous. As a matter of fact, there is similarity between \gls{XL}-\gls{MIMO} and pixel-based antenna.\footnote{Evidently, it is worth pointing out that a key difference in pixel-based antenna systems is that optimization in electromagnetic properties and signal processing is jointly done, which is not possible in \gls{XL}-\gls{MIMO}.} This suggests that key enabling hardware designs in \gls{XL}-\gls{MIMO} may be applicable to \gls{FAS}, and vice versa. However, \gls{FAS} typically operates with a fewer number of \gls{RF} chains and can be seen as \gls{XL}-\gls{MIMO} with antenna selection, the size of which may not be restricted. The latter design is referred to as \gls{CAP}-\gls{MIMO} \cite{5707050}, also known as holographic \gls{MIMO} \cite{9374451} or large intelligent surface \cite{9139999}. Unlike \gls{FAS}, the entire aperture in \gls{CAP}-\gls{MIMO} is fully used for communications. Recent works in this area have explored pattern optimization \cite{10158997}, channel estimation \cite{9716880}, antenna design \cite{Yurduseven:18} and an approximation via discrete antennas \cite{10303285}. Although \gls{CAP}-\gls{MIMO} has shown promising gains, much of the theory is still not fully understood. In terms of implementation, the holographic concept offers a promising approach to optimize precoding. However, the technology bottleneck arises from the necessity to fully manipulate the current distribution of a continuous aperture. Therefore, \gls{FAS} can be interpreted as a simplified \gls{CAP}-\gls{MIMO} architecture. Specifically, \gls{FAS} can activate several regions of ports instead of requiring control over the current distribution across an entire radiating surface. The number of regions depends on the number of \gls{RF} chains of the transceiver. The advantage of such architecture is that there is no need to fully manipulate the current distribution over any position of a continuous aperture antenna. Furthermore, in this architecture, the impact of transmit power and noise can be better understood \cite{10303285}. Also, this architecture opens the new possibility of implementing a simple and scalable multiple access scheme, as demonstrated in \cite{10318083,wong2022extra}.

\subsubsection{Integrated Computing and Communication}
Cutting-edge information processing technologies like AI and \gls{ML} have the potential to deliver ubiquitous computing and intelligent services, facilitating the efficient analysis and processing of massive data from wireless devices. However, leveraging these approaches encounters significant challenges due to constraints such as limited radio resources, the need for ultra-low latency, ultra-high reliability, and high capacity in the next generation of wireless communications, i.e, \gls{6G} technology \cite{she2020deep}. Hence, the traditional \textit{communication-before-computing} technique is not able to handle such massive data computation from smart wireless devices, e.g., \gls{IoT}, wearable, and sensor devices, due to excessively high latency and low spectrum efficiency. 

One intelligent solution to tackle this is to integrate computing into communication, referred to as \textit{over-the-air computation} (AirComp) \cite{abari2016over,zhu2021over,qi2020integration}. Contrary to conventional wireless communication via multiple access channels, which needs data transmission and decoding, AirComp utilizes the signal-superposition feature of wireless multiple access channels to compute a class of nomographic functions \cite{goldenbaum2014nomographic} of distributed data from smart devices via concurrent transmission. Hence, \textit{communication-while-computing} is possible \cite{nazer2007computation}. 

Incorporating \gls{FAS} into AirComp has potential to provide a pioneering fusion of dynamic adaptability and distributed intelligence \cite{zhang2023fluid,cheng2023movable,zuo2024fluid}. This synergy can be achieved by exploiting the ``fluidic'' electromagnetic wave properties of \gls{FAS}, while AirComp harnesses distributed computational resources for real-time data processing and optimization. More precisely, \gls{FAS}, when integrated with AirComp, can not only transmit and receive data but also actively engage in computational tasks. In this regard, fluid antennas that are equipped with computational capabilities will become active nodes in the network that are able to perform signal processing, data fusion, and optimization tasks in real-time. Crucially, channel strength variation over the \gls{FAS} ports at the transmitter side can provide high-resolution computation in the analogue domain without the need of power control. Such distributed intelligence improves the energy and spectral efficiency of wireless communications for mobile edge computing, i.e., reducing latency and maximizing bandwidth utilization. 


Nonetheless, coordinating the dynamic adjustments of \gls{FAS} alongside the computational tasks performed by AirComp can be intricate, and ensuring that the computational resources by AirComp are synchronized with the changing configurations of \gls{FAS} requires sophisticated algorithms and communication protocols. In addition, constraints on computational resources, e.g., CPU, memory, available within AirComp can complicate the optimization of \gls{FAS}. Besides, given dynamic environmental conditions, both \gls{FAS} and AirComp will need to quickly make real-time adaptations to maintain desirable performance, which is not straightforward. Most importantly, optimizing the allocation of computational resources based on the dynamic characteristics of the \gls{FAS} requires formulating the problem as a multi-objective optimization problem, which is an NP-hard combinatorial optimization problem with intricate coupling amongst optimization variables. Hence, designing mathematical models and optimization algorithms that can dynamically allocate computational resources within AirComp to maximize the performance of such an integrated system while adapting to the changing characteristics of the \gls{FAS} is critical.

\subsubsection{Others}
There are evidently more areas deserving discussion that can benefit from the application of \gls{FAS} but are not covered in this article. For instance, it is speculated that coding that imposes time correlation of data could help a \gls{FAS}-equipped UE figure out the best port on a per-symbol basis, for realizing fast \gls{FAMA}. Furthermore, the results in Section \ref{ssec:fama4ss} revealed that \gls{FAS} could greatly simplify precoding in the case of \gls{LoS}-dominant channels. Therefore, Li-Fi (light fidelity) and cell-free \gls{MIMO} may be ideal scenarios for applying \gls{FAS}. In addition, \gls{FAMA} has a unique capability of dealing with co-channel interference without precoding and should find applications in cognitive radio networks. On the other hand, \gls{CUMA} may be regarded as performing hybrid beamforming without phase shifters in the analogue domain. Thus it is reasonable to expect that certain hybrid signal processing techniques may be useful to enhance the \gls{CUMA} architecture. Last but not least, while semantic communication is gathering much attention in recent years, the operating conditions that do need semantic communication are usually hostile and narrowband, e.g., not permitting heavy channel coding and interference-rich. In this case, \gls{FAS} can be an essential tool to provide some immunity. In summary, there are many uncharted territories that involve \gls{FAS} but are not mentioned in this article.


\section{Conclusion}\label{sec:conclude}
This paper provided a comprehensive tutorial on \gls{FAS}, an emerging shape-flexible position-flexible antenna technology poised to redefine the landscape of \gls{6G} and beyond networks. Specifically, it was pointed out that it is essential to consider the fluid antenna architecture, circuit and system configuration, spatial correlation, and environmental factors to develop physics- and electromagnetic-compliant models for \gls{FAS}. Also, we demonstrated that with \gls{ML} and advanced mathematical methods, \gls{FAS} could obtain the full \gls{CSI} by only observing on a few ports (or positions) due to the strong spatial correlation or channel sparsity. Also, the extreme number of channels that can be accessed by \gls{FAS} contributes to its superiority over \gls{TAS}, not only in diversity gain but also energy efficiency. Moreover, \gls{FAS} not only liberates the fundamental tradeoff between signal maximization and null steering but facilitates rapid channel hardening. Intriguingly, \gls{FAS} opens the door for novel multiple access schemes to increase the network multiplexing gain in which \gls{CSI} is not required at the transmitter and \gls{SIC} is also not necessary at the receiver. This becomes possible because the interference signal may experience deep fade in the spatial domain while the desired signal remains strong. In addition, the hardware designs are left open for researchers to determine although we reviewed several emerging hardware designs for realizing \gls{FAS}. This open-ended approach allows researchers to define their own hardware architecture, pushing the envelope of what is possible in \gls{FAS}-assisted wireless communications. On the other hand, we discussed how \gls{FAS}-assisted wireless communications can be potentially useful to various state-of-the-art technologies. While we acknowledge the challenges of \gls{FAS} and its fast evolving nature, it is hoped that this tutorial will serve as a guiding light. We envisage this paper being not just a culmination of existing information but a catalyst for future breakthroughs. As the journey to \gls{6G} has already begun, we invite researchers to explore new horizons and unlock the full potential of \gls{FAS} for unprecedented performance. 

\bibliographystyle{ieeetr}

\begin{IEEEbiography}[{{\includegraphics[clip,width=1in,height=1.25in]{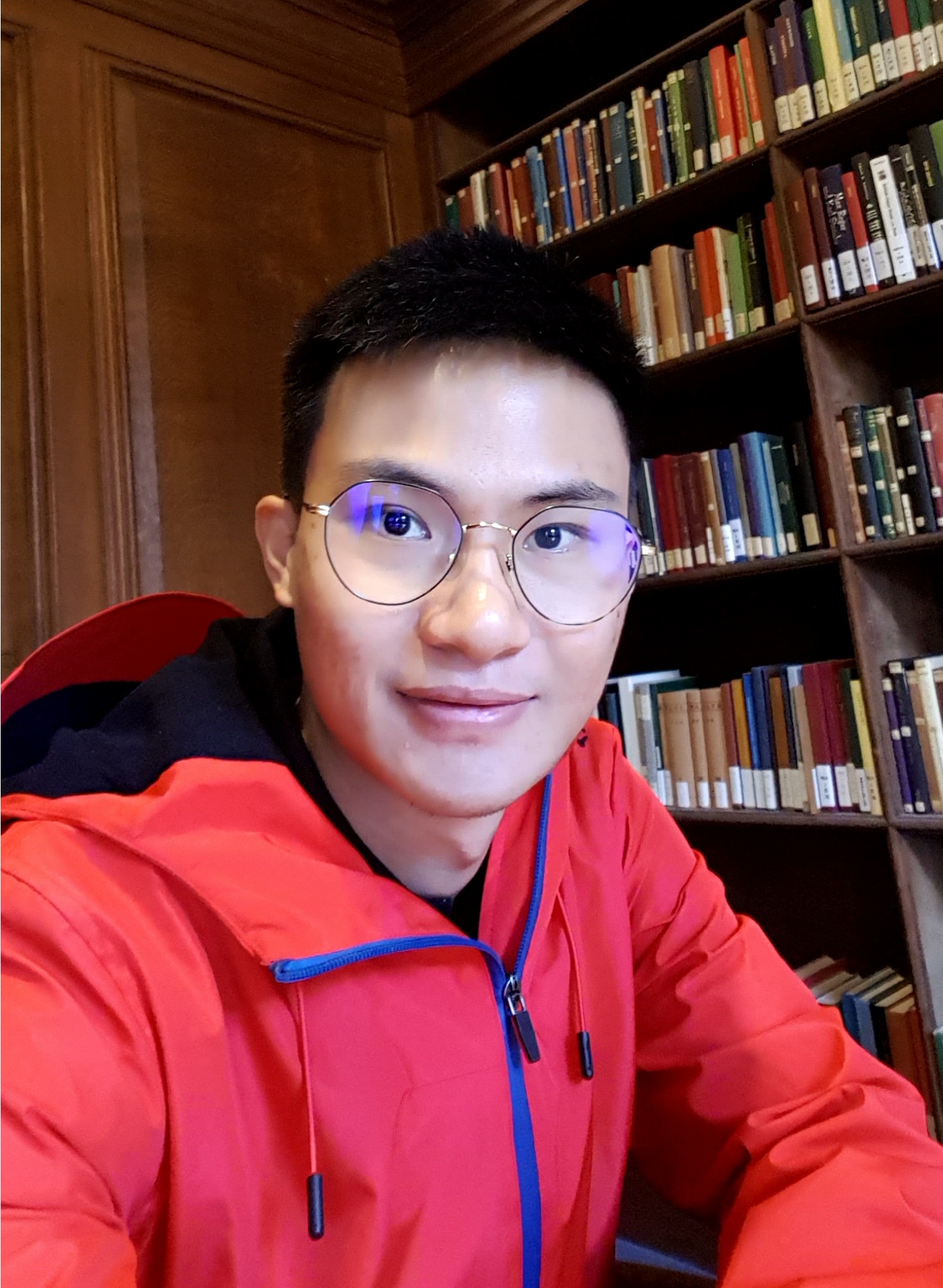}}}]{(Aven) Wee Kiat~New}
received his Ph.D in Electrical Engineering from Universiti Teknologi Malaysia, M.Eng.Sc in Electrical Engineering from University of Malaya and B.IT in Data Communications and Networking from Multimedia University. He was a visiting researcher at Lancaster University and University of Cyprus. He is currently a Research Fellow at the Department of Electronic and Electrical Engineering, University College London, UK. His research interests include information theory, optimization, stochastic processes, machine learning, and their applications in emerging areas of communications. He serves as an Associate Editor for  IEEE Transactions on Vehicular Technology and as a Guest Editor for the IEEE Journal on Selected Areas in Communications on Fluid Antenna System and Other Next-Generation Reconfigurable Antenna Systems for Wireless Communications. He was also the TPC co-chair for the 2024 ICC Workshop on Fluid Antenna Systems for 6G. He was the recipient of the 2021 IEEE Malaysia Comsoc/VTS Best Paper Award, the IEEE Malaysia AP/MTT/EMC Best Paper Awards in 2020, 2021, and 2022, and the 2024 IEEE ISTT Best Paper Award.
\end{IEEEbiography}

\begin{IEEEbiography}[{\includegraphics[width=1in,height=1.25in,clip,keepaspectratio]{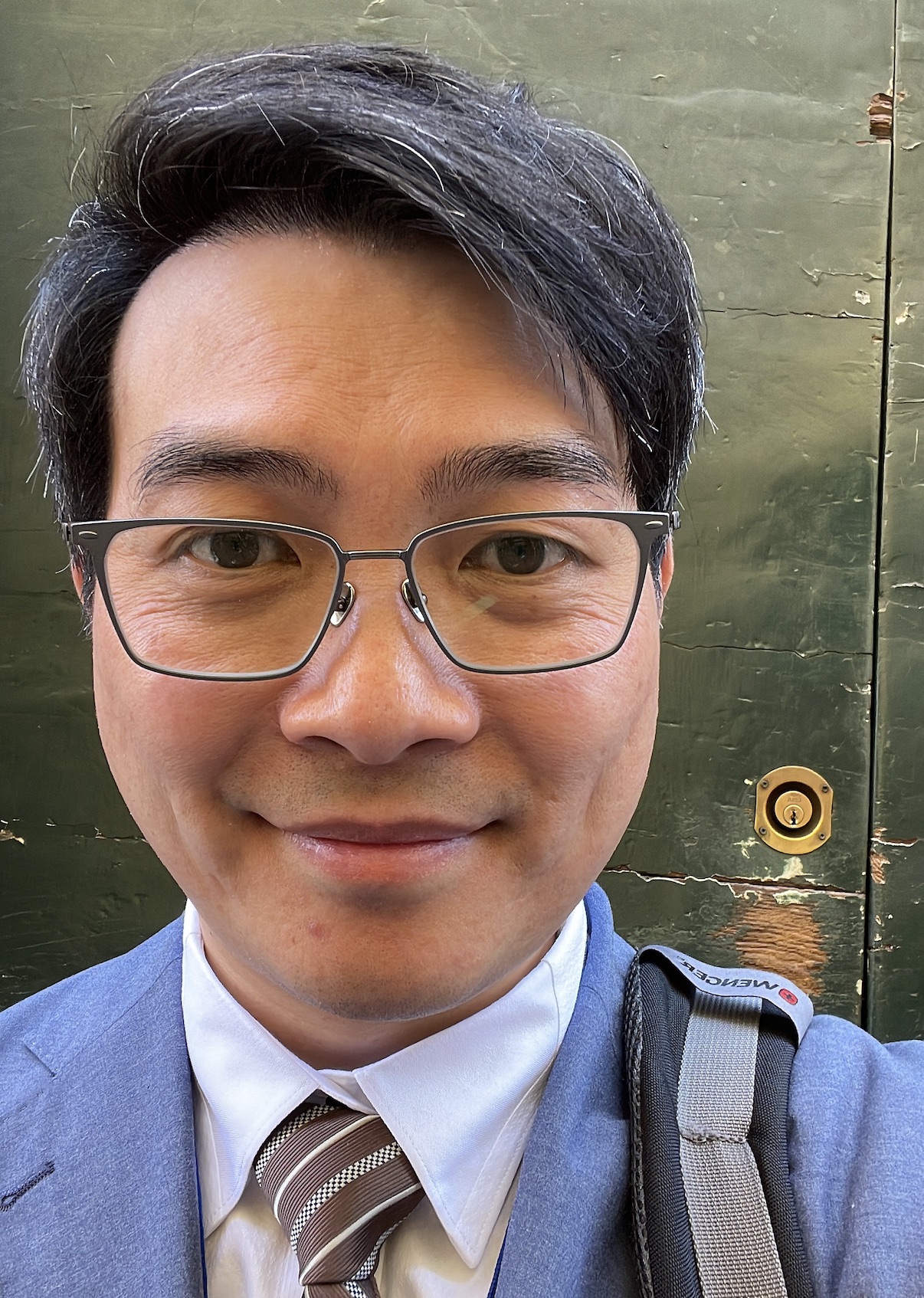}}]{(Kit) Kai-Kit~Wong} (M'01-SM'08-F'16) was born in Hong Kong in 1973. He received the BEng, the MPhil, and the PhD degrees, all in Electrical and Electronic Engineering, from the Hong Kong University of Science and Technology, Hong Kong, in 1996, 1998, and 2001, respectively. After graduation, he took up academic and research positions at the University of Hong Kong, Hong Kong, Lucent Technologies, Bell-Labs, Holmdel, the Smart Antennas Research Group of Stanford University, USA, and the University of Hull, U.K. Presently he is Chair Professor of Wireless Communications at the Department of Electronic and Electrical Engineering, University College London, U.K. He is Fellow of IEEE and IET. He served as the Editor-in-Chief for IEEE Wireless Communications Letters between 2020 and 2023. He is currently the Subject Editor-in-Chief for Wireless Communications for IET Electronics Letters and also on the Advisory Board of IEEE Communications Letters and the Steering Committee of IEEE Wireless Communications Letters.
\end{IEEEbiography}

\begin{IEEEbiography}[{\includegraphics[width=1in,height=1.25in,clip,keepaspectratio]{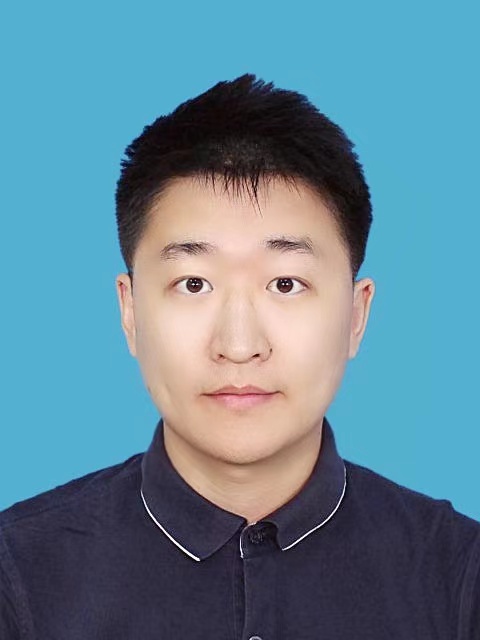}}]{Hao~Xu} (S'15-M'19) received the B.S. degree in communication engineering from Nanjing University of Science and Technology, Nanjing, China, in 2013, and the Ph.D. degree in information and communication engineering with the National Mobile Communications Research Laboratory, Southeast University, Nanjing, China, in 2019. From 2019 to 2021, he was an Alexander von Humboldt (AvH) Post-Doctoral Research Fellow with the Faculty of Electrical Engineering and Computer Science at the Technical University of Berlin, Germany. He is currently a Marie Sklodowska-Curie Actions (MSCA) Individual Fellow with the Department of Electronic and Electrical Engineering, University College London, UK. His research interests mainly include information theory, mathematical optimization, MIMO systems, and physical layer security in wireless networks.
\end{IEEEbiography}

\begin{IEEEbiography}[{\includegraphics[width=1in,height=1.25in,clip,keepaspectratio]{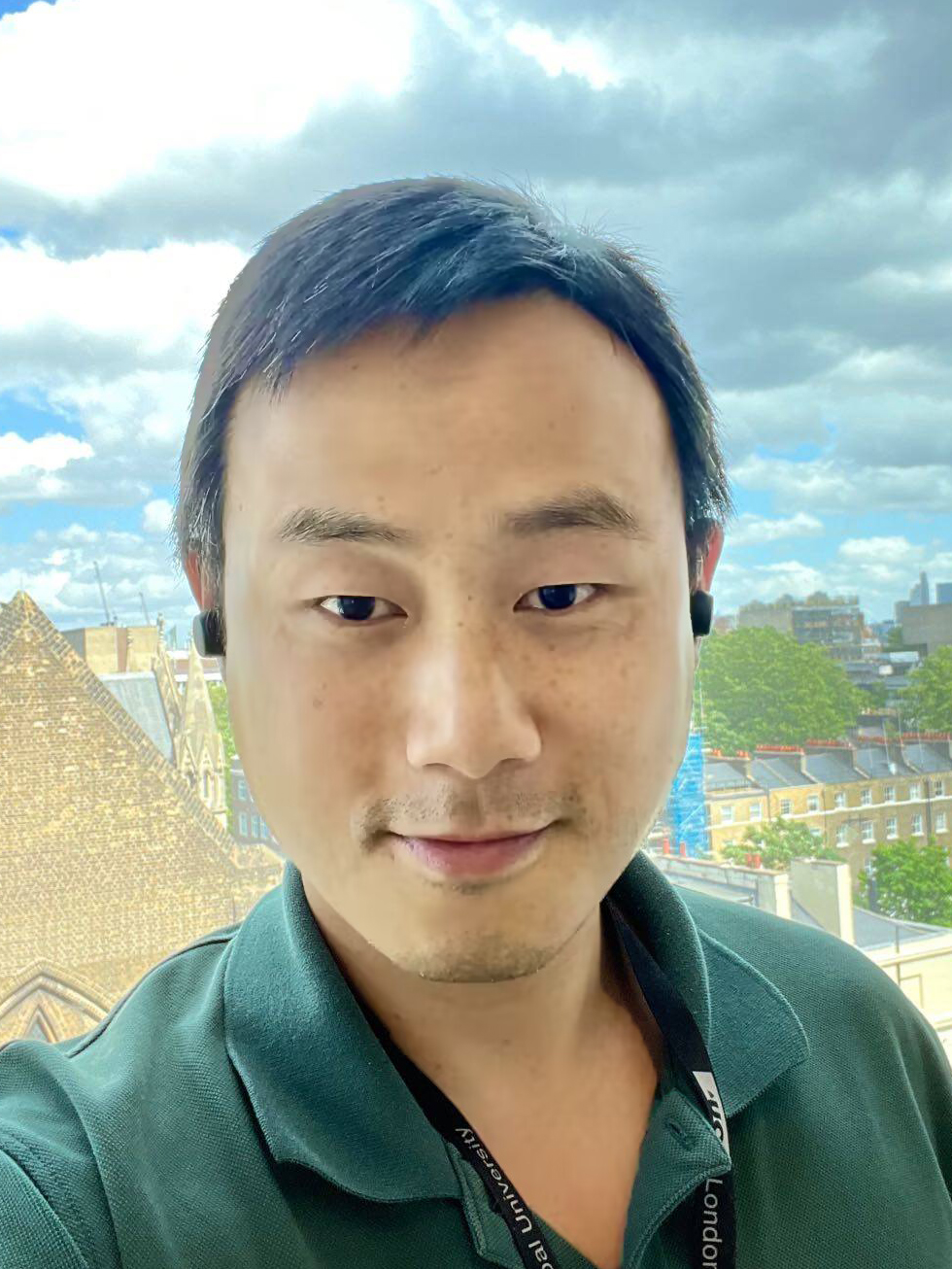}}]{Chao~Wang} (M'20) received the B.S., M. S. and Ph.D. degrees, in information and communication engineering from Xi?an Jiaotong University, Xi?an, China, in 2008, 2013, and 2016, respectively. He is currently an Associate Professor with the State Key Laboratory of Integrated Services Networks, School of Telecommunications Engineering, Xidian University. His current research interests include 5G \& 6G wireless networks and key technologies, physical-layer security, covert communications, and deep learning and its application in the optimization of wireless communications . He was a recipient of the Excellent Doctoral Dissertation Award of Shaanxi Province in 2018 and China Institute of Communications in 2016, the Best Paper Awards at the IEEE ICCC 2014 and IEEE SAGC 2020, respectively. 
\end{IEEEbiography}

\begin{IEEEbiography}[{\includegraphics[width=1in,height=1.25in,clip,keepaspectratio]{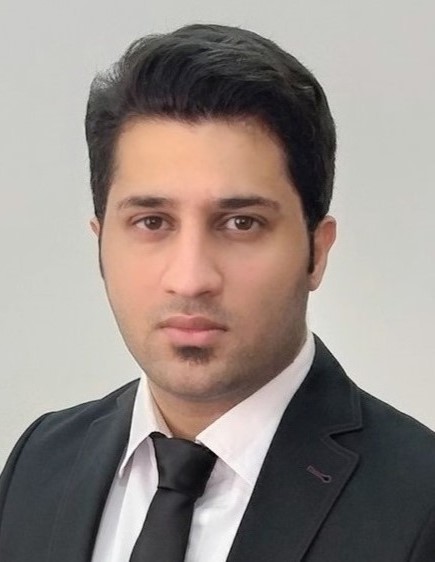}}]{Farshad~Rostami~Ghadi} (Member IEEE) received his Ph.D. degree (Hons.) in electrical communication systems engineering from the Ferdowsi University of Mashhad, Mashhad, Iran, in 2021. He was a Postdoctoral Research Fellow with the Department of Communication Engineering, University of Malaga, Malaga, Spain, in 2021. He is currently a Research Fellow with the Department of Electronic and Electrical Engineering, University College London, London, UK. His main research interests include analyzing wireless communication systems, network information theory, and copula theory, with an emphasis on wireless channel modeling and physical layer security. He has received the Best Paper Award at the IEEE ISTT 2024.
\end{IEEEbiography}

\begin{IEEEbiography}[{\includegraphics[width=1in,height=1.25in,clip,keepaspectratio]{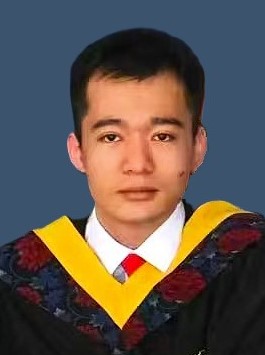}}]{Jichen~Zhang} (Graduate Student Member, IEEE) received his B.Eng. degree in communication engineering from the Harbin Institute of Technology, Shenzhen, China, in 2023. He is currently pursuing the Ph.D. degree at the Department of Electronic and Computer Engineering, the Hong Kong University of Science and Technology, Hong Kong. His research interests include fluid antenna systems, reconfigurable antennas, MIMO systems, and lens antennas. He won the Best Student Paper Award in IEEE APCAP 2023.
\end{IEEEbiography}

\begin{IEEEbiography}[{\includegraphics[width=1in,height=1.25in,clip,keepaspectratio]{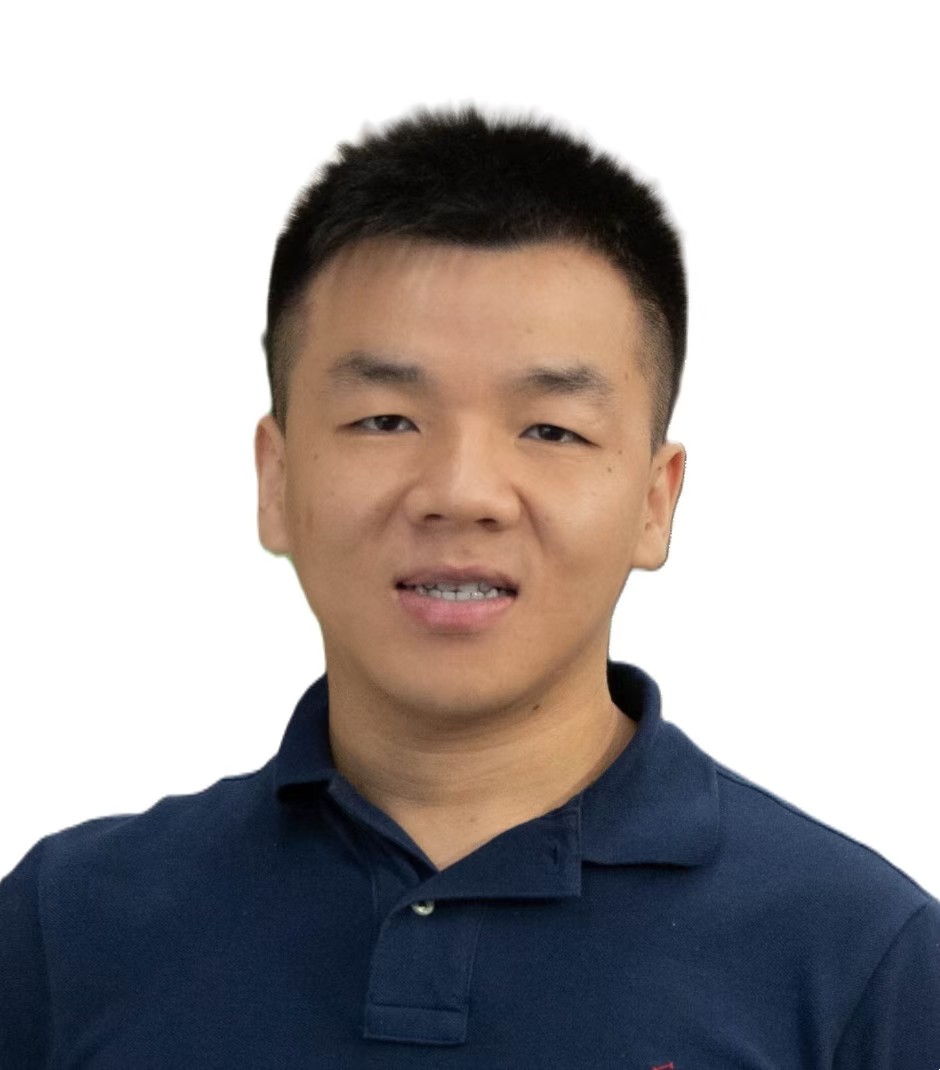}}]{Junhui~Rao} (Member, IEEE) received his B.Eng. degree from the University of Electronic Science and Technology of China in Chengdu, China, in 2020 and his Ph.D. degree from the Department of Electronic and Computer Engineering at The Hong Kong University of Science and Technology in 2024, where he is currently a Post-Doctoral Fellow. His research interests include reconfigurable intelligent surfaces, microwave circuits, MIMO systems, millimeter waves, and 6G technology. Dr. Rao has been recognized with several prestigious fellowships and awards at national, international, and university levels, including the distinguished IEEE Antennas and Propagation Society Fellowship (IEEE APSF Award 2023) and HONORABLE MENTION award in IEEE AP-S/URSI 2024.
\end{IEEEbiography}

\begin{IEEEbiography}[{\includegraphics[width=1in,height=1.25in,clip,keepaspectratio]{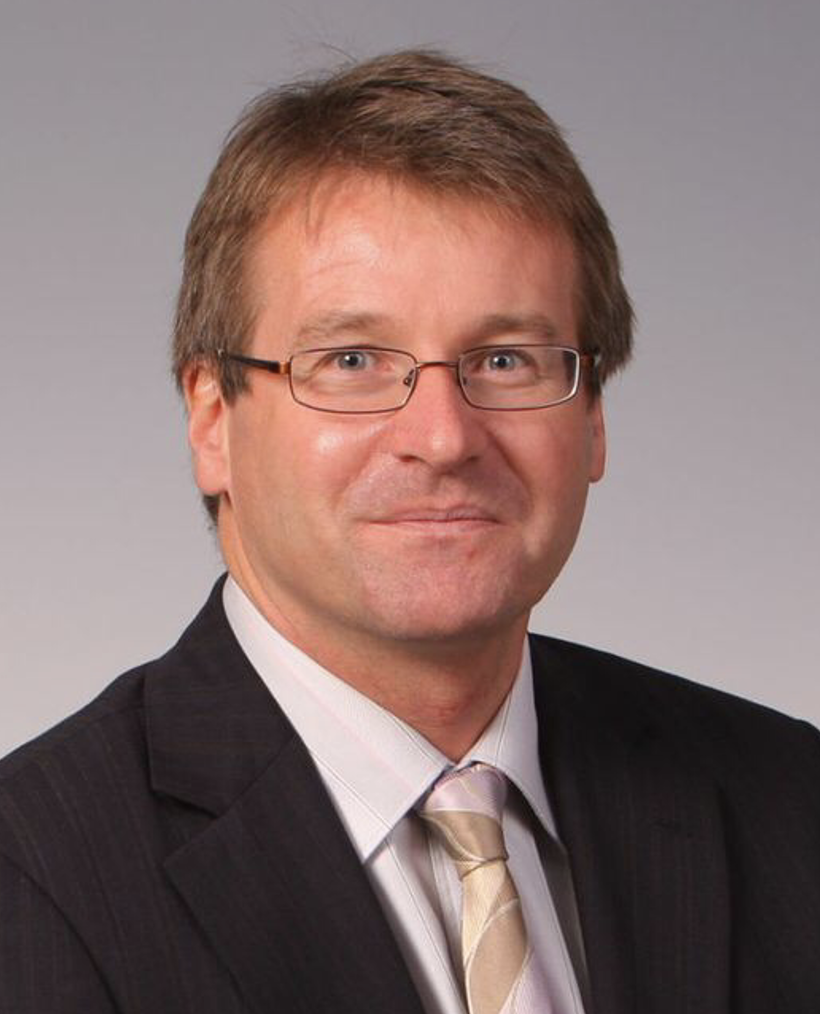}}]{Ross~Murch} (S'84-M'90-SM'98-F'09) received the bachelor's and Ph.D. degrees in Electrical and Electronic Engineering from the University of Canterbury, New Zealand.  

He is currently a Chair Professor in the Department of Electronic and Computer Engineering and a Senior Fellow at the Institute of Advanced Study both at the Hong Kong University of Science and Technology (HKUST). He is known for his research on multiple antenna technology including multiuser-MIMO, compact multiport antennas and multiport energy harvesting. His current research focus is creating new RF wave technology for making a better world and this includes RF imaging, energy harvesting, electromagnetic information theory, 6G, IoT, multiport antenna systems and reconfigurable intelligent surfaces. His unique expertise lies in his combination of knowledge from both wireless communication systems and electromagnetics and he publishes in both areas. For example, he has more than 50 journal papers in IEEE Transactions on Antennas and Propagation, 40 journal papers in IEEE Transactions on Wireless Communications from over 200 journal paper publications and 20 patents that have attracted 20,000+ citations. Prof. Murch has also successfully supervised more than 50 research students, enjoys teaching and has won five teaching awards. 

Prof. Murch was Department Head at HKUST from 2009-2015, is an IEEE, IET, HKAE, HKIE and NASI Fellow. He has been a David Bensted Fellow, Simon Fraser University, Canada, an HKTIIT fellow at Southampton University, U.K and has spent sabbaticals at MIT, USA; AT\&T, USA; Allgon Mobile Communications, Sweden; Imperial College London. He was awarded the IEEE Communications Society Wireless Communications Technical Committee Recognition Award in 2023. He has served IEEE in various positions including IEEE area editor, technical program chair, distinguished lecturer and Fellow evaluation committee. 

Professor Ross Murch joined HKUST in 1992 as an Assistant Professor and has remained at HKUST in Hong Kong since then, where he is now a Chair Professor. From 1990--1992 he was a Post-Doctoral Fellow at the Department of Mathematics and Computer Science, University of Dundee, UK. 
\end{IEEEbiography}

\begin{IEEEbiography}[{\includegraphics[width=1in,height=1.25in,clip,keepaspectratio]{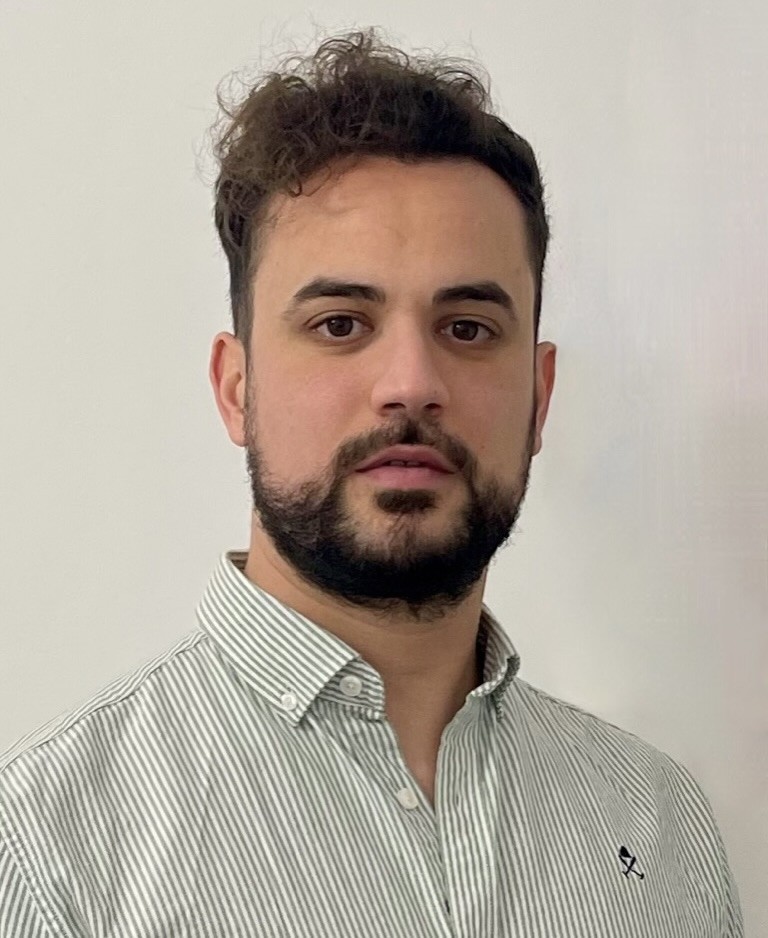}}]{Pablo Ram\'irez-Espinosa} received the M.Sc. and Ph.D. degrees in telecommunication engineering from the University of M\'{a}laga, Spain, in 2017 and 2020, respectively. From 2020 to 2022, he was a Post-Doctoral Researcher with the Connectivity Section, Department of Electronic Systems, Aalborg University, Denmark. From 2022 to 2024, he was a ``Maria Zambrano'' Post-Doctoral Fellow (National-Funded) with the University of Granada, Spain. In 2018, he was a Visiting Researcher with Queen?s University Belfast. He is currently an MSCA Post-Doctoral Fellow with the Telecommunications Research Institute (TELMA), University of M\'{a}laga. His main research interests include wireless communications, particularly dynamic metasurface antennas, fluid antennas, ultra-reliable low-latency communications, and channel modeling.
\end{IEEEbiography}

\begin{IEEEbiography}[{\includegraphics[width=1in,height=1.25in,clip,keepaspectratio]{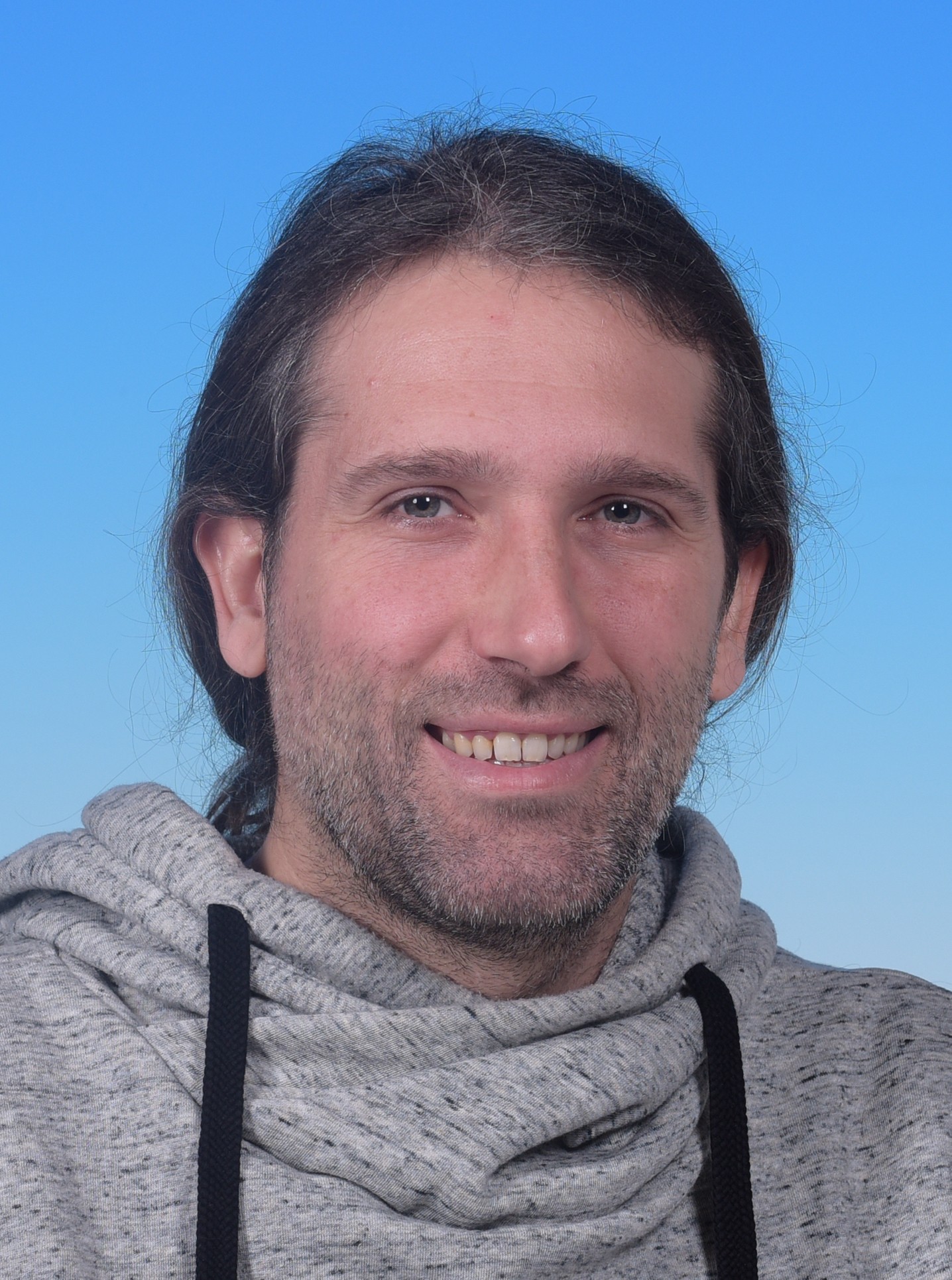}}]{David Morales-Jimenez} (M'13-SM'19) is an RyC Research Professor with the Department of Signal Theory, Networking and Communications at University of Granada (Spain). He received the M.Sc. and Ph.D. degrees in Telecommunication Technologies from University of Malaga (Spain) in 2008 and 2011, respectively. Between 2011 and 2013 he was a Postdoctoral Fellow at Universitat Pompeu Fabra (Barcelona, Spain). He then joined the Hong Kong University of Science and Technology (HKUST), first as Visiting Scholar (2014--2016) and then as Research Assistant Professor (2016--2018) with the Department of Electronic and Computer Engineering. He was a Lecturer (Assistant Professor) at Queen?s University Belfast (2018--2021) and an Associate Professor at University of Malaga (2021--2022). He also held visiting appointments at University College London (Electronic and Electrical Engineering, 2010) and at Stanford University (Statistics Department, 2015). His research interests include statistical signal processing, random matrix theory, and high-dimensional statistics, with multidisciplinary applications to wireless communications and computational biology.

Prof. Morales is a Senior Area Editor of the IEEE Transactions on Signal Processing and an Elected Member of the IEEE Technical Committee on Signal Processing for Communications and Networking (SPCOM). He received the Best Ph.D. Thesis Award in Electrical and Computer Engineering by the University of Malaga. He and his coauthors received the Best `Statistica Sinica' paper award at Joint Statistical Meetings 2020. He was a Poster Co-Chair of the IEEE Communication Theory Workshop 2022 and a General Co-Chair of the IEEE Spanish Workshop on Signal Processing, Information Theory and Communications 2022.
\end{IEEEbiography}

\begin{IEEEbiography}[{\includegraphics[width=1in,height=1.25in,clip,keepaspectratio]{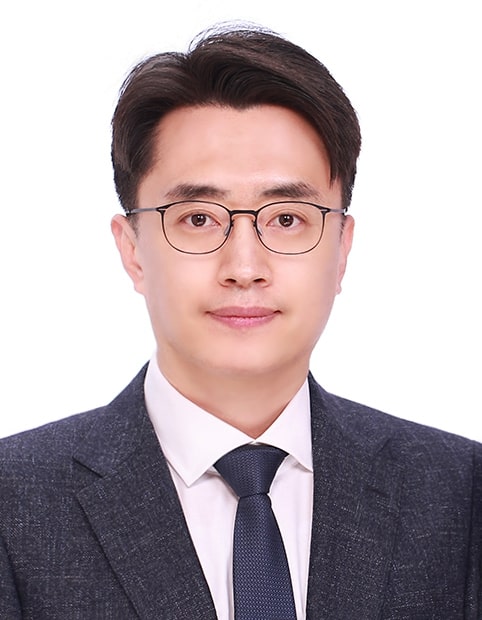}}]{Chan-Byoung~Chae}	(S'06-M'09-SM'12-F'21)  is an Underwood Distinguished Professor in the School of Integrated Technology, Yonsei University, Korea. Before joining Yonsei University, he was with Bell Labs, Alcatel-Lucent, Murray Hill, NJ, USA from 2009 to 2011, as a Member of Technical Staff, and Harvard University, Cambridge, MA, USA from 2008 to 2009, as a Postdoctoral Research Fellow. He received his Ph.D. degree in Electrical \& Computer Engineering from The University of Texas at Austin in 2008. Prior to joining UT, he was a research engineer at the Telecommunications R\&D Center, Samsung Electronics, Suwon, Korea, from 2001 to 2005.
	
He has been an Editor-in-Chief of the IEEE Trans. Molecular, Biological, and Multi-scale Communications (2019-2022) and a Senior Editor of the IEEE Wireless Communications Letters (2020-present). He has served/serves as an Editor for the IEEE Communications Magazine (2016-present), the IEEE Trans. on Wireless Communications (2012-2017), and the IEEE Wireless Communications Letters (2016-present). He is an IEEE ComSoc Distinguished Lecturer for the term 2020-2021 and 2022-2023.

He was the recipient/co-recipient of the CES Innovation Award in 2023, the IEEE ICC Best Demo Award  in 2022, the IEEE WCNC Best Demo Award in 2020, the Best Young Engineer Award from the National Academy of Engineering of Korea (NAEK) in 2019, the IEEE DySPAN Best Demo Award in 2018, the IEEE/KICS Journal of Communications and Networks Best Paper Award in 2018, the IEEE INFOCOM Best Demo Award in 2015, the IEIE/IEEE Joint Award for Young IT Engineer of the Year in 2014, the KICS Haedong Young Scholar Award in 2013, the IEEE Signal Processing Magazine Best Paper Award in 2013, the IEEE ComSoc AP Outstanding Young Researcher Award in 2012, the IEEE VTS Dan. E. Noble Fellowship Award in 2008.
\end{IEEEbiography}

\begin{IEEEbiography}[{\includegraphics[width=1in,height=1.25in,clip,keepaspectratio]{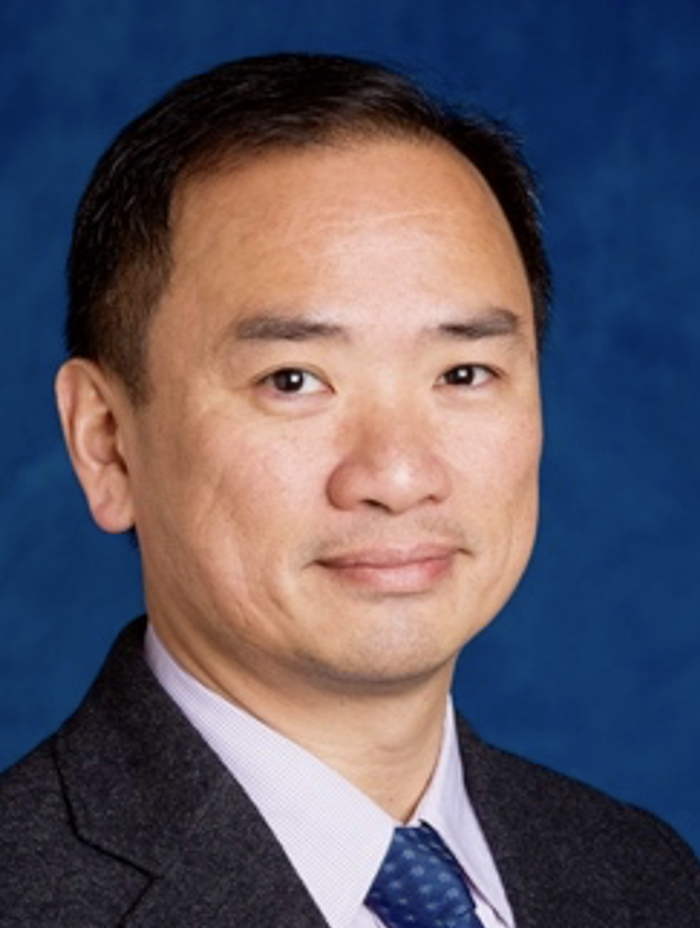}}]{Kin-Fai Tong} is a Full Professor of Antennas and Applied Electromagnetics at the Department of Electronic and Electrical Engineering of UCL. During his PhD research, he was credited with being one of the first to introduce the idea of embedding microstrip patch antennas into mobile phone handsets. As an Expert Researcher in the Photonic and Millimetre-wave Devices Group of the National Institute of Information and Communications Technology, Japan, he worked on novel wideband photonic antennas at 38 GHz and 60 GHz. Prof. Tong is a Fellow of IEEE, Chartered Engineer of UK Engineering Council, Fellow of Electromagnetic Academy US and Fellow of Higher Education Academy UK. His Innovate UK project was graded as ?OUTSTANDING?, i.e., the top 5\%, amongst all the funded projects. Recently, his AgriTech Internet-of-Thing (IoT) Hub project supported by EPSRC has resulted in two start-up companies and winning the UCL Provost?s Spirit of Enterprise Award. Prof Tong was the general chairman of the IEEE iWEM 2017 held in UK, the Lead Guest Editor of the IEEE OJAP Special Section, the Subject Editor of IET Electronics Letters, Associate Editor of IEEE AWPL.
\end{IEEEbiography}

\end{document}